\documentclass{iopart_mod}

\usepackage[bookmarks=false]{hyperref}
\usepackage{graphicx,epsfig}
\usepackage{longtable}
\usepackage{color}
\usepackage[sort,compress,numbers]{natbib}
\usepackage{doi}

\usepackage{bm}
\usepackage{capt-of}
\usepackage{relsize}
\usepackage{xcolor}
\usepackage{amssymb}
\usepackage[normalem]{ulem}
\usepackage{mathptmx}
\usepackage[export]{adjustbox}
\usepackage{url}

\def\al{\alpha}
\def\be{\beta}
\def\ga{\gamma}
\def\de{\delta}
\def\ep{\epsilon}

\def\th{\theta}

\def\si{\sigma}

\def\ta{\tau}

\def\om{\omega}

\def\De{\Delta}

\def\Ph{\Phi}

\def\nue{\nu_e}
\def\numu{\nu_\mu}
\def\nutau{\nu_\ta}

\def\nuebar{\bar\nu_e}
\def\numubar{\bar\nu_\mu}

\newcommand{\beq}{\begin{eqnarray}}
\newcommand{\eeq}{\end{eqnarray}}
\def\to{\rightarrow}

\def\mix{\leftrightarrow}

\def\no{\nonumber}
\def\fr#1#2{\frac{#1}{#2}}

\def\uGeV{\mbox{GeV}}
\def\uMeV{\mbox{MeV}}

\def\uGeVt{\mbox{GeV}^2}

\def\piz{\pi^{\circ}}
\def\pip{\pi^{+}}
\def\pim{\pi^{-}}
\def\pipm{\pi^{\pm}}
\def\CfA{C_5^A}

\newcommand{\np}{{\bf p}}

\newcommand{\nh}{{\bf h}}

\newcommand{\nq}{{\bf q}}

\begin{document}

\topical{Neutrino-Nucleus Cross Sections for Oscillation Experiments}

\author{Teppei Katori$^1$ and Marco Martini$^{2,3}$}

\address{$^1$School of Physics and Astronomy, Queen Mary University of London, London, UK}
\address{$^2$ESNT, CEA, IRFU, Service de Physique Nucl\'eaire, Universit\'e de Paris-Saclay, F-91191 Gif-sur-Yvette, France}
\address{$^3$Department of Physics and Astronomy, Ghent University, Proeftuinstraat 86, B-9000 Gent, Belgium}

\ead{t.katori@qmul.ac.uk, marco.martini@cea.fr}

\begin{abstract}
Neutrino oscillations physics is entered in the precision era. 
In this context accelerator-based neutrino experiments 
need a reduction of systematic errors to the level of a few percent.
Today one of the most important sources of systematic errors are neutrino-nucleus cross sections
which in the hundreds-MeV to few-GeV energy region are known with a precision not exceeding 20\%.
In this article we review the present experimental and theoretical knowledge of the neutrino-nucleus interaction physics. 
After introducing neutrino oscillation physics and accelerator-based neutrino experiments,
we overview general aspects of the neutrino-nucleus cross sections, both theoretical and experimental views. 
Then, we focus on these quantities in different reaction channels. 
We start with the quasielastic and quasielastic-like cross section, putting a special emphasis on
multinucleon emission channel which attracted a lot of attention in the last few years.
We review the main aspects of the different microscopic models for this channel by discussing analogies and differences among them.
The discussion is always driven by a comparison with the experimental data.
We then consider the one pion production channel where data-theory agreement remains very unsatisfactory.
We describe how to interpret pion data,
then we analyze in particular the puzzle related to the impossibility of theoretical models and Monte Carlo
to simultaneously describe MiniBooNE and MINERvA experimental results. 
Inclusive cross sections are also discussed,
as well as the comparison between the $\nu_\mu$ and $\nu_e$ cross sections, relevant for the CP violation experiments.
The impact of the nuclear effects on the reconstruction of neutrino energy and
on the determination of the neutrino oscillation parameters is reviewed.
A window to the future is finally opened by discussing projects and efforts in future detectors, beams, and analysis.

\end{abstract}
\submitto{\JPG}

\tableofcontents

\newpage
\section{Introduction \label{sec:intro-gen}}
Neutrino oscillation physics is one of the most flourishing fields in particle physics.
The visibility of the active community is further increased by the 
the 2015 Nobel Prize and the 2016 Breakthrough Prize in Fundamental Physics both given to neutrino
oscillations~\cite{Fukuda:1998ah,Ahmad:2002jz,Ahn:2002up,Eguchi:2002dm,Abe:2011sj,An:2012eh}.

After the discovery of the neutrino oscillations from the atmospheric neutrinos and solar neutrinos,
neutrino oscillations have been further studied in the long baseline accelerator and reactor experiments. 
Neutrino masses and mixing, a first evidence of a new particle physics beyond the Standard Model (SM),
are now well-accommodated in the standard framework of three-neutrino mixing,
the so called ``Neutrino Standard Model ($\nu$SM)'',
where the three active neutrinos $\nu_e$, $\nu_\mu$, $\nu_\tau$ are super-positions of three massive neutrinos
$\nu_1$, $\nu_2$, $\nu_3$ with respective masses $m_1$, $m_2$, $m_3$.
With the high precision measurement of the small parameter
$\sin^22\theta_{13}$~\cite{Abe:2011sj,An:2012eh,Ahn:2012nd,Abe:2011fz} 
we can affirm that neutrino physics is definitely entered the precision era.
Beyond a better and better determination of the five known oscillation parameters
(two squared-mass differences and three mixing angles) the determination of two unknown parameters,
the Dirac CP-violating phase and the neutrino mass ordering (NMO) is motivating the present
and future neutrino oscillation experiments. 
In parallel oscillation experiments aiming at the investigation of
the existence of additional massive neutrinos (the sterile neutrinos) are also pursued. 

In the present review article we focus on accelerator-based neutrino experiments which,
in the precision era, needs a reduction of systematic errors to the level of a few percent.
These experiments measure the rate of neutrino interactions,
which is the convolution of three factors: the neutrino flux, the interaction cross section and the detector efficiency.
Here we discuss all three aspects but we pay particular attention to
the neutrino-nucleus cross sections which in the hundreds-MeV to few-GeV energy region are one of
the most important sources of systematic errors, being known with a precision not exceeding 20\%.
Although a majority of interaction systematic errors can be canceled by
the internal measurement of oscillation experiments mainly by the near detectors, 
without improving the interaction models the limitations of internal constrains remain.

After an introduction on neutrino oscillation physics and accelerator-based experiments in Section \ref{sec:intro-gen},
we discuss in Section \ref{sec:xs-gen} some general theoretical
and experimental aspects of the neutrino cross sections on nuclei since the detectors of modern neutrino experiments
are composed of complex nuclei ($^{12}$C, $^{16}$O, $^{40}$Ar, $^{56}$Fe...)
which are more than a simple assembly of protons and neutrons. 
In accelerator-based experiments the neutrino beams
(at difference with respect to electron beams, for example)
are not monochromatic but they span a wide range of energies.
Several reaction channels can be open and the incomplete
lepton kinematic information prevents to compare theories with data in the same way as in the electron scattering.
At this moment, the flux-integrated differential cross sections are the golden observables
for the theory-experiment comparisons in neutrino scattering. 

We focus on these quantities in different reaction channels.
We start, in Section \ref{sec:QE} with the quasielastic (QE) and quasielastic-like cross section,
putting a special emphasis on multinucleon emission channel which 
attracted a lot of attention in the last few years, 
after the suggestion~\cite{Martini:2009uj,Martini:2010ex} of the inclusion of this channel
as possible explanation of the MiniBooNE quasielastic cross section on carbon
with unexpectedly large normalization~\cite{Katori:2009du,AguilarArevalo:2010zc}. 
Several theoretical calculations agree today on its crucial role to reproduce MiniBooNE,
as well as more recent MINERvA and T2K data.
However important quantitative differences remain between the calculations.
These differences largely contribute to the systematic error of the neutrino experiments,
depending on the way the multinucleon emission channel is inserted in the Monte Carlo generators used for the neutrino experiments. 
This channel was totally ignored in the generators,
and the effort to include these contributions in several Monte Carlo simulations started after 2010 and it is far from conclusion. 
A treatment of the multinucleon emission channel
(related to nucleon-nucleon correlations and meson exchange current contributions)
without approximations is particularly difficult and computationally very demanding. Therefore different
approximations are employed by the different theoretical approaches and by the Monte Carlo implementations.
We review the main aspects of the different microscopic models by discussing analogies and differences among them.
The discussion is always driven by a comparison with data (MiniBooNE, T2K, MINERvA, ArgoNeuT) and often respects the chronological
order of the theoretical and experimental results, which allows,
in our opinion, to better follow the rapid evolution of the field.     

The single pion production is the largest misidentified background
for both $\numu$-disappearance and $\nue$-appearance experiments.
However, data-theory agreement remains very unsatisfactory.
In particular there is no model which describes MiniBooNE and MINERvA simultaneously, the so called ``pion puzzle''.
In Section~\ref{sec:pion}, we introduce pion data and describe their interpretations.
The complications of pion data analyses lay not only on their primary production models,
but also on the fact that all hadronic processes have to be modeled correctly.
Combination of data from different channels and different experiments hope to entangle and constrain all processes,
however, such an approach has been started very recent and currently we are still struggling against pion puzzle.

Inclusive cross sections are the subject of Section~\ref{sec:cc} where not only the $\nu_\mu$ scattering case,
but also the $\nu_e$ one is presented. The $\nu_\mu$ \textit{vs} $\nu_e$ comparison,
relevant for the CP violation experiments is also discussed.

Since the neutrino beams are not monochromatic but wide-band, 
the incoming neutrino energy is reconstructed from the final states of the reaction.
The determination of the neutrino energy is crucial since it enters the expression of the neutrino oscillation probability.
This determination is typically done through the charged current quasielastic events. 
The reconstructed energy hypothesis used to obtain the neutrino energy from the measured charged lepton variables 
(energy and scattering angle) via a two-body formula is that the neutrino interaction in the nuclear target takes place on a nucleon at rest.
The identification of the reconstructed neutrino energy with the real one is too crude. 
Several nuclear effects, such as multinucleon emission need to be taken into account.
A review on the impact of the nuclear effects on the neutrino energy reconstruction is given in Section \ref{sec:enrec}.

A window to the future is opened in Section \ref{sec:future}
where we start the discussion from two flagship future accelerator-based long-baseline neutrino oscillation experiments,
DUNE (argon target)~\cite{DUNE_CDR1} and Hyper-Kamiokande (water target)~\cite{HK_2015}.
This clearly shows that argon and oxygen are two of the most important nuclear targets to study.
The high precision measurements can be achieved by a number of new approaches
mainly focusing on hadronic system information which was previously ignored.
Further reductions of systematics could be possible by improving the neutrino beam quality.
We discuss the effort to the future both detectors, analyses, and the beam.

We retain that the choices and the emphasis we have put on the different subjects of
the present manuscript render the present review a complement of other recent
articles~\cite{Gallagher:2011zza,Morfin:2012kn,Formaggio:2013kya,Alvarez-Ruso:2014bla,Garvey:2014exa,Mosel:2016cwa}.

\subsection{Neutrino oscillation physics\label{sec:nuosc}}

Neutrinos are peculiar particles within the Standard Model (SM), 
because their flavor states $|\nu_\alpha\rangle$ (productions and detections)
are superpositions of their Hamiltonian eigenstates, $|\nu_i\rangle$,
and they are related with unitary transformation $V$,
\beq
  |\nu_\al\rangle= \sum_i V_{\alpha i}^*(E) |\nu_i\rangle.
\eeq
In the vacuum, Hamiltonian eigenstates can be identified with
mass eigenstates, {\it i.e.},
\beq
  |\nu_\al\rangle= \sum_i U_{\alpha i}^* |\nu_i\rangle.
\eeq
Here, the unitary matrix $U$, so-called PMNS matrix,
diagonalize the mass matrix in the flavor basis
\beq
  m=
\left(\begin{array}{ccc}
m_{ee}     & m_{e\mu}     & m_{e\ta} \\
m_{e\mu}^* & m_{\mu\mu}   & m_{\mu\ta} \\ 
m_{e\ta}^* & m_{\mu\ta}^* & m_{\ta\ta}
\end{array}\right)
=
U
\left(\begin{array}{ccc}
m_1 & 0 & 0 \\
0 & m_2 & 0 \\ 
0 & 0 & m_3
\end{array}\right)
U^{\dagger}~.~
\eeq

Over the past years, the neutrino experiment community has tried to
understand the structures of neutrino masses and the PMNS matrix.
The PMNS matrix is usually written in terms of three Euler angle-like matrices and phase terms
\beq
U=
\left(\begin{array}{ccc}
1 & 0      & 0    \\
0 & c_{23} & s_{23}\\
0 &-s_{23} & c_{23}
\end{array}\right)
\left(\begin{array}{ccc}
c_{13}            & 0& s_{13}e^{-i\de_{CP}}\\
0                & 1& 0\\
-s_{13}e^{i\de_{CP}}& 0& c_{13}
\end{array}\right)
\left(\begin{array}{ccc}
c_{12} & s_{12} & 0\\
-s_{12}& c_{12} & 0\\
0     & 0      & 1
\end{array}\right)&&\no\\
\sim
\left(\begin{array}{ccc}
  c_{12} c_{13} & s_{12} c_{13} & s_{13} e^{-i\de_{CP}} \\
  - s_{12} c_{23} - c_{12} s_{23} s_{13} e^{i\de_{CP}} & c_{12} c_{23} - s_{12} s_{23} s_{13} e^{i\de_{CP}} & s_{23} c_{13} \\
  s_{12}s_{23} -c_{12}c_{23}s_{13}e^{i\de_{CP}} & - c_{12} s_{23} - s_{12} c_{23} s_{13} e^{i\de_{CP}} & c_{23} c_{13}
\end{array}\right)&&~,
\label{eq:U}
\eeq
where $s_{ij}$ and $c_{ij}$ are the sine and cosine of the $\theta_{ij}$ angle, respectively.
The right matrix represents the rotation between $|\nu_1\rangle$ and $|\nu_2\rangle$, or $1\mix 2$ mixing, 
and this is usually measured by solar neutrino oscillation
experiments~\cite{Homestake,KamiokandeII,SK_solar,SAGE,Gallex,GNO,SNO,Borexino}
and long-baseline reactor experiment~\cite{Eguchi:2002dm}.
Because of the matter effect in the Sun,
$\nu_2$ is known to be heavier than $\nu_1$. 

The left matrix represents $2\mix 3$ mixing, which is measured through
long-baseline accelerator neutrino oscillations~\cite{K2K_osc,T2K_numu2012,Agafonova:2010dc,NOvA_numu2016,Adamson:2014vgd}
or atmospheric neutrino oscillations~\cite{Adamson:2014vgd,SK_LoverE,MACRO,ANTARES_osc,IC_osc2013}.
Although this is the first established neutrino oscillation sector,
uncertainty in $\th_{23}$ is the largest among three rotation angles~\cite{nufit},
and precise value of $\th_{23}$ is the key for both neutrino mass ordering and $\de_{CP}$ measurements.

Finally, the middle matrix is for  $1\mix 3$ mixing.
Since $\th_{12}$ and $\th_{23}$ are nonzero,
nonzero $1\mix 3$ mixing implies nonzero Dirac CP phase, {\it i.e.},
possible leptonic CP violation in neutrino oscillations.
After the discovery of the nonzero $\th_{13}$ both by accelerator neutrino experiments
~\cite{Abe:2011sj,MINOS_osc1,T2K_nue2013,Adamson:2016tbq}
and reactor neutrino experiments
~\cite{An:2012eh,Ahn:2012nd,Abe:2011fz},
the focus of the neutrino oscillation community is mainly the determination of neutrino mass ordering (NMO) 
and the Dirac CP phase. 
Note, in the case of Majorana neutrinos, there are two Majorana CP phases, however,
they do not contribute to neutrino oscillations so we do not discuss them at here.

\begin{figure}[t!]
  \begin{center}
    \includegraphics[width=6.5cm,valign=m]{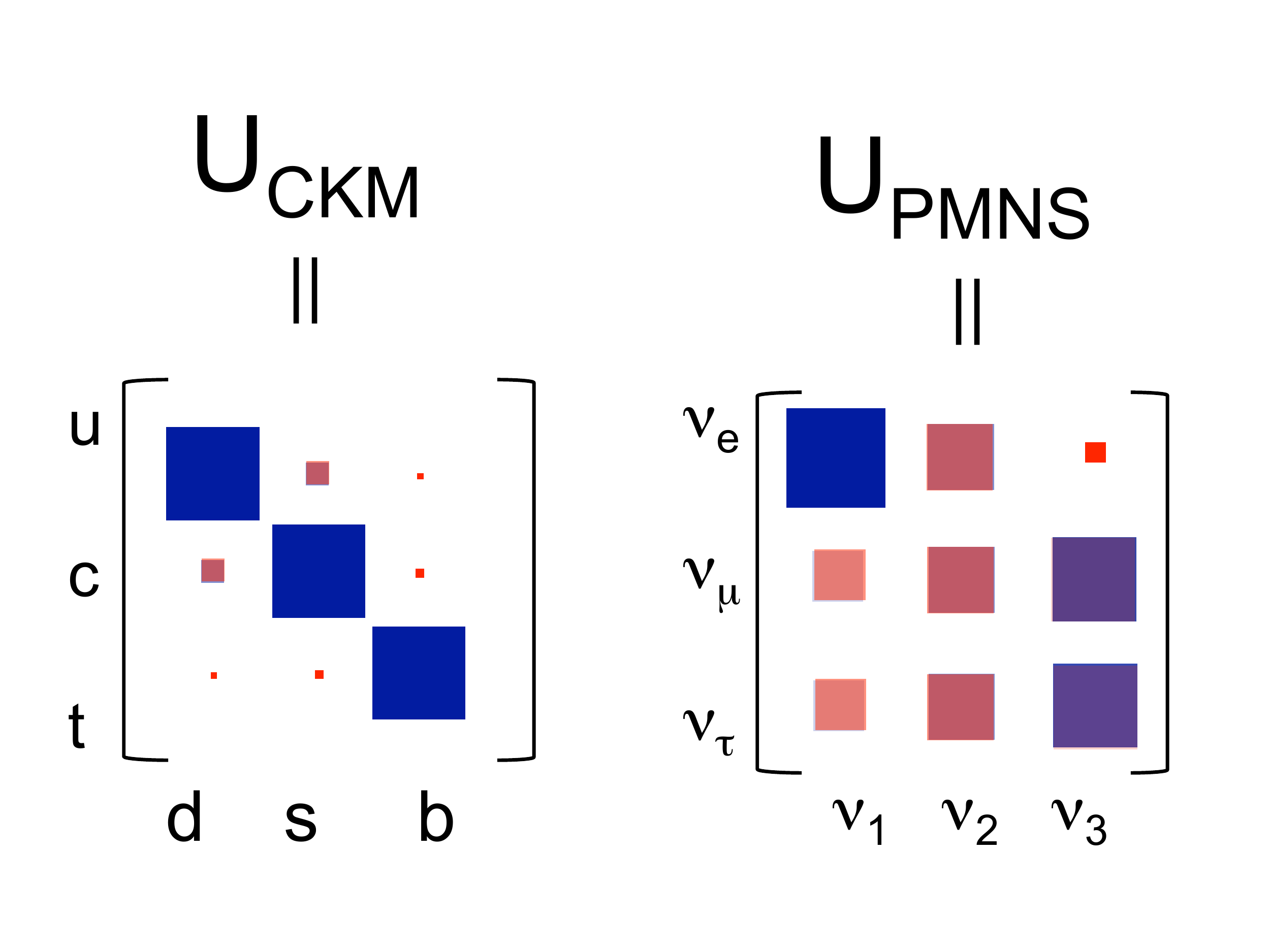}
    \includegraphics[width=5.5cm,valign=m]{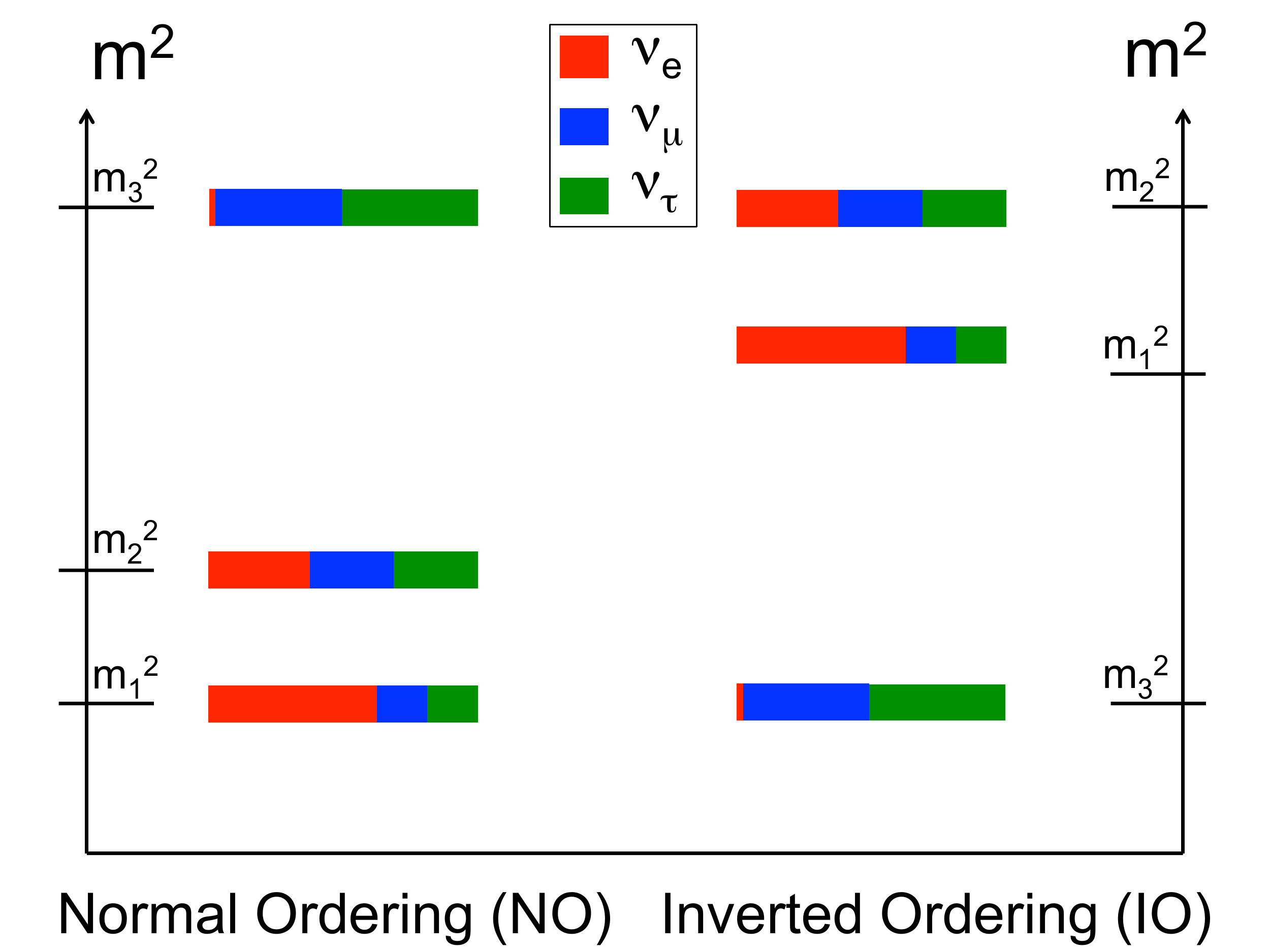}
    \end{center}
\vspace{-2mm}
\caption{
Graphical representations of CKM matrix and PMNS matrix (left)
and neutrino mass ordering (right).
}
\label{fig:PMNS}
\end{figure}

The structure of PMNS matrix~\cite{nufit,Capozzi:2013csa} is strikingly different from CKM matrix~\cite{PDG}. 
Figure ~\ref{fig:PMNS} left graphically shows the size of matrix elements.
CKM matrix is dominated by diagonal terms, which allows only a small mixing between different flavors.
On the other hand, PMNS matrix has large off-diagonal terms,
and mixing between different generations are large.
However, we do not know which is more ``natural'',
and the structures of these matrices are big interests of
the particle theory community (see for example Ref.~\cite{King_summary}). 


Another mystery is the structure of neutrino mass ordering (NMO).
Figure ~\ref{fig:PMNS} right shows the order of neutrino masses.
Each bar shows fraction of neutrino flavors.
Currently, there are two candidates of NMO,
so called ``normal ordering'', where $\nu_3$ comes to the top,
or ``inverted ordering'', where $\nu_3$ comes to the bottom.
The goal of current and future neutrino oscillation experiments is to find the aforementioned unknowns.
On top of this, the absolute neutrino mass scale is not known but this is not measurable by oscillation experiments
and we do not discuss here.


The completeness of the 3 neutrino mixing paradigm sketched above is challenged by the so called
short-baseline anomalies, including 
reactor \cite{Mention:2011rk,Mueller:2011nm,Huber:2011wv},
Gallium \cite{Abdurashitov:2005tb,Laveder:2007zz,Giunti:2010zu,Giunti:2006bj,Giunti:2012tn}, 
LSND~\cite{Aguilar:2001ty},
and MiniBooNE results~\cite{AguilarArevalo:2008rc,Aguilar-Arevalo:2013pmq}
which could indicate that the neutrino mixing framework
need an extension in order to accommodate short-baseline oscillations.
The additional squared-mass difference required to explain these anomalies with neutrino oscillations
necessitates the existence of at least an additional massive neutrino at the $\sim$1~eV scale.
Since from the LEP measurement of the invisible width of the $Z$ boson~\cite{ALEPH:2005ab},
we know that there are only three active neutrinos, in the flavor basis the additional massive neutrinos correspond
to sterile neutrinos~\cite{Pontecorvo:1968fh},
which do not have standard weak interactions.
The search for sterile neutrinos is another strong dynamo of neutrino experimental programs
in the world~\cite{Giunti:2006bj,Giunti:2012tn,Teppei_SBL,Christina_sterile,sterile,Giunti:2015wnd}.

Next, we take a look on standard neutrino oscillations more closely. 
The non-trivial part of the vacuum Hamiltonian
in the flavor basis can be written as

\beq
  H(E)\sim \frac{1}{2E}U
\left(\begin{array}{ccc}
m_1^2  & 0 & 0 \\
0 &m_2^2 & 0 \\ 
0 & 0 &m_3^2
\end{array}\right)
U^{\dagger}~.
\label{eq:hmass}
\eeq
From this Hamiltonian, the evolution of an initial flavor
state $|\nu_\al\rangle$ over a distance $L$ is solved. 
The probability of measuring a flavor state $|\nu_\be\rangle$ of energy $E$ 
after traveling in vacuum with distance $L$ is

\beq
&&P_{\nu_{\al} \to \nu_{\be}}(L,E)=\de_{\al \be}
-4  \sum_{i>j}Re(U_{\al i}^{*}U_{\be i}U_{\al j}U_{\be j}^{*})
\sin^2\left(\fr{\De m_{ij}^2}{4E}L\right)\no \\
&+&2\sum_{i>j}Im(U_{\al i}^{*}U_{\be i}U_{\al j}U_{\be j}^{*})
\sin\left(\fr{\De m_{ij}^2}{2E}L\right)~,~\De m_{ij}^2\equiv m_i^2-m_j^2~.
\label{eq:osc}
\eeq
To discuss further details of neutrino oscillations, we reduce this equation to a simpler form
by using the two neutrino oscillation approximation, which is successfully used in past years. 
We define $\th$ to be the mixing angle of two mass eigenstates, $|\nu_1\rangle$ and $|\nu_2\rangle$.
Then, for example, flavor state $|\nu_\al\rangle$ to $|\nu_\be\rangle$ oscillation is 

\beq
P_{\nu_{\al} \to \nu_\be}(L,E)=\sin^22\th \sin^2\left(\fr{\De m_{21}^2}{4E}L\right)~.~
\label{eq:osc2nu}
\eeq

First, the neutrino oscillation formula is now reduced to a simple one sinusoidal function with $L/E$ dependence.
Super-Kamiokande was first to show the evidence of this $L/E$ dependence of oscillations~\cite{SK_LoverE}, 
which is strong evidence that neutrino oscillations are
actually caused by neutrino masses within the measured energy region by the Hamiltonian described by Eq.~(\ref{eq:hmass}),
and not any other exotic physics which do not have this $L/E$ dependence.

Second, Since the oscillations generated by a squared-mass difference
$\Delta{m}^2$ is observable for $\Delta{m}^2 L / 4 E \gtrsim 1$, 
long-baseline neutrino oscillation experiments are characterized by a ratio $L/E \gtrsim 100 \, \textrm{m} \, \textrm{MeV}^{-1}$
which make them to be sensitive to $\Delta{m}^2 \lesssim 10^{-2} \, \textrm{eV}^2$.
On the other hand, short-baseline neutrino oscillation experiments
are characterized by a ratio $L/E \lesssim 10 \, \textrm{m} \, \textrm{MeV}^{-1}$
to explore $\Delta{m}^2 \gtrsim 10^{-1} \, \textrm{eV}^2$. 

Third, the imaginary part is zero and the oscillation is insensitive to complex CP phases.
This is equivalent to the quark sector which requires three flavors for CP violation 
because there is no Dirac CP phase for two quark flavor mixing. 
Likewise for neutrino oscillations, 
leptonic Dirac phase appears only under three neutrino oscillation framework. 

Fourth, the mixing angle is involved in terms of $\sin^22\th$,
and this causes a degeneracy in $\th$,
for example $\th=40^\circ$ and $\th=50^\circ$
give the same oscillation amplitudes within two neutrino oscillation framework.
However, they give different results in full three neutrino oscillation framework where
a dependence on $\sin^2\th$ also appears.
In fact, this degeneracy is the biggest systematic to measure neutrino mass ordering
through atmospheric neutrino oscillations~\cite{Patterson_NMH}.

Finally, now there is only one $\De m_{21}^2$ involved in sine square, 
which means that there is no sensitivity of the sign of $\De m_{21}^2$. 
The situation changes if the neutrinos propagate in matter, 
where the cross section is different between electron neutrinos and others.
For the electron neutrinos, both charged and neutral current interactions are possible with electrons in matter,
but for other flavors only neutral current interactions are possible.
This makes flavor asymmetric potential, the so-called Wolfenstein term~\cite{Wolfenstein}, in the Hamiltonian
\beq
H(E)&=&
\fr{1}{2E}
U
\left(\begin{array}{cc}
m_1^2 & 0     \\
0     & m_2^2 
\end{array}\right)
U^{\dagger}+
\left(\begin{array}{cc}
\sqrt{2}G_F n_e & 0     \\
0     & 0 
\end{array}\right)\no\\
&=&
V(E)
\left(\begin{array}{cc}
m_1^2(E)^* & 0     \\
0     & m_2^2(E)^* 
\end{array}\right)
V(E)^{\dagger}.
\eeq
The Wolfenstein term adds different energy dependence in the Hamiltonian
and the $\fr{1}{E}$ term cannot be factored out any more, 
and both mixing matrix elements and effective neutrino mass terms become energy dependent.
This makes the oscillation equation sensitive to the sign of $\De m^2$. 
This technique is used to fix the mass ordering of $\nu_1$ and $\nu_2$ in solar neutrino mixing, 
and long-baseline neutrino oscillation experiments are planning to use this to find
the mass ordering of $\nu_3$~\cite{T2K_nue2013,Adamson:2016tbq,PINGU,KM3NeT,HK_2015,DUNE_CDR1,INO}.
On the other hand, neutrino mass ordering can be measured through the precise measurement of
neutrino oscillations within three neutrino mixing framework. 
The reactor experiments use that approach to measure neutrino mass ordering~\cite{JUNO_CDR,Reno50}.

\subsection{Accelerator-based neutrino experiments\label{sec:nuexp}}

Neutrinos can be generated by natural sources --this is the case of solar neutrinos,
atmospheric neutrinos, geo-neutrinos, supernova neutrinos, galactic or extra-galactic neutrinos--
or by artificial sources such as reactors and accelerators. 
In this article we especially focus on accelerator-based neutrino experiments. 
Here we give a brief overview of these past, present and future experiments which,
in parallel with neutrino oscillation program, 
measure neutrino-nucleus cross sections.

K2K (KEK to Kamioka) experiment was the first long-baseline
accelerator-based neutrino oscillation experiment~\cite{K2K_osc},
designed to measure $\numu\to\numu$ disappearance oscillation.
The neutrino beam ($\sim$1.3~GeV) aims 250~km away at the Super-Kamiokande detector. 
The near detector complex is consisted of multiple detectors,
including 1 kton water Cherenkov detector and the vertex detector of
the plastic scintillation fiber tracker
``SciFi''~\footnote{This is still the best name for the particle detector in the world.}~\cite{K2K_SciFi}
which was later replaced with the extruded plastic scintillator tracker ``SciBar''~\cite{K2K_SciBar}. 
Although data from the water Cherenkov detector (H$_2$O) was important for the oscillation physics~\cite{K2K_NCpi0},   
most of the neutrino interaction data are from tracker measurements~\cite{Gran:2006jn,K2K_CCpip,K2K_CCmpi0}.
Among them, SciBar analysis demonstrates that the measurement of energy deposits around the interaction vertex, the 
so called ``vertex activity'', is a useful variable, especially to select coherent scattering events~\cite{K2K_CCcohpi}.

MiniBooNE (mini-Booster Neutrino Experiment) was a mineral-oil based Cherenkov detector~\cite{MB_detec}
located on the Booster neutrino beamline ($E_\nu\sim$800~MeV, $E_{\bar\nu}\sim$600~MeV)~\cite{MB_flux}, 
designed to test the LSND neutrino oscillation results~\cite{Aguilar:2001ty}, 
namely the goal of MiniBooNE was to measure $\numu\to\nue$($\numubar\to\nuebar$)
appearance short-baseline oscillation signals~\cite{AguilarArevalo:2008rc,Aguilar-Arevalo:2013pmq}.
The Cherenkov detector was chosen as a crude way to count neutrino interactions with large fiducial volume, 
but it turns out that the 4$\pi$ coverage detector can produce excellent neutrino cross section data.  
To overcome the disadvantage of not having the near detector unlike other accelerator-based oscillation experiments,
MiniBooNE utilized measured $\numu$($\numubar$) interaction to
control systematics of $\nue$($\nuebar$) appearance oscillation analysis. 
This made MiniBooNE to try to understand detailed kinematics and backgrounds of
$\numu$($\numubar$) interactions~\cite{MB_CCQEPRL,MB_NCpi0PLB,MB_qepipratio,MB_WS},
and they became the series of first flux-integrated differential cross-section
measurements~\cite{MB_NCpi0,AguilarArevalo:2010zc,MB_NCEL,MB_CCpip,MB_CCpi0,AguilarArevalo:2013hm,MB_antiNCEL}.

SciBooNE (SciBar Booster Neutrino Experiment) uses the SciBar detector from K2K at the Booster neutrino beamline at
Fermilab~\footnote{Extruded scintillators were made by Fermilab. They were shipped to Japan for K2K, and shipped back to Fermilab for SciBooNE, 
and later shipped to Puebla, Mexico for a solar neutron measurement.
Similarly, EC~\cite{SB_EC} was originally constructed in Italy and used for CHORUS and HARP at CERN,
before joining SciBar for K2K, then SciBooNE.}. 
SciBooNE measured many aspects of neutrino interactions which MiniBooNE cannot measure very well, 
for example it measured charged current quasielastic (CCQE) from 1 and 2-track samples~\cite{AlcarazAunion:2009ku},
direction reconstruction for protons below Cherenkov threshold~\cite{SB_NCEL}, 
and the coherent pion production was measured utilizing
the vertex activity~\cite{SB_antiCCpip,SB_NCpi02,SB_NCpi01,SB_CCpi0,SB_CCpip}.
SciBooNE is a nice cross-check of MiniBooNE results, 
because Cherenkov and tracker detectors are complimentary in neutrino interaction measurements:
\begin{itemize}
\item Cherenkov detector, isotropic 4$\pi$ coverage, but it's hard to measure more than 1 track.
\item Tracker detector, relatively smaller angular acceptance, but excellent performance for multi-track events.
\end{itemize}
\begin{figure}[t!]
  \begin{center}
    \includegraphics[width=5cm,valign=m]{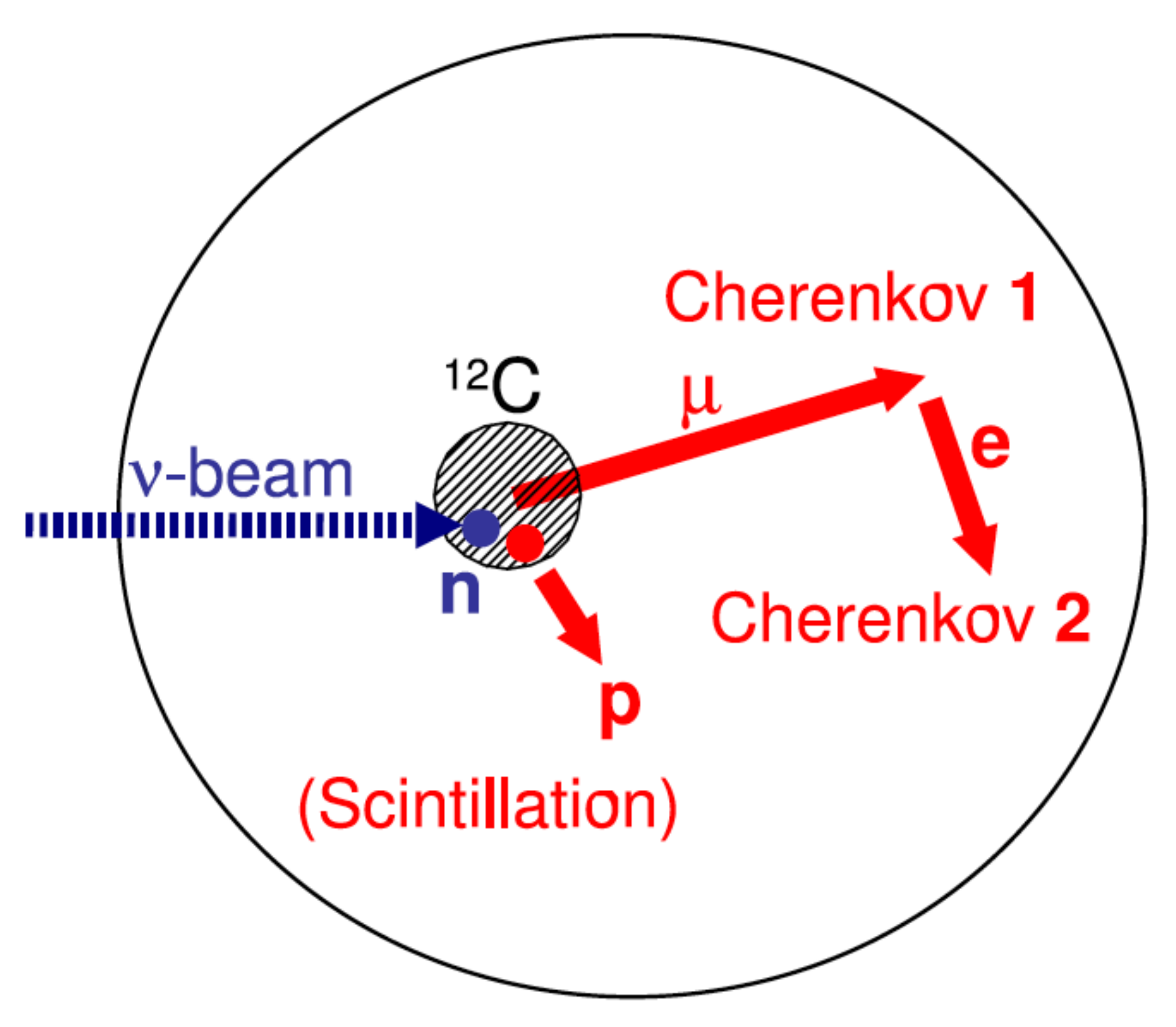}
    \includegraphics[width=7cm,valign=m]{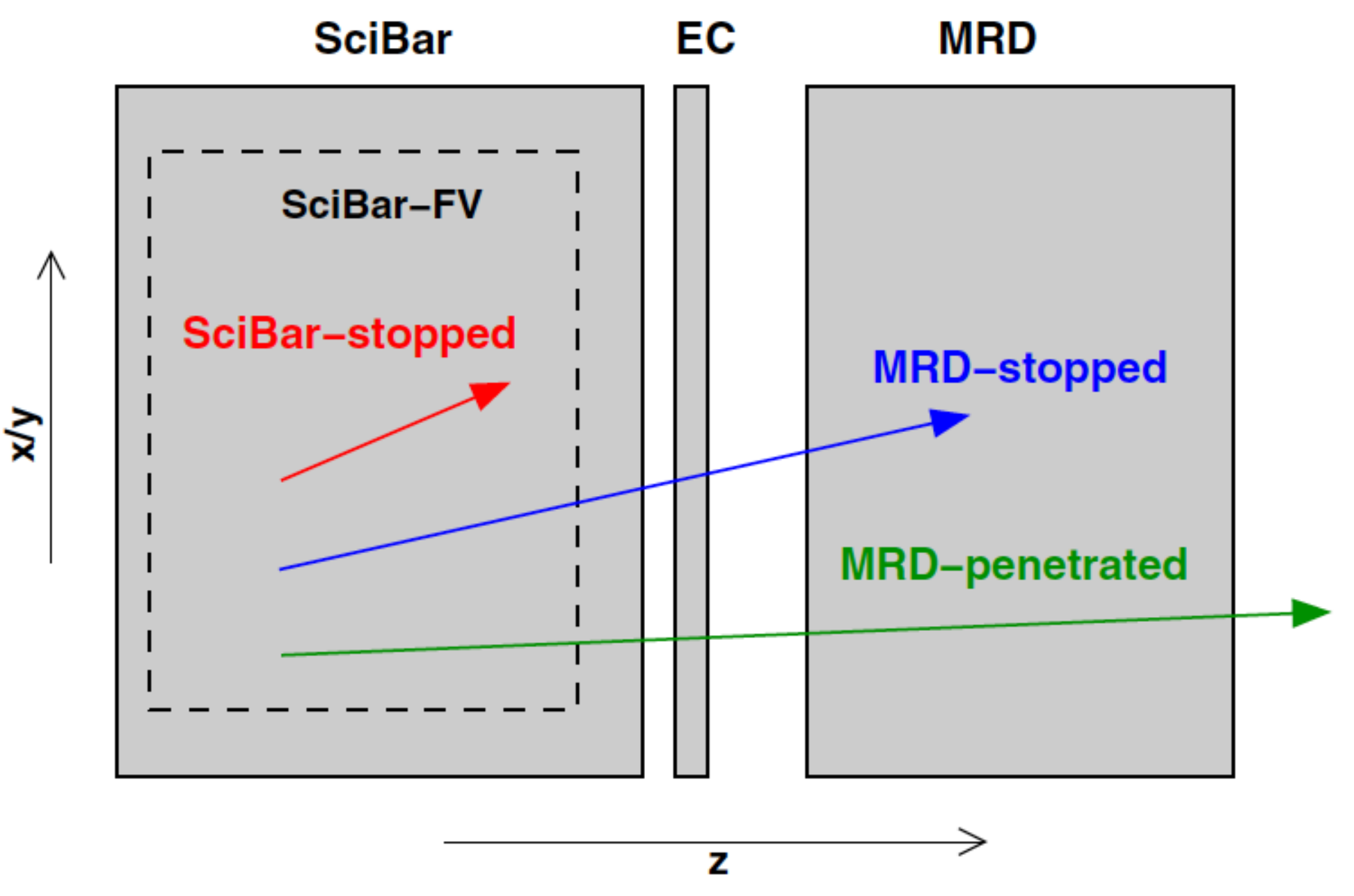}
    \end{center}
\vspace{-2mm}
\caption{
Event selection cartoons from MiniBooNE CCQE analysis~\cite{AguilarArevalo:2010zc} and SciBooNE CC inclusive analysis~\cite{Nakajima:2010fp}.
The MiniBooNE detector is the spherical mineral oil-based Cherenkov detector.
The SciBooNE detector consisted on three sub-detectors;
extruded plastic scintillator tracker ``SciBar''~\cite{K2K_SciBar},
electromagnetic shower radiator ``EC'' (electron catcher)~\cite{SB_EC}, and muon range detector ``MRD''~\cite{SB_MRD}.
}
\label{fig:int_cartoon}
\end{figure}
Figure~\ref{fig:int_cartoon} shows a comparison of a Cherenkov detector selection and a tracker detector selection.
The left figure describes the CCQE sample event selection in MiniBooNE~\cite{AguilarArevalo:2010zc}. 
Isotropic detector contains outgoing muons in 4$\pi$ direction with relatively constant efficiency.
In this analysis a muon decay electron is also tagged,
but again the efficiency of this is uniform across the detector. 
Although kinematic coverage of muons is excellent, the detector has a hard time reconstructing multiple tracks,
especially, it can see the low energy protons only through the isotropic scintillation light because
most protons are below Cherenkov threshold.
The right figure describes the CC inclusive sample event selection in SciBooNE~\cite{Nakajima:2010fp}.
Notice that the event samples are further classified depending on the event topologies,
such as muons stopped inside of the SciBar (``SciBar-stopped''), muons stopped inside of the MRD (``MRD-stopped''),
and muons which penetrate the MRD (``muon penetrated'').
This is often inevitable because the acceptance of each sub-detector is different. 
In general, this makes detector efficiency correction more complicated,
and it makes it harder to access the true particle kinematics. 
On the other hand, the detector can analyze more than one particle track,
and also the segmented tracker allows measurement of the vertex activity. 

MINOS (Main Injector Neutrino Oscillation Search) long-baseline oscillation experiment
started shortly after K2K~\cite{MINOS_osc1,Adamson:2014vgd}.
It utilizes both accelerator-based neutrinos and atmospheric neutrinos, however,
in this article we focus on former.
The goals of MINOS were to measure both $\numu\to\numu$($\numubar\to\numubar$) disappearance oscillation
and $\numu\to\nue$($\numubar\to\nuebar$) appearance oscillation signals~\cite{Adamson:2014vgd}. 
The experiment uses the on-axis NuMI neutrino beam~\cite{NuMI,Aliaga:2016oaz},
which has the ability to change the energy spectrum by configuring target and horn locations, 
but most of the data are taken with the low energy configuration, $\sim$3~GeV at the flux peak.
MINOS has similar near and far detectors, this means the MINOS near detector is also a magnetized iron-scintillator sandwich tracker,
and muon momentum is measured by the curvature and the range.
A magnetic field allows a charge separation of muon neutrino and muon anti-neutrino interactions, 
but on the other hand calorimetric reconstruction works fine to measure energies from electromagnetic and hadronic showers. 
There is a handful of neutrino interaction data published~\cite{MINOS_CC,Adamson:2014pgc,MINOS_NCpi0}.
Currently, NuMI is running with medium energy configuration ($\sim$7~GeV flux peak on on-axis),
and further data are expected from the MINOS extension run, called MINOS+~\cite{MINOSp}.

ArgoNeuT is the liquid argon time projection chamber (LArTPC),
which we further discuss in Sec.~\ref{sec:lartpc}.
It is a dream detector, because it has almost all the features of all detectors we have discussed,
such as 4$\pi$ coverage, 
multi-particle tracking,
vertex activity measurement,
and calorimetric energy reconstruction. 
Although ArgoNeuT is only 180~L volume, 
it was located in front of the MINOS near detector to use
it as a muon range detector
~\footnote{Before ArgoNeuT, emulsion detector PEANUT~\cite{PEANUT}
was located in front of the MINOS near detector.
Indeed, MINOS near detector is serving as a muon range detector for someone more than 10 years,
and contributed a lot of cross section measurements!},
and they produced number of interesting
results~\cite{ArgoNeuT_CCcohpi,Acciarri:2014gev,Acciarri:2014isz,Anderson:2011ce,Acciarri:2015ncl,Acciarri:2016sli}. 

MINERvA (Main Injector Experiment for v-A) is the tracker detector located in front of the MINOS near detector after ArgoNeuT.
MINERvA uses the extruded plastic scintillator for the active material.
However, MINERvA also has passive targets which are used to produce
a neutrino interaction target dependence results~\cite{MINERvA_TRatio,MINERvA_DISRatio}.
The charge separation at the down stream MINOS near detector can distinguish neutrino and antineutrino
interactions~\cite{Fiorentini:2013ezn,Fields:2013zhk,Eberly:2014mra,Higuera:2014azj,
Walton:2014esl,Aliaga:2015wva,Rodrigues:2015hik,McGivern:2016bwh,DeVan:2016rkm},
and the high segmentation and timing of the detector allows various
particle identifications~\cite{Wolcott:2015hda,Park:2015eqa,Wolcott:2016hws,Marshall:2016rrn,Wang:2016pww}.
Data are taken from MINOS period to NOvA period of NuMI,
which means the averaged neutrino energy of earlier MINERvA data are with NuMI low energy configuration ($\sim$3~GeV),
and later data are taken with NuMI medium energy configuration ($\sim$~7GeV).

NOMAD (Neutrino Oscillation Magnetized Detector) is a rather higher energy ($\sim$17~GeV) experiment
originally designed for $\numu\to\nue$($\numubar\to\nuebar$) appearance oscillation measurements. 
Superior flux systematics and fine grained detector with a magnetic field produced very important data,
relevant for context of this review~\cite{Lyubushkin:2008pe,NOMAD_NCgamma}. 

CHORUS (CERN Hybrid Oscillation Research ApparatUS) and
OPERA (Oscillation Project with Emulsion-tRacking Apparatus)
are $\numu\to\nutau$ appearance oscillation experiments.
Both experiments, as well as DONUT (Direct Observation of NU Tau),
use emulsion as an active target material for the main neutrino vertex detector.
It has the highest resolution and is useful for the $\nutau$ appearance measurements in DONUT and OPERA~\cite{DONUT,Agafonova:2010dc}, 
but it is also useful to measure high charged hadron multiplicity from neutrino interactions~\cite{CHORUS_had}.
The main chemical elements of the emulsion is the hydrocarbon ($\sim$60\% of mass), however,
there are non-negligible amounts of heavy elements (silver, bromine, etc),
and this makes interpretation of high precision emulsion data difficult. 

T2K (Tokai to Kamioka) experiment is one of two flagship long-baseline neutrino oscillation experiments in the world to
date~\cite{Abe:2015awa,T2K_nue2013}. It uses the J-PARC off-axis neutrino beam ($\sim$600~MeV)~\cite{T2K_flux}. 
The Super-Kamiokande detector located 295~km away is the far detector for the oscillation measurement.
These designs are for
$\numu\to\nue$($\numubar\to\nuebar$) appearance oscillation measurements
to find nonzero $\th_{13}$ and leptonic Dirac CP phase,
as well as precise $\th_{23}$ measurement through $\numu\to\numu$($\numubar\to\numubar$) disappearance measurements. 
Although the far detector (Super-Kamiokande) can provide some interesting neutrino cross section results~\cite{T2K_SK_NCQE}, 
the majority of cross section data are provided from the  near detector complex. 
The ND280 near detector complex is consisted of two tracker-style near detectors. 
The on-axis INGRID detector~\cite{INGRID} is the iron-scintillator sandwich trackers 
designed to measure the neutrino beam profile.
INGRID itself can provide interesting data~\cite{Abe:2014nox,Abe:2015biq}, 
but the later installed fully active ``proton module'' (full scintillator tracker)
can measure multi-track events and provide further high precision data~\cite{Abe:2015oar,Kikawa_thesis}.
The off-axis ND280 detector consists of five sub-detectors: 
pi-zero detector (P0D)~\cite{P0D}, fine-grained detector (FGD)~\cite{FGD},
gas argon TPC~\cite{TPC}, electromagnetic calorimeter (ECal)~\cite{ECal}, and
side muon range detector (SMRD)~\cite{SMRD}.
The analyses combine all of these sub-detectors,
and typical analyses are based on either
P0D (water target)~\cite{T2K_P0D_nueCC,DanRuterbories_thesis,Shamil_thesis} or
FGD (CH and water target)~\cite{Abe:2016fic,Abe:2013jth,Abe:2014usb,Abe:2014agb,Abe:2014iza,MattLawe_thesis,Abe:2016aoo}
as vertex detectors.
However, some analyses also use different targets,
such as argon gas (TPC)~\cite{PipHamilton_thesis}
and lead (ECal)~\cite{DomBrailsford_thesis}. 

NOvA (Numi Off-axis Neutrino Appearance) experiment is the other
flagship long-baseline neutrino oscillation experiment located at Fermilab.
Although NuMI is running with higher energy on on-axis ($\sim$7~GeV),
NOvA near and far detectors are located off-axis from the beam center,
making $\sim$2~GeV narrow band beam available. 
This configuration maximizes the sensitivity of 
$\numu\to\nue$($\numubar\to\nuebar$) appearance oscillation measurements~\cite{Adamson:2016tbq} and
$\numu\to\numu$($\numubar\to\numubar$) disappearance measurements~\cite{NOvA_numu2016}  
to find both leptonic Dirac CP phase and NMO as well as $\th_{23}$~\cite{Patterson_NMH}. 
Like MINOS, identical design is accepted for both near and far detectors. 
The liquid scintillator tracker detector (CH$_2$) may be too coarsely instrumented 
to perform the precise interaction measurement, however,
unique beam profile still allows us to measure useful
flux-integrated differential cross sections~\cite{NOvA_NDOS_CCQE,Bu:2016grw}.

\subsection{Neutrino fluxes\label{sec:flux}}

One of the key ingredients in neutrino physics is the neutrino flux which has to be known with the maximal precision. 
As already mentioned,
in this article we focus only on accelerator-based neutrino experiments in the neutrino energy range of 1-10 GeV.
Also in this case understanding of the neutrino flux is crucial, as it will be discussed in the following.
Below we give a brief overview of current and future neutrino beams. 

Modern accelerator-based neutrino beams around 1 to 10 GeV are made in several steps.
First, primary protons hit the target to produce secondary meson beams.
The target is usually located inside of the magnetic focusing horn,
and in the neutrino mode (antineutrino mode) it focuses positive mesons (negative mesons). 
Then, decay-in-flight (DIF) of mesons make tertiary muon neutrino (neutrino mode)
or muon antineutrino (antineutrino mode) dominated beams. 
These beams are so-called ``super beam'', this is contrast with other types of
accelerator-based neutrino beam, discussed later (Sec.~\ref{sec:futurebeam}). 

It is common to place the detector not on the beam center (on-axis),
but off from the beam center (off-axis), 
this makes the neutrino beam narrower and make it more convenient for oscillation analyses,
comparing with on-axis configuration~\cite{offaxis}. 
Understandings of production processes of neutrino beams are crucial for accelerator-based neutrino experiments.
Detailed descriptions of them are published for major neutrino beamlines,
including the Booster Neutrino beamline (BNB)~\cite{MB_flux} ---
used for the MiniBooNE/SciBooNE experiments and the host of future Fermilab short baseline programs~\cite{ICARUS_FNAL,ANNIE},  
J-PARC neutrino beamline~\cite{T2K_flux}---
being used for the T2K experiment,
and the NuMI neutrino beamline~\cite{NuMI,Aliaga:2016oaz} ---
the host neutrino beamline for MINOS, ArgoNeuT, MINERvA, and NOvA.
The biggest systematics of the neutrino beam prediction
is the correct simulation of the secondary meson kinematics.
This is often difficult to simulate,
and the most reliable method is to use data by measuring meson production directly 
from the replica target at hadron measurement facilities. 
These experiments include HARP at CERN (with BNB and K2K neutrino beam replica target)~\cite{HARP},
MIPP at Fermilab (with NuMI replica target)~\cite{MIPP},
and SHINE at CERN (with J-PARC neutrino beamline replica target)~\cite{SHINE}. 
For more discussions on neutrino beams, see for example Ref.~\cite{Kopp_flux}.


\begin{figure}[t!]
  \begin{center}
    \includegraphics[width=12cm]{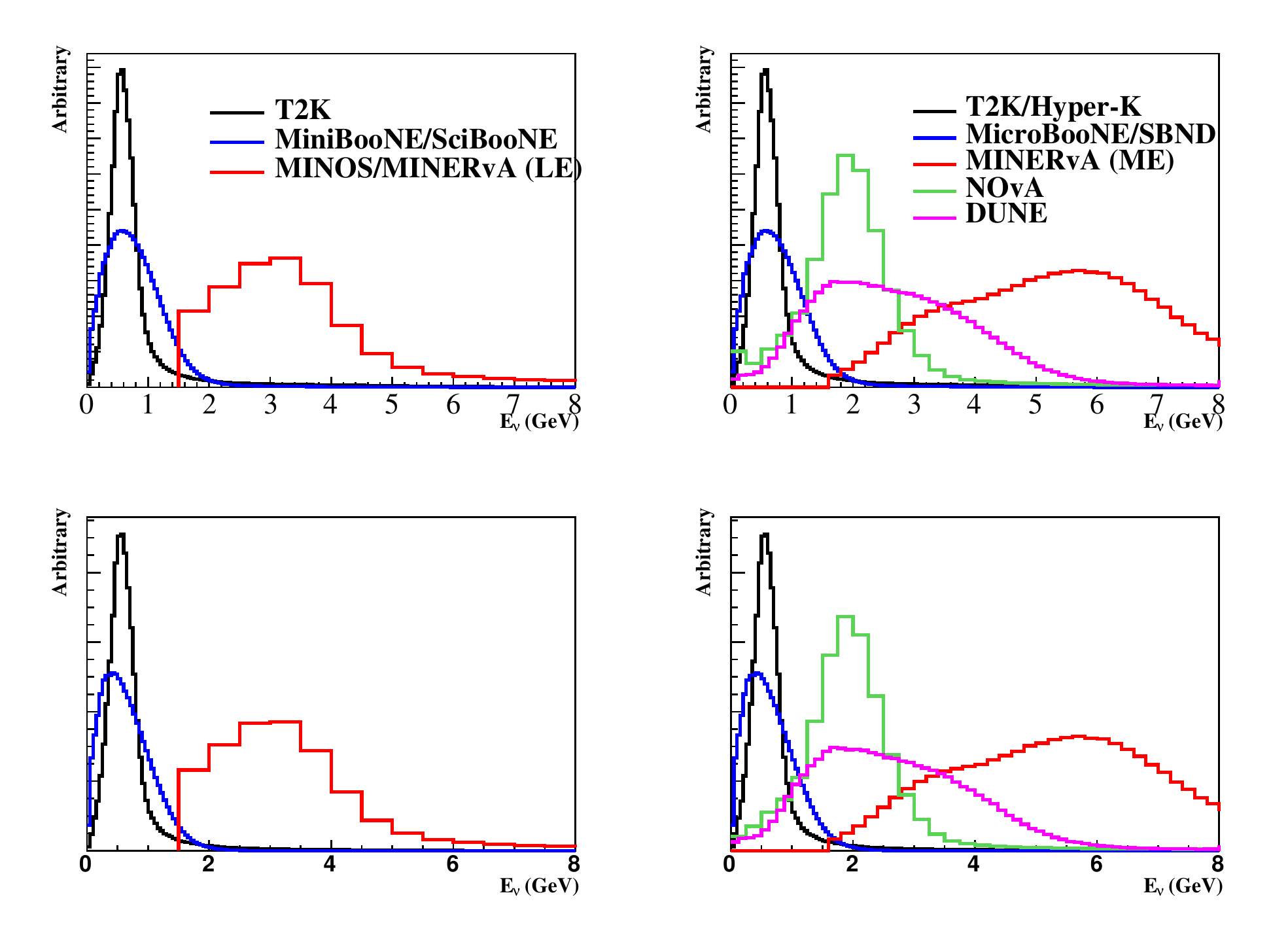}
    \end{center}
\vspace{-2mm}
\caption{
Muon neutrino and muon anti-neutrino flux predictions from current and future accelerator based neutrino experiments.
Here, the top two plots are neutrino mode beam muon neutrino flux predictions,
where the bottom two plots are anti-neutrino mode beam muon anti-neutrino flux predictions.
Predictions are all arbitrary normalized.
Left plots are current experiments (T2K, MiniBooNE, MINERvA with low energy NuMI),
and right plots are current to future experiments (Hyper-Kamiokande, MicroBooNE, NOvA, DUNE, MINERvA with medium energy NuMI).
}
\label{fig:flux_all}
\end{figure}

After incorporating the hadron production information in the neutrino beamline simulation,
neutrino flux at the detector site is predicted.
In this article we mainly focus on the measurements 
with neutrino baseline less than $<1$~km,
since that is the typical baseline for the neutrino cross section experiments and
near detectors of long-baseline oscillation experiments. 
Figure~\ref{fig:flux_all} shows flux predictions from current and future accelerator based neutrino experiments.
Note, some of them, especially flux predictions of future experiments~\cite{HK_2015,DUNE_CDR3} are preliminary results. 
The top two plots are for neutrino mode muon neutrino flux predictions,
and the bottom two plots are muon antineutrino flux prediction for antineutrino mode.
Left two plots are past experiments, and right two plots are current to future experiments.

\begin{itemize}
\item MiniBooNE used BNB, which also provided the beam for the SciBooNE experiment. 
Future experiments, such as Fermilab short baseline programs~\cite{ICARUS_FNAL,ANNIE} will also use it.
\item MINERvA, MINOS, and NOvA use NuMI neutrino beamline.
The two important flux configurations are low energy (LE) mode and medium energy (ME) mode.
Also, detector configurations can be on-axis or off-axis.
Here, MINOS and MINERvA are both LE and ME on-axis experiments,
and NOvA is a ME off-axis experiment,
and their flux predictions are quite different.
Note MINERvA does not provide neutrino flux below 1.5~GeV where flux systematic errors
have not been evaluated yet.
\item DUNE will use a dedicated beamline, which will have a wide-band beam to measure
neutrino oscillations not only the first maximum,
but also the second oscillation maximum~\cite{DUNE_CDR3}.
\item Hyper-Kamiokande uses higher power J-PARC off-axis neutrino beam~\cite{HK_2015},
and here we simply assumed the same shape with current T2K J-PARC off-axis neutrino beam.
\end{itemize}

The on-axis beam experiments,
such as MiniBooNE, MINERvA, and DUNE have a wider beam spectrum,
and off-axis beam experiments, such as T2K and NOvA have narrower spectrums. 
Although spectra are narrower for off-axis beams,
they have long tails going to higher energy. 
This is a standard feature for off-axis beams.
Therefore understanding of neutrino interactions are important in all 1-10 GeV spectrum
for both on-axis and off-axis beam experiments.


\vspace{0.5cm}

\begin{figure}[t!]
  \begin{center}
    \includegraphics[width=10cm]{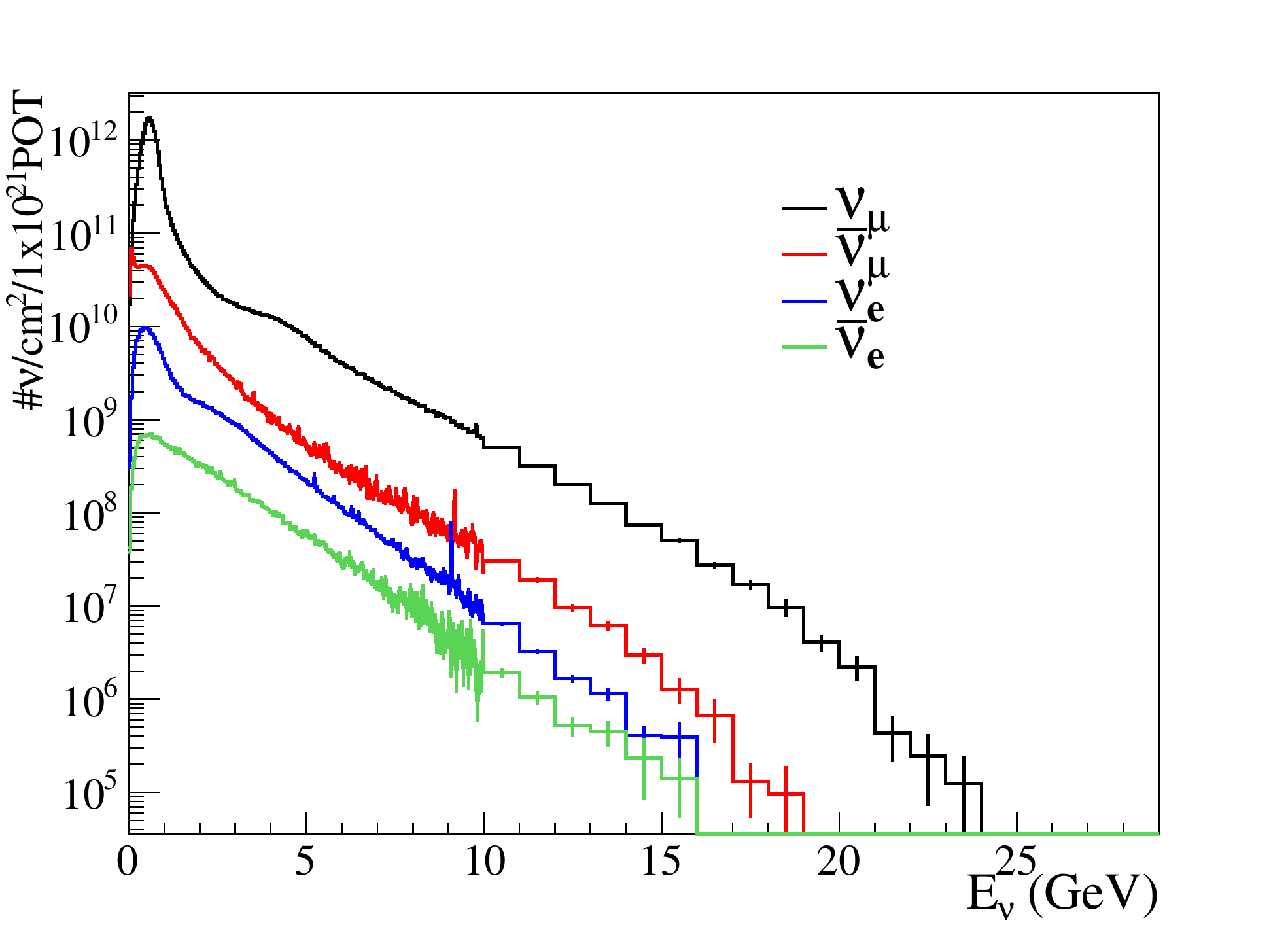}
    \end{center}
\vspace{-2mm}
\caption{
Details of T2K neutrino mode neutrino flux predictions at the near detector site (off-axis). 
}
\label{fig:flux_T2K}
\end{figure}

Figure~\ref{fig:flux_T2K} shows more detailed neutrino flux predictions.
Here, we use T2K neutrino mode flux prediction as an example to demonstrate the common features
of the off-axis super beam.
First, we see high energy tails in these fluxes, even though the off-axis beam peak is tuned around 600~MeV. 
For the neutrino mode flux, positive mesons are focused to enhance muon neutrino ($\numu$) components.
However, there are always inevitable contaminations of muon antineutrinos ($\numubar$)
due to inefficiency to reject negative (wrong sign) mesons and muon decays.
Such background ($\numubar$ in neutrino mode beam) is called ``wrong sign'' (WS) background. 
There are also tiny amounts of electron neutrinos ($\nue$)
and electron antineutrinos ($\nuebar$) from muon and kaon decays. 
All of them are considered to be the intrinsic backgrounds
of both $\numu$ disappearance and $\nue$ appearance oscillation measurements in neutrino mode.
Large far detectors of oscillation experiments often lack magnetic fields
to perform a sign separation of positive and negative leptons. 
Also, resolutions are sacrificed to maximize the fiducial volume,
and this makes rather poorer particle ID ability to reject background events. 
Therefore, for future experiments, understanding of these tiny contaminations is crucial.

These beam contaminations are also major backgrounds of cross section measurements. 
The experiments often encounter the problem of predicting small contaminations correctly. 
For example, many WS backgrounds are originated from the forward going mesons
which are not rejected by the magnetic horn,
and distributions are not measured by hadron production experiments
due to the lack of forward direction coverage of detector arrays~\cite{MB_WS}. 
Instead, experimentalists often try to measure beam contaminations in neutrino detectors
by applying cuts to make a background control sample,
and correct their distributions in the simulation. 
Such a technique was demonstrated by several experiments~\cite{MB_WS,SB_kaon}.
This is a powerful way to correct intrinsic beam background distributions and to constrain their errors. 
However, this technique also needs care.
The ``backgrounds'' are by definition unwanted events in the signal sample. 
Since measured background events in background control sample never pass the signal selection 
(if so, they are not the ``background'' from the beginning!),
one needs to be very careful of how to relate measured beam intrinsic background events from background control sample
and background events contaminated in the signal sample. 

Detailed predictions of the neutrino flux is always difficult,
and it is also common to correct flux predictions based on
interactions with known cross sections in the neutrino detectors.
NOMAD~\cite{Lyubushkin:2008pe} checked the flux normalization in two ways: 
by utilizing DIS and inverse muon decay (IMD) cross sections.
For lower energy neutrino experiments which we focus on this review,
neither DIS nor IMD are not very practical for such a purpose. 
MINERvA~\cite{Park:2015eqa} measures $\numu-e$ elastic scattering to constrain the flux.
This process also has a theoretically well-known cross section,
and distinctive experimental signature (forward going electromagnetic shower) allows them to be selected efficiently. 
By measuring $\numu-e$ elastic scattering events,
MINERvA can effectively measure the neutrino flux even though the cross section is rather small, 
and MINERvA found predictions of NuMI flux are consistently higher than the measurement. 
Later the NuMI flux simulation was improved~\cite{Aliaga:2016oaz} and this disagreement no longer exists.

\section{Neutrino cross section generalities \label{sec:xs-gen}} 
In this Section we discuss some general aspects of the neutrino-nucleus cross sections. We omit to treat the neutrino-nucleon scattering, reviewed for example in Ref. \cite{Alvarez-Ruso:2014bla}. Our choice is driven by the fact that, 
as discussed in Sec. \ref{sec:nuexp}, the detectors of modern neutrino experiments are composed of complex nuclei.

\subsection{Theory\label{sec:theo}}

The double differential cross-section  for the reaction
\mbox{$ \nu_l \, (\bar{\nu}_l) + A \longrightarrow l^- \, (l^+) + X $}
is given by
\begin{equation}
\label{m_eq_1}
\frac{d^2 \sigma}{d\Omega_{k'} d \omega} =  
\frac{G_F^2 \cos^2\theta_C}{32\pi^2}\frac{|{\bf{k}}'|}{|{\bf{k}}|}L_{\mu\nu}W^{\mu\nu}({\bf{q}},\omega). 
\end{equation}
Here $d\Omega_{k'}$ is the differential solid angle in the direction specified
by the charged lepton momentum ${\bf{k}}'$ in the laboratory frame,
$\omega=E_\nu-E_l'$ is the energy transferred to the nucleus,
the zero component of the four momentum transfer $q=k-k^\prime\equiv(\omega,{\bf{q}})$,
with $ k \equiv (E_\nu,{\bf{k}})$ and $ k^\prime \equiv (E_l',{\bf{k}}')$, being the initial and final lepton four momenta.
In Eq.~(\ref{m_eq_1}), $ G_F $ is the weak coupling constant,
$ \theta_c $ is the Cabbibo angle, and $L$ and $W$ are the leptonic and hadronic tensors, respectively. 

The leptonic tensor is 
\begin{equation}
L_{\mu\nu} = 8 (k_\mu {k'}_\nu + k_\nu {k'}_\mu - g_{\mu\nu} k.k' \mp i \varepsilon_{\mu\nu\alpha\beta} k^\alpha {k'}^\beta)
\end{equation}
where the metric is $g_{\mu\nu}=(+,-,-,-)$ and the convention for the fully anti-symmetric Levi-Civita tensor is $\varepsilon_{0123}=+1$.
The sign $-$ ($+$) before the Levi-Civita tensor refers to neutrino (antineutrino) interaction.
This basic asymmetry which follows from the weak interaction theory has important consequences on
the differences between neutrino and antineutrino cross sections, as it will be illustrated later.  

The hadronic tensor describes the hadronic part. It is defined as
\begin{equation}
W^{\mu\nu}({\bf{q}},\omega)=\sum_f
\langle \Psi_i |J^{\mu}(q)|\Psi_f\rangle
\langle \Psi_f |J^{\nu}(q)|\Psi_i\rangle
\delta^{(4)}(P_i+q-P_f) \,,
\end{equation}
where $|\Psi_i\rangle$ and $|\Psi_f\rangle$ are the initial and final hadronic states with four momenta $P_i$ and $P_f$. $J^{\mu}$ 
is the electroweak nuclear current operator which can be expressed as a sum of
one-body $J^{\mu}_{OB}$ and two-body $J^{\mu}_{TB}$ contributions. 
The sum over final states can be decomposed as the sum of
one-particle one-hole (1p-1h) plus two-particle two-hole (2p-2h) excitations plus additional channels 
\begin{equation}
 W^{\mu\nu} ({\bf{q}},\omega)= W^{\mu\nu}_{1p1h} ({\bf{q}},\omega)+ W^{\mu\nu}_{2p2h} ({\bf{q}},\omega)+ \cdots
\end{equation}
According to Eq.~(\ref{m_eq_1}), the same decomposition holds for the cross section. 

The different components of the hadronic tensor can be combined allowing a reformulation of Eq.~(\ref{m_eq_1})
in terms of projections with respect to the momentum transfer direction.
The charged current cross section is a linear combination of five response functions
\begin{equation}
\label{m_eq_general}
\frac{d^2 \sigma}{d\Omega_{k'} d \omega} =  
\sigma_0\left[L_{CC}R_{CC}+L_{CL}R_{CL}+L_{LL}R_{LL}+L_TR_{T}\pm L_{T'} R_{T'} \right],
\end{equation}
where the kinematical factors come from the contraction with the leptonic tensor and
the plus (minus) sign applies to neutrinos (antineutrinos).
The letters $C$, $L$ and $T$ stay for Coulomb,
longitudinal and transverse respectively\footnote{The notation $\{CC,CL,LL,T,T'\}$ is often replaced
by $\{00,0z,zz,xx,xy\}$ or $\{00,03,33,11,12\}$.}. 
We omit here to give the explicit expression of the kinematical factors and of the responses,
which can be found in many books (e.g. Ref.~\cite{Walecka:1995mi}) and articles
(e.g. Refs.~\cite{O'Connell:1973gus,Amaro:2005dn,Nieves:2004wx,Martini:2009uj,Shen:2012xz}).

Below we give instead a simplified expression which ignores the final lepton mass contributions and
which is obtained keeping only the leading terms
for the hadronic tensor in the development of the hadronic current in $p/M_N$ \cite{O'Connell:1973gus},
where $p$ denotes the initial nucleon momentum and $M_N$ the nucleon mass.
In this case the response functions entering into the expression of the cross section are reduced to three
\begin{equation} \label{eq:_risposte_alpha}
R_\alpha({\bf{q}},\omega) = \sum_f \, 
\langle f | \sum_{j=1}^A \, O_\alpha(j) \, 
e^{ i \, {\bf{q}} \cdot {\bf{x}}_j } | 0 \rangle 
\langle f | \sum_{k=1}^A \, O_\alpha(k) \, 
e^{ i \, {\bf{q}} \cdot {\bf{x}}_k } | 0 \rangle^* \, \delta (\omega - E_f + E_0 )  
\end{equation}
with 
\begin{equation}
O_\alpha(j) = \tau_j^\pm, \,\,\, ( {\bf{\sigma}}_j \cdot \widehat{q} ) \, 
\tau_j^\pm, \,\,\, 
( {\bf{\sigma}}_j \times \widehat{q} )^i
\, \tau_j^\pm, 
\end{equation}
for $ \alpha = \tau $,  $ \sigma\tau (L) $, $ \sigma\tau (T) $
\footnote{The $ \sigma\tau$ operators are replaced by the usual 1/2 to 3/2 transition operators $ST$ in the case of coupling to the $\Delta$.}. We have thus the \textit{isospin} ($ R_\tau $), 
the \textit{spin-isospin longitudinal} ($ R_{\sigma\tau (L)} $) and the \textit{spin-isospin transverse} ($ R_{\sigma\tau (T)} $)
responses (the longitudinal and transverse character of these last two responses refers to the direction of the spin operator with respect to the direction of the transferred momentum ${\bf{q}}$). The explicit expression of the cross section in terms of these three responses is 
\begin{eqnarray}
\label{eq1:cross_section}
\frac{d^2\sigma}{d \cos\theta d \omega} & = & \frac{G_F^2 \, 
\cos^2\theta_c}{\pi} |{\bf{k}}'|E_l' \, \cos^2\frac{\theta}{2} \, 
\left[ \frac{({\bf{q}}^2-\omega^2)^2}{{\bf{q}}^4}\,G_E^2 \, R_\tau({\bf{q}},\omega) \right.\nonumber \\  
& + &  \left. \frac{\omega^2}{{\bf{q}}^2} \, G_A^2 \, R_{\sigma\tau (L)}({\bf{q}},\omega)  \right.\nonumber \\ 
& + & \left.
 2 \left( \tan^2\frac{\theta}{2}+ \frac{{\bf{q}}^2-\omega^2}{2 {\bf{q}}^2} \right ) 
\left( G_M^2 \,
\frac{{\bf{q}}^2}{4M_N^2}
+ G_A^2 \right)  R_{\sigma\tau (T)}({\bf{q}},\omega)\right.\nonumber \\ 
& \pm& \left. 2 \,  \frac{E_\nu + E_l'}{M_N} \, 
\tan^2\frac{\theta}{2} \, G_A \, G_M \, R_{\sigma\tau (T)}({\bf{q}},\omega) \right]. 
\end{eqnarray}
This expression is particularly useful for illustration since \textit{i)} the different kinematic variables
(related to the leptonic tensor),
\textit{ii)} the nucleon electric, magnetic,
and axial form factors (that contain the information about the nucleon properties),
and \textit{iii)} the nuclear response functions (that contain the information about the nuclear dynamics)
explicitly appear~\footnote{In the generic expression of Eq.(\ref{m_eq_general})
the nucleon form factors are implicitly included in the response functions.}.
It is important to stress that Eqs.~(\ref{m_eq_1}), (\ref{m_eq_general}) and (\ref{eq1:cross_section}) 
are totally general and apply to different excitations channels: 1p-1h QE, 2p-2h, 1p-1h 1$\pi$ production,
coherent $\pi$ production, etc. 

Concerning the form factors, they depend on the square of the 4-momentum transfer $Q^2=-q^2$. 
The conserved vector current hypothesis allows to apply the vector (electric and magnetic) form factors measured in electron scattering to neutrino scattering. The axial form factor is usually described by a dipole parameterization
\begin{equation}
G_A(Q^2)=\frac{g_A}{(1+Q^2/M_A^2)^2}.
\label{eq:axialff}
\end{equation}
The $g_A$ coupling is well known from neutron $\beta$ decay, $g_A=G_A(0)=1.26$.
The value of the axial mass parameter $M_A$ extracted 
from charged current quasielastic experiments on deuterium bubble
chambers~\cite{Mann:1973pr,Barish:1977qk,Baker:1981su,Kitagaki:1983px} 
is $M_A=1.026 \pm 0.021$ GeV~\cite{Bernard:2001rs}. The value of this axial mass parameter attracted
a lot of attention in connection with the MiniBooNE CCQE result, as it will be discussed in Sec.~\ref{sec:QE}. 

\begin{figure}[t!]
  \begin{center}
    \includegraphics[width=8cm]{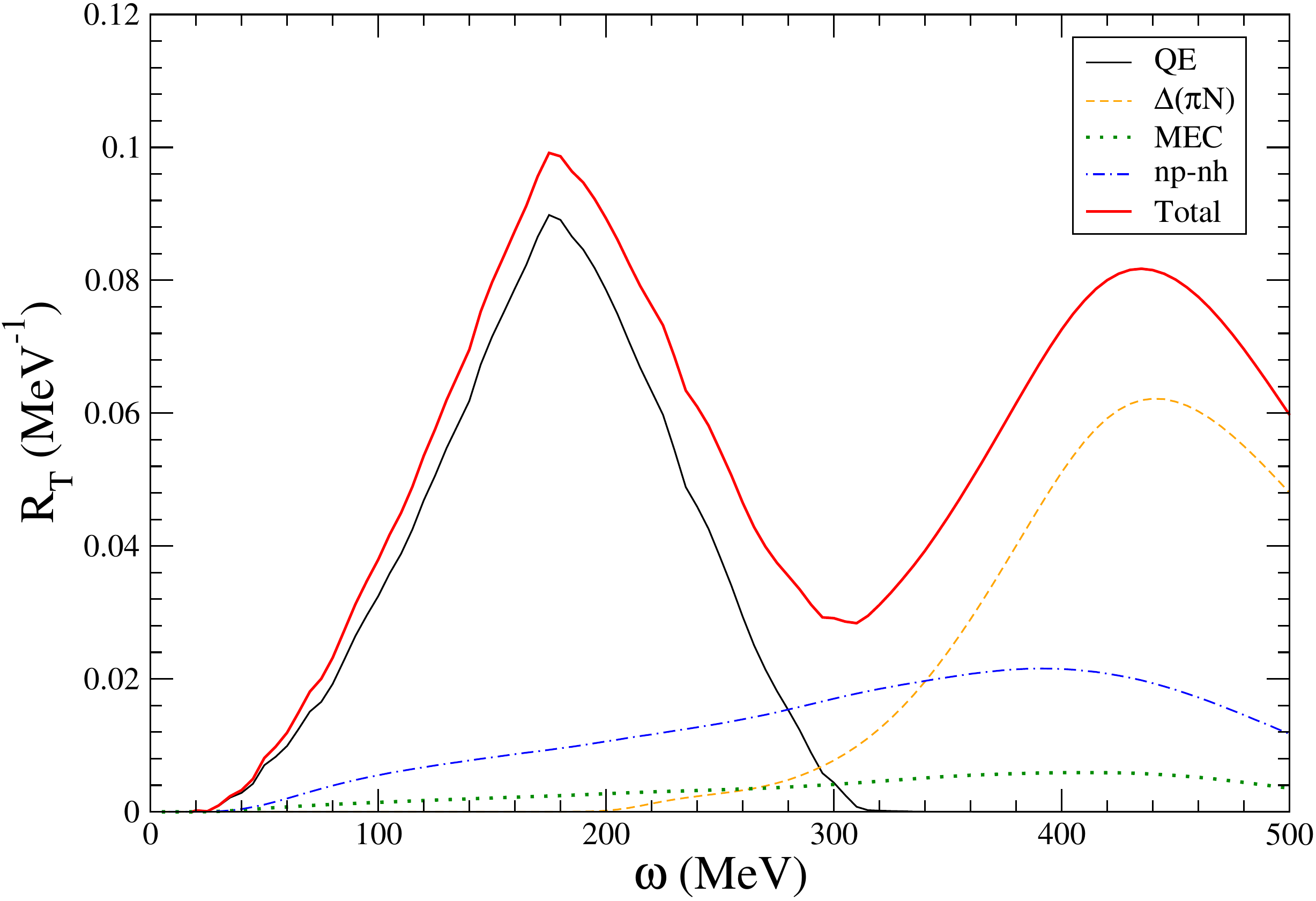}
    \end{center}
\caption{
spin-isospin transverse response $R_{\sigma\tau (T)}({\bf{q}},\omega)$
for $^{12}$C calculated in Random Phase Approximation (RPA) according to the approach of Ref.~\cite{Martini:2009uj} 
at $|{\bf{q}}|$= 600 MeV/c as a function of the energy transfer $\omega$ with its different components,
quasielastic, pion emission and np-nh.
The MEC contribution to np-nh is also shown. 
}
\label{fig_risp_q600_diversicanali}
\end{figure}

Turning to the  nuclear responses entering in Eq.~(\ref{eq1:cross_section}) an example of these is given
in Fig.~\ref{fig_risp_q600_diversicanali} 
where the different $^{12}$C spin-isospin transverse responses 
$R_{\sigma\tau (T)}({\bf{q}},\omega)$, calculated in Random Phase Approximation (RPA) according
to the approach of Ref.~\cite{Martini:2009uj},
are plotted for fixed values of the momentum transfer $|{\bf{q}}|$, as a function of the energy transfer $\omega$. 
One can easily distinguish the quasielastic response, which corresponds to one nucleon knockout.  
It is peaked around  
\begin{equation}
\label{qe_disp_rel}
\omega=\sqrt{{\bf{q}}^2+M_N^2}-M_N=\frac{Q^2}{2M_N}=\frac{{\bf{q}}^2-\omega^2}{2M_N}, 
\end{equation}
the value corresponding to the quasielastic scattering with a free nucleon at rest. 
RPA collective effects can shift the position of this quasielastic peak~\cite{Alberico:1981sz}
with respect to the one given by Eq.~(\ref{qe_disp_rel}). 
The broadening of the quasielastic response is due to the Fermi motion.
The quasielastic response can be distorted by Pauli correlations or by the collective nature of the response, as described in RPA. 

The curve characterized by a bump at higher $\omega$ corresponds to the $\Delta$ resonance excitation,
which decays via the pionic channel $\Delta \to \pi N$. 
It is peaked around  
\begin{equation}
\label{delta_disp_rel}
\omega=\sqrt{{\bf{q}}^2+M_{\Delta}^2}-M_N=\frac{Q^2}{2M_N}+\frac{M_{\Delta}^2-M_N^2}{2M_N}.
\end{equation}
This second curve is related to the $\Delta$ pionic decay $\Delta \to \pi N$
in the nuclear medium (hence to the 1$\pi$ production channel). 
Pauli blocking of the nucleon and the distortion of the pion are taken into account. 
In the nuclear medium non pionic $\Delta$ decay channels are also possible
such as the two-body (2p-2h) and three-body (3p-3h) absorption channels, 
which leads to np-nh excitations. 

The total np-nh excitations channel, which also includes other 2p-2h excitations
which are not reducible to a modification of the $\Delta$ width is also shown.
The part of the np-nh excitations related to the two-body meson exchange currents (MEC) contributions is separately plotted. 
We postpone a detailed discussion of these np-nh excitations to the Section~\ref{sec:npnh}.
By comparing with data, one can notice that it is crucial to fill the dip between the quasielastic and $\Delta$ excitations,
as first observed by Van Orden and Donnelly~\cite{VanOrden:1980tg} and Alberico \textit{et al.}~\cite{Alberico:1983zg}
in the studies of the electron scattering cross sections and transverse responses.  

\begin{figure}
\begin{minipage}[c]{80mm}
\begin{center}
\includegraphics[width=80mm]{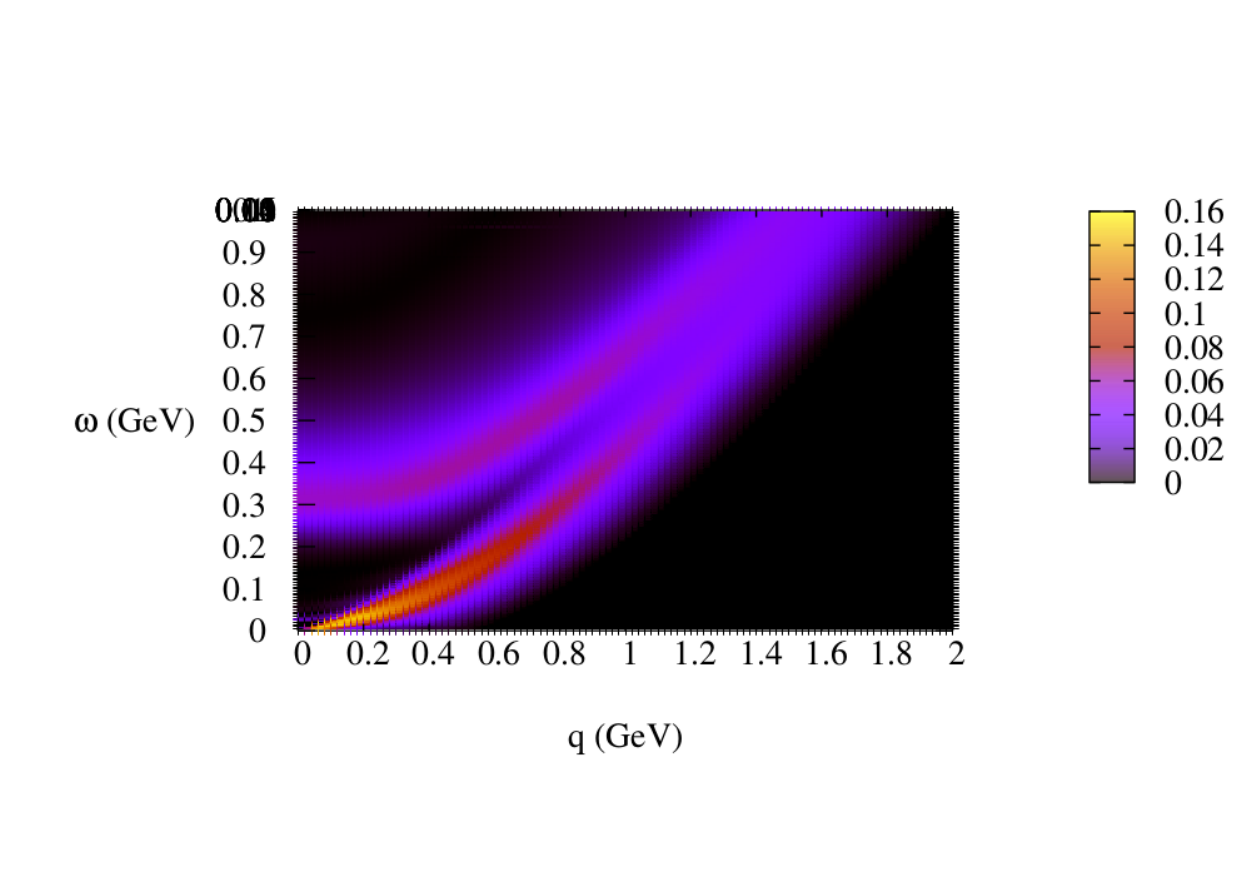}
\caption{\label{fig_qe_Delta}Quasielastic and Delta spin-isospin transverse response function (and region)}
\end{center}
\end{minipage}
\hspace{1mm}
\begin{minipage}[c]{80mm}
\begin{center}
      \includegraphics[width=80mm]{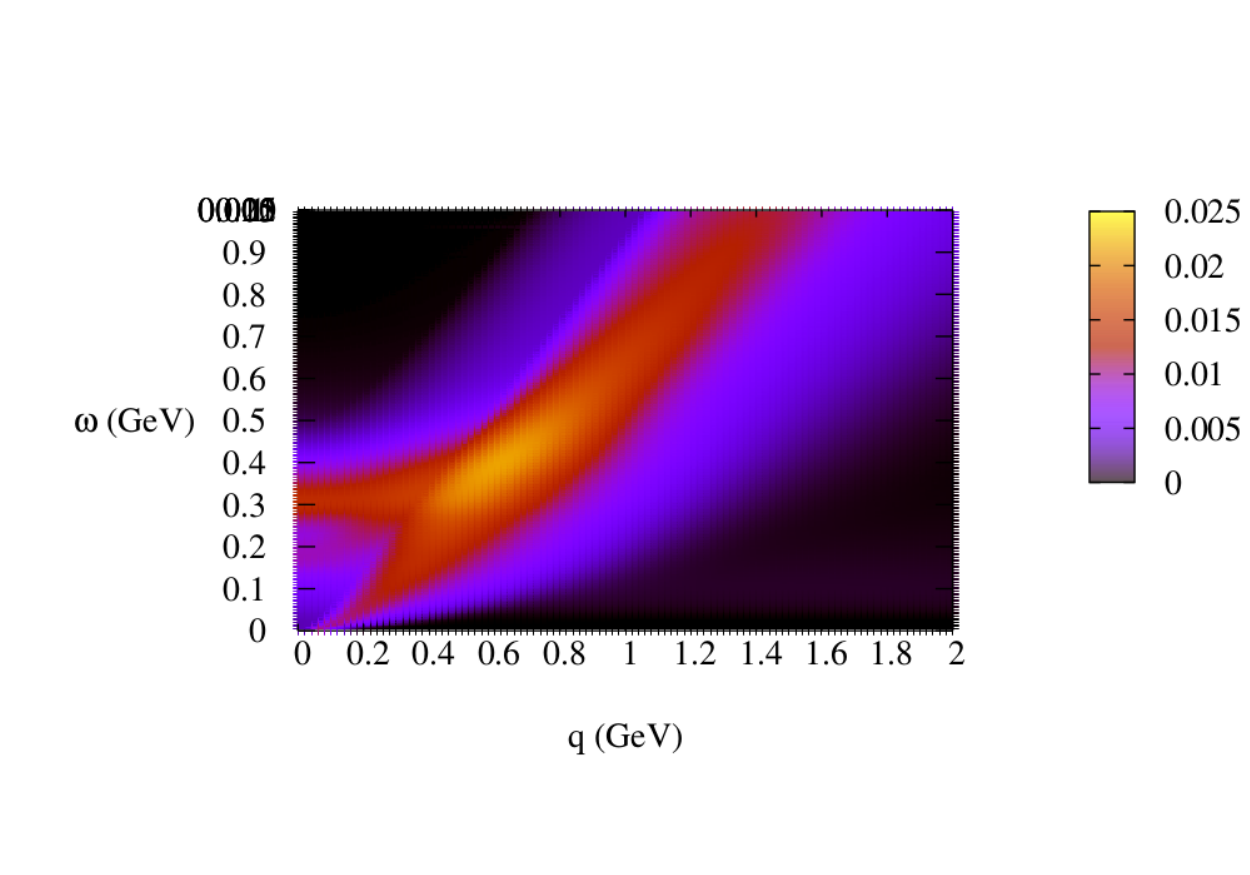}
\caption{\label{fig_npnh}np-nh spin-isospin transverse response function (and region)}
\end{center}
\end{minipage}
\hspace{1mm}
\begin{minipage}[c]{80mm}
\begin{center}
      \includegraphics[width=80mm]{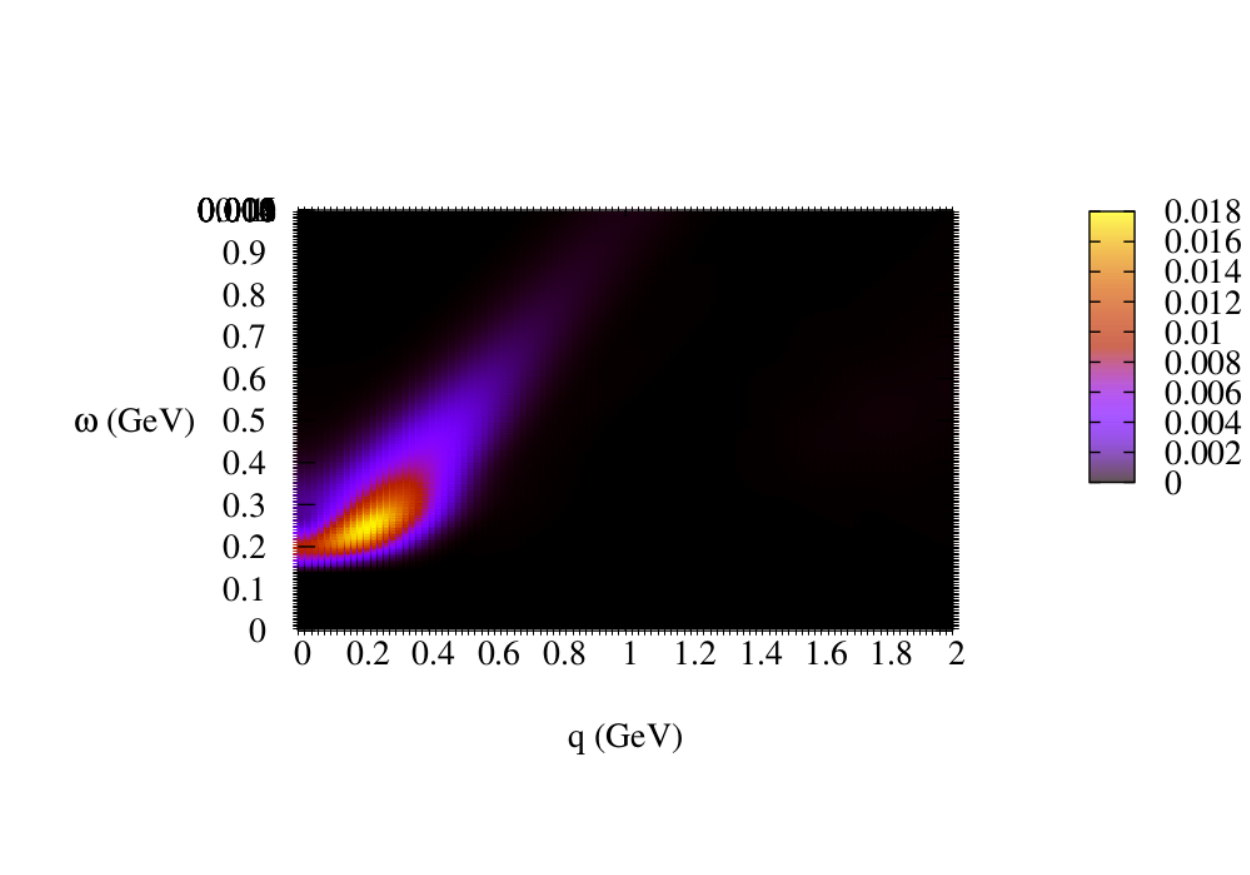}
\caption{\label{fig_cohpi}Coherent spin-isospin longitudinal response function (and region)}
\end{center}
\end{minipage}
\end{figure} 

The different spin-isospin transverse responses $R_{\sigma\tau (T)}({\bf{q}},\omega)$
calculated in the approach of Ref.~\cite{Martini:2009uj}
are shown in three-dimensional plots in Figs.~\ref{fig_qe_Delta} and~\ref{fig_npnh}. 
These figures well illustrate the response regions \textit{i.e.}
the regions of the $\omega$ and $|{\bf{q}}|$ plane where the responses are nonzero.   
The sum of quasielastic and $\Delta$ contributions is given in Fig.~\ref{fig_qe_Delta}.
This figure allows to easily distinguish the quasielastic and $\Delta$ region
as well as the position of the quasielastic and $\Delta$ peaks,
which for a non-interacting system would follow the nucleon and $\Delta$ dispersion relations of
Eqs.~(\ref{qe_disp_rel}) and (\ref{delta_disp_rel}), respectively.
The quasielastic response region is delimited by the two lines
$\omega_{\pm}=\sqrt{{\bf{q}}^2\pm 2|{\bf{q}}|k_F +M^2_N} - M_N$, where $k_F$ is the Fermi momentum. 
The spreading of the $\Delta$ response region is due to the nucleon Fermi motion and to the $\Delta$ decay width
which is modified by the interaction of the $\Delta$ with surrounding nucleons. 
The np-nh response function (and region) is separately plotted in Fig.~\ref{fig_npnh}. 
One can observe that it covers mostly the whole $\omega$ and $|{\bf{q}}|$ plane,
in particular it is non vanishing in the dip region between the nucleon-hole and $\Delta-hole$ domains.
The two substructures appearing in Fig.~\ref{fig_npnh} reflects the different origins of the np-nh excitations,
such as the nucleon-nucleon correlation contributions (lower $\omega$ part)
and the non pionic $\Delta$ decay contributions (higher $\omega$ part). 

For completeness, the main contribution to the coherent pion production response,
represented by spin-isospin longitudinal coherent response 
$R_{\sigma\tau (L)}^{coh}({\bf{q}},\omega)$ is also shown in Fig.~\ref{fig_cohpi}.
Other examples of nuclear responses calculated in the same approach can be found in Ref.~\cite{Martini:2009uj}.
For example, Figs. 5 and 6 of Ref.~\cite{Martini:2009uj} illustrate the reshaping effects of the nuclear responses
due to the role of the effective interaction between particle-hole excitations as well as collective features of the spin-isospin longitudinal response,
in particular in the coherent channel.

\begin{figure}[tb]
  \begin{center}
    \includegraphics[width=12cm,valign=m]{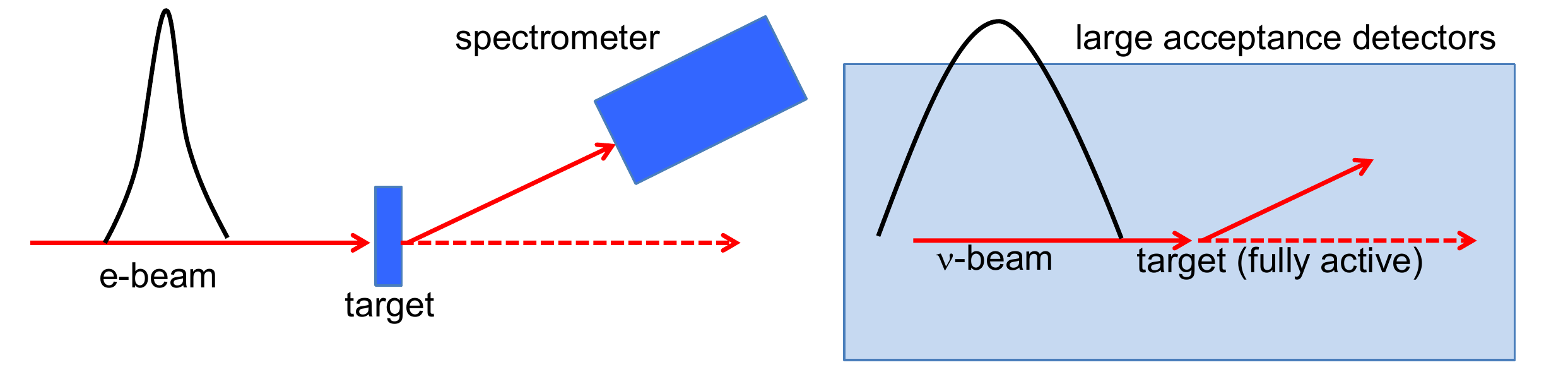}
    \end{center}
\vspace{-2mm}
\caption{
Cartoons for a typical electron scattering experiment (left) and a typical neutrino experiment (right).
}
\label{fig:garvey}
\end{figure}

As illustrated, nuclear cross sections are naturally expressed in terms of the nuclear responses,
functions of the energy and momentum transferred to the nuclear system.
Figure~\ref{fig:garvey} shows a comparison of a typical electron scattering experiment and
a typical accelerator-based neutrino experiment~\cite{garvey}. 
In the electron scattering experiment (left), the beam energy is precisely known,
and experimentalists measure energy and angle of scattered electron.
In this way, both $\om$ and ${\bf q}$ are determined from given interactions and kinematics are fully fixed.
On the other other hand, modern accelerator-based neutrino experiments (right)
are performed with a wide-band beam with a fully active detector to maximize the interaction rate. 
Since the neutrino energy of a given interaction is not know {\it a priori},  
experimentalists do not know the neutrino energy of the given interaction,
and $\om$ and ${\bf q}$ cannot be determined from the measured lepton distribution
\footnote{On the other hand, the fully active detector can records all other tracks and activities from hadrons, 
and such information can be used to reconstruct the kinematics, as demonstrated by MINERvA~\cite{Rodrigues:2015hik} and discussed in Sec.\ref{subsubsec_minerva_evail}.} 
The measured variables in neutrino scattering are only the charged lepton energy $E_l$ (or kinetic energy $T_l$)
and its scattering angle $\theta$, related to the energy and momentum transfer by 
\begin{equation}
\label{eq:w}
\omega=E_\nu- E_l
\end{equation}
and
\begin{equation}
\label{eq:q}
{\bf{q}}^2 = E^2_\nu +  {\bf{k}}_l'^2 - 2  E_\nu |{\bf{k}}_l'| \cos{\theta}.
\end{equation}
The experimental measured quantity is then the flux-integrated double differential cross section 
in terms of the measurable variables $T_l$ and $\theta$:  
 \begin{equation}
 \label{cross}
\frac{d^2 \sigma}{dT_{l}~d~\mathrm{cos}\theta}=
\frac{1}
{ \int \Phi(E_{\nu})~d E_{\nu}}
 \int ~d E_{\nu}
\left[\frac{d^2 \sigma}{d \omega  ~d\mathrm{cos}\theta}\right]_{\omega=E_{\nu}-E_{l}} \Phi(E_{\nu}), 
\end{equation}
where $\Phi(E_{\nu})$ is the neutrino flux. 

\begin{figure}
\begin{center}
  \includegraphics[width=10cm]{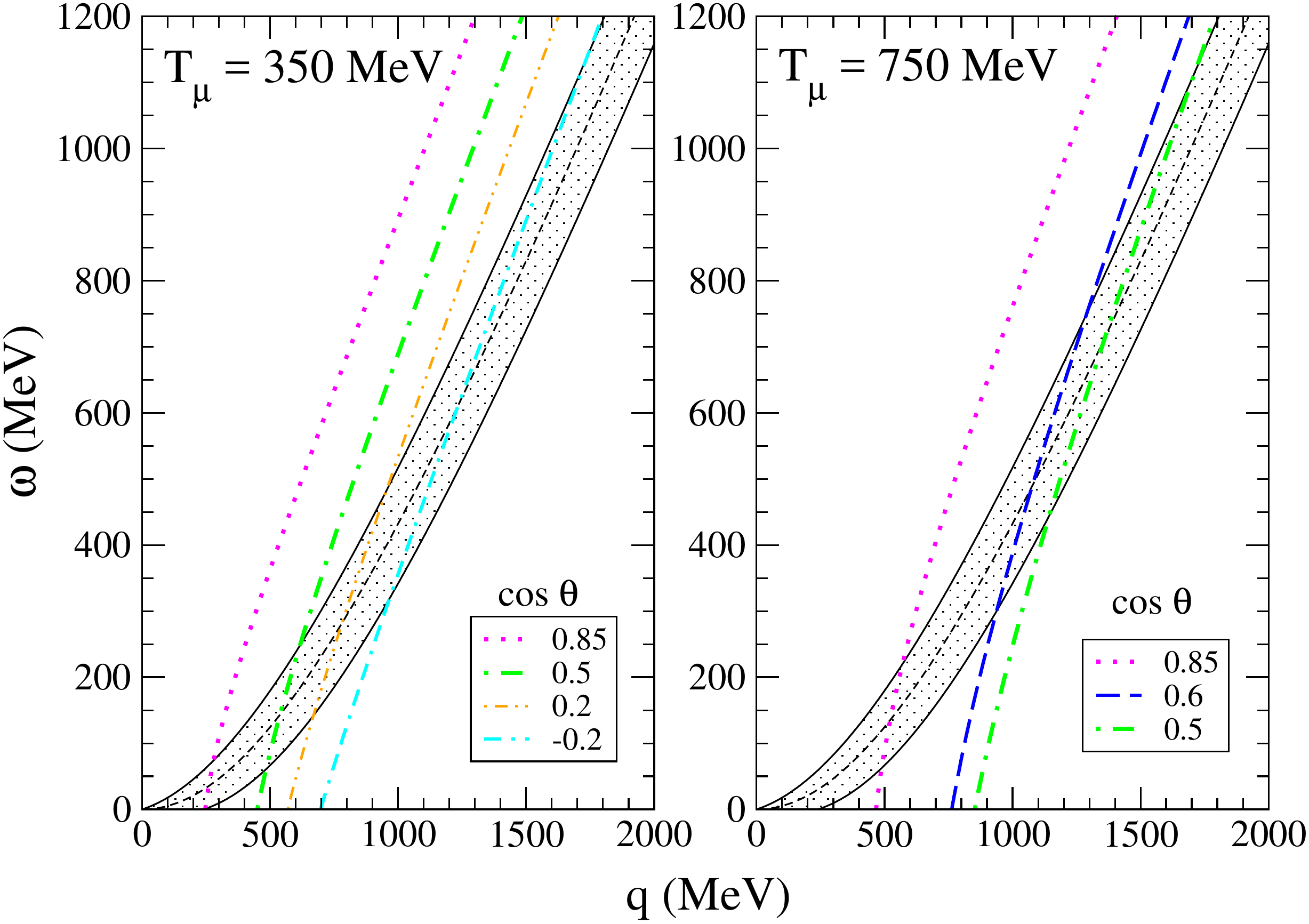}  
\caption{(Color online) The neutrino 
hyperbolas defined by Eq.~(\ref{eq_hyp}) for a muon kinetic energy $T_\mu$=350 MeV,
$T_\mu$=750 MeV and several muon scattering angles indicated in the figure.
The shaded area delimited by the two corresponding continuous lines represents
the region of the quasielastic response of a Fermi gas.
The central dashed lines show the position of the quasielastic peak. The figure is taken from Ref.~\cite{Martini:2012fa}.}
\label{fig_hyperbolas}
\end{center}
\end{figure}

The cross section of the r.h.s. of Eq.~(\ref{cross}), as expressed in terms of the nuclear responses,
according to Eqs.~(\ref{m_eq_1}) and~(\ref{eq1:cross_section}), 
is non vanishing in the regions of the $\omega$  and $|{\bf{q}}|$ plane where the responses are non-zero, 
regions shown in Figs.~\ref{fig_qe_Delta},~\ref{fig_npnh} and~\ref{fig_cohpi}. 
To illustrate how these regions are explored in neutrino reactions we repeat an argument of
Refs.~\cite{Delorme:1985ps,Martini:2011wp,Martini:2012fa}. We write the squared four momentum transfer 
in terms of the lepton observables (for illustration we take the example of an ejected muon)
\begin{equation}
\label{eq_hyp}
Q^2={\bf{q}}^2 - \omega^2= 4 (E_\mu+\omega) E_\mu ~\mathrm{sin} ^2 \frac{\theta}{2}-m_{\mu}^2+2 (E_\mu+\omega) (E_\mu-|{\bf{k}}_\mu'|) ~\mathrm{cos} {\theta}.
\end{equation} 
For a given set of observables $E_\mu$ and $\theta$, this relation defines a hyperbola
in the  $\omega$ and $|{\bf{q}}|$ plane~\cite{Delorme:1985ps}. 
The asymptotic lines are parallel to the $\omega=|{\bf{q}}|$ line
($|{\bf{q}}| > \omega$ always, the hyperbolas lie entirely in the space-like region)
and the intercept of the curves with the $\omega=0$ axis occurs at a value of the momentum
\begin{equation}
{\bf{q}}^2_{\omega=0}= 4 E_\mu^2 ~\mathrm{sin} ^2 \frac{\theta}{2}-m_{\mu}^2+2 E_\mu (E_\mu-|{\bf{k}}_\mu'|) ~\mathrm{cos} {\theta}
\simeq  4 E_\mu^2 ~\mathrm{sin} ^2 \frac{\theta}{2},
\end{equation}
where the second expression is obtained by neglecting the muon mass. 
With increasing $E_\mu$ or increasing  angle, this point shifts away from the origin. 
The neutrino cross section for a given $T_\mu$ and $\theta$ explores the nuclear responses along the corresponding hyperbola. 
Some examples of hyperbolas are shown in Fig.~\ref{fig_hyperbolas},
together with the region of the quasielastic response of a Fermi gas. 
For simplicity we have omitted to show the $\Delta$, the np-nh and the coherent pion response regions,
already illustrated in Figs.~\ref{fig_qe_Delta},~\ref{fig_npnh} and~\ref{fig_cohpi}. 
The intercept of the hyperbola with the region of response of the nucleus, whatever its nature (quasielastic, $\Delta$, np-nh, coherent pion),  fixes the possible $\omega$ and $q$ values for a given value of $T_\mu$ and $\theta$. 

In the case of a dilute Fermi gas where the region of quasielastic response is
reduced to the quasielastic line given by Eq.~(\ref{qe_disp_rel})
(the black dashed line of Fig.~\ref{fig_hyperbolas}), 
the intercept values $\omega =\omega_{\textrm{int.}}$ and $|{\bf{q}}|= q_{\textrm{int.}}$ are completely fixed.
Hence the neutrino energy is also determined,  
$\overline{E_\nu} = E_\mu + \omega_{\textrm{int.}}$, which leads to the well known expression of the reconstructed energy for a fixed 
set of muon observables, $E_\mu$ and $\theta$:
\begin{equation}
\label{enubar_muon}
\overline{E_\nu} = \frac{E_\mu-m^2_\mu/(2M_N)}{1-(E_\mu-|{\bf{k}}_\mu'| \cos \theta)/M_N}. 
\end{equation}
The corresponding value of the reconstructed squared four momentum transfer is 
\begin{eqnarray}
\label{eq_Q_rec}
\overline{Q}^2&=&{\bf{q}}_{\textrm{int.}}^2 - \omega_{\textrm{int.}}^2\nonumber \\
&=& 4 (E_\mu+\omega_{\textrm{int.}}) E_\mu ~\mathrm{sin} ^2 \frac{\theta}{2}-m_{\mu}^2+2 (E_\mu+\omega_{\textrm{int.}}) (E_\mu-|{\bf{k}}_\mu'|) ~\mathrm{cos} {\theta}\nonumber \\
&=& -m_{\mu}^2+2 \overline{E_\nu} (E_\mu-|{\bf{k}}_\mu'| ~\mathrm{cos} {\theta}).
\end{eqnarray}
The quantities $\overline{E_\nu}$ and $\overline{Q}^2$ are often equivalently called $E_\nu^{QE}$ and $Q^2_{QE}$ in literature. 
The set of ($T_\mu$,$\cos\theta$) points satisfying Eqs.~(\ref{enubar_muon}) and~(\ref{eq_Q_rec})
for various values of $\overline{E_\nu}$ and $\overline{Q}^2$ 
is shown in the $\cos\theta$ and $T_\mu$ plane in Fig.~\ref{fig:kin_all} of Sec.~\ref{sec:kinematics}. 

A similar procedure could be repeated by considering the intersection of the hyperbolas shown in Fig.~\ref{fig_hyperbolas} 
with the $\Delta$ dispersion relation of Eq.~(\ref{delta_disp_rel})
instead of the free nucleon dispersion relation of Eq.~(\ref{qe_disp_rel}),
when interested in pion production instead of quasielastic.

The complexity of the nuclear physics implies a more subtle and delicate situation. 
As already discussed, first, the nuclear region of response is not restricted to a line for the QE and $\Delta$, 
but it spreads around these lines (see Fig.~\ref{fig_qe_Delta}) and second,
very important, it covers the whole $\omega$ and $q$ plane due to multinucleon emission (see Fig.~\ref{fig_npnh}). 
As a consequence, for a given set of values of $E_\mu$ and $\theta$,
moving along a hyperbola one explores the whole $\omega$ and $q$ plane hence, 
all values of the energy transfer $\omega$ 
contribute to the cross sections. In other words, \textit{for a given set of values of $E_\mu$ and $\theta$
one explores the full energy spectrum of neutrinos} above the muon energy since $E_\nu=E_\mu+\omega$. 
This fact has fundamental consequences on the determination of the neutrino energy in the neutrino oscillation experiments,
as it will be illustrated in Section~\ref{sec:enrec}.
Another aspect related to this point is that all the reaction channels (QE, np-nh, pion production,...)
are entangled and isolating a primary vertex process from the measurement of
neutrino flux-integrated differential cross section is much more difficult than in the cases of
monochromatic (such as electron) beams.
This is illustrated for example in Ref.~\cite{Benhar:2010nx} where a theoretical model based on
the impulse approximation scheme and the nuclear spectral function turns to successfully reproduce
the quasielastic peak of electron scattering double differential cross section data on carbon
as a function of the transferred energy $\omega$ for fixed scattering angle but not
the MiniBooNE neutrino flux integrated quasielastic-like double differential cross section as
a function of the muon kinetic energy at the same scattering angle.

\subsection{Experiment\label{sec:introexp}}

Let's now analyze the neutrino cross sections from an experimental perspective. 
Neutrino experiments measure the rate of neutrino interactions, $R$.
The interaction rate is the convolution of three factors: 
neutrino flux, $\Ph$, interaction cross section, $\si$, and the detector efficiency, $\ep$~\cite{Teppei_CCQE}
\beq
R\sim\Ph(E_\nu)\otimes\si(k,k')\otimes\ep(\mbox{observed particle kinematics}). 
\eeq
Here, neutrino flux is a function of neutrino energy, $E_\nu$,
and neutrino interaction cross section is a function of initial and final lepton kinematics.
The detector efficiency can be a function of any kinematic variables of observables,
such as energy deposit, scattering angle, etc. 
To compare the experimental data with predictions,
experimentalists simulate the neutrino flux, the interaction cross section, and the detector model,
and convolute them with the Monte Carlo (MC) method. 
Every experiment strives to understand the detector performance for given kinematics,
and this allows experimentalists to unfold detector responses.
Then, the measured quantity is the detector effect unfolded rate,
and it is proportional to the flux-integrated cross section.

\beq
R'\sim\Ph(E_\nu)\otimes\si(k,k')~.
\label{eq:diffxs}
\eeq

Reconstruction of neutrino energy is non-trivial in this energy region as we discuss in Sec.~\ref{sec:enrec}.
This essentially makes it impossible to unfold neutrino flux term
without introducing a model dependence in the final result. 
Therefore, modern neutrino interaction measurements focus on
producing flux-integrated differential cross-sections of direct observable kinematics
(lepton energy, lepton scattering angle, total hadron energy deposit, etc.) from topology-based signals.
Here, we show how to relate measured event distribution to theoretically calculable quantities. 

A histogram of observed neutrino interaction events is distributed in a vector $d_j$,
which may be the functions of lepton kinetic energy and scattering angle, {\it i.e.}, $d_j=d_j(T_l$,$\cos\th)$. 
The index $j$ of the data vector $d_j$ emphasizes this vector is a function of observed variables
(such as measured muon energy and direction). 
After subtracting the background contribution $b_j$,
detector bias is corrected.
This detector bias unfolding process is often separated into two processes: 
unsmearing and efficiency correction.
A proper unsmearing method transforms the background subtracted data (or sometimes data including backgrounds)
from the function of observables to the function of true variables.
This corresponds to applying unsmearing matrix $U_{ij}$ to change the index 
from $j$ to $i$, which represents true variables (such as muon energy and direction without any detector bias).
Then the estimated detection efficiency $\ep_i$ is inversely applied to recover the true distribution. 
Finally, by correcting all normalizations, such as total exposed flux $\Ph$, total target number, T,
and bin width $\De T_l$ and $\De \cos\th$,
the double differential cross section is obtained  
\beq
\left(\frac{d^2\si}{dT_l \cos\th}\right)_i=\frac{\sum_jU_{ij}(d_j-b_j)}{\Ph\cdot T\cdot\ep_i\cdot(\De T_l,\De \cos\th)_i}~.
\label{eq:ddexp}
\eeq
As one can see, this is equivalent to Eq.~(\ref{cross}),
therefore, {\it flux-integrated differential cross section is the point where theorists
and experimentalists meet for neutrino interaction physics.}
Note, details of Eq.~(\ref{eq:ddexp}) may depend on experiments since there are a number of ways
to remove backgrounds, unsmear distribution, and correct the efficiency.
Furthermore, one could compare experimental data with a theory without unfolding, instead,
``fold'' detector efficiency and smear and add backgrounds on a theoretical model (forward folding)
which may be a favored method from a statistical point of view~\cite{Cousins:2016ksu}.
However, currently the community standard for experimentalists is to
unfolded detector bias to present data.

\subsection{Matching Theory and Experiment\label{sec:kinematics}}

Experimentalists strive to measure the flux-integrated differential cross section, 
and theorists calculate that by convoluting their models with flux from each experiment.
Because of the flux-dependence of the measured cross sections unlike flux-unfolded total cross sections,
every single experiment with different locations or different neutrino beams 
would measure different differential cross sections.
Therefore, comparisons of data sets from different neutrino beams are non-trivial,
and they can only be related through the theoretical interaction models.
We will come back to this in the next sections, in particular
in Sec.~\ref{sec:ccpip} in connection with the one pion production cross sections.
Here we give an example of analysis which can be performed
by using a Monte Carlo events generator, the tool bridging theory and experiment. 

\begin{figure}[t!]
  \begin{center}
    \includegraphics[width=6.4cm]{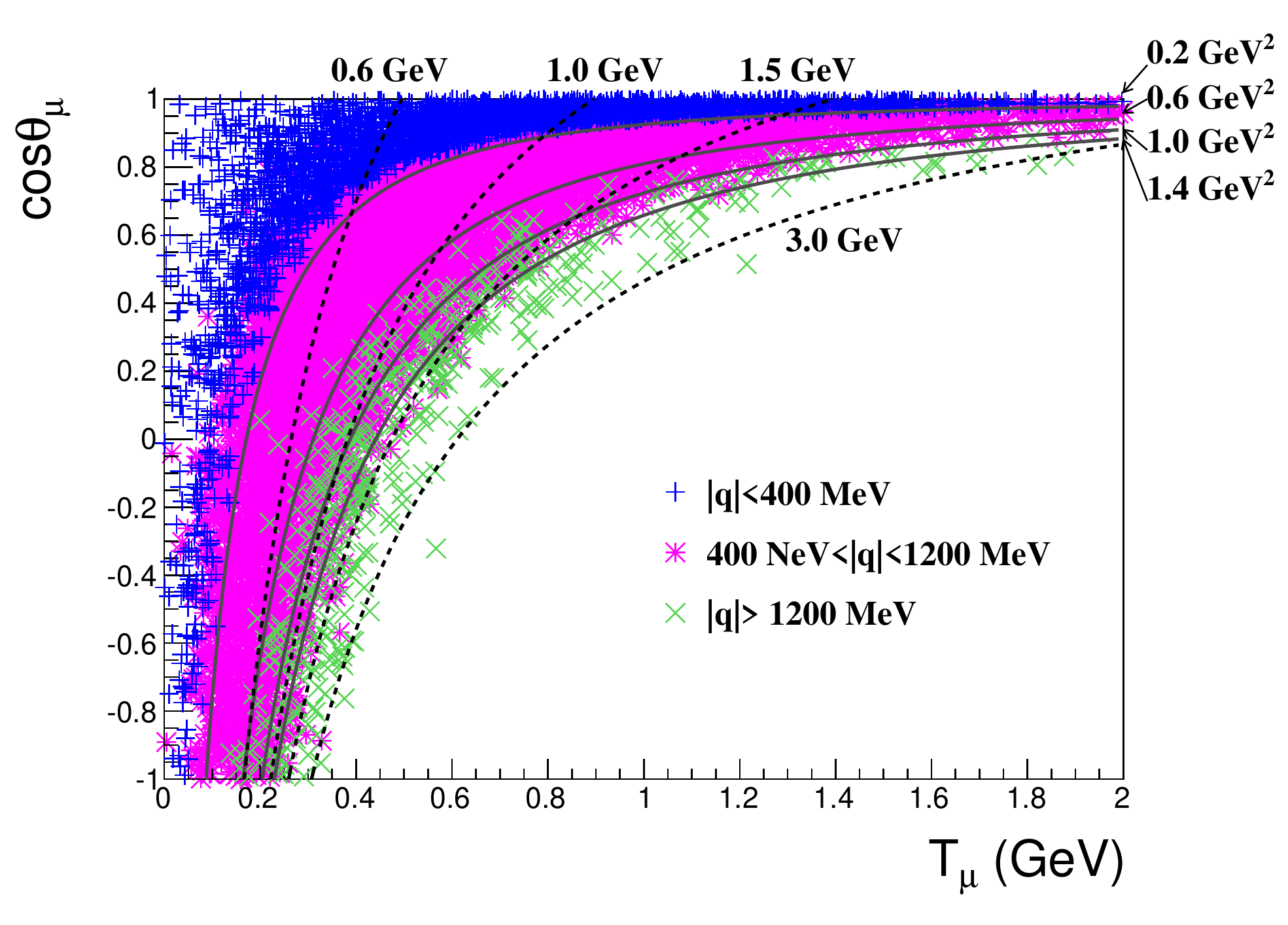}
    \includegraphics[width=6.4cm]{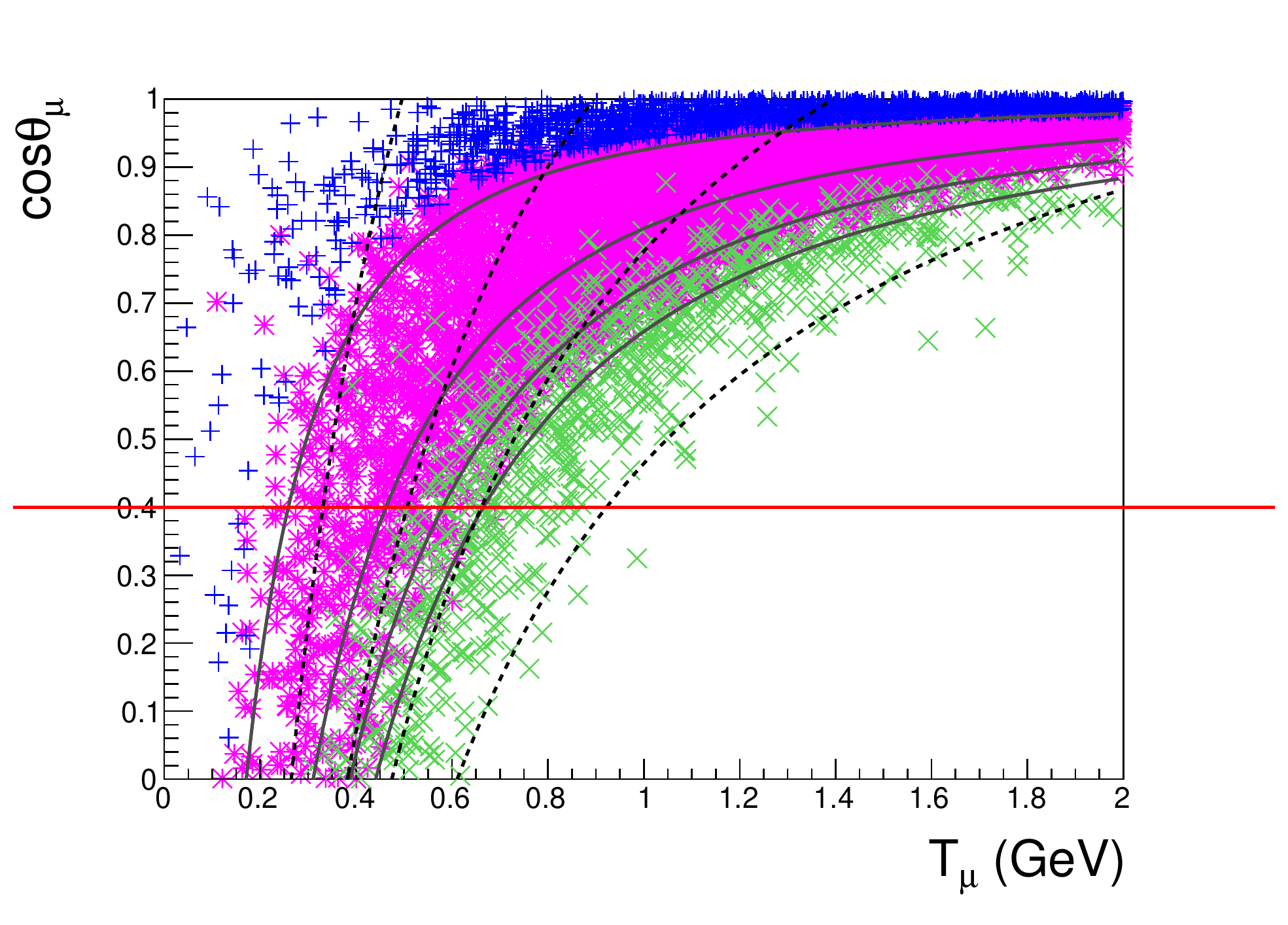}
    \includegraphics[width=6.4cm]{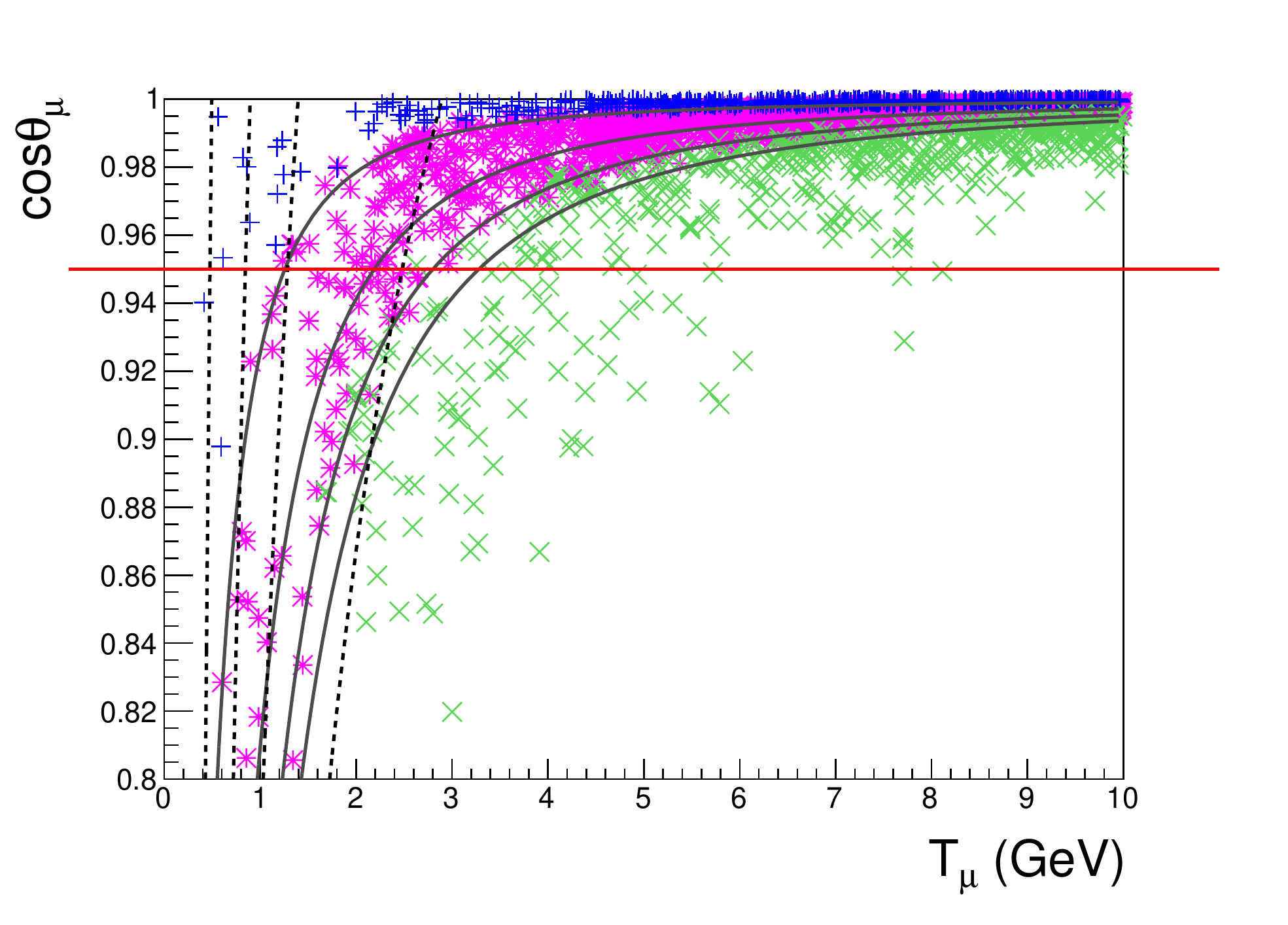}
    \includegraphics[width=6.4cm]{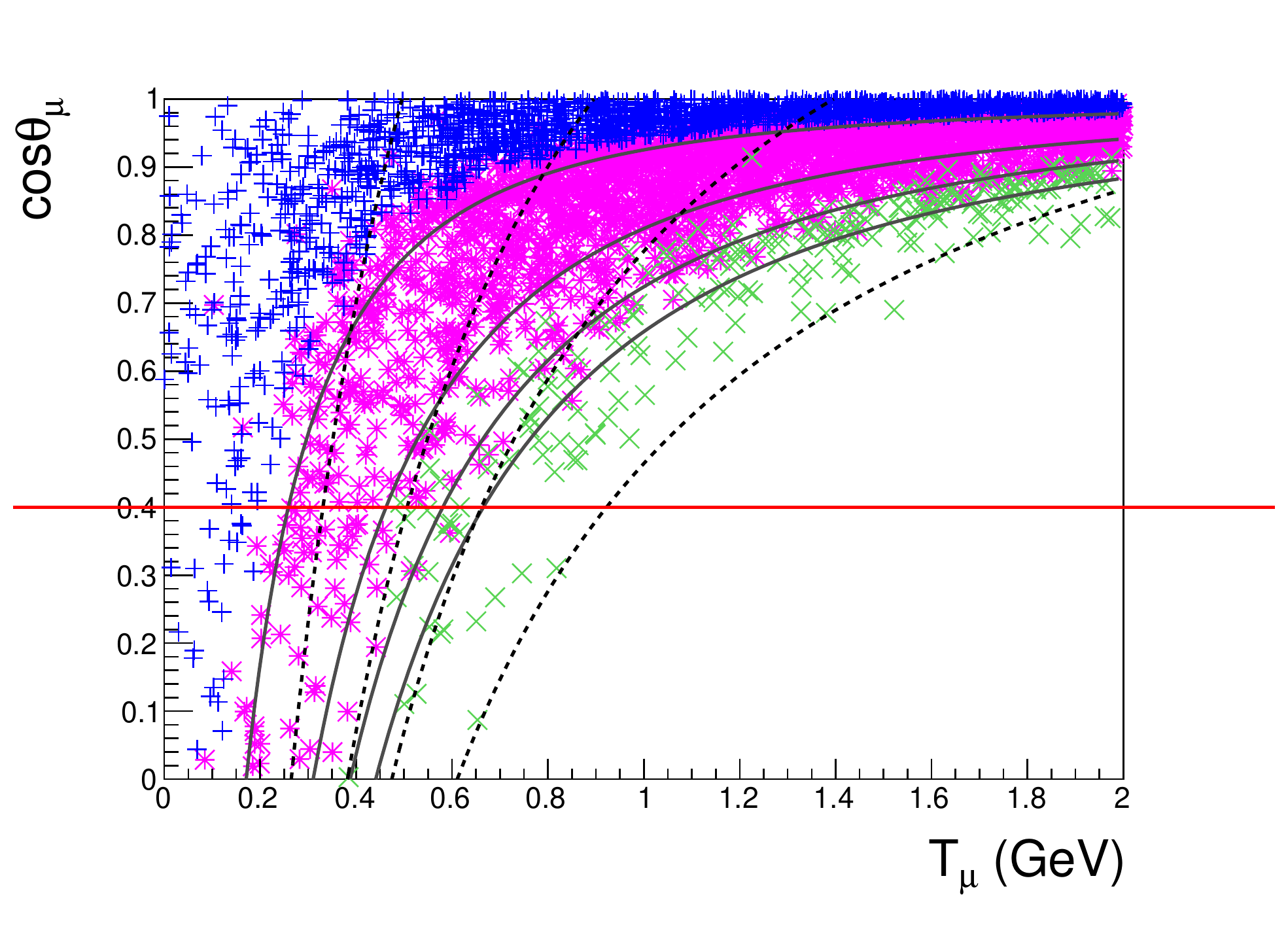}
    \end{center}
\vspace{-2mm}
\caption{
Flux-integrated double-differential cross section for CCQE interactions on carbon target
function of muon kinetic energy, $T_\mu$, and scattering angle, $\cos\th_\mu$.
Top left panel is for MiniBooNE neutrino mode, bottom left panel is for MINERvA neutrino mode,
and right two panels are for T2K neutrino and anti-neutrino mode.
The red horizontal line indicates the angular acceptance of the detector (see text).
Note the different scales on the two axes in the different panels.  
}
\label{fig:kin_all}
\end{figure}

We have seen in Sec.~\ref{sec:theo} that the  $\omega$ and $|{\bf{q}}|$ plane of Figs.~\ref{fig_hyperbolas} 
are useful to appreciate where the nuclear responses lie and which region is
explored for fixed values of $T_\mu$ and $\theta$. 
Since the flux-integrated double differential cross sections are function of $T_\mu$ and $\theta$, 
to analyze what happens in the $\cos\theta$ and $T_\mu$ plane
is also very illuminating and allows to bridge theoretical properties of neutrino interactions
and nuclear models with the experimental situation. To illustrate this point we consider a simple situation:
genuine CCQE events generated by GENIE neutrino interaction generator (version 2.8.0)~\cite{GENIE}.
This generator uses the relativistic Fermi gas (RFG) model, the simplest model for
the nuclear structure and the only one considered in Monte Carlos for many years. 
Figure~\ref{fig:kin_all} shows CCQE flux-integrated double differential cross sections on carbon target for MiniBooNE (top left),
T2K ND280 near detector complex (right), and MINERvA (bottom left).
For T2K, both muon neutrino in neutrino mode (top right) and muon antineutrino in antineutrino mode (bottom right) are calculated.
For MiniBooNE and MINERvA, only muon neutrino CCQE results are shown. 
Each marker represents an event.
The events can be thought of as the intersection between the nuclear response 
region and the hyperbolas of Figs.~\ref{fig_hyperbolas},
weighted by the different neutrino fluxes shown in Sec.~\ref{sec:flux}.
The differences in these fluxes (in particular the one of MINERvA with respect to the MiniBooNE and T2K cases) are reflected
in the different behavior in the $\cos\theta$ and $T_\mu$ plane. 
Events are classified by three colored marker types, depending on their three-momentum transfer.

The blue (dark gray) ``${\color{blue}+}$'' markers are events with $|\nq|<400~\uMeV$.
This is smaller than roughly the twice of Fermi motion of typical nuclei,
and it is the kinematic region where impulse approximation (interaction with one single nucleon in the nucleus) 
starts to be violated, as illustrated by comparisons with inclusive electron scattering 
data~\cite{Ankowski:2010yh,Leitner:2008ue}. Therefore if many events are classified in here,
one should consider models beyond the impulse approximation (IA), such as the RPA, or 
IA-based models with necessary corrections.
As one can see in Fig.~\ref{fig:kin_all}, all experiments considered here
include sizable amount of low momentum transfer events (roughly 20\% of all CCQE interactions).

The magenta (gray) ``${\color{magenta}\ast}$'' markers have $400~\uMeV<|\nq|<1200~\uMeV$.
In this context, this is the ``safe region'' where most models work fine for the genuine CCQE. 
We remind however that multinucleon interaction contributes also in this region,
so the correct description of the experimental data requires care on models.

Finally, the green (light gray) ``${\color{green}\times}$'' are for $|\nq|>1200~\uMeV$.
This is the other delicate region. The validity of the nucleon-nucleon effective interaction employed in the RPA-based 
calculations for genuine quasielastic, such as the ones of Refs.~\cite{Nieves:2004wx,Martini:2009uj}, 
at high momentum transfer is delicate~\cite{Chanfray:1985zz}. This is true also for other channels, 
for example $|\nq|=1200$ MeV is the limit up to which 2p-2h contributions are included 
in the model of Nieves \textit{et al.} ~\cite{Nieves:2011pp} even
when they investigate neutrino interactions up to 10 GeV~\cite{Gran:2013kda}.

Several lines are overlaid on Fig.~\ref{fig:kin_all} to clarify some kinematics discussions.  
Four dashed lines are for constant $\overline{E_\nu}$ (0.6, 1.0, 1.5, 3~$\uGeV$), according to Eq.~(\ref{enubar_muon}), 
and four solid lines are for constant $\overline{Q}^2$ (0.2, 0.6, 1.0, 1.4~$\uGeVt$),
according to Eq.~(\ref{eq_Q_rec}). 
The angular acceptance in the different experiments is represented by a red horizontal line. 
Below, we summarize what we learn from each plot.

\vspace{0.5cm}

{\it MiniBooNE} ---
There are many events with $|\nq|<400~\uMeV$ ($\sim$27\% of all CCQE events) and they are all below $\overline{Q}^2<0.2~\uGeVt$.
Because the detector has a 4$\pi$ coverage, angular acceptance is $\cos\th_\mu=+1$ (forward scattering)
to $\cos\th_\mu=-1$ (back scattering). This large acceptance helps to understand underlying interaction physics.
The K2K experiment~\cite{Gran:2006jn} showed a deficit of CCQE candidate events at very forward scattering region.
It wasn't understood until MiniBooNE analyzed full 4$\pi$ kinematic space to show the deficit is
not only forward scattering, that could be an inefficiency of the detector, 
but low $Q^2$ region, which must be some physics~\cite{MB_CCQEPRL}. 
As it will be illustrated in Sec.~\ref{sec:QE}, 
RPA collective effects give a reduction of the CCQE cross section at low $Q^2$. 

\vspace{0.5cm}

{\it T2K (neutrino mode)} ---
Narrower J-PARC off-axis beam (see Fig.~\ref{fig:flux_all})
makes a very small number of interactions below $\overline{E_\nu}<0.6~\uGeV$ unlike MiniBooNE. 
The tracker nature of T2K near detectors has small angular acceptance,
and here we defined the current acceptance as $\cos\th_\mu>0.4$ and draw a red line above
where $\sim$95\% of CCQE candidates are observed by the T2K ND280 CCQE analysis of 2014~\cite{Abe:2014iza}. 
In this limited kinematic space, there is a similar fraction of $|\nq|<400~\uMeV$ events as in MiniBooNE
($\sim$21\% of all CCQE interactions). This means that T2K and MiniBooNE
should follow similar interaction physics in terms of RFG model. 
Although current analysis can accept majority of total CCQE events,
larger angular events may be also very interesting since they are related
to transverse response (the response most affected by the two-body current contributions). 
We note the potential acceptance of T2K ND280 near detector is larger,
because the detector itself has an ability to measure leptons with higher scattering angle thanks to
the ECal and the SMRD surrounding the fiducial volume. 
In principle, $|\cos\th_\mu|\ge 0.15$ can be measured and in this case
the analysis can accept more than 95\% of CCQE events.

We note the problem of the detector for the total cross section measurements. 
If the acceptance is small, one needs to ``guess'' unmeasured number of events to estimate the total cross section,  
and this often requires the model dependent correction. 
To overcome this model dependency,
neutrino interaction model systematics error must be included and the total error increases,
as shown in T2K on-axis CCQE cross section measurements~\cite{Abe:2015oar}.
On top of that, flux-unfolded total cross section requires the reconstruction of
neutrino energy (Sec.~\ref{sec:enrec}) which adds additional errors.
This is a common problem for any cross section measurements.
One solution is to present ``fiducial cross sections'',
that are defined in restricted phase space or limited acceptance.
By applying those restrictions in theoretical models,
a theory and data are comparable without adding bias in the data. 

\vspace{0.5cm}

{\it T2K (antineutrino mode)} ---
The distribution of events is similar with its neutrino mode,
except that more events are concentrated (as expected and discussed also in Sec.~\ref{subsec:QEintro}
in connection with the MiniBooNE results) in the forward angle region, 
and lower momentum transfer events ($|\nq|<400~\uMeV$) are around 33\%, hence higher than in the neutrino mode. 
This implies physics beyond IA is more important for antineutrino mode than neutrino mode,
and models working for neutrino mode, if based on IA, may not work properly in antineutrino mode. 
Since the antineutrino mode measurement is an important part of CP violation measurement, 
one should keep in mind this kinematics difference.
A quantitative analysis of RPA and multinucleon effects for neutrino
and antineutrino can be found for example in Ref.~\cite{Martini:2010ex}  

\vspace{0.5cm}

{\it MINERvA} ---
Higher energy NuMI beam makes very different kinematics with the previous two experiments.
The angular acceptance of MINERvA experiment is considerably smaller if the MINOS matching is required.  
However, in this energy region momentum transfer can be well separated by small angles,
and MINERvA also covers a similar kinematic space with MiniBooNE and T2K,
for example, the fraction of events with $|\nq|<400~\uMeV$ is $\sim$21\%.
Not surprisingly, there are also many events with $|\nq|>1200~\uMeV$ ($\sim$18\%),
because of the higher energy beam.
Majority of them are higher energy, forward going events.
A similar analysis is performed in Ref.~\cite{Megias:2014kia}
where the Super-Scaling approach is used to evaluate the different 
$|\nq|$ and $\omega$ contributions to the MINERvA $Q^2$ distribution.

\section{Quasielastic \label{sec:QE}}
\subsection{CCQE, CCQE-like, and CC0$\pi$ \label{subsec:QEintro}}

\begin{figure}
\begin{minipage}[c]{70mm}
\begin{center}
      \includegraphics[width=70mm]{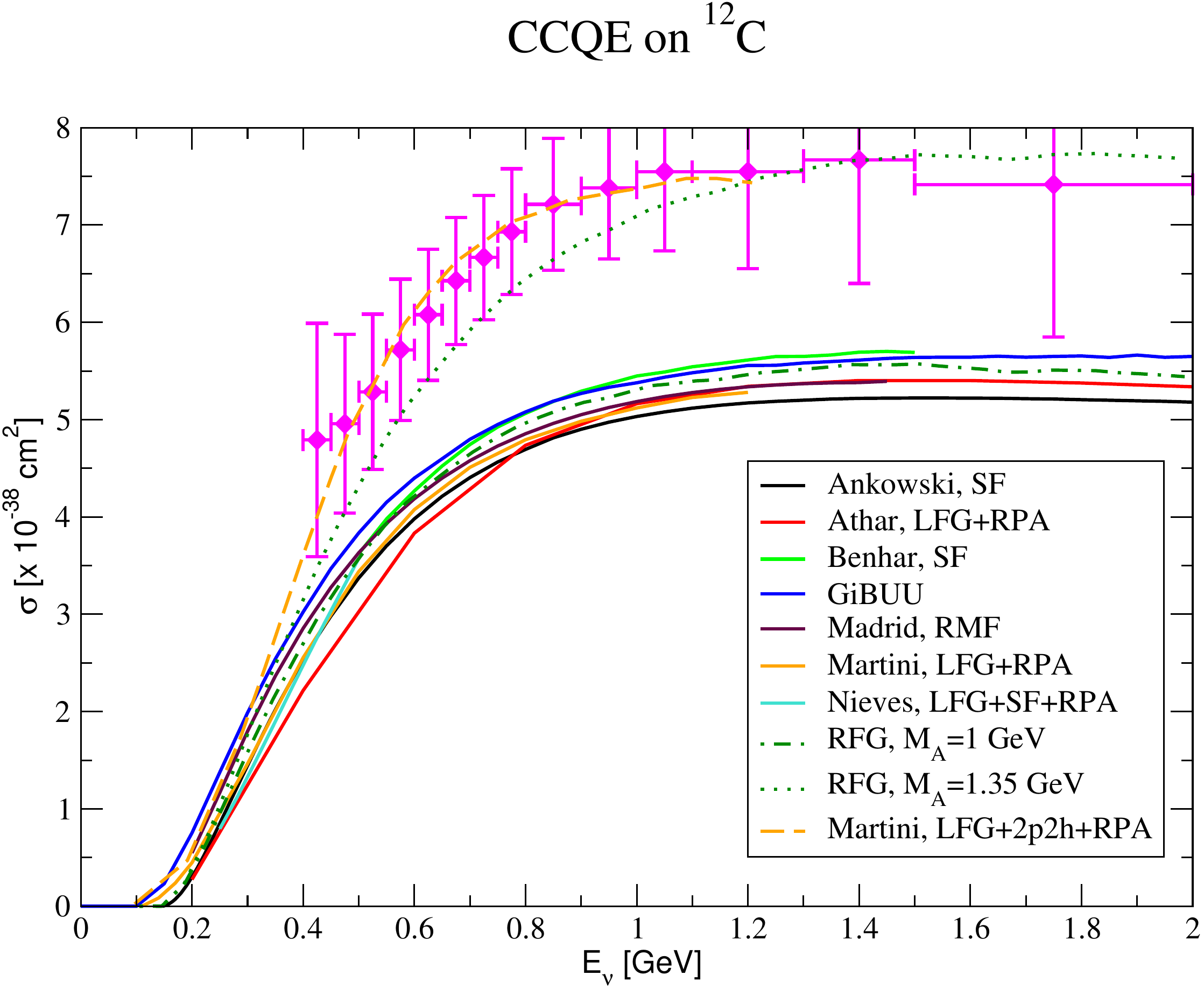}
\caption{\label{fig_luis}Charged current quasielastic  $\nu_\mu$-$^{12}$C cross section  as a function of neutrino energy calculated
within several models collected and published in Ref.~\cite{Boyd:2009zz}.
The MiniBooNE data are taken from Ref. ~\cite{AguilarArevalo:2010zc} The figure is published in Ref.~\cite{AlvarezRuso:2010ia}.}
\end{center}
\end{minipage}
\hspace{1mm}
\begin{minipage}[c]{70mm}
\begin{center}
   \includegraphics[width=70mm]{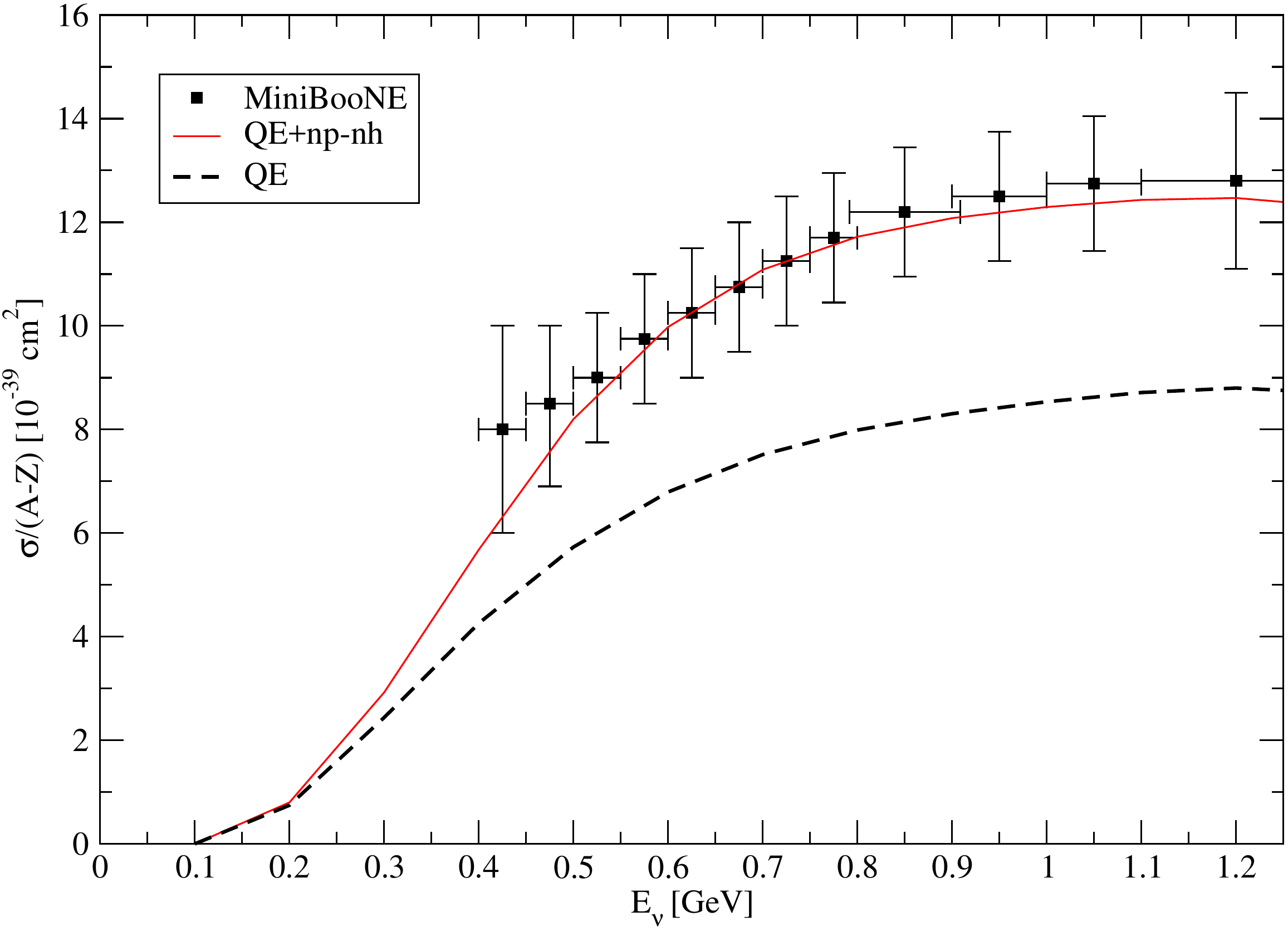} 
\caption{\label{fig_sig_tot_miniboone}``Quasielastic-like'' $\nu_\mu$-$^{12}$C cross sections measured by MiniBooNE~\cite{AguilarArevalo:2010zc}
compared to Martini \textit{et al.} calculations.
The figure is taken from Ref.~\cite{Martini:2009uj}.}
\end{center}
\end{minipage}
\end{figure}

In the discussion of the CCQE cross section, the MiniBooNE measurement, 
obtained using a high-statistics sample of $\nu_\mu$ CCQE events on $^{12}$C, plays a central role.
The results were presented for the first time at the NuInt09 conference~\cite{Katori:2009du}
and then published in Ref.~\cite{AguilarArevalo:2010zc}.
In this work  the quasielastic cross section is defined as the one for processes 
in which only a muon is detected in the final state, but no charged pions.
However it is possible that in the neutrino interaction,
a pion produced via the excitation of the $\Delta$ resonance escapes detection,
for instance because it is reabsorbed in the nucleus. 
In this case it imitates a quasielastic process. 
The MiniBooNE analysis of the data corrected for this possibility via
a data driven correction based on the simultaneous measurement of CC1$\pip$ sample.
The net effect amounted to a reduction of the observed quasielastic cross section. 
After applying this correction,
the quasielastic cross section thus defined still displayed an anomaly. 
The comparison of these results with a prediction based on
the relativistic Fermi gas model using in the axial form factor (cf. Eq.~(\ref{eq:axialff}))
the standard value of the axial cut-off mass $M_A=1.03$ GeV$/c^2$,
consistent with the one extracted from bubble chamber experiments, reveals a substantial discrepancy. 
The introduction of more realistic theoretical nuclear models,  
assuming the validity of the hypothesis that the neutrino interacts with a single nucleon in the nucleus,
does not alter this conclusion.
This is illustrated in Fig.~\ref{fig_luis}.
This figure, published in Ref.~\cite{AlvarezRuso:2010ia} shows
the CCQE $\nu_\mu$-$^{12}$C cross section as a function of neutrino energy calculated
within several models already applied in electron scattering studies where they provided satisfactory agreement with data.
These models are the spectral function approaches of Refs.~\cite{Benhar:2006nr,Ankowski:2007uy}, 
the local Fermi gas plus RPA approaches of Refs.~\cite{Nieves:2004wx,Martini:2009uj,SajjadAthar:2009rd},
the relativistic mean field of Ref.~\cite{Maieron:2003df} and GiBUU~\cite{Leitner:2006ww,Leitner:2008ue}.
All these theoretical results were collected for the NuInt09 conference and published in Ref.~\cite{Boyd:2009zz} where a synthetic description of these different theoretical models is also given.
From Fig.~\ref{fig_luis} it clearly appears that the mentioned theoretical predictions,
all using the standard value for the axial mass, underestimate the MiniBooNE data.
In Fig.~\ref{fig_luis} is also shown that in the relativistic Fermi gas model an increase of
the axial mass from $M_A=1$ GeV$/c^2$ to the larger value of $M_A=1.35 (\pm 0.17)$ GeV$/c^2$ can reproduce the MiniBooNE data,
as discussed in Refs.~\cite{Katori:2009du,AguilarArevalo:2010zc}.

\begin{figure}
\begin{center}
  \includegraphics[width=150mm]{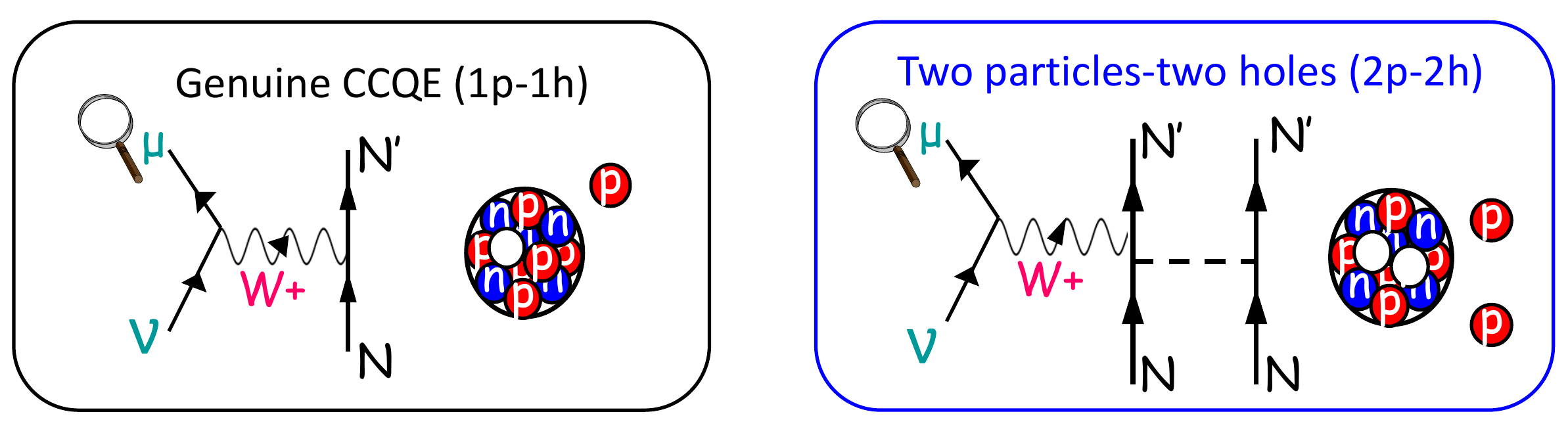}  
  \caption{Schematic and pictorial representation of the 1p-1h and 2p-2h excitations.}
\label{fig_pic_1p1h_2p2h}
\end{center}
\end{figure}

A possible solution of this apparent puzzle was suggested by Martini \textit{et al.}~\cite{Martini:2009uj}
which drew the attention to the existence of additional mechanisms
beyond the interaction of the neutrino with a single nucleon in the nucleus,
which are susceptible to produce an increase of the quasielastic cross section.
The absorption of the $W$ boson by a single nucleon,
which is knocked out,
leading to 1 particle - 1 hole (1p-1h) excitations, is only one possibility.
In addition  one must consider coupling to nucleons belonging to correlated pairs (NN correlations)
and two-nucleon currents arising from meson exchange (MEC). 
This leads to the excitation of two particle -two hole (2p-2h) states.
3p-3h excitations are also possible. 
Together they are called np-nh (or multinucleon) excitations.
A schematic and pictorial representation of the 1p-1h and 2p-2h excitations is shown in Fig.~\ref{fig_pic_1p1h_2p2h}.
As shown in Ref.~\cite{Martini:2009uj} and in Fig.~\ref{fig_sig_tot_miniboone} the addition of the np-nh excitations 
to the genuine quasielastic (1p-1h) contribution leads to an agreement with the MiniBooNE data without any increase of the axial mass. 
Isolating a genuine quasielastic event in electron scattering experiments where
the kinematics are fixed by the knowledge of the energy and momentum of incoming and outgoing electron beams is relatively easy.
In the double differential cross sections, or in the nuclear responses,
one can isolate the bump centered at $Q^2/(2M_N)$ corresponding to single nucleon knockout,
shown for example in Fig.~\ref{fig_risp_q600_diversicanali}.
This is not the case in neutrino scattering experiments. 
Due to the broadening of the incoming neutrino flux as illustrated in Sec. \ref{sec:theo}, 
one explores the whole energy- and momentum-transfer plane, 
hence the multinucleon excitations are strictly entangled with the single nucleon knockout events.
This is particularly true for Cherenkov detectors. 
The importance of 2p-2h excitations in neutrino scattering processes was suggested for
the first time by Delorme and Ericson in Ref.~\cite{Delorme:1985ps}. 
The confusion between one nucleon knock out processes and multinucleon excitations
in the Cherenkov detectors was stressed as first by Marteau \textit{et al.}~\cite{Marteau:1999kt,Marteau:2002sh}
in connection with the atmospheric neutrino measurements at Super-Kamiokande. Today one generally refers to single nucleon knockout processes as true or genuine quasielastic. Processes in which only a final charged lepton is detected, hence including multinucleon excitations,
but pion absorption contribution is subtracted, 
are usually called {\bf quasielastic-like}, or {\bf QE-like}.
Thus, what MiniBooNE published was not CCQE data, but CCQE-like data.
To avoid the confusion of the signal definition,
it is increasingly more popular to present the data in terms of the final state particle,
such as ``1 muon and 0 pion, with any number of protons''.  
This corresponds to the CCQE-like data without subtracting any intrinsic backgrounds
(except beam and detector related effects) and it is called {\bf CC0$\pi$}. 
We will discuss the advantage of such topology-based signal definition in Sec.~\ref{sec:ccpip}.

\begin{figure}
\begin{center}
  \includegraphics[width=12cm]{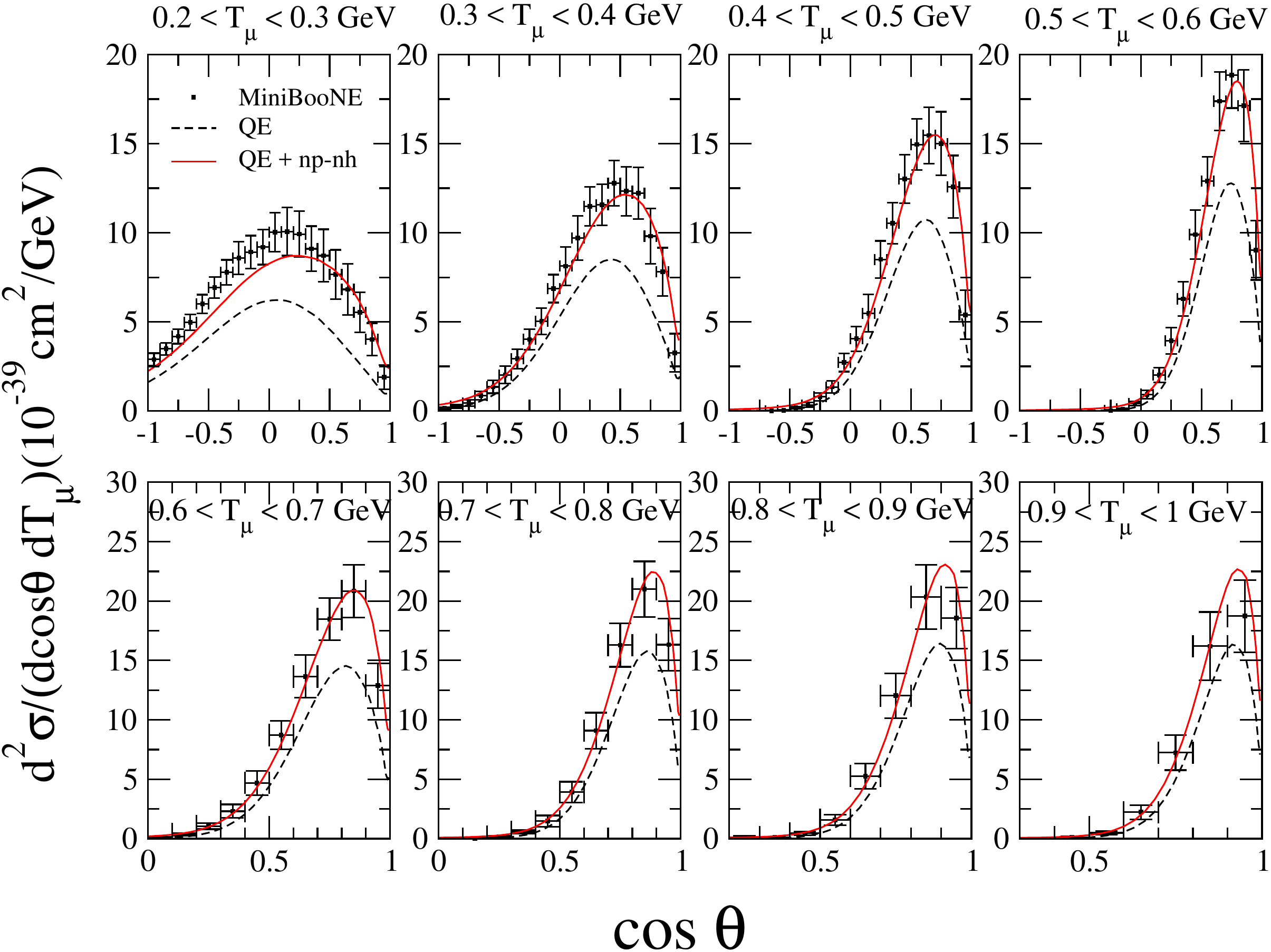}  
\caption{(color online). MiniBooNE flux-integrated CCQE-like $\nu_\mu$-$^{12}$C
double differential cross section per neutron for several values of muon kinetic energy
as a function of the scattering angle. Dashed curve: pure quasielastic (1p-1h) cross section calculated in RPA; 
solid curve: with the inclusion of np-nh component. 
The experimental MiniBooNE points 
are taken from~\cite{AguilarArevalo:2010zc}. The figure is taken from Ref.~\cite{Martini:2011wp}.}
\label{fig_minib_d2s}
\end{center}
\end{figure}

The results presented in Figs.\ref{fig_luis} and \ref{fig_sig_tot_miniboone} 
relate to cross sections as a function of the neutrino energy.
Nevertheless the experimental points shown in 
these figures are affected by the energy reconstruction problem, see Section~\ref{sec:enrec}. 
For a comparison between theory and experiment,
the most significant quantities are the neutrino flux-integrated double differential cross sections,
as defined in Eq.~(\ref{cross}) (theory) and Eq.~(\ref{eq:ddexp}) (experiment),
which are functions of two measured variables:
the muon energy and the scattering angle.
The comparison between the experimental MiniBooNE results~\cite{AguilarArevalo:2010zc}
and the theoretical calculations of Martini \textit{et al.},
as published in Ref.~\cite{Martini:2011wp} is given in Fig.~\ref {fig_minib_d2s}. 
A very good agreement with data is obtained once the multinucleon component is included. 
Similar conclusions have been obtained by Nieves \textit{et al.} in Ref.~\cite{Nieves:2011yp}.

\begin{figure}
\begin{center}
  \includegraphics[width=10cm]{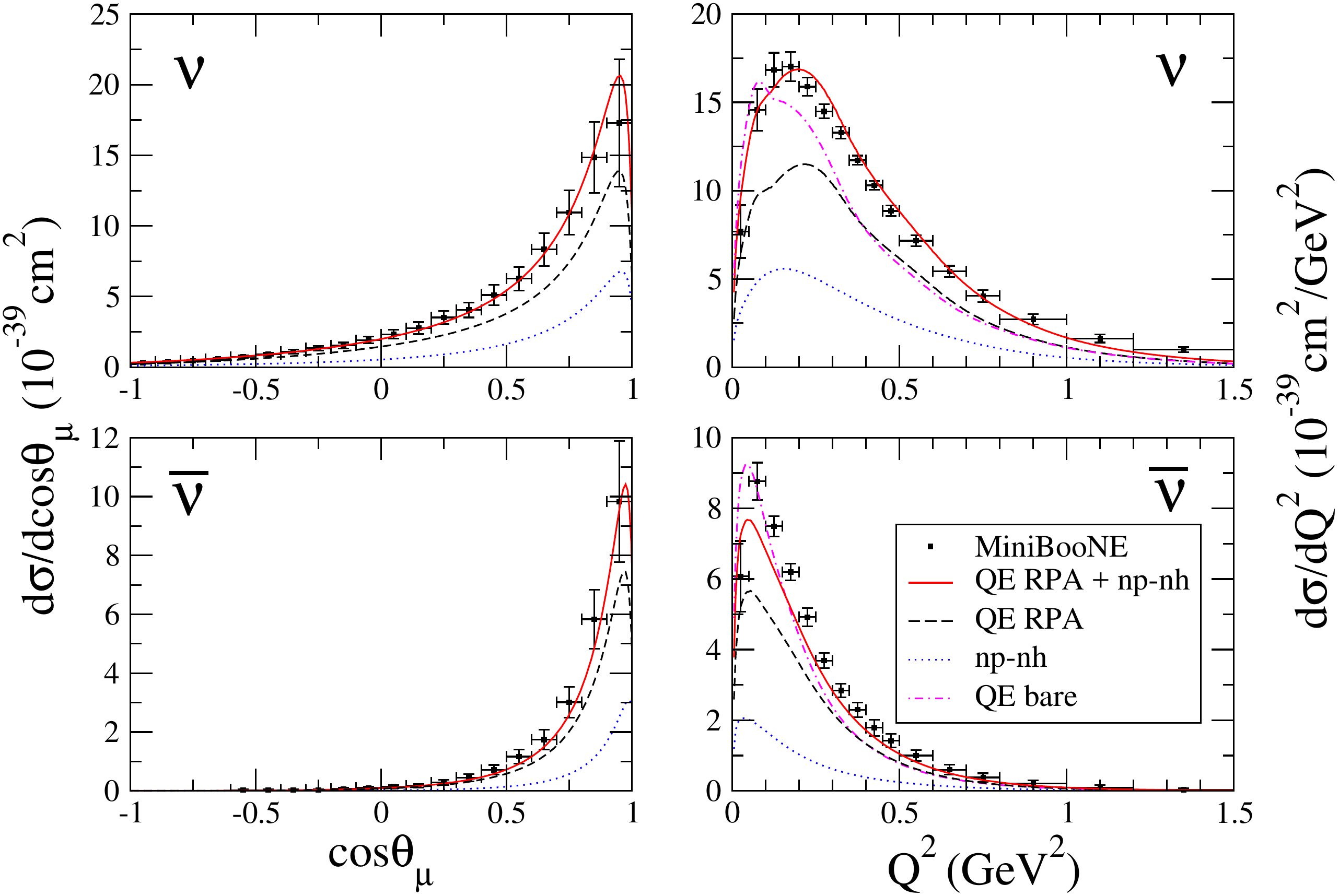}  
\caption{(color online). MiniBooNE flux-integrated differential cross sections
$d\sigma/d\cos\theta$ (left panels) and $Q^2$ distributions (right panels) 
for neutrino (upper panels) and antineutrino (lower panels) CCQE-like scattering on carbon. 
The experimental MiniBooNE points 
are taken from~\cite{AguilarArevalo:2010zc} and~\cite{AguilarArevalo:2013hm}. 
The theoretical results are the ones of Refs.~\cite{Martini:2011wp} and~\cite{Martini:2013sha}.}
\label{fig_minib_dsdcos_dQ2}
\end{center}
\end{figure}

In 2013 the MiniBooNE collaboration published the measurements of
the antineutrino CCQE-like cross section on carbon~\cite{AguilarArevalo:2013hm}. 
Similar agreements between theory and experiments for the flux-integrated double differential cross sections
and similar conclusions on the crucial role of np-nh excitations have been obtained by Nieves \textit{et al.}~\cite{Nieves:2013fr}
and by Martini and Ericson~\cite{Martini:2013sha}. 
A full calculation of the MiniBooNE flux-integrated neutrino and
antineutrino double differential cross section was also given by Amaro \textit{et al.} 
in Ref.~\cite{Amaro:2010sd,Amaro:2011aa} in the super-scaling analysis (SuSA) approach. 
In this context for the neutrino scattering the inclusion of
the vector MEC gives a relatively small contribution which reduces the discrepancy between
the MiniBooNE results and the theoretical predictions,
but it is not enough to reproduce data. 
The situation is different for antineutrino cross section where the vector MEC contribution turns to be large. 
These results have been updated by Megias~\textit{et al.} in Ref.~\cite{Megias:2014qva} and further updated in Ref.~\cite{Megias:2016fjk} by considering the SuSAv2 approach and by including the axial MEC contributions. 
We postpone the discussion on the comparison among models and on the
relative role of np-nh excitation in neutrino and antineutrino scattering to the next subsection. 
For the moment we show in Fig.~\ref{fig_minib_dsdcos_dQ2} the neutrino
and antineutrino differential cross section $d\sigma/d\cos\theta$ and $d\sigma/dQ^2$, 
as measured in Refs.~\cite{AguilarArevalo:2010zc} and~\cite{AguilarArevalo:2013hm}
and as calculated in Refs.~\cite{Martini:2011wp} and ~\cite{Martini:2013sha}. 
From the $d\sigma/d\cos\theta$ panels,
one can observe that the antineutrino cross sections falls more rapidly with angle than the neutrino one. 
This also reflects in the $Q^2$ distributions. 
It is a consequence of the difference of sign in front of the vector-axial interference term of the cross section, 
cf. Eq.~(\ref{eq1:cross_section}). 
This different $\cos \theta$ and $Q^2$ behavior between neutrino and antineutrino cross section
is due first to the kinematic factor in front of the nuclear responses.
It would survive even if one considered,
as in the Fermi gas model, that all the nuclear responses would be the same. 
In reality, due to the nuclear interaction, the nuclear responses are different in the different spin isospin channels. 
As a consequence, the vector-axial interference term introduce an additional asymmetry between neutrino and antineutrino since
the various nuclear responses weigh differently in the neutrino and antineutrino cross sections.
This point was analyzed in details in Refs.~\cite{Martini:2010ex,Amaro:2011aa}. 
The asymmetry of the nuclear effects for neutrino and antineutrino is important for CP violation studies.
The nuclear cross-section difference for neutrinos and antineutrinos stands as a potential obstacle in the interpretation of experiments
aimed at the measurement of the CP violation angle, hence has to be fully mastered.
It will be further discussed in subsection~\ref{sec:npnh}.
In subsections~\ref{sec:npnh} and \ref{sec:hadron},
other experiments following the MiniBooNE one will be also discussed.

\subsection{np-nh excitations: theory \textit{vs} experimental $d\sigma$ \label{sec:npnh}}

After the suggestion~\cite{Martini:2009uj} of the inclusion of np-nh excitations mechanism as the likely explanation of the MiniBooNE anomaly, 
the interest of the neutrino scattering and oscillation communities on the multinucleon emission channel rapidly increased. 
Indeed this channel was not included in the generators used for the analyses of the neutrino cross sections and oscillations experiments. 
It can be inferred also from the large values of the axial mass deduced from other neutrino scattering experiments on nuclei: 
K2K on oxygen  $M_A=1.20 \pm 0.12$ GeV$/c^2$~\cite{Gran:2006jn}; K2K on carbon  $M_A=1.14 \pm 0.11$ GeV$/c^2$~\cite{Espinal:2007zz};
MINOS on iron $M_A=1.23^{+0.13+0.12}_{-0.09-0.15}$ GeV$/c^2$~\cite{Adamson:2014pgc}. 
The only exception is NOMAD,
a higher-energy experiment on carbon who obtains a value of $M_A=1.05 \pm 0.02 \pm 0.06$ GeV$/c^2$~\cite{Lyubushkin:2008pe}.
The high values of the axial mass indicate the presence of a np-nh component not considered in the analysis.
Today there is an effort to include this np-nh channel in several Monte Carlo~\cite{Sobczyk:2012ms,Katori:2013eoa,Schwehr:2016pvn,Wilkinson:2016wmz}. 

Concerning the theoretical situation, nowadays several calculations agree on the crucial role of the multinucleon emission in order to explain the 
MiniBooNE neutrino~\cite{AguilarArevalo:2010zc} and antineutrino~\cite{AguilarArevalo:2013hm} data as well 
as the SciBooNE~\cite{Nakajima:2010fp} and T2K inclusive ~\cite{Abe:2013jth,Abe:2014agb} and CC$0\pi$ ~\cite{Abe:2016tmq} cross sections. 
Nevertheless there are some differences on the results obtained for this np-nh channel by the different theoretical approaches. 
The aim of this section is to review the actual theoretical status on this subject.

The theoretical calculations of np-nh excitations contributions to neutrino-nucleus cross sections are actually performed 
essentially by three groups. There are the works of
Martini~\textit{et al.}~\cite{Martini:2009uj,Martini:2010ex,Martini:2011wp,Martini:2012fa,Martini:2012uc,Martini:2013sha,Martini:2014dqa,Ericson:2015cva,Martini:2016eec}, 
the ones of Nieves~\textit{et al.}~\cite{Nieves:2011pp,Nieves:2011yp,Nieves:2012yz,Nieves:2013fr,Gran:2013kda}
and the ones of Amaro~\textit{et al.}~\cite{Amaro:2010sd,Amaro:2011qb,Amaro:2011aa,Simo:2014wka,Simo:2014esa,Megias:2014qva,Ivanov:2015aya,Simo:2016ikv,RuizSimo:2016ikw,Megias:2016fjk}. 

The np-nh channel is taken into account through more phenomenological approaches by Lalakulich, Mosel~\textit{et al.} 
\cite{Lalakulich:2012ac,Lalakulich:2012hs,Mosel:2014lja,Gallmeister:2016dnq} 
in GiBUU and by Bodek~\textit{et al.}~\cite{Bodek:2011ps} in the so called Transverse Enhancement Model (TEM). 
In the case of GiBUU in Refs.~\cite{Lalakulich:2012ac,Lalakulich:2012hs,Mosel:2014lja}
the size of the squared matrix element of the neutrino-induced two-nucleon knock-out cross section 
is obtained by fitting the neutrino charged current quasielastic MiniBooNE cross section on carbon. 
This is a pure two-nucleon phase-space model. Recently a more realistic np-nh contribution, 
where an empirical response function deduced from electron scattering data is used as a basis, 
has been implemented in GiBUU \cite{Gallmeister:2016dnq}. 
It allows a simultaneous description of neutrino and antineutrino MiniBooNE CCQE-like data, as well as $\nu_\mu$ and $\nu_e$ T2K CC inclusive cross sections. 
In the TEM model~\cite{Bodek:2011ps} the magnetic form factor is enhanced with respect to the standard dipole parameterization 
according to the formula $G_M^{TEM}(Q^2)=G_M^{dipole}\sqrt{1+AQ^2e^{-Q^2/B}}$. 
The parameters $A$ and $B$ are fitted to reproduce inclusive electron scattering data on carbon. 
This is is the effective way to include the meson exchange currents contribution in electron and neutrino scattering. 
In the same spirit of the modification of the axial mass, instead to modify
the nuclear responses contribution, which requires elaborated many-body calculations,
a simple modification of the magnetic form factor, easy to implement in the Monte Carlo is proposed. 
Two-body current contributions to the axial part of the cross sections are not taken into account in this TEM model. 
Recent~\textit{ab initio} many-body calculations of 
neutral-weak responses and sum rules performed by Lovato~\textit{et al.}~\cite{Lovato:2014eva,Lovato:2015qka} 
have separated these axial contributions, showing their relevance. 
Also very recently,
fully relativistic calculations of MEC contributions to the weak nuclear responses 
and neutrino cross sections, performed by Simo~\textit{et al.} ~\cite{Simo:2016ikv} and Megias~\textit{et al.} ~\cite{Megias:2016fjk} respectively, show the role of the axial contribution. 
This important point will be analyzed in more detail later. 

Beyond all the theoretical approaches and models mentioned above, 
other interesting calculations discussing the 2p-2h excitations in connection with the neutrino scattering 
appeared in 2015 and 2016 ~\cite{Benhar:2015ula,Rocco:2015cil,VanCuyck:2016fab}. 
Since, for the moment no comparison with neutrino flux-integrated differential cross sections are shown,  
in the following we will focus essentially on the results obtained by the three theoretical approaches 
which calculate these quantities: the ones of Amaro~\textit{et al.}, Martini~\textit{et al.} and Nieves~\textit{et al.}

Considering these three different models, it is important to remind that there exist some differences already at level of genuine quasielastic,  
which can be particularly important when one compares the double differential cross sections.
Amaro~\textit{et al.} considered the relativistic super-scaling approach (SuSA)~\cite{Amaro:2004bs} based on
the super-scaling behavior exhibited by electron scattering data. 
It has been recently extended by Gonzalez-Jimenez~\textit{et al.}~\cite{Gonzalez-Jimenez:2014eqa}
in order to take into account the different behavior of the longitudinal and transverse nuclear responses due to relativistic mean field effects.
This new version of the model is called SuSAv2. The models of Martini~\textit{et al.} and Nieves~\textit{et al.} are more similar:
they start from a local Fermi gas picture of the nucleus. 
They consider medium polarization and collective effects through the random phase approximation (RPA) 
including $\Delta$-hole degrees of freedom, $\pi$ and $\rho$ meson exchange and $g'$ Landau-Migdal parameters in the effective $p-h$ interaction.

Turning to the np-nh sector, let's remind,
as first, the sources and the references of the calculations of the three groups. 
The np-nh contributions in the papers of Martini~\textit{et al.} are obtained starting 
from the microscopic calculations of the transverse response in electron scattering performed by Alberico~\textit{et al.}~\cite{Alberico:1983zg},
from the results of pion and photon absorption of Oset and Salcedo~\cite{Oset:1987re} and
from the results of pion absorption at threshold of Shimizu and Faessler~\cite{Shimizu:1980kb}.
The 2p-2h contributions considered by Amaro~\textit{et al.} 
are taken from the full relativistic model of De Pace~\textit{et al.}~\cite{DePace:2003xu} 
related to the electromagnetic transverse response. 
The extension of this full relativistic model to the weak sector by the addition of the axial MEC 
has been given very recently by Simo~\textit{et al.} ~\cite{Simo:2016ikv}. 
The results of Amaro~\textit{et al.} of Refs. ~\cite{Amaro:2010sd,Amaro:2011qb,Amaro:2011aa,Megias:2014qva,Ivanov:2015aya} 
do not include these last contributions, while the very recent results of Refs.~\cite{RuizSimo:2016ikw,Megias:2016fjk} do it. 
The approach of Nieves~\textit{et al.} can be considered as a generalization of the work of Gil~\textit{et al.}~\cite{Gil:1997bm},
developed for the electron scattering, to the neutrino scattering. 
The contributions related to the non-pionic $\Delta$ decay are taken,
as in the case of Martini ~\textit{et al.} from Oset and Salcedo~\cite{Oset:1987re}. 

We remind that there exist several contributions to two-body currents $J^{\mu}_{TB}$,
see for example Refs.~\cite{Alberico:1983zg,DePace:2003xu,Nieves:2011pp,Simo:2016ikv}. 
In the electromagnetic case, there are the so called pion-in-flight term $J^{\mu}_{\pi}$,
the contact term $J^{\mu}_{\textrm{contact}}$ and the $\Delta$-intermediate state or $\Delta$-MEC term $J^{\mu}_{\Delta}$. 
At level of terminology,
in the past some authors refer just to the first two terms as Meson Exchange Currents contributions
(like in~\cite{Martini:2009uj}) but actually the most current convention consists of including the $\Delta$-term into MEC.
Here we follow this convention. 
In the electroweak case another contribution, the pion-pole term $J^{\mu}_{\textrm{pole}}$, appears. 
It has only the axial component and therefore it is absent in the electromagnetic case. 

In the 2p-2h sector, the three microscopic models that we are discussing now are based on the Fermi gas,
which is the simplest independent particle model. 
In other words, the calculations are performed in a basis of uncorrelated nucleons. 
If also in the 1p-1h sector a basis of uncorrelated nucleons is used, 
one needs to consider also the nucleon-nucleon (NN) correlations contributions
since the protons and the neutrons in the nucleus are correlated and 
the short range correlated (SRC) pairs act as a unique entity in the nuclear response to an external field.
In the framework of independent particle models, like Fermi gas based models or mean field based models, 
these NN correlation are included by considering an additional two-body current,
the correlation current $J^{\mu}_{\textrm{NN-corr}}$. 
Detailed calculations and results for these NN correlation current contributions are given for example 
in Refs.~\cite{Alberico:1983zg,Alberico:1990fc,Amaro:2010iu,VanCuyck:2016fab}. 
In other approaches, like the one of Lovato~\textit{et al.}~\cite{Lovato:2014eva,Lovato:2015qka} 
the NN correlations are included in the description of the nuclear wave functions. 
With the introduction of the NN correlation contributions,
also the NN correlations-MEC interference contributions to the 2p-2h excitations 
(the terms called $N\Delta$ in the works of Martini~\textit{et al.}) naturally appear.
In the correlated-basis based approach, these contributions are referred as one nucleon-two nucleon currents interference.
This point will be further analyzed later.   

Focusing now on the Fermi gas based models,
it is important to stress that even in this simple model an exact relativistic calculation is difficult for several reasons. 
Let's start from the general expression of the 2p-2h hadronic tensor
\begin{eqnarray}
W^{\mu\nu}_{2p-2h}(\nq,\omega)
&=&
\frac{V}{(2\pi)^9}\int
d^3p'_1
d^3p'_2
d^3h_1
d^3h_2
\frac{m_N^4}{E_1E_2E'_1E'_2}
\nonumber \\ 
&&
\theta(p'_2-k_F)
\theta(p'_1-k_F)
\theta(k_F-h_1)
\theta(k_F-h_2)
\nonumber \\ 
&&
\langle 0 | J^\mu|\nh_1\nh_2\np'_1\np'_2 \rangle 
\langle \nh_1\nh_2\np'_1\np'_2| J^\nu| 0 \rangle 
\nonumber \\ 
&&
\delta(E'_1+E'_2-E_1-E_2-\omega)\delta(\np'_1+\np'_2-\nh_1-\nh_2-\nq) \,,
\label{hadronic}
\end{eqnarray}
where $\np'_1$ and $\np'_2$ are the momenta of the two nucleons ejected out of the Fermi sea, 
leaving two hole states in the daughter nucleus with moments $\nh_1$ and $\nh_2$. 
Using energy and momentum conservation, for fixed values of $\omega$ and $q$,
the 2p-2h calculations involves the computation of 7-dimensional integrals $\int d^3h_1 d^3h_2~d\theta'_1$.  
The first difficulty is that one needs to perform these 7-dimensional integrals for a huge number of 2p-2h response Feynman diagrams. 
Second, divergences in the NN correlations sector and in the angular distribution of the ejected nucleons~\cite{Simo:2014wka,Simo:2014esa}
may appear and need to be regularized.
Furthermore, as illustrated in Sec.~\ref{sec:theo} the neutrino cross section calculations should
be performed for all the kinematics compatible with the experimental neutrino flux
(and not only for some fixed values of the momentum- or energy-transfer,
as in the case of the electron scattering 
where the incoming and outgoing electron energies and momenta are known). 
For these reasons an exact relativistic calculation is very demanding with respect to computing,
and as a consequence different approximations are employed by the different groups
in order to reduce the dimension of the integrals, and to regularize the divergences. 
The choice of subsets of diagrams and terms to be calculated also presents important differences. 
In this connection Amaro \textit{et al.} only explicitly add the MEC contributions and not the NN correlations-MEC interference terms
(these last terms were analyzed for electron scattering in Ref.\cite{Amaro:2010iu})
to the genuine quasi-elastic.  
MEC contributions, NN correlations and NN correlations-MEC interference are present both in Martini \textit{et al.} and Nieves \textit{et al.} 
Martini~\textit{et al.} consider only the $\Delta$-MEC 
\footnote{The main reason for Martini~\textit{et al.} to discard the other contributions from the explicit calculation of MEC
is that they are peculiar to the external probe.
They want a ``universal'' spin-isospin 2p-2h response, to use in different processes, like in Ref.~\cite{Alberico:1983zg} where this response was used to study electron scattering and pion absorption. 
However MEC contributions to the time component of the axial current (due to the seagull and pion-pole terms) are taken into account in an effective way by introducing in the time component of the axial current a renormalization factor $G_A^*=G_A(1+\delta)$, 
see Appendix A of Ref.~\cite{Martini:2009uj}. The impact of this renormalization on the cross section is small.}.
This is the dominant contribution, as shown for example by De Pace~\textit{et al.} (Ref.~\cite{DePace:2003xu}, Fig.~9). 
The interference between direct and exchange diagrams is neglected by Martini~\textit{et al.} and Nieves~\textit{et al.} 
The treatment of Amaro~\textit{et al.} is fully relativistic as well as the one of Nieves~\textit{et al.} 
(even if the non pionic $\Delta$ decay contribution of $\Delta$-MEC are taken from
the non-relativistic work~\cite{Oset:1987re}, as in the case of Martini~\textit{et al.}) 
while the results of Martini~\textit{et al.} are related to a non-relativistic reduction of the two-body currents. 
Interestingly, Simo~\textit{et al.} have shown (Ref.~\cite{Simo:2014wka}, Fig.12)
that for the 2p-2h phase-space integral (obtained from Eq.(\ref{hadronic}) 
by setting to one the current matrix elements) the differences between a non relativistic and a fully relativistic calculation are relatively small.
The two results are close to each other in particular if compared with the calculation implementing only relativistic kinematics and not 
the Lorentz-contraction factor $M_N/E$.

Beyond these differences, there are also other differences 
related to the terms of the neutrino-nucleus cross sections affected by 2p-2h contribution. 
Amaro~\textit{et al.} in Refs. ~\cite{Amaro:2010sd,Amaro:2011qb,Amaro:2011aa,Megias:2014qva,Ivanov:2015aya} 
consider the 2p-2h contribution only in the vector sector while
Martini~\textit{et al.} and Nieves~\textit{et al.} also consider the axial vector.
Fully relativistic calculations of Amaro~\textit{et al.} 
for the axial sector have been recently presented in the paper of Simo~\textit{et al.}~\cite{Simo:2016ikv}
focusing on the nuclear responses and in the 
in the paper of Megias~\textit{et al.} ~\cite{Megias:2016fjk} focusing on the neutrino flux integrated differential cross sections.

Entering in further details, 
in the case of Martini~\textit{et al.} the coupling of the weak current to the nucleon 
or $\Delta$ inducing 2p-2h excitations is the spin isospin one ($\sigma\tau$). 
\footnote{We remind that other nuclear processes where the 2p-2h are relevant and are excited 
by the same excitation operator $\sigma\tau$ are the pion absorption,
the photon absorption and the magnetic excitations in electron scattering (\textit{i.e.} the transverse response).} 
In this case the 2p-2h term only affects the magnetic and axial responses which enter the neutrino cross section. 
In expression (\ref{eq1:cross_section}) these are the terms in $G_A^2, G_M^2$, and the interference term in $G_AG_M$. 
In the version of GiBUU discussed in Ref.~\cite{Gallmeister:2016dnq} the 2p-2h contributions enter in the same terms via 
an empirical spin-isospin transverse response deduced from electron scattering data. 
The isovector response (term in $R_{\tau}$) is not affected by 2p-2h in Martini~\textit{et al.} neither in GiBUU. 
On the contrary, as discussed in Ref.~\cite{Nieves:2013fr}, in the case of Nieves~\textit{et al.}
2p-2h mechanisms that affect the isovector response $R_{\tau}$ are also included
(beyond, obviously, to the contributions to the spin-isospin responses $R_{\sigma\tau}$).
In the case of Amaro~\textit{et al.}, referring always to Eq.~(\ref{eq1:cross_section}) for sake of illustration, in Refs. ~\cite{Amaro:2010sd,Amaro:2011qb,Amaro:2011aa,Megias:2014qva,Ivanov:2015aya} the vector MEC contributions affect only the nuclear response $R_{\sigma\tau}$ which is multiplied by the factor $G_M^2$, while in Ref.~\cite{RuizSimo:2016ikw,Megias:2016fjk} the vector and axial MEC contributions affect all the nuclear responses. 
We remind also that, as already mentioned the TEM is based on an effective enhancement of $G_M$,
hence the MEC contributions enter, in an effective way in the terms in $G_M^2$ and in $G_AG_M$, due to the modification of $G_M$.
A graphic illustration of all these analogies and differences can be found for example in the presentation ~\cite{marco_nuint14} at the NuInt14 workshop or in the presentations at CEA-ESNT 2016 2p-2h workshop~\cite{esnt_2p2h}.

\subsubsection{MiniBooNE}

\begin{figure}
\begin{center}
  \includegraphics[height=.7\textheight,angle=90]{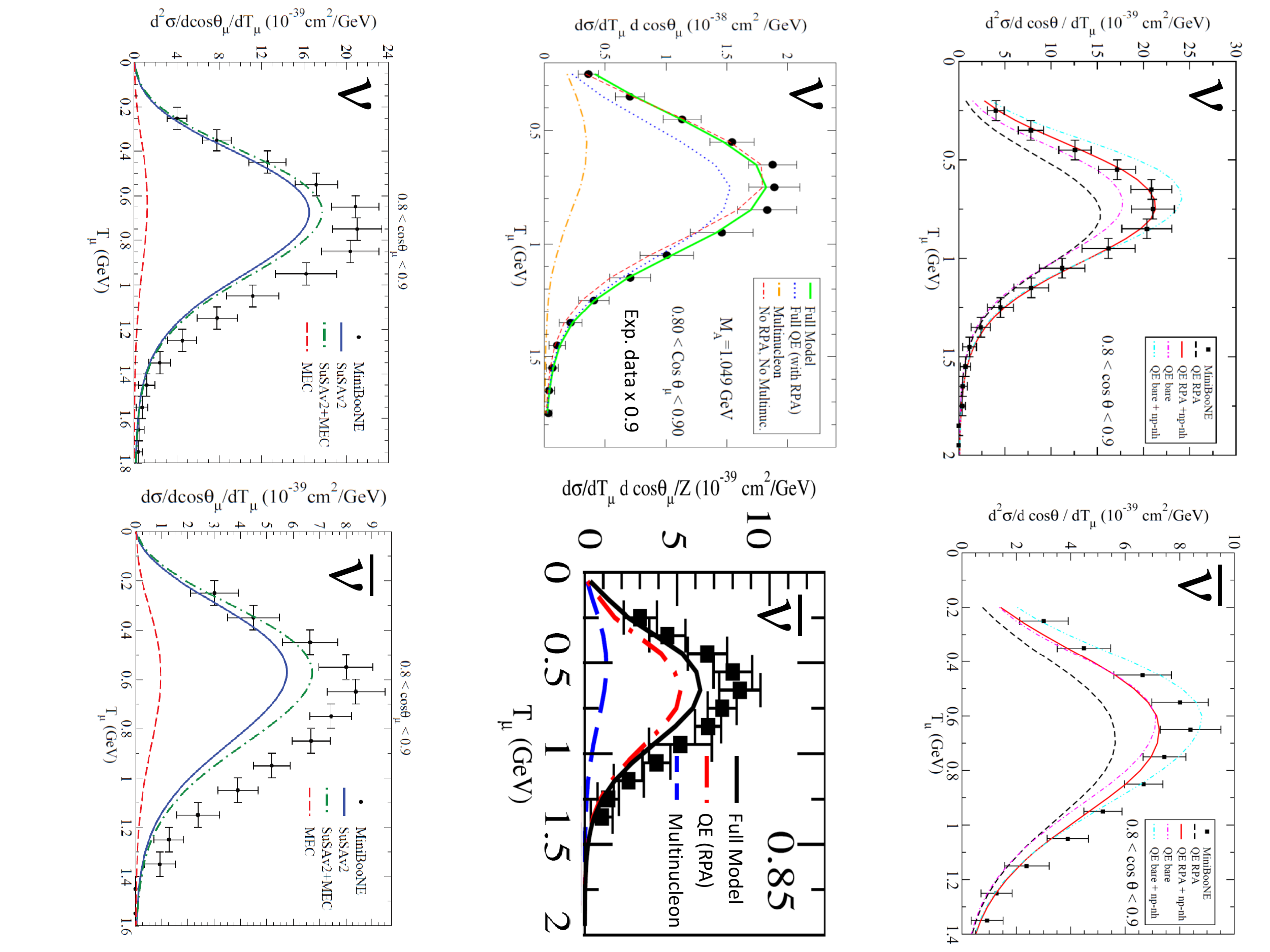}
  \caption{MiniBooNE flux-integrated neutrino (left panels) and antineutrino (right panels) CCQE-like double differential cross section
  on carbon per active nucleon for $0.8<\cos\theta<0.9$ as a function of the muon kinetic energy.
  Top panels: Martini~\textit{et al.}~\cite{Martini:2011wp,Martini:2013sha} results.
  Middle panels: Nieves~\textit{et al.}~\cite{Nieves:2011yp,Nieves:2013fr} results.
  Bottom panels: Megias~\textit{et al.}~\cite{Megias:2014qva} results representing
  an update of the Amaro~\textit{et al.}~\cite{Amaro:2010sd,Amaro:2011aa} results.}
\label{fig_confronto}
\end{center}
\end{figure}

\begin{figure}
\begin{center}
  \includegraphics[height=.7\textheight,angle=90]{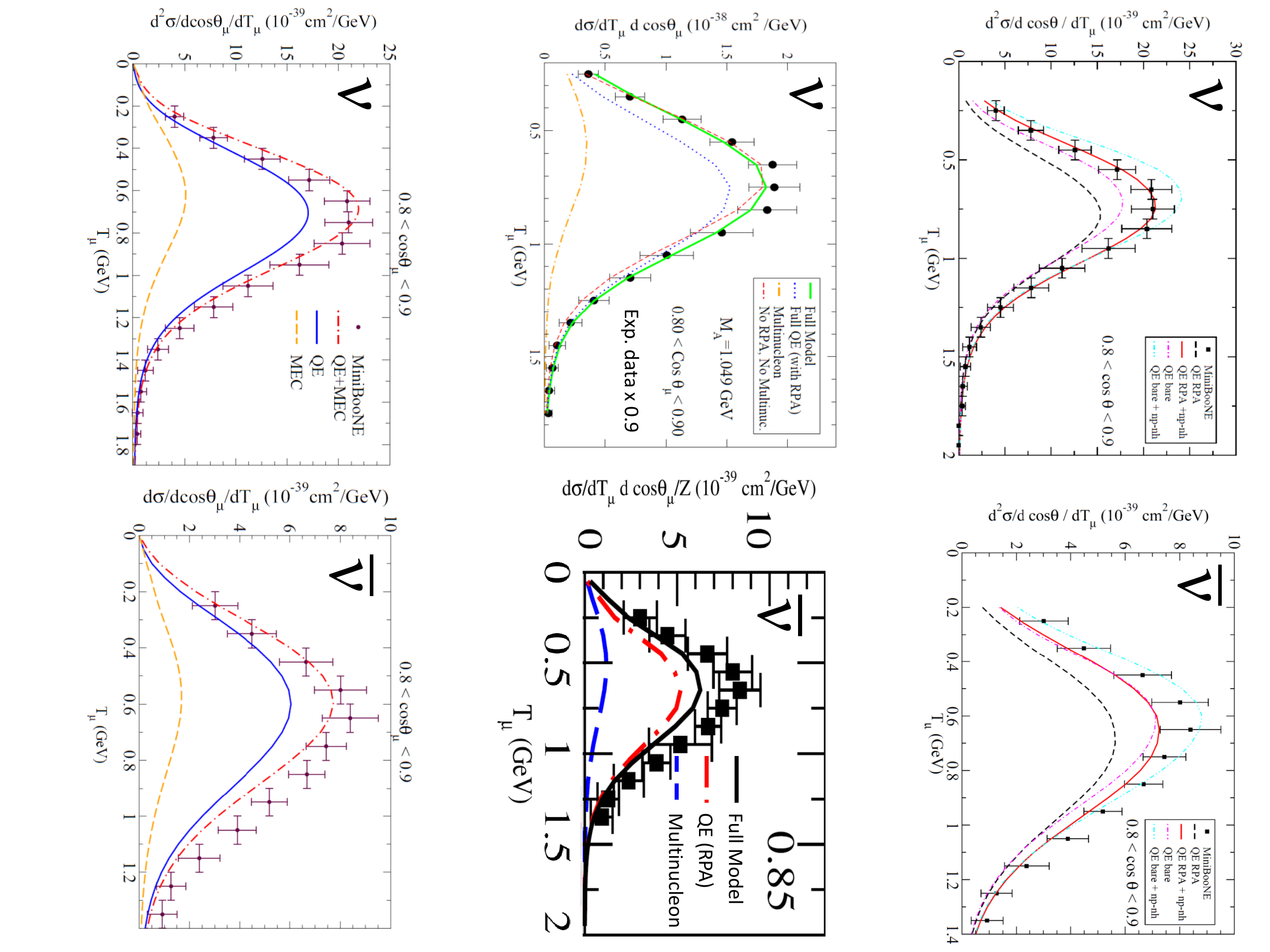}
  \caption{The same as Fig.~\ref{fig_confronto} but with the updated Megias~\textit{et al.} results of Ref.~\cite{Megias:2016fjk}.}
\label{fig_confronto_2}
\end{center}
\end{figure}

Taking into account the existence of all the differences discussed in detail above,
it is not surprising that the models produce different final results.
This point is illustrated in Fig.~\ref{fig_confronto} where the MiniBooNE neutrino and antineutrino
flux-integrated double differential CCQE-like cross sections calculated in the different approaches are displayed. 
For the sake of illustration the results are given for $0.8< \cos\theta_\mu<0.9$ as a function of the muon kinetic energy.
The complete theoretical results in the different bins for neutrinos (see Fig.~\ref{fig_minib_d2s})
and antineutrinos are given in Refs.~\cite{Martini:2011wp,Martini:2013sha} for Martini~\textit{et al.},
in Refs.~\cite{Nieves:2011yp,Nieves:2013fr} for Nieves~\textit{et al.} and in Refs.~\cite{Amaro:2010sd,Amaro:2011aa} for Amaro~\textit{et al.}
An updated version of these last results,
but still considering MEC contributions only in the vector sector,
is given by Megias~\textit{et al.} in Ref.~\cite{Megias:2014qva} from which
we take the results reported in the last two panels of Fig.~\ref{fig_confronto}. 
A further update of these last results, now including MEC contributions also
in the axial sector is given by Megias~\textit{et al.} in Ref.~\cite{Megias:2016fjk}. 
Figure~\ref{fig_confronto_2} is the analog of Fig.~\ref{fig_confronto} containing
the latest results of Megias~\textit{et al.}~\cite{Megias:2016fjk}. 
We prefer to show both in order to better illustrate the (recent and rapid) chronological evolution of the research in this field.  
 
Let's start the discussion with the, in some sense obsolete, Fig.~\ref{fig_confronto}.
As one can observe, the results of Martini~\textit{et al.} are in agreement with the experimental data. 
In the case of Nieves~\textit{et al.} and Amaro~\textit{et al.} a tendency to underestimate the MiniBooNE data appears. 
Nevertheless also these theoretical results are compatible with MiniBooNE since
data points do not show the overall normalization errors, and neutrino and antineutrino cross section data
have additional 10\% and 17\% errors. This is the reason why Nieves~\textit{et al.} multiplied the neutrino MiniBooNE data by 0.9 in their figure. 
One can also notice from the two panels of Martini~\textit{et al.} and from the neutrino panel of Nieves~\textit{et al.} that
the agreement between the theoretical predictions and the MiniBooNE data is given by
a delicate balance between RPA suppression and np-nh enhancement effects. 
This point is evident also in the $Q^2$ distributions, as shown in figure~\ref{fig_minib_dsdcos_dQ2}. 
However there are regions of the $Q^2$ distributions ($Q^2\geq$0.3 GeV$^2$) and
of the double differential cross sections which are not affected by RPA quenching hence there the multinucleon contributions are singled out. 

An other important point is that, as it appears from Fig.~\ref{fig_confronto} the relative role of
the multinucleon contribution is different for neutrinos and antineutrinos in the different approaches.
In the case of Martini~\textit{et al.} the relative role of the multinucleon contribution is larger for neutrinos,
even if it remains important also for antineutrinos,
in the case of Nieves~\textit{et al.} it is more or less the same for neutrinos and antineutrinos while
for Amaro~\textit{et al.} the relative np-nh contribution is larger for antineutrinos with respect to neutrinos.
This difference was even more pronounced in the previous version of the Amaro~\textit{et al.} results~\cite{Amaro:2011aa}.
As discussed in Refs.~\cite{Martini:2010ex,Amaro:2011aa,Ericson:2015cva} the difference between
the neutrino and antineutrino results is due to the presence in the neutrino-nucleus cross section expression of
the vector-axial interference term, which changes sign between neutrino and antineutrino (see Eq.~(\ref{eq1:cross_section})),
the basic asymmetry which follows from the weak interaction theory.
Due to this vector-axial interference term,
the relative weight of the different nuclear responses is different for neutrinos and antineutrinos. 
As a consequence also the relative weight of the np-nh contributions is different for neutrinos and antineutrinos.
For example the fact that np-nh contributions are larger for antineutrinos with respect to neutrinos
in the case of Amaro~\textit{et al.} is due to the fact that Amaro~\textit{et al.}
in Refs.~\cite{Amaro:2010sd,Amaro:2011aa,Megias:2014qva} consider the np-nh contribution only in the vector sector,
hence not in the vector-axial interference term, as already discussed. 

\begin{figure}
\begin{minipage}[c]{70mm}
\begin{center}
      \includegraphics[width=70mm]{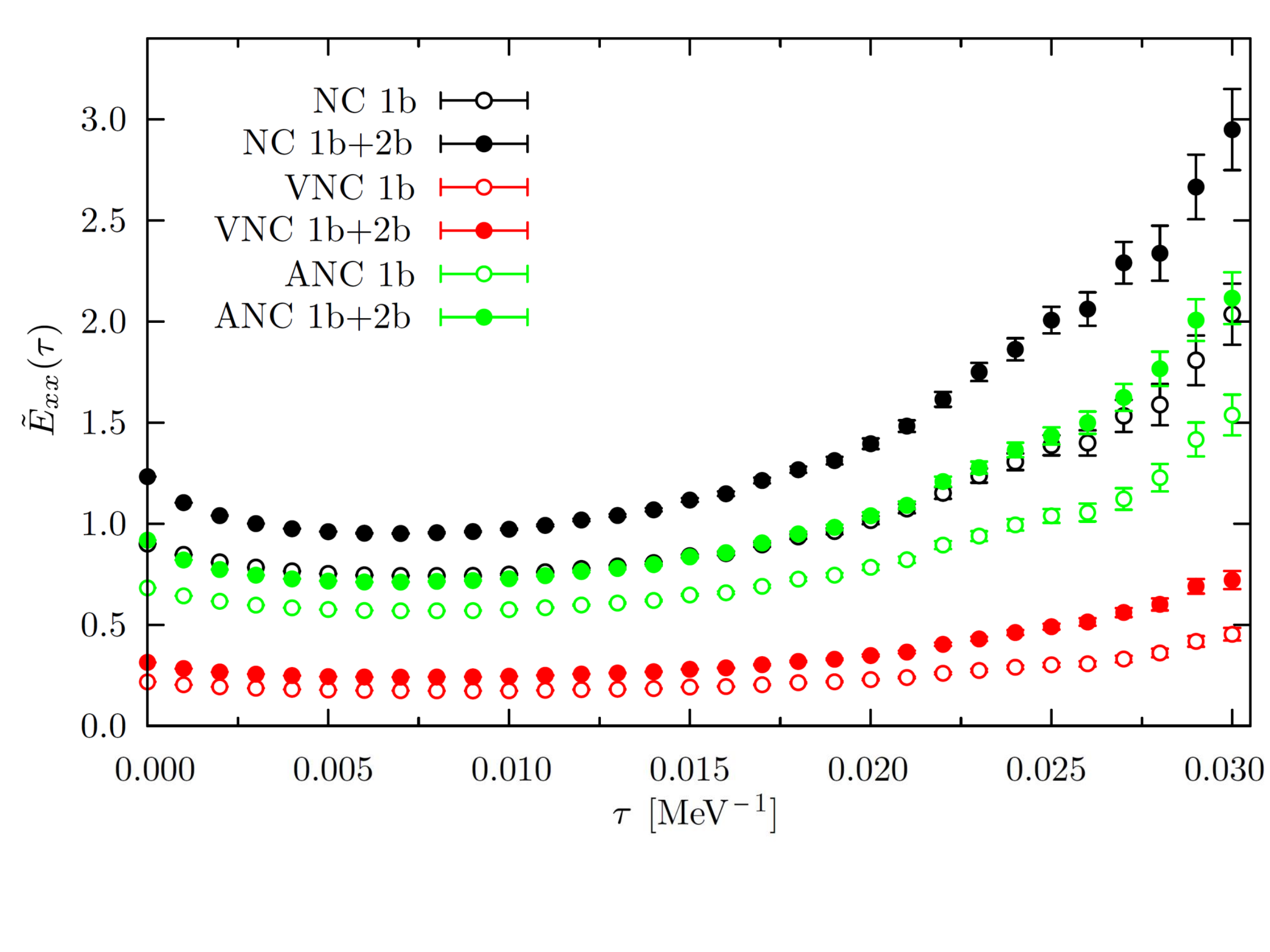}
\caption{\label{fig_lovato}Euclidean neutral-weak transverse response functions of $^{12}$C at $q=570$ MeV. 
The figure is taken from Ref.~\cite{Lovato:2015qka} of Lovato~\textit{et al.}}
\end{center}
\end{minipage}
\hspace{1mm}
\begin{minipage}[c]{70mm}
\begin{center}
  \includegraphics[width=70mm]{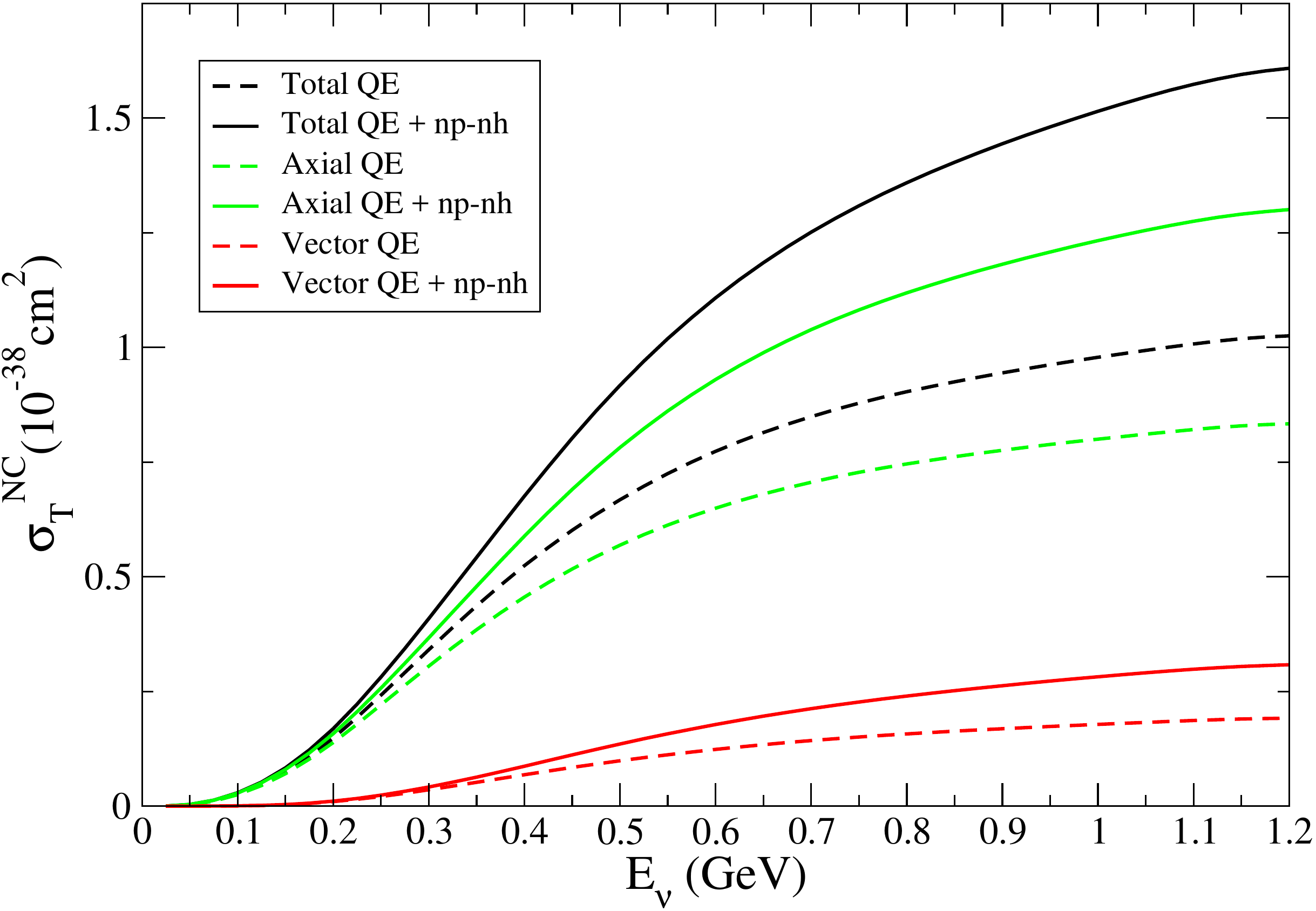}  
\caption{\label{fig_xtot_NC_11_tot_ax_vect} 
Transverse contribution to the neutral current $\nu_\mu$-$^{12}$C QE-like cross section
calculated in the approach of Martini~\textit{et al.}}
\end{center}
\end{minipage}
\end{figure} 

\begin{figure}
\begin{minipage}[c]{70mm}
\begin{center}
      \includegraphics[width=70mm]{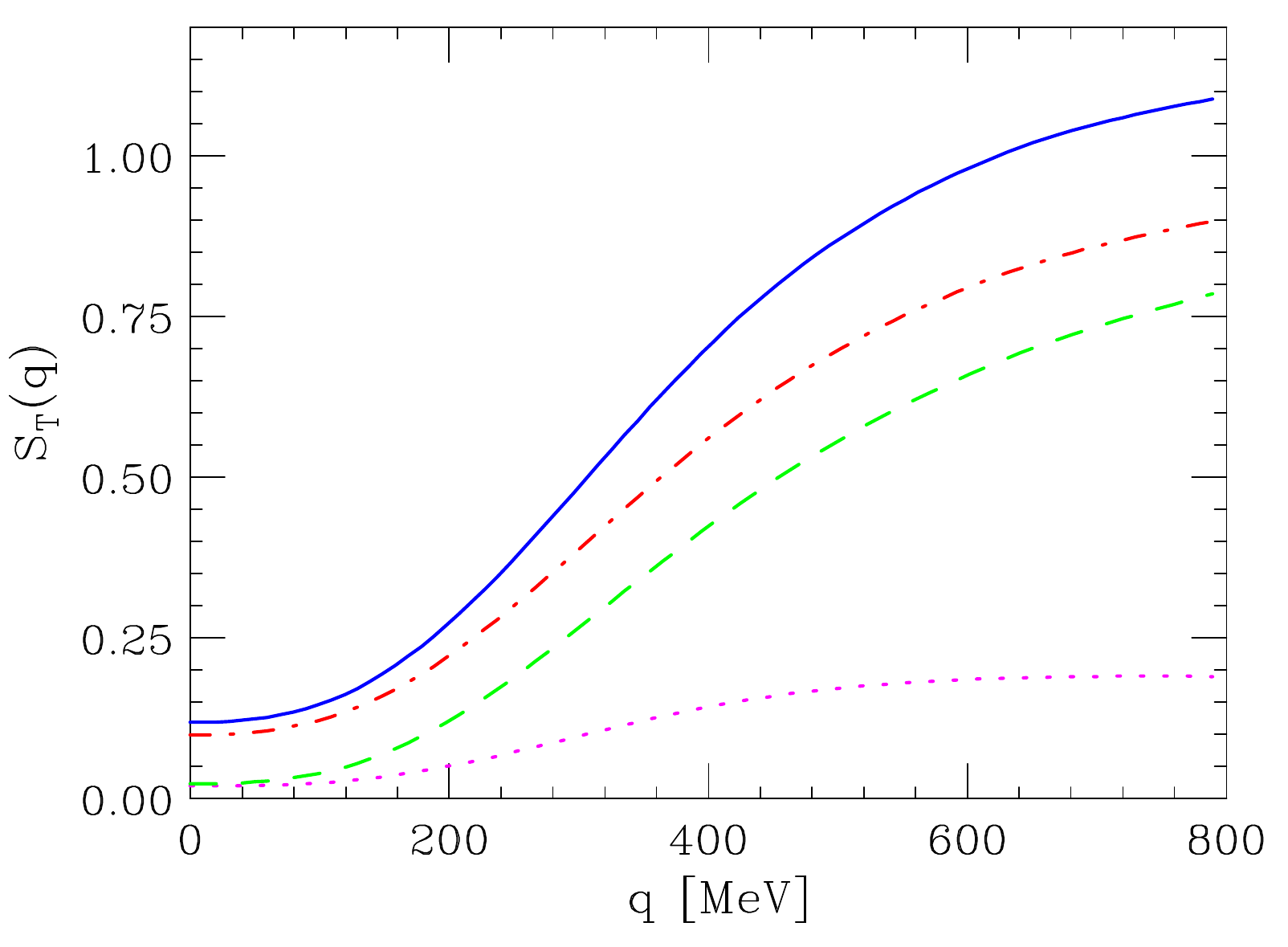}
\caption{\label{fig_intf_sr_benh} Sum rule of the $^{12}$C electromagnetic transverse response,
as calculated in the approach of Lovato~\textit{et al.}~\cite{Lovato:2013cua}. 
The figure is taken from Ref.~\cite{Benhar:2015ula}. The dashed line shows the results
obtained including the one-nucleon current only, while the solid line
corresponds to the full calculation. The dot-dashed line represents the
sum rule computed neglecting the NN correlations-MEC interference term, the contribution
of which is displayed by the dotted line. }
\end{center}
\end{minipage}
\hspace{1mm}
\begin{minipage}[c]{70mm}
\begin{center}
  \includegraphics[width=70mm]{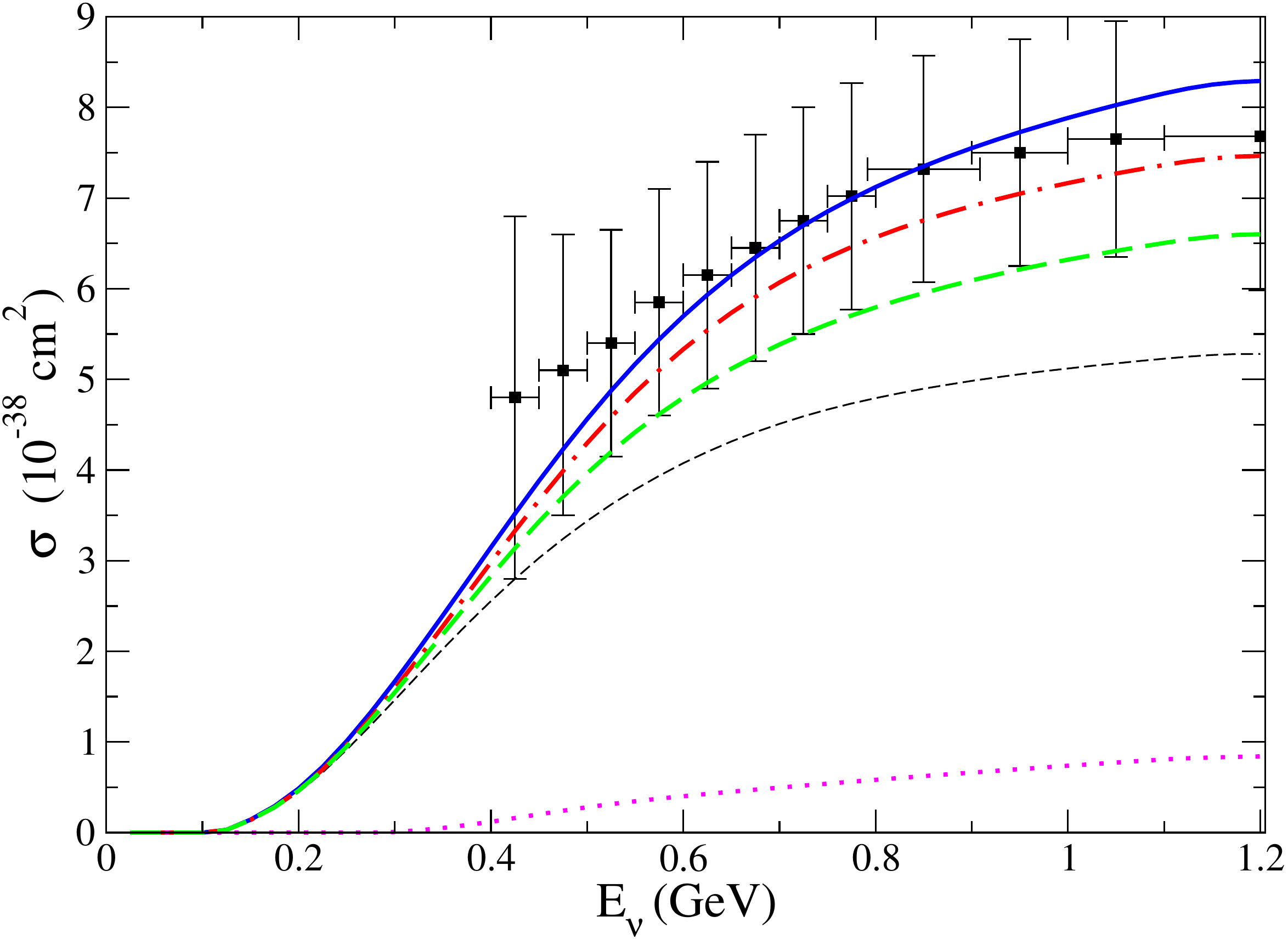}  
\caption{\label{fig_intf_xs_mart} Charged current $\nu_\mu$-$^{12}$C QE-like
cross section calculated in the approach of Martini~\textit{et al.} The short-dashed line shows the results
obtained including the one-nucleon current only (genuine QE). The long-dashed line shows the results
obtained including the one-nucleon current and the NN correlation two-body current contributions. The solid line
corresponds to the full calculation. The dot-dashed line represents the
cross section computed neglecting the NN correlations-MEC interference term, the contribution
of which is displayed by the dotted line.}
\end{center}
\end{minipage}
\end{figure} 

In order to investigate the multinucleon content of the vector-axial interference term,
Ericson and Martini have considered in Ref.~\cite{Ericson:2015cva} the difference between the neutrino and antineutrino 
MiniBooNE CCQE-like double differential cross sections.
These quantities depend on the neutrino or antineutrino normalized energy flux profiles.
In the case of identical ones, the difference provides a direct access to the vector-axial interference term. 
For the MiniBooNE fluxes, Ericson and Martini have tested how much the flux difference influences the combination of the two cross sections, 
showing that this influence is small.
The difference between the MiniBooNE quasielastic-like double-differential neutrino and
antineutrino cross sections is rather pure with respect to the vector-axial interference term, which remains dominant.
This allows more specific tests of theoretical models on the vector-axial interference term. 
The model of Martini~\textit{et al.}, which includes the np-nh excitations in the vector-axial interference term,
gives a good fit for the difference  of the MiniBooNE cross sections reproducing well the data in the full range of muon energy and emission angle.
This result represents an important test for the presence of the multinucleon component in the vector-axial interference term.

A similar conclusion on a relevant two-body current contribution in the vector-axial interference term
has been obtained by Lovato~\textit{et al.}~\cite{Lovato:2014eva,Lovato:2015qka}
who calculated the neutral weak current two-body contributions to sum rules and Euclidean responses in $^{12}$C.
The main advantage of the calculation of Lovato~\textit{et al.} is that it is an~\textit{ab initio}
microscopic approach which considers a full realistic nuclear interaction, fitted on nucleon-nucleon data,
with a simultaneous treatment of the two-body current.
It can be considered as a state-of-the-art description of nuclear ground state and correlations.
The disadvantages are that the currents are non-relativistic,
the pion production channel is not included and, most important, these calculations are computationally very demanding. 
Within this approach an evaluation of $^{12}$C responses in the whole $(\omega, |{\bf{q}}|)$ plane and of neutrino cross sections is beyond the present computational capabilities
\footnote{The longitudinal and transverse electromagnetic responses of $^{12}$C have been calculated in the same \textit{ab initio} approach
for the first time in Ref.~\cite{Lovato:2016gkq} for $|{\bf{q}}|$ in the range 300-570 MeV. They are compared with the ones obtained from the spectral function approach in Ref.~\cite{Rocco:2016ejr}. }
. 
But the results obtained with this approach offer a benchmark for more phenomenological methods.
An example of results that can be obtained within this approach is the already mentioned conclusion of
an important two-body current contribution in the vector-axial interference term.
Another very interesting result is that two-body current contribution is also important in the axial part of the transverse response.
These results support the important 2p-2h contribution in the axial sector of Martini~\textit{et al.} 
This point can be noticed also by a qualitative comparison between two different quantities,
\textit{i.e.} the Euclidean neutral weak transverse response function of $^{12}$C, 
as shown in Fig.~\ref{fig_lovato} taken from the Lovato~\textit{et al.}~\cite{Lovato:2015qka} and
the transverse contribution (the $R_{\sigma\tau}(T)$ terms) to the neutral current $\nu_\mu$-$^{12}$C QE-like cross section
calculated in the approach of Martini~\textit{et al.}, as shown in Fig.~\ref{fig_xtot_NC_11_tot_ax_vect}. 
Another interesting qualitative comparison between two different quantities is the one between Fig.~\ref{fig_intf_sr_benh} 
(which corresponds to Fig.~2 of the paper of Benhar~\textit{et al.}~\cite{Benhar:2015ula}) and Fig.~\ref{fig_intf_xs_mart}. 
Figure~\ref{fig_intf_sr_benh} shows the sum rule of the $^{12}$C electromagnetic transverse response,
as calculated in the approach of Lovato~\textit{et al.}~\cite{Lovato:2013cua}. 
Figure~\ref{fig_intf_xs_mart} shows the charged current $\nu_\mu$-$^{12}$C
QE-like cross section calculated in the approach of Martini~\textit{et al.}. 
In these figures the 2p-2h NN correlations-MEC interference contributions 
(referred as one nucleon-two nucleon currents interference contributions in Refs.~\cite{Lovato:2014eva,Lovato:2015qka,Benhar:2015ula}) 
are separately plotted. One can notice the quantitative relevance of this interference contribution. 

With the recent inclusion of the axial two-body currents contributions to the responses entering in the neutrino cross sections 
via a fully relativistic calculation of Simo~\textit{et al.}~\cite{Simo:2016ikv}, also the group of Amaro~\textit{et al.} 
agrees now on the fact that 2p-2h are important in the axial part of the transverse response 
and in the vector-axial interference term. 
In Ref.~\cite{Simo:2016ikv} Simo~\textit{et al.} calculated the different responses for fixed values of $q$.  
In Ref.~\cite{Megias:2016fjk} Megias~\textit{et al.} calculated the MiniBooNE,
T2K and MINERvA flux integrated differential cross sections as well as the total cross sections compared with the MiniBooNE and NOMAD results.
As already mentioned, two examples of the new Megias~\textit{et al.} results are shown in Fig.~\ref{fig_confronto_2}. 
Figure~\ref{fig_confronto_2} is the analog of Fig.~\ref{fig_confronto} containing
the latest results of Megias~\textit{et al.}~\cite{Megias:2016fjk}. 
These results are now closer to the ones of Martini~\textit{et al.} for neutrinos and for antineutrinos.
One of the major difference between the results of Amaro~\textit{et al.} on one hand and Martini~\textit{et al.} and Nieves~\textit{et al.}
on the other hand, related to the presence or not of 2p-2h contributions in the axial sector and in the vector-axial interference term,
and as a consequence,
on the relative role of 2p-2h contributions for neutrinos and antineutrinos seems now to be disappeared. 
The MEC contributions to neutrino-nucleus cross sections in the three different microscopic approaches seem now to be compatible among
them~\footnote{We mention that the 2p-2h MEC contribution to $R_{\sigma\tau}(T)$, as evaluated by 
Martini~\textit{et al.} and shown in Fig.~\ref{fig_risp_q600_diversicanali} for $|{\bf{q}}|$= 600 MeV/c,
is similar to the one reported in the paper of Megias~\textit{et al.} (Ref.~\cite{Megias:2014qva}, Fig.~1). 
Furthermore as shown by Megias~\textit{et al.} in Fig.~2 of Ref.~\cite{Megias:2016fjk}
the following three MEC contributions to the neutrino cross section: transverse-vector,
transverse-axial and transverse-vector-axial are very similar up to a neutrino energy of $E_\nu\simeq 2$ GeV
justifying the approximation of Martini~\textit{et al.}
and GiBUU of obtaining these contributions from a unique response, $R_{\sigma\tau}(T)$.}.
Probably the major differences that still remain, 
are related to the treatment of the NN correlations and NN correlations-MEC interference terms.

\subsubsection{T2K}

\begin{figure}
\begin{center}
  \includegraphics[height=\textheight]{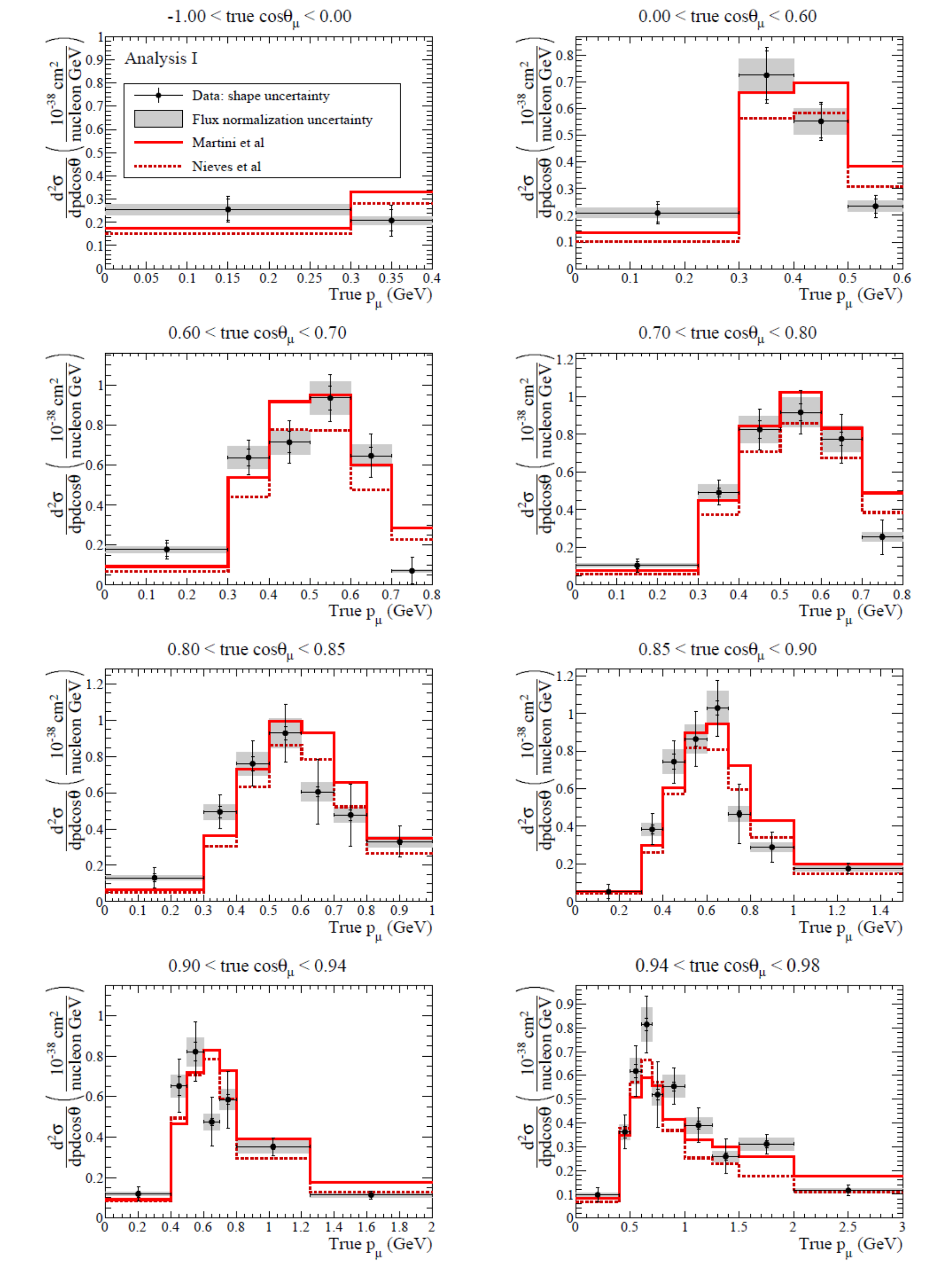}
  \caption{Double-differential muon neutrino charged-current interactions on carbon 
without pions in the final state (\textbf{CC$0\pi$}) performed by T2K using
the off-axis near detector ND280 compared with the two theoretical calculations of Martini~\textit{et al.} and Nieves~\textit{et al.}}
\label{fig_comp_t2k_cc0pi}
\end{center}
\end{figure}

A good illustration of the amount of the differences between the results of 
Martini~\textit{et al.} and Nieves~\textit{et al.} is also given 
in Fig.~\ref{fig_comp_t2k_cc0pi}, 
taken from Ref.~\cite{Abe:2016tmq}, where the \textbf{CC$0\pi$} flux-integrated double-differential cross section on carbon 
performed by T2K using the off-axis near detector ND280 is compared with these two theoretical calculations. 
In Fig.~\ref{fig_comp_t2k_cc0pi} the T2K results are compared to the ones of Martini~\textit{et al.} and Nieves~\textit{et al.}
obtained as a sum genuine quasielastic contribution calculated in RPA and np-nh excitations.
Pion absorption contribution due to pion FSI is not taken into account in these approaches but,
as already discussed, the pion-less Delta decay is included in the np-nh contributions. 
As shown in Ref~\cite{Morfin:2012kn} in connection with the MiniBooNE results and in Ref.~\cite{Abe:2016tmq}
for the T2K flux integrated double-differential cross sections,
the two theoretical approaches give very similar results for the genuine quasielastic calculated in RPA.
The major differences are related to the np-nh channel. At the present level of experimental accuracy 
quantifying the agreement between the T2K data and the two models is not evident; the uncertainties are too large for any conclusive statement.
For the moment, from Ref.~\cite{Abe:2016tmq} one can only conclude that both models agree with the data,
and the data seems to suggest the presence of np-nh with respect to pure CCQE RPA predictions.
This is an important conclusion,
since these results represent a successful test of the necessity of the multinucleon emission channel
in RPA based models in order to reproduce the data of an experiment with another
neutrino flux (but in the same neutrino energy domain) with respect to the one of MiniBooNE.
The same conclusion holds for the CC inclusive cross sections of T2K, see Sec.\ref{sec:cc}.  
A comparison with the T2K CC$0\pi$ (and CC inclusive) has been now also performed by Megias~\textit{et al.} in Ref.~\cite{Megias:2016fjk}. 
As shown in Fig.~14 of Ref.~\cite{Megias:2016fjk},
even if in this case, a better agreement with data seems to be obtained by including the MEC contributions,
genuine theoretical CCQE results obtained in the SuSAv2 approach are very often in the error bars.
Interestingly the authors observe that in this T2K case the relative contribution of the 2p-2h MEC compared with the pure QE ($\sim 10\%$)
is significantly smaller than in the MiniBooNE case ($\sim 25-35 \%$). They connect this difference with the shape of the T2K neutrino flux that,
although with an averaged neutrino flux similar to MiniBooNE, shows a much narrower distribution.

\subsubsection{MINERvA\label{sec:qe_minerva}}

\begin{figure}
\begin{center}
  \includegraphics[height=\textheight]{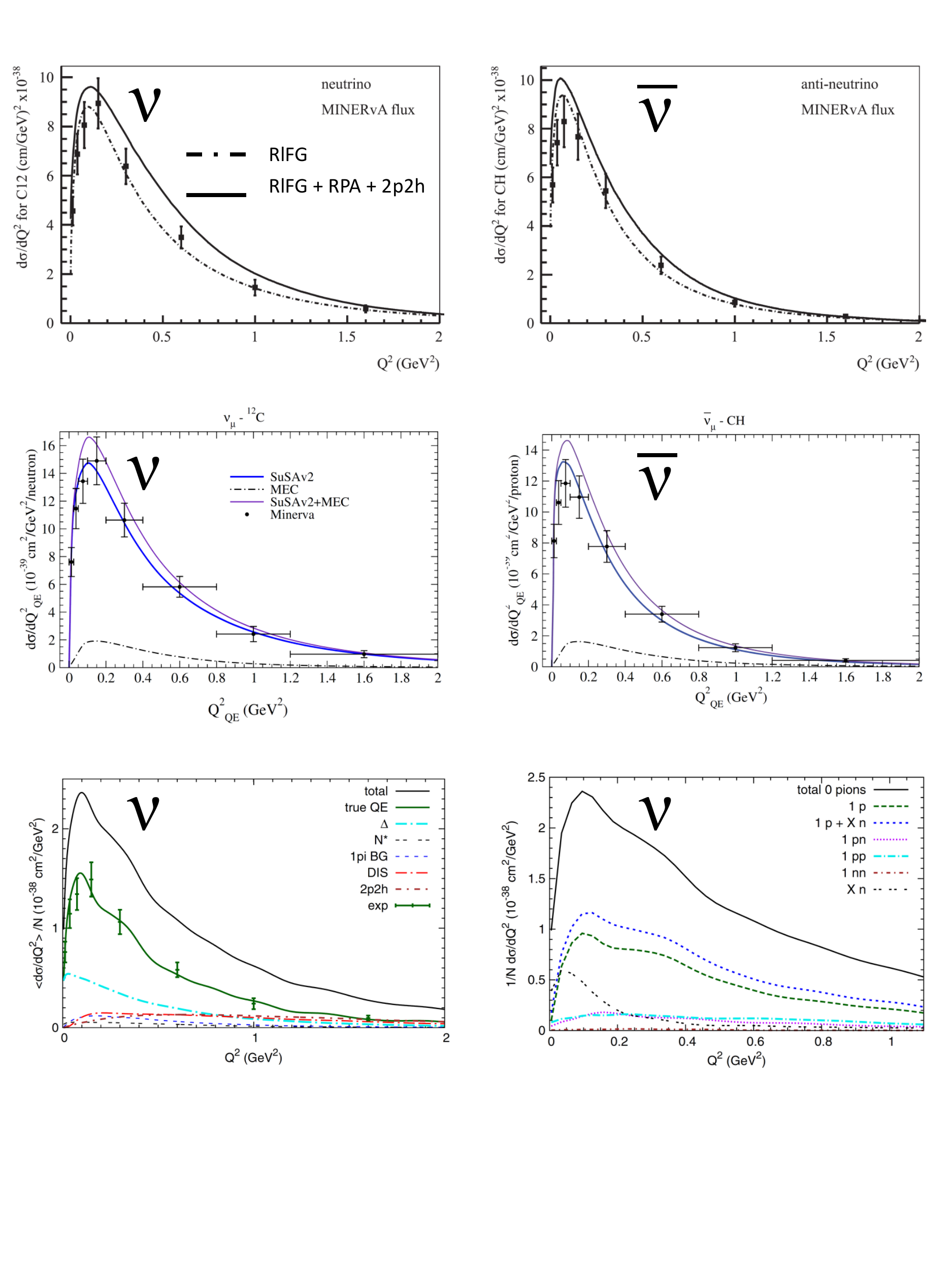}
  \caption{The MINERvA flux-integrated $Q^2$ distributions calculated by Gran~\textit{et al.}~\cite{Gran:2013kda} 
(top panels),
Megias~\textit{et al.}~\cite{Megias:2014qva} (middle panels), and Mosel~\textit{et al.}~\cite{Mosel:2014lja} (bottom panels) 
compared with the MINERvA neutrino~\cite{Fiorentini:2013ezn} and antineutrino~\cite{Fields:2013zhk} CCQE-like data.}
\label{fig_comp_minerva_q2}
\end{center}
\end{figure}

Up to now we have discussed the theoretical models in connection with the MiniBooNE and T2K cross sections. 
For the moment most of the theoretical calculations for the np-nh excitations are restricted to
the relatively small energy and momentum transfer, prevalent in the MiniBooNE and T2K experiments.
How the np-nh processes behave at large energy- and momentum transfer is now under investigation.
In Ref.~\cite{Megias:2014qva} Megias~\textit{et al.} applied the model SuSAv2+vector MEC to neutrino energies of up to 100 GeV 
and compared their predictions with NOMAD~\cite{Lyubushkin:2008pe} and
MINERvA neutrino~\cite{Fiorentini:2013ezn} and antineutrino~\cite{Fields:2013zhk} 
CCQE-like data. In Ref.~\cite{Megias:2016fjk} Megias~\textit{et al.} repeated the same study by including axial MEC contributions. 
Gran~\textit{et al.}~\cite{Gran:2013kda} applied the model of 
Nieves~\textit{et al.} to neutrino energies of up to 10 GeV.
However, with thinking of the kinematic limitations of their model,
they placed a cut on  the  three-momentum transfer of 1.2 GeV.
They compared their results with the MINERvA neutrino and antineutrino CCQE $Q^2$ distribution.
A similar comparison has been performed also by Mosel~\textit{et al.}~\cite{Mosel:2014lja} using GiBUU.
The special implementation of 2p-2h used in GiBUU is also subject to the same uncertainty at higher energies as in the other approaches. 
The MINERvA flux-integrated $Q^2$ distributions calculated by Gran~\textit{et al.}~\cite{Gran:2013kda},
Megias~\textit{et al.}~\cite{Megias:2014qva}, and Mosel~\textit{et al.}~\cite{Mosel:2014lja} are given in Fig.~\ref{fig_comp_minerva_q2}. 

As a general remark, by comparing these results with MINERvA data, 
one can observe that the MINERvA $Q^2$ distributions can be reproduced also without the inclusion of np-nh excitations. 
This is not the case of the MiniBooNE $Q^2$ distributions, as shown in
Refs.~\cite{Martini:2011wp,Martini:2013sha,Megias:2014qva} and in Fig.~\ref{fig_minib_dsdcos_dQ2}. 
As stressed by Mosel~\textit{et al.}, in the case of MINERvA the sensitivity to details of
the treatment of np-nh contributions is smaller than the uncertainties introduced by
the $Q^2$ reconstruction and our insufficient knowledge of pion production.
The MINERvA experiment being at higher energies with respect to the MiniBooNE one,
the pion production channel becomes in this case more important hence the background subtraction
to isolate genuine CCQE and 2p-2h events is delicate,
in particular when information related only to muon variables are considered.  
Furthermore a recent reanalysis of the  MINERvA flux results~\cite{Aliaga:2016oaz} seems to lead
to an increase of the normalization of the MINERvA CCQE-like cross sections discussed here. 
This new result will probably invalidate the conclusion that the MINERvA $Q^2$ distributions
can be reproduced also without the inclusion of np-nh excitations.

It could also reduce the strong tension between the MiniBooNE and MINERvA quantitatively analyzed by the 
T2K’s Neutrino Interaction Working Group in Ref.~\cite{Wilkinson:2016wmz}. 
In this paper it is shown how the published neutrino and antineutrino CCQE-like  data sets from the MiniBooNE
and MINERvA are used to test the models implemented in the NEUT neutrino interaction generator~\cite{NEUT}.
The results from this global fit show that none of the models that are currently available in
NEUT describe all of the CCQE data adequately,
and the fit returns poor goodness-of-fit with unreasonable parameters. 
For example, the value of the axial mass obtained from the fit using 
the NEUT implementation of the RPA+2p2h model of Nieves \textit{et al.} turns to be $M_A$= 1.15 GeV.
This value is lower than that obtained from past fits of the RFG model to MiniBooNE data alone,
but it is still inconsistent with that obtained in global fits to light target bubble chamber data or high energy heavy target data. 
Additionally, the data require a large suppression of the 2p-2h channel.
At the best-fit point the 2p-2h contribution is suppressed to 27\% of the Nieves \textit{et al.} nominal value.
This poor fit result is 
due to several reasons. 
First, MINERvA data do not require any additional strength,
where the fit completely suppress the 2p-2h component if the MiniBooNE data were not included in the fit.
Second, current 2p-2h model in NEUT does not have realistic systematic errors and it does not have any freedoms to change the shape,
and the fit just suppress it to avoid a tension.
This would be a source of a poor fit.
On top of them, MiniBooNE data do not have a full covariance matrix
and this makes statistical interpretation of the global fit result impossible.
Theorists should investigate systematic errors and alternative 2p-2h models,
and experimentalists should publish covariance matrix in the future. 

\begin{figure}
\begin{center}
  \includegraphics[width=150mm]{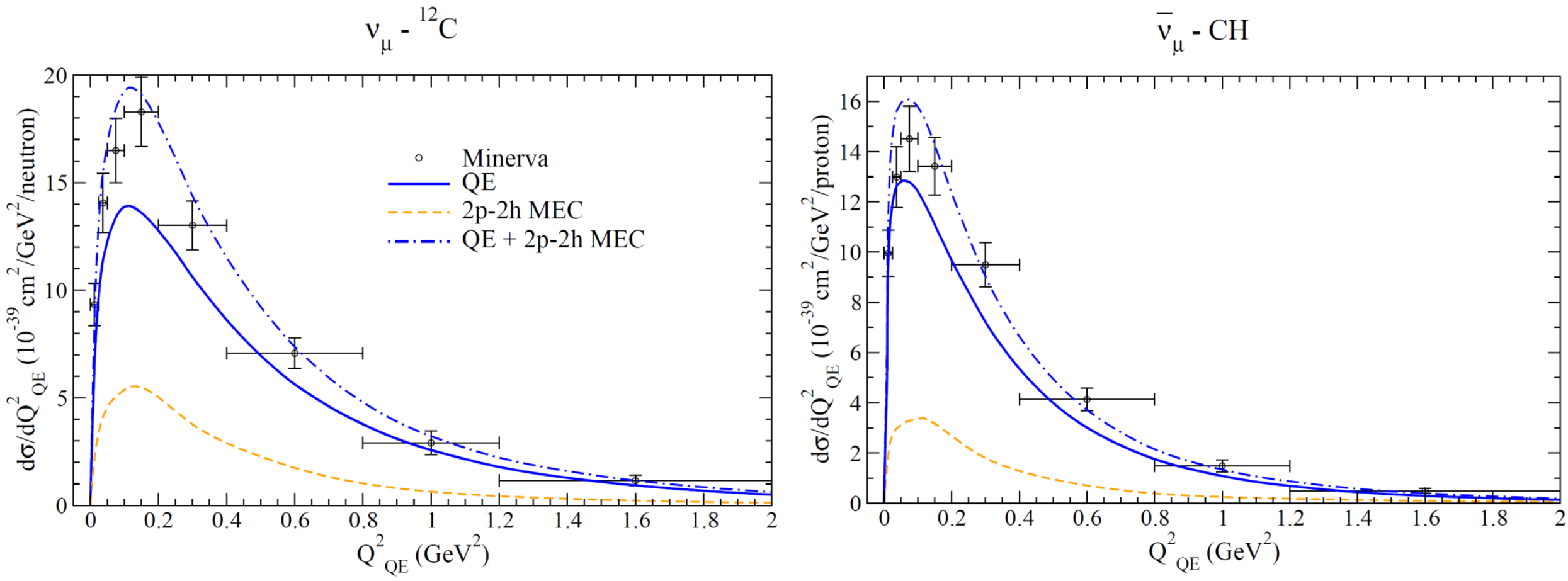}  
  \caption{The MINERvA flux-integrated $Q^2$ distributions calculated by 
Megias~\textit{et al.}~\cite{Megias:2016fjk}
compared with the new preliminary MINERvA neutrino and antineutrino  CCQE-like data.}
\label{fig_megias_minerva}
\end{center}
\end{figure}

New preliminary MINERvA results are shown in Fig.~\ref{fig_megias_minerva} that we take from Ref.~\cite{Megias:2016fjk} of Megias~\textit{et al.} 
The data correspond to the new analysis performed by the MINERvA collaboration
which takes into account the reevaluation of the  MINERvA flux~\cite{Aliaga:2016oaz}. 
These data exceed the ones published in Refs.\cite{Fiorentini:2013ezn,Fields:2013zhk} and shown in Fig.~\ref{fig_comp_minerva_q2} by $\sim 20 \%$. 
These new experimental results are compared with the theoretical evaluations of
Megias~\textit{et al.}~\cite{Megias:2016fjk} including vector and axial MEC contributions. 
The comparison leads to major differences with respect to the previous conclusions related to the MiniBooNE - MINERvA discrepancies. 
Now significant contributions of the 2p-2h MEC, of the order of $\sim 35$ - $40\%$ ($\sim 25\%$) at the
maximum for $\nu_\mu$ ($\bar{\nu}_\mu$), are needed in order to reproduce the experimental data. 
In spite of the different neutrino energy fluxes, the np-nh contributions are finally crucial in order to fit not only the MiniBooNE (and T2K) CCQE-like data, but also the MINERvA data. 

\subsection{Hadron information \label{sec:hadron}}

\subsubsection{one-track vs. two-track}

In past years, all information of lepton kinematics are exploited. 
Any more lepton kinematics differential cross section data from CCQE-like or CC0$\pi$ sample is
unlikely to contribute to further understanding of this channel, 
unless systematic error is squeezed down to order few \% which is unlikely to happen in next few years. 
Because of this, there is a growing interest to utilize hadron information to understand further structures of np-nh nature. 

Naively, genuine quasielastic scattering involves in one outgoing nucleon.
However, if QE-like data is really a combination of genuine QE and np-nh contribution as suggested,
we expect more nucleons in the final state from QE-like data. 
We have an indication of this from MINERvA QE-like vertex activity data,
where neutrino mode QE-like data prefer more visible energy around the vertex,
which can be interpreted as a contribution of low energy protons created by interactions with correlated nucleon pairs, 
but not antineutrino mode QE-like data~\cite{Fiorentini:2013ezn,Fields:2013zhk}. 
This is in agreement with the fact that the strongly correlated initial state nucleon pairs are essentially neutron-proton pairs,
leading predominantly to proton-proton (neutron-neutron) pairs in neutrino (antineutrino)
charged current reactions, as discussed in
Refs.~\cite{Martini:2009uj,Gran:2013kda,Megias:2014qva,Lovato:2014eva,VanCuyck:2016fab,RuizSimo:2016ikw}.
The GiBUU calculations of Mosel~\textit{et al.}~\cite{Mosel:2014lja} (which include final state interaction for the emitted particles)
of the MINER$\nu$A neutrino CCQE $Q^2$ distribution lead to different conclusions. 
Mosel~\textit{et al.} observe that the channels with a pp or a pn pair are very similar, quite flat, and suppressed.
They also observed an interesting pileup at small $Q^2$ in the Xn channel (see bottom right panel of Fig.~\ref{fig_comp_minerva_q2}).
This is entirely due to the final state interaction (FSI) of the nucleons.

Although FSIs change the number of outgoing nucleons,
Lalakulich {\it et al.} shows that CC interaction with one outgoing nucleon and no pions
in the final state is dominated by genuine CCQE interaction, 
where one outgoing nucleon and no pions can be by genuine CCQE and 2p-2h contributions~\cite{Lalakulich:2012ac}. 
In fact, utilizing two-body kinematics from a muon and a proton final state is
a common technique to select genuine CCQE interactions~\cite{Gran:2006jn,AlcarazAunion:2009ku,Lyubushkin:2008pe}.
However, there is always a question how to use such two-track sample in the analyses,  
because high statistics one-track sample (one outgoing muon and no protons detected) also contains genuine CCQE,  
and experimentalists often try to merge one-track and two-track samples to present the CCQE cross sections.
The data-MC disagreement is interpreted as a mis-modeling of FSI, 
for example NOMAD fits the one- and two-track samples to tune formation time to
have better data-MC agreement in one- and two-track sample simultaneously~\cite{Lyubushkin:2008pe}. 
However, the justification of this is recently revised~\cite{Garvey:2014exa}. 

\begin{figure}[t!]
  \begin{center}
    \includegraphics[width=6cm,valign=m]{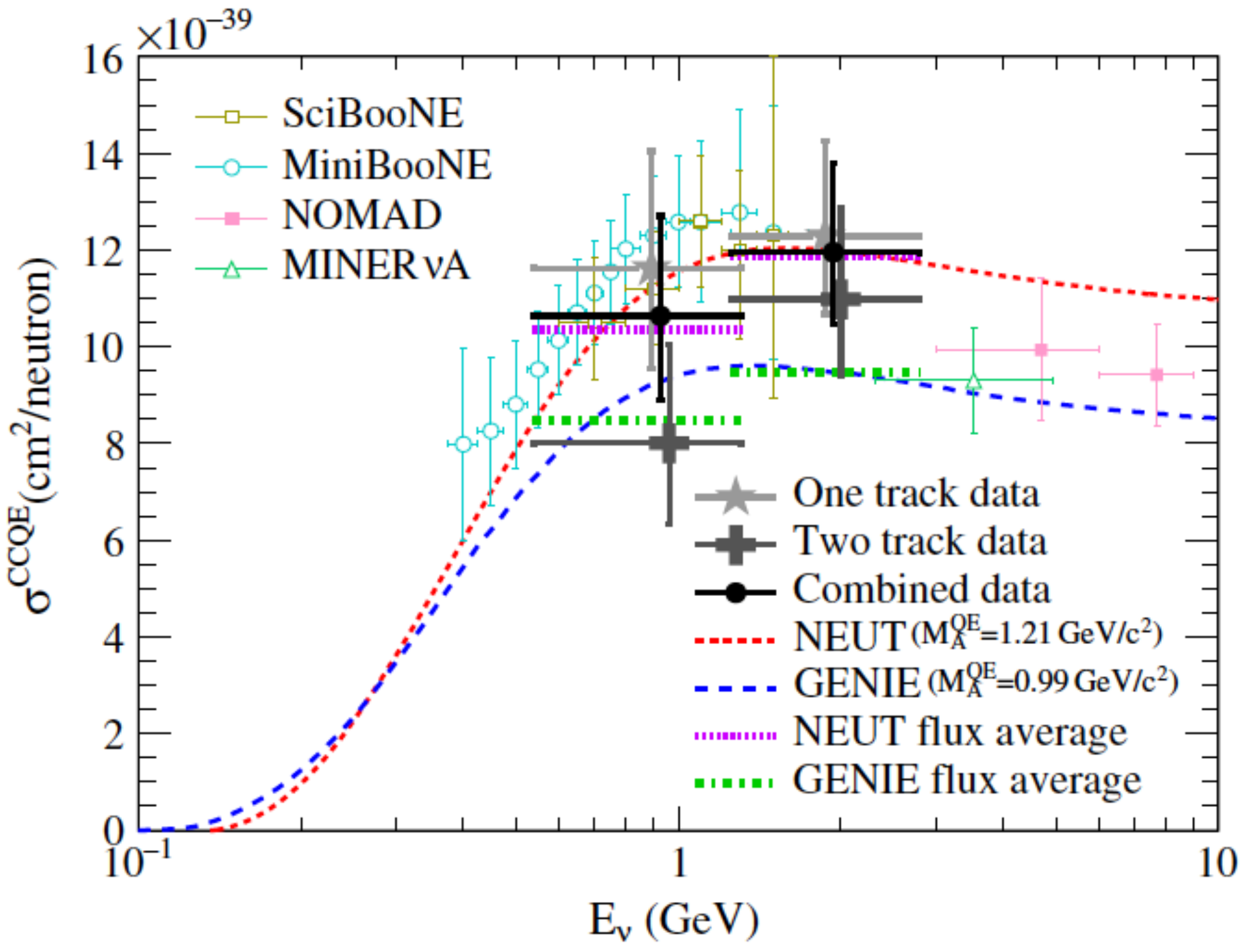}
    \includegraphics[width=6cm,valign=m]{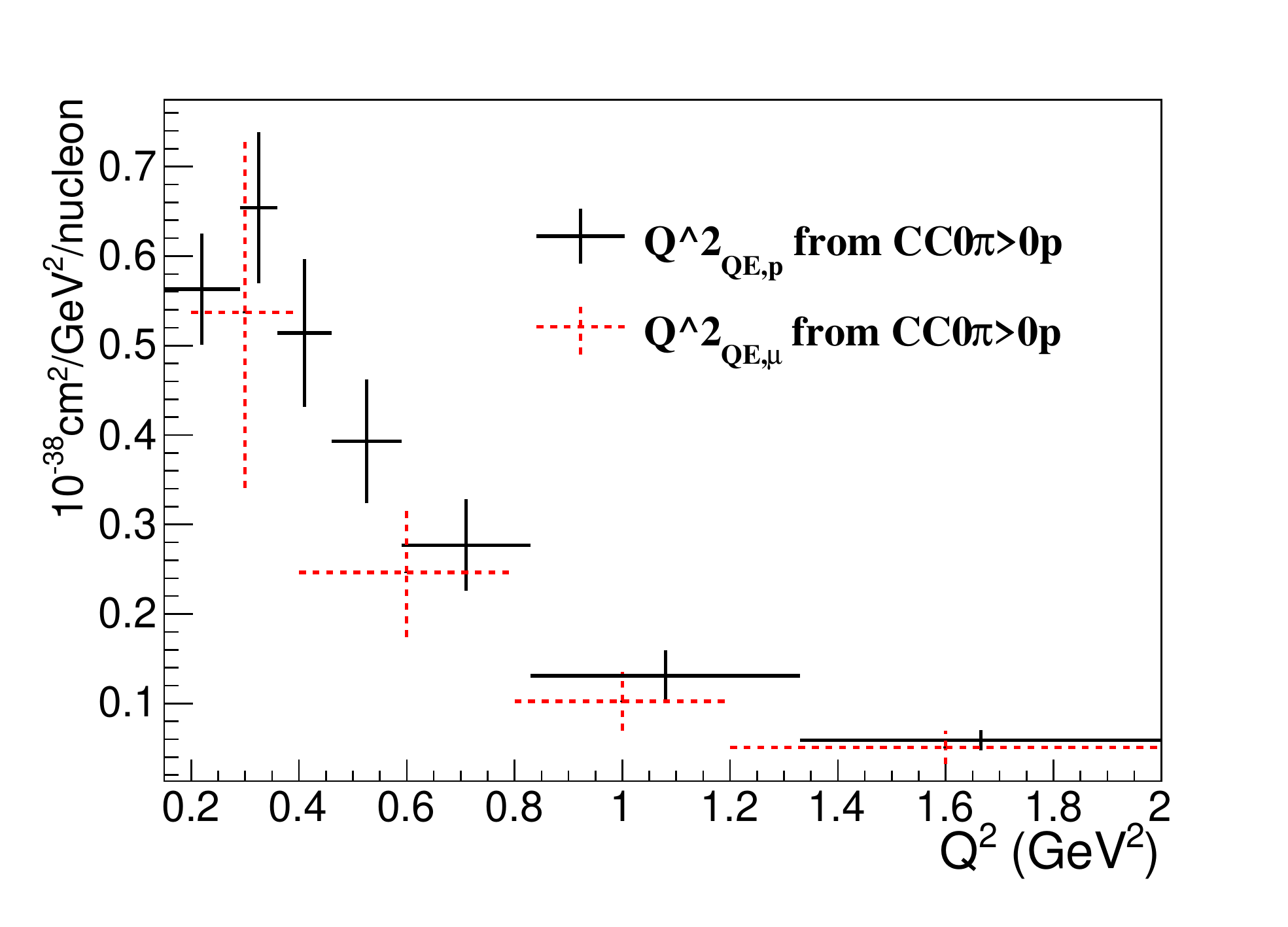}
    \end{center}
\vspace{-2mm}
\caption{
  Left, flux-unfolded total cross section of one- and two-track samples from T2K~\cite{Abe:2015oar}.
  The total cross sections are extracted by assuming RFG model with $M_A=1.21~\uGeV$.
  Right, flux-integrated differential cross section of one- and two-track samples from MINERvA~\cite{Walton:2014esl}.
  Note, two-track sample allows to reconstruct $Q^2$ in two ways, either from muon kinematics or proton kinematics.
}
\label{fig:twotrack}
\end{figure}

T2K INGRID proton module analysis took a different path.
They presented one- and two-track total cross sections separately~\cite{Abe:2015oar}. 
This may be a more honest way to report data, 
because potential disagreements of measured total cross sections
from one- and two-track samples could be an important information to understand hadronic system of neutrino interactions.
If so, merging them by interpreting the difference as FSI within an experimental simulation could bias the data and
erase a potentially important information of hadronic system. 
INGRID proton module analysis indeed discovered something very interesting. 
As you see from Figure~\ref{fig:twotrack}, left,
they found total cross section obtained from two-track sample is
lower than that from the one-track sample,
and the total cross section estimated from
the two-track sample indeed agrees with genuine CCQE cross section.

Two tracks are selected by assuming two-body kinematics of a muon and a proton,
{\it i.e.}, genuine CCQE interaction is assumed. 
In T2K INGRID proton module analysis, this is done by two angular cuts.
The first cut is about the opening angle of a muon and a proton candidate tracks,
and that is required to be $>60^{\circ}$. 
Since genuine CCQE interaction is back-to-back in the center of mass system,
it tends to have larger opening angle comparing to background interactions which does not follow two-body kinematics. 
The second cut is about coplanarity angle. 
Coplanarity angle is defined from projections of a muon and a proton candidate tracks on to
the plane perpendicular to the neutrino beam direction,
and the selection requires this to be $>150^{\circ}$
where genuine CCQE interaction with the rest target without FSIs has $180^{\circ}$.

To measure the total cross section, it is required to correct the acceptance. 
In other words, we need to rely on the MC to calculate how many events missed the detector, 
and such correction is, unfortunately, largely dependent on the assumed interaction model. 
To overcome this situation, the analysis was repeated with three different nuclear models.
Here we show results from RFG model as an example.
Interestingly, regardless of the choice of interaction models, 
the total cross sections extracted from the two-track sample is always lower than that from one-track sample
and the two-track sample total cross sections are more consistent with genuine CCQE models.
This naively shows two-track sample succeeded to select genuine CCQE correctly.  
Then the remaining task is the interpretation of one-track sample..., 
why is this larger than genuine CCQE cross section?  
According to Lalakulich {\it et al.}~\cite{Lalakulich:2012ac},
higher nucleon multiplicity data contains a higher fraction of 2p-2h components, after taking into account FSIs.
This indicates that we would measure a higher cross section for the two-track sample, not the one-track sample,
because the chance that events with multinucleon emissions end up as two-track sample by FSI (=losing one proton by FSI)
is higher and potential 2p-2h contributions should contaminate in two-track sample, not one-track sample. 

What is missing from above arguments is the proton detection efficiency,
or other words, the experimental performance of proton track reconstruction. 
In the case of an interaction with a correlated nucleon pair,
an energy-momentum transferred to the hadronic system is shared with two or more nucleons,
and the detection efficiency of outgoing protons from such interaction may be lower than protons from genuine CCQE. 
If this is the case, the 2p-2h contribution would end up in one-track sample, not two-track sample, 
simply because no protons can exceed the detection threshold and reconstructed.
The problem is such detection threshold is seldom quoted in experimental papers.
It is relatively easier to estimate the detection threshold for a Cherenkov detector,
where Cherenkov thresholds of particles can be interpreted a detection threshold.
For the tracker experiment, such as T2K INGRID,
proton detection efficiency or proton track reconstruction efficiency
is a function of both energy and angles.
Nevertheless, MINERvA and ArgoNeuT quote 110 and 21 MeV, respectively,
as the detection threshold of the proton kinetic energy for their $\mu+p$ two-track sample~\cite{Walton:2014esl}
and $\mu+2p$ three-track sample~\cite{Acciarri:2014gev}. 

As we mentioned, to overcome the interaction model dependence of the acceptance correction,
T2K used several nuclear models for the detector effect unfolding process to evaluate
the size of interaction model dependencies. 
One result comes with the unfolding including np-nh model from Nieves {\it et al.}~\cite{Nieves:2011pp}.
In this case np-nh contribution is treated explicitly as a background,
and one would expect both one- and two-track samples should provide the same cross sections,
and more or less agree with RFG model with $M_A=0.99 \uGeV$. 
Unfortunately, INGRID data do not support this golden scenario. 

\begin{figure}[t!]
  \begin{center}
    \includegraphics[width=4cm,valign=m]{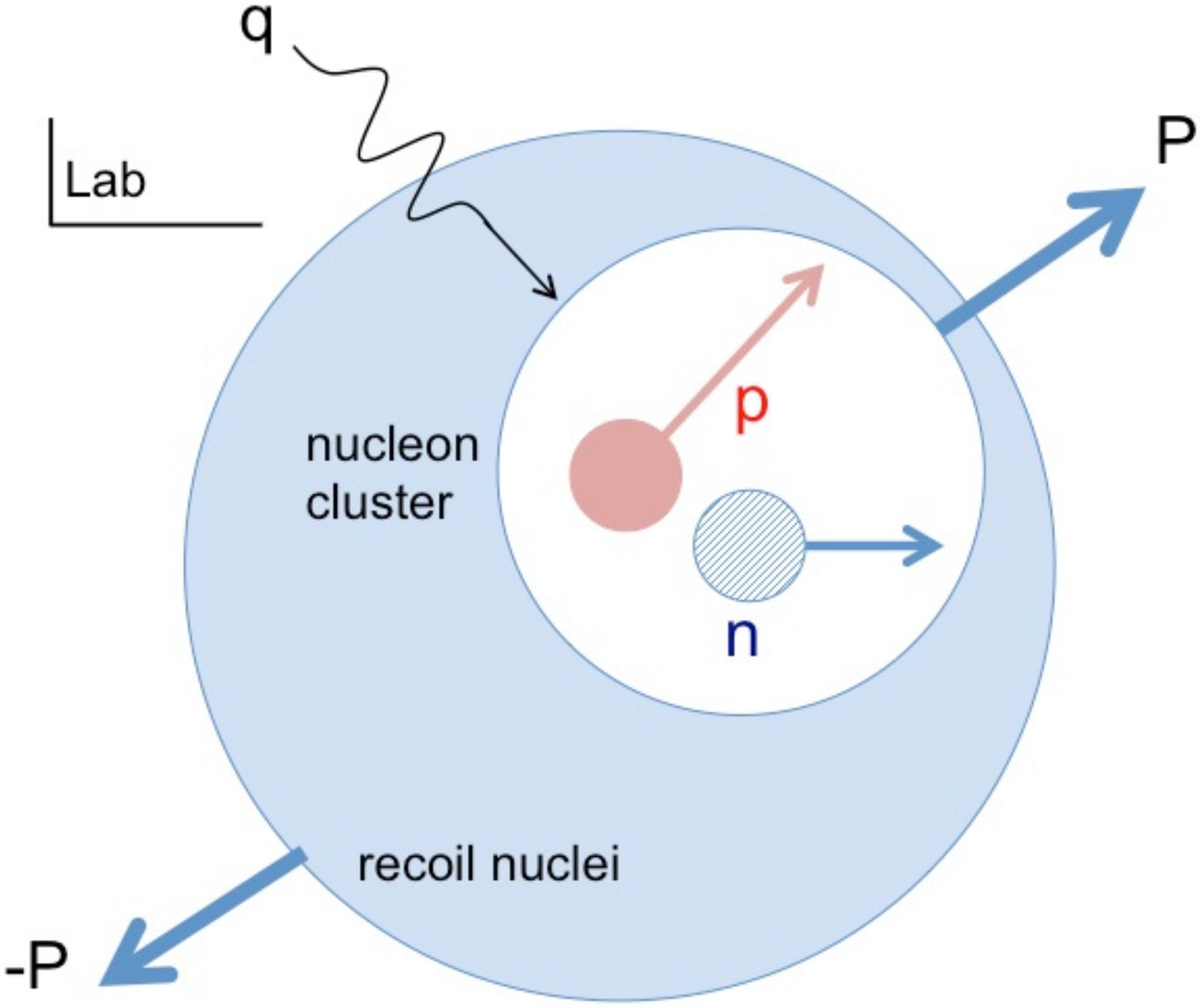}
    \includegraphics[width=4cm,valign=m]{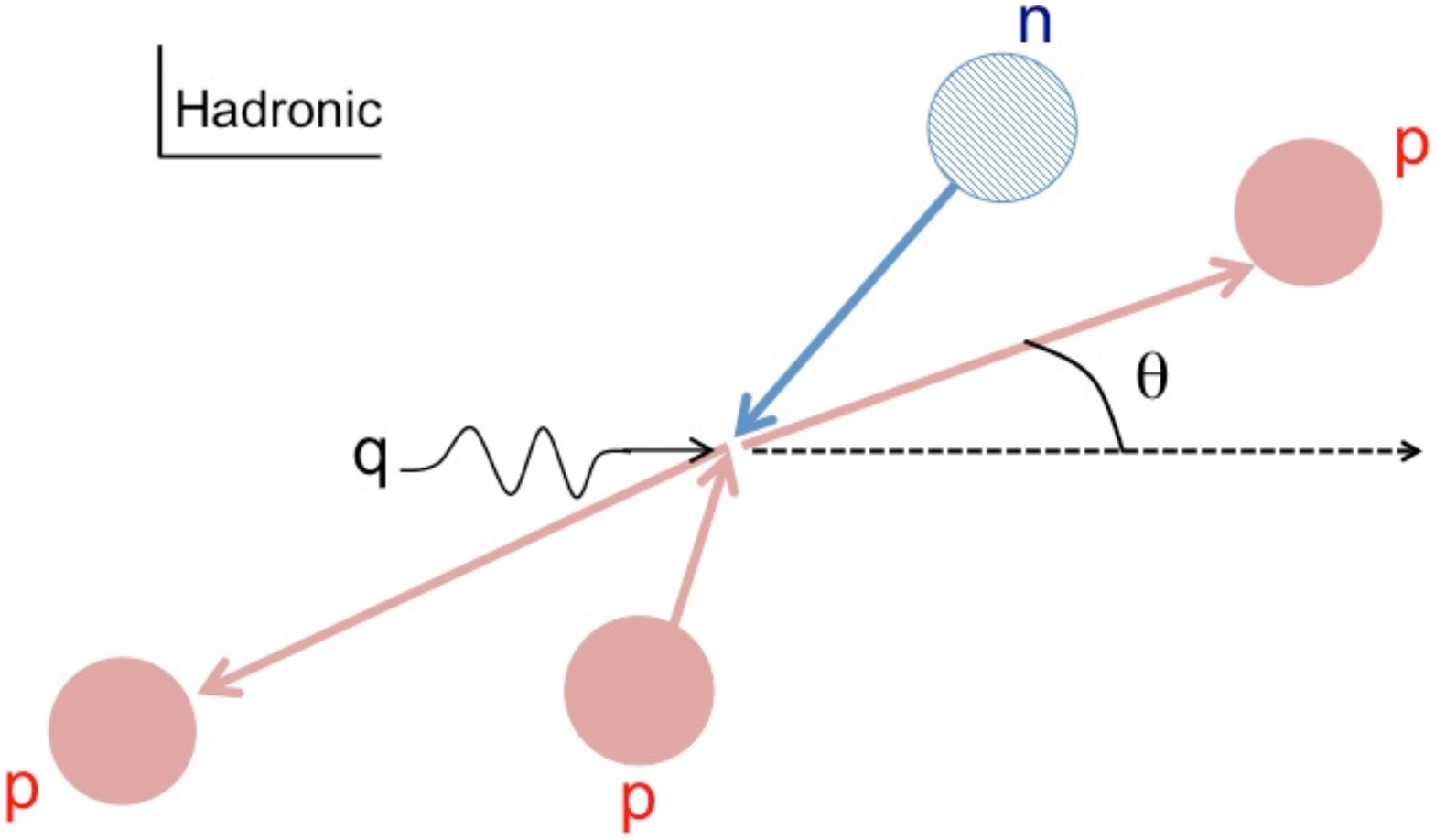}
    \includegraphics[width=4cm,valign=m]{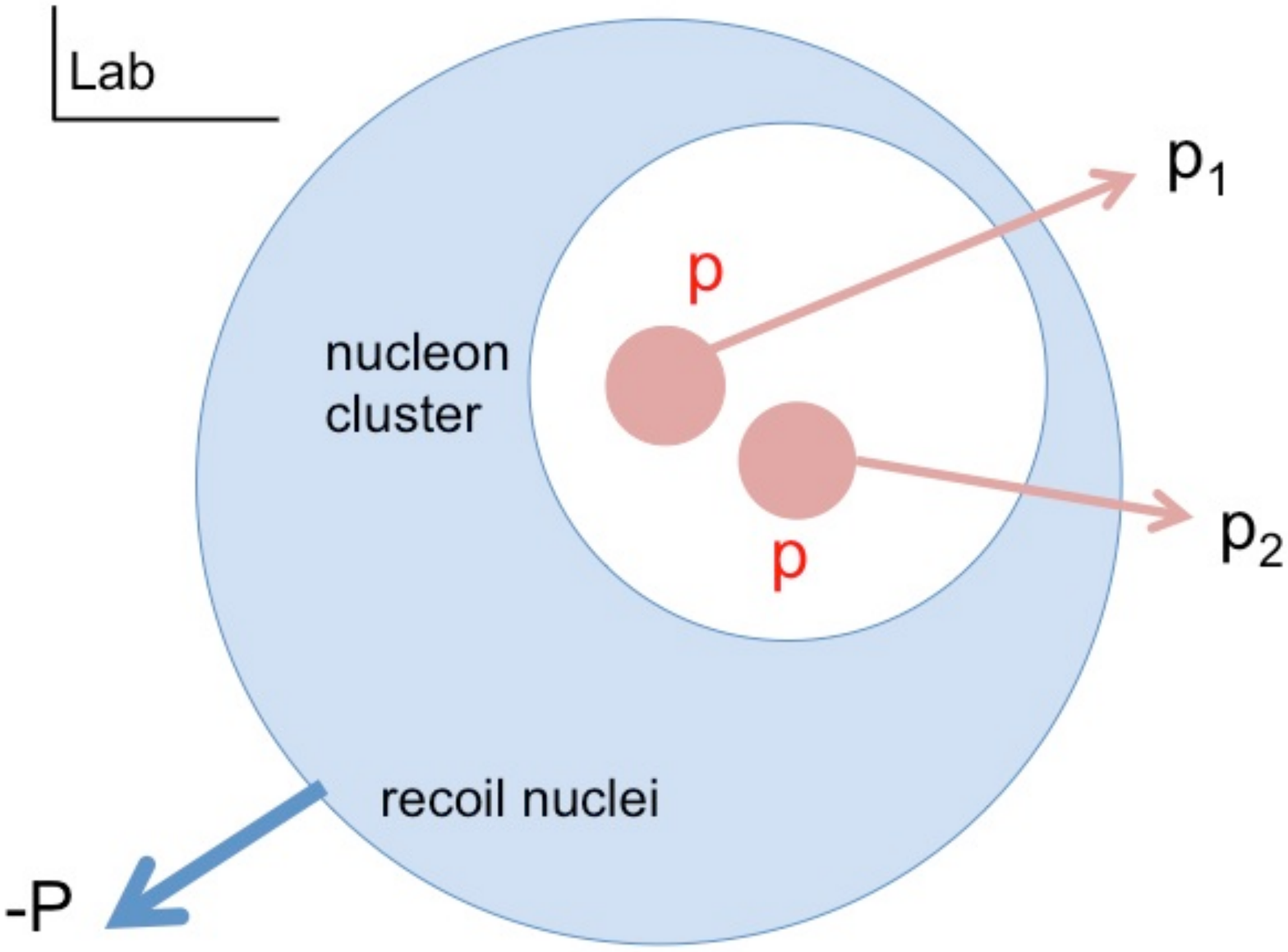}
    \end{center}
\vspace{-2mm}
\caption{
A cartoon to describe the nucleon cluster model.
First, nucleon pair is chosen from the Fermi sea and the leftover system is recoiled (left),
then 4 vectors of energy-momentum transfer and 2 nucleons make the center-of-mass system (hadronic system),
and it decays isotropic in this system (middle).
Finally, they are boosted in the lab frame to simulate outgoing 2 nucleons (right)~\cite{Katori:2013eoa}.
}
\label{fig:ncm}
\end{figure}

On top of proton detection efficiency and proton FSI errors,  
at this moment any neutrino interaction generators do not even have
a reliable model of outgoing nucleons from correlated nucleon pairs. 
No aforementioned theoretical models provide final nucleon distributions, 
and to simulate outgoing hadrons in the experiment, 
na\"{i}ve ``nucleon cluster model'' is used,
independently developed by Sobczyk (NuWro) and Andreopoulos and Dytman (GENIE) ~\cite{Katori:2013eoa,Sobczyk:2012ms}.
Figure~\ref{fig:ncm} shows a graphic representation of the model. 
In this model, two nucleons (nucleon cluster) and energy-momentum transfer four vectors make a hadronic system, 
and the nucleon cluster decays to two nucleons in this frame,
then boost back to the lab frame to simulate outgoing nucleons. 
This model was accepted in NuWro, NEUT, and GENIE. 
Although this is a reasonable approach for an initial guess,
validity of this model is unknown.
For the state of the art of theoretical studies on multinucleon emission
in terms of hadronic variables see Sec.\ref{subsubsec_theonppp}.
The NOvA oscillation analysis presented in Neutrino 2016~\cite{NOvA_Nu2016}
took into account multinucleon excitations as implemented in GENIE
with final state nucleons produced by the nucleon cluster model. 
The experiment successfully reduced the error associated to the hadronic energy deposit. 
This suggests, although the details of the nucleon kinematics needs to be investigated,
the current empirical model is not too off from the data
as long as the experiment only focus on the total hadron energy deposit.

MINERvA published two-track sample differential cross sections.
In MINERvA two-track sample analysis, Figure~\ref{fig:twotrack}, right,
$Q^2$ is reconstructed from both muon kinematics, $Q^2_{QE}(CC0\pi>0p)$,
and proton kinematics, $Q^2_{QE,p}(CC0\pi>0p)$.
The differential cross sections of them agree well.
Here, we add  ``$>0p$'' to clarify that MINERvA sample is defined as ``1 muon and 0 pion and at least 1 proton''.
The main difference of T2K INGRID and MINERvA analyses is the definition of the signal.
In the former, the signal is defined as ``genuine CCQE'', and CCQE-like intrinsic background is subtracted. 
In the latter, the signal is defined from the final state particles (CC0$\pi>$0p). 
We will discuss the advantage of such definition more in Sec.~\ref{sec:ccpip}.
There are two immediate consequences from such signal definition;
(i) intrinsic backgrounds,
mainly $\De$-resonance with pion absorption, are counted as signal and not subtracted,
and (ii), unlike T2K INGRID analysis, selection does not assume genuine CCQE interaction 
and the cuts are not applied to the angular distribution of a muon and a proton. 

The MINERvA two-track data indeed agree well with GENIE prediction (Fig.~4 of Ref.~\cite{Walton:2014esl}).
This is a striking difference comparing with Fig.~\ref{fig_comp_minerva_q2},
where all models (including GENIE) over-estimate differential cross section for one-track sample.
This looks contrary with the T2K two-track analysis where total cross section obtained from
two-track sample is smaller than one-track sample. 
However, as noted above those two analyses have different concept about what they measured,
and the simple comparison is not easy.

Note, in this MINERvA analysis, pion absorption in the nuclei is signal,
but pion absorption in the detector (so called secondary interaction)
is considered inefficiency of the detector and background,
even though the final state particles is one muon and one proton for both case. 
Such intrinsic background is carefully tuned based on data before the subtraction, 
and depending on $W$ region it is tuned down as large as 50\%.
On the other hand, similar treatment in one-track sample modify the intrinsic background
at most 15\%~\cite{Fiorentini:2013ezn,Fields:2013zhk}.
This large reduction of background is a worry of this analysis, in fact,
this is common for several MINERvA analyses with hadron final states
(this will be discussed more in Sec.~\ref{sec:pion}). 
All in all, it is too early to draw any conclusions from these results,
but T2K, MINERvA, and ArgoNeuT (see subsec. \ref{subsec_argoneut}) collaborations certainly initiate new type of analyses
and a clear ``path to the forward'' for the neutrino interaction physics community.

\subsubsection{MINERvA $\frac{d^2 \sigma}{dE_{avail} dq}$\label{subsubsec_minerva_evail}}

Another charged-current $\nu_\mu$-$^{12}$C measurement which clearly goes in this direction is
the one presented by the MINERvA collaboration in Ref.~\cite{Rodrigues:2015hik} 
where the observed hadronic energy is combined with muon kinematics allowing to give
the results in terms of a pair of variables which separate genuine QE and $\Delta$ resonance events,
like in Fig.~\ref{fig_risp_q600_diversicanali} and in inclusive electron scattering experiment.
In the case of Ref.~\cite{Rodrigues:2015hik} these two variables are the magnitude of three-momentum transfer $q=|{\bf{q}}|$ and
the hadronic energy available to produce activity in the detector $E_{avail}$,
which is the sum of proton and charged pion kinetic energy,
plus neutral pion, electron, and photon total energy.
The $E_{avail}$ observable is closely related to the transferred energy $\omega$.
The reason why $E_{avail}$ is preferred to $\omega$ is that 
the reconstruction of the energy transfer $\omega$ requires additional
MC dependent corrections to correct nucleon removal energy and unobserved neutrons. 
The reconstructed $E_{avail}$ is estimated using just the 
calorimetric sum of energy (not associated with the muon) 
in the central tracker region of the MINERvA detector and the electromagnetic 
calorimeter region immediately downstream of the tracker. An unfolding procedure is 
applied to translate the data from reconstructed quantities to true $(E_{avail},|{\bf{q}}|)$.
Note, this unfolding involves a MC dependent correction to estimate $\om$ from $E_{avail}$, to obtain $|{\bf{q}}|$. 
Giving the double differential cross sections in terms of these two variables, \textit{i.e.} 
$\frac{d^2 \sigma}{dE_{avail} dq}$, two different nuclear-medium effects are isolated: 
the RPA suppression which has a significant effect on the lowest  $E_{avail}$ bins of $\frac{d^2 \sigma}{dE_{avail} dq}$
and the necessity of the np-nh excitations to fill the dip region between the quasielastic and the $\Delta$ peaks.
This np-nh contribution is simulated via a GENIE implementation of the model of Nieves \textit{et al.}
with the nucleon cluster model.
The insertion of this contribution mitigates some of the discrepancy between the simulation and the data in the dip region,
however does not fully describe the data.
Another result of Ref.~\cite{Rodrigues:2015hik},
obtained using the Bragg peak technique to effectively count protons,
is that the data have more events with two or more observable protons in the final state,
compared to a default model which includes only genuine quasielastic, $\Delta$ resonance and coherent pion production channels.
The addition of np-nh excitations via a GENIE implementation of the model of Nieves \textit{et al.}
with the nucleon cluster model
reduces once again the discrepancy but more multinucleon events would further improve the agreement with data.

\subsubsection{ArgoNeuT\label{subsec_argoneut}}

\begin{figure}[t!]
  \begin{center}
    \includegraphics[width=6cm,valign=m]{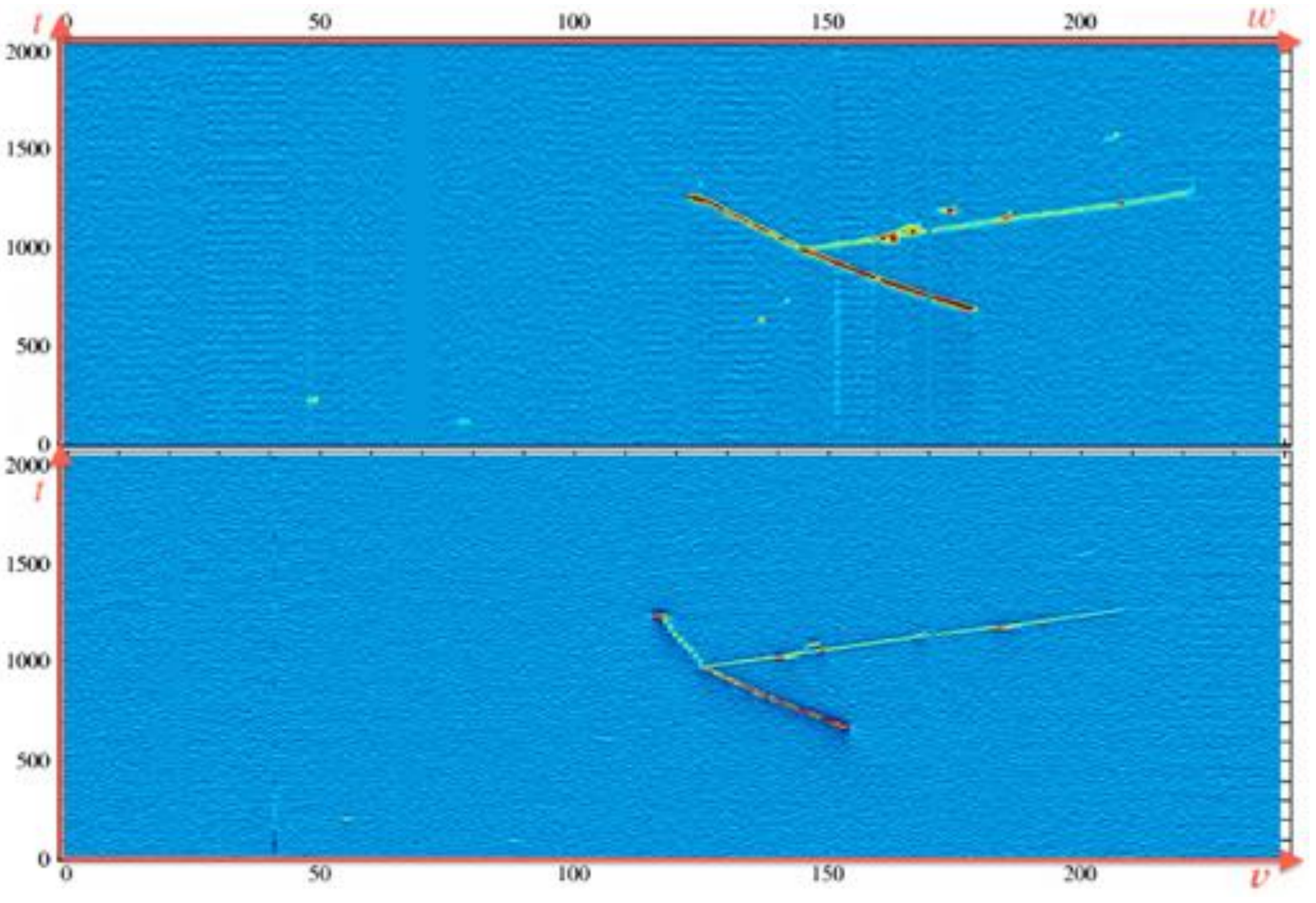}
    \includegraphics[width=6cm,valign=m]{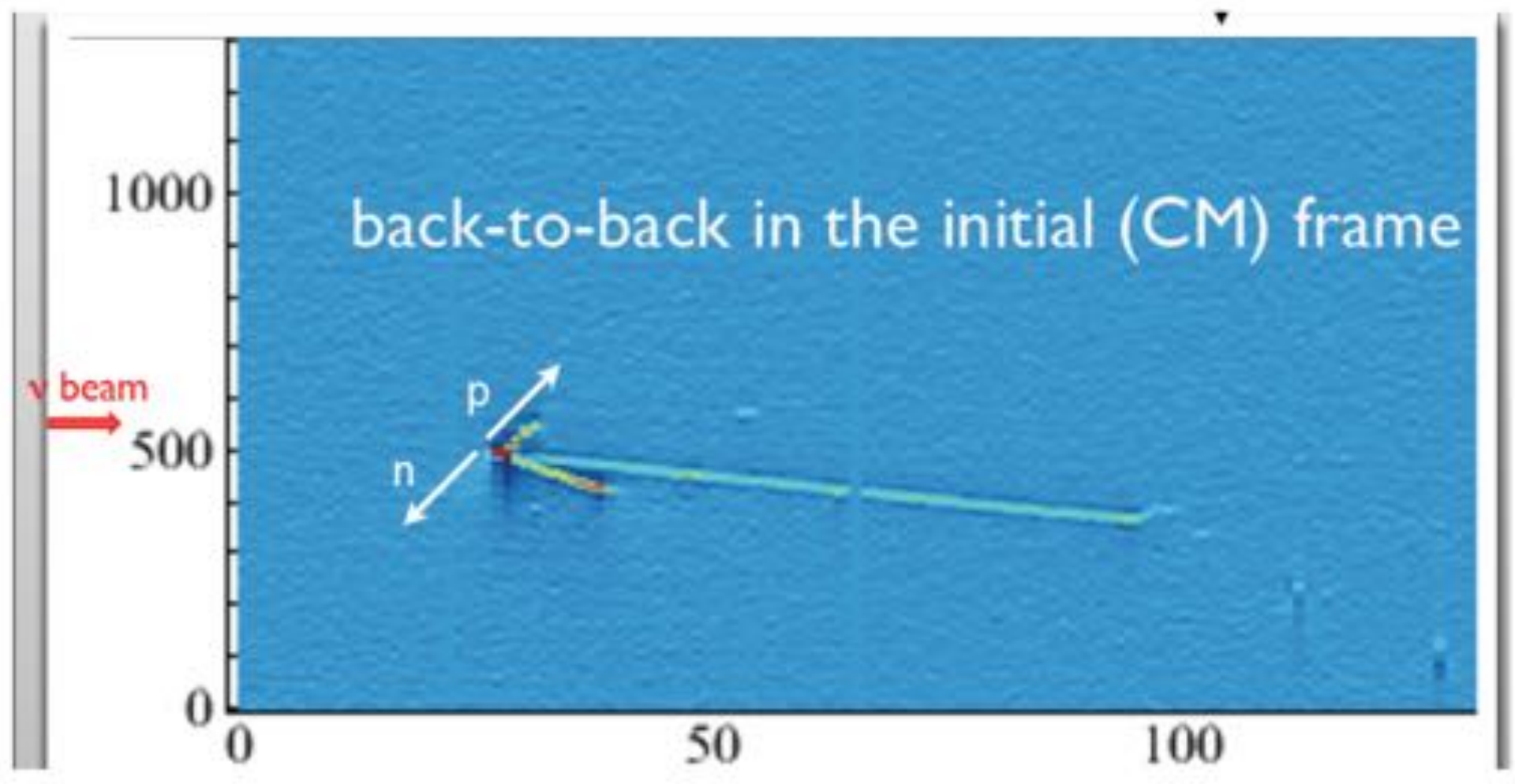}
    \end{center}
\vspace{-2mm}
\caption{
Left: two-dimensional  views  of  one of the hammer events, as measured by  ArgoNeuT~\cite{Acciarri:2014gev} with  a  forward  going muon  and  a  back-to-back proton  pair. Right: two-dimensional views of one of the ArgoNeuT events with a reconstructed back-to-back
np pair in the initial state. The figure is taken from Ref.~\cite{Cavanna:2015sla}.
}
\label{fig:argoneutdisplay}
\end{figure}

A very interesting study on proton pair emission was published by ArgoNeuT ~\cite{Acciarri:2014gev}.  
This experiment is based on Liquid Argon Time Projection
chamber (LArTPC) technique which we discuss in details in Sec.~\ref{sec:lartpc}. 
The reconstruction of the individual proton kinematics (kinetic energy and three momentum) is determined
with good angular resolution and down to a low proton
kinetic energy threshold of 21 MeV or 200 MeV/c of momentum, below the argon Fermi momentum ($k_F \simeq$ 220 MeV/c). 
The capability to detect neutrons emerging from the interaction vertex is very limited in ArgoNeuT because the
detector size is too small to have a significant chance to
allow neutrons to convert into visible protons in the
LArTPC volume before escaping. Taking advantage of the reconstruction capabilities of
LArTPCs, individual events are categorized in terms of
exclusive topologies. 
In Ref.~\cite{Acciarri:2014gev} the ArgoNeuT collaboration search for possible
hints of nucleon-nucleon correlations in the data by specifically
looking for exclusive $\nu_\mu$ CC 0-pion events with 2 protons in the final state, i.e., the ($\mu^- + 2 p$) triple coincidence topology,
as an analogy of JLab Hall A triple coincidence measurement~\cite{Shneor:2007tu}.
Here, neutrino CC interactions make $\mu^- + 2 p$ topology from more abundant np pair ($\numu+(np)\to\mu^-+p+p$),
but electron scatterings make  $e^- + 2 p$ topology from less abundant pp pair ($e^-+(pp)\to e^-+p+p$).


They collected 30 events of this type, 19 of which have both protons above the $^{40}$Ar Fermi momentum. 
Out of these 19 events, 4 are found with the pair in a back-to-back configuration ($\cos \gamma<0.95$) in the \textit{laboratory frame}. 
Visually, the signature of these events gives
the appearance of a hammer, with the muon forming
the handle and the back-to-back protons forming the
head. As an example, the two-dimensional views
from the two wire planes of the LArTPC for one of
these hammer events are reported in the left panel of Fig.~\ref{fig:argoneutdisplay}. 
Another fraction (4 events) of the remaining 15 events was found compatible with a reconstructed back-to-back
configuration of np pair in the \textit{initial state} (CM frame), right panel of Fig.~\ref{fig:argoneutdisplay}.

The ArgoNeuT collaboration argued that the hammer events are most likely due to pionless (positive charged) resonance (RES) mechanisms
involving a pre-existing short range correlated np pair in the nucleus. More precisely they suggested two mechanisms: 
(i) nucleon RES excitation and subsequent two-body absorption of the decay $\pi^+$ by a SRC pair,
leading hence to 3p-3h in the final state and (ii) RES formation inside a SRC pair (hit nucleon in the pair) and
de-excitation through multi-body collision within the A-2 nuclear system,
leading to 2p-2h in the final state.
We could rephrase these two mechanisms as (i) FSI pion absorption and (ii) $\Delta$-MEC contribution. 

In Ref.~\cite{Niewczas:2015iea} Niewczas and Sobczyk analyzed these hammer events by using the NuWro Monte Carlo event generator~\cite{NuWro}.
The result of this study is that most of these hammer events come from the RES and MEC mechanisms.
However NuWro cannot explain the fact that so many (4 or more) hammer events are contained in the samples of 30 or 19 ArgoNeuT events.
According to NuWro,
the probability to have 4 or more hammer events out of 30 and 19 events is of ∼3\% and ∼1\% respectively. 

In Ref.~\cite{Weinstein:2016inx} Weinstein \textit{et al.} modeled these hammer events with two semi-classical models. 
The first model describes pion production on a nucleon followed by pion absorption on an NN pair and the second model 
describes resonance excitation (primarily $\Delta (1232)$) followed by de-excitation via the reaction $\Delta N \to pp$. 
The first model -- pion production and re-absorption process -- results in events with two back-to-back protons
each with momentum of about 500 MeV/c and moderate transverse missing momentum, very similar to that of
the observed ArgoNeuT events. Contrary to the claims of Ref.~\cite{Acciarri:2014gev} the final state distribution of pp pairs
is relatively insensitive to the details of the ``initial pair configuration'' 
(relative momentum of the NN pair and its center of mass momentum distribution). Furthermore this model predicts that a third
nucleon is emitted from the nucleus and that about half the time this third nucleon is an easily detectable proton. 
The results of the second model ($\Delta N \to pp$) agree less well with the four observed hammer events: 
the protons are significantly less back-to-back, have higher momentum, and have less missing transverse momentum 
than the pion production and re-absorption model. One should be able to decisively distinguish between the two models
by the fraction of hammer events with a third emitted proton. In summary, the conclusion of  Ref.~\cite{Weinstein:2016inx} 
is that ArgoNeuT hammer events can be described by a simple pion production and re-absorption model. 
Another conclusion is that these events can be used to determine the incident neutrino energy, but cannot teach us anything significant about short range correlated NN pairs.

As already mentioned, beyond the hammer events, ArgoNeuT detected also 4 events compatible with a reconstructed back-to-back
configuration of np pair in the \textit{initial state} (CM frame). These events have been suggested to correspond  
to the $^{12}$C$(e,e'np)$ events observed in the JLab experiment~\cite{Subedi:2008zz} where the projectile scatters from 
one nucleon in the correlated pair and its correlated partner then emerges from the nucleus. Sometimes in literature 
one refers to these events as one-body quasi-elastic interaction on a neutron in a SRC pair. 
From a theoretical point of view these events would correspond to the  
NN correlated pair contribution included by considering the NN correlation two-body current $J^{\mu}_{\textrm{NN-corr}}$ 
in the independent particle approaches. 
In the correlated basis approaches these NN correlations are included in the description of the nuclear wave functions, 
and hence are referred as one-body contributions, see Sec. \ref{sec:npnh}. 
The ArgoNeuT results on np pair in the \textit{initial state} was 
obtained with an approach similar to the electron scattering triple coincidence analysis~\cite{Shneor:2007tu}:
the initial momentum of the struck neutron was determined by transfer-momentum
vector subtraction to the higher proton momentum ($\vec{p}^i_n$=$\vec{p}_{p1} - \vec{q}_{\textrm{rec}}$) and the lower momentum
proton ($\vec{p}_{p2}$) was identified as the recoil spectator nucleon from within SRC,
as shown in the right panel of Fig.~\ref{fig:argoneutdisplay}.
The momentum transfer $\vec{q}_{\textrm{rec}}$ is calculated from the reconstructed neutrino energy and
the measured muon kinematics, hence it is a quantity less precisely determined than in electron-scattering experiments. 
In Ref.~\cite{Niewczas:2015iea} Niewczas and Sobczyk  analyzed also these 4 back-to-back in the initial state events. 
They show that if one defines $\vec{p}^i_n$=$\vec{p}_{p1} - \vec{q_{\textrm{rec}}}$, as done by ArgoNeuT,
the shape of the distribution of the cosine of the reconstructed angle between the two nucleon in
the initial state is universal and does not depend much on the dynamical mechanism (genuine QE, SRC, MEC)
behind the appearance of the two-proton final state. 
Hence it is not directly related with existence of SRC nucleon pairs.
Nevertheless they show also that the details of the distribution shape is sensitive to SRC pairs. 
To summarize, the event statistics from ArgoNeuT was very limited, hence not enough to lead to definitive conclusions,
by their analysis as well as by the the two subsequent ones \cite{Niewczas:2015iea,Weinstein:2016inx}.  
However examples of important observable, to be further investigate in future experiments, was done for the first time in the context of neutrino scattering experiments.

\subsubsection{Theoretical studies\label{subsubsec_theonppp}}

Beyond the MC and semi-classical studies, from a theoretical point of view only few,
and very recent, calculations have been performed focusing on hadronic information in connection with the neutrino-nucleus scattering.
We can essentially mention and discuss two studies, one of Ruiz Simo \textit{et al.} ~\cite{RuizSimo:2016ikw}
and one of Van Cuyck \textit{et al.} ~\cite{VanCuyck:2016fab}, related to the emission of nucleon pairs induced by MEC and SRC, respectively. 

In Ref.~\cite{RuizSimo:2016ikw} Ruiz Simo \textit{et al.} investigate the relative effects of MEC
on the separate proton-proton (i.e., neutron-proton in the initial state) and neutron-proton emission channels.
For this purpose they studied the cross sections and the nuclear responses for 
the semi-inclusive $^{12}$C$(\nu_\mu,\mu^−pp)$ and $^{12}$C$(\nu_\mu,\mu^−np)$ reactions integrated over
the two emitted nucleons. For all the nuclear responses, as well as
for the cross section they obtained that the pp channel (i.e., neutron-proton in the initial state) clearly dominates. 
The pp/np ratio is around 5-6 near the maximum, but its precise value depends on the $\omega$ and 
${\bf{q}}$ variables, i.e. on the kinematics. The np distribution is shifted towards
higher muon energies and smaller lepton scattering angles with respect to the pp one. 
The pp/np ratio, with its $(\omega,|{\bf{q}}|)$ dependence, 
critically depends on the treatment of the interference between the direct and exchange matrix elements,
the so called ``exchange'' contribution. 
For pp pair emission this contribution is almost negligible.
On the other hand, for np emission it is of the same order as the direct contribution. Hence, although the net effect of the
interference (or ``exchange'')  is less than 20\% of the total MEC contribution and often it can be safely disregarded --
for example when one studies cross sections only as a function of lepton variables
(this is the case of the approaches of Martini \textit{et al.} and Nieves \textit{et al.} who discard  these terms) --
the situation is different when one want to separate the pp and pn contributions and calculate their ratio.
The pp/np ratio obtained by Ruiz Simo \textit{et al.} ~\cite{RuizSimo:2016ikw} for the neutrino scattering neutrino
can be compared to the np/pp ratio in the $(e,e')$ reaction studied by the same authors in Ref.~\cite{Simo:2016imi},
because they correspond to the same pairs in the initial state.
For the transverse response that ratio for neutrino scattering is roughly a factor of two smaller than for the electron case. 
For the electron scattering cross sections this ratio was found to be roughly between twelve
and six depending on the kinematics. 
For the specific kinematics of the electron scattering measurement of Ref.\cite{Subedi:2008zz} the value 
obtained by Ruiz Simo \textit{et al.} ~\cite{Simo:2016imi} for the np/pp ratio due to MEC is 6. 
This is not sufficient to explain the factor 18$\pm$5 found in the
experiment \cite{Subedi:2008zz} and attributed to SRC, coming mainly from
the tensor nuclear force. However, it does suggest that in order
to understand in depth the size of SRC effects, the MEC
contributions should also be included.

Theoretical calculations in neutrino scattering for the separate
pp and np contributions due to SRC
have been presented by Van Cuyck \textit{et al.} in Ref. ~\cite{VanCuyck:2016fab} for the $^{12}$C transverse and charge response.
The results are given for a fixed value of the momentum transfer, $q = 400$ MeV/c. 
In this case the contribution of initial np pairs to the transverse response
(the one who dominates the neutrino cross section) is about twice that of the initial nn pairs.
In Ref. ~\cite{VanCuyck:2016fab} Van Cuyck \textit{et al.} 
investigate also exclusive cross sections, giving for the first time the SRC contribution of 2 nucleon knockout process 
for exclusive $^{12}$C$(\nu_\mu,\mu^- N N)$ and semi-exclusive $^{12}$C$(\nu_\mu,\mu^- N)$ reactions.
Some cases of differential cross sections for fixed values of lepton variables are given as a function of hadronic variables.
As an example of exclusive cross section, they consider in-plane kinematics, with both nucleons emitted in the lepton scattering plane. 
They obtain that the cross section is dominated by back-to-back nucleon knock-out, 
the most strength residing in a region with initial center-of-mass momentum $P_{12}<300$ MeV/c. 
Also for the examples of semi-exclusive cross sections they consider, they obtain that these cross sections 
are dominated by pairs with small initial center-of-mass momentum.
As already mentioned Van Cuyck \textit{et al.} consider also inclusive quantities,
such as nuclear responses and double differential cross sections for fixed values of transfer momentum and
lepton kinematics as a function of the transferred energy $\omega$. Beyond the already discussed isospin content of the initial NN SRC pairs,
another interesting message of Ref. ~\cite{VanCuyck:2016fab} is that
SRC similarly affect the vector and axial parts of the two-body currents.  

All the theoretical results discussed up to now refer to $^{12}$C. Since also other nuclear targets, 
such as $^{16}$O and $^{40}$Ar, are used in present and future neutrino experiment,
the mass dependence of np-nh excitations require important investigations.
The mass dependence is strictly related to the range of the pairs interaction. 
For zero-range interactions the mass dependence should go $\propto A$ , 
whereas for the extreme of long-range interaction it should go $\propto A^{2}$. 
No explicit calculations of mass dependence of 2p-2h excitations in neutrino-nucleus scattering have been performed up to now.
However theoretical calculations have been performed in connection with electron scattering reactions. 
We mention for examples the works of Refs.~\cite{Vanhalst:2011es,Vanhalst:2012ur,Vanhalst:2014cqa,Colle:2015ena,Mosel:2016uge}
related to the SRC pairs and the works of Refs.~\cite{VanOrden:1980tg,DePace:2004cr} related to the vector MEC. 
Without entering in details, the $A$ dependence of SRC pairs is found to be close to
linear~\cite{Vanhalst:2014cqa,Colle:2015ena,Mosel:2016uge}. 
Concerning the MEC, in Refs.~\cite{VanOrden:1980tg,DePace:2004cr} it is shown that the one-body (QE)
response scales as $A/k_F$ (scaling of second kind) and that,
even if an exact $k_F$ dependence of the 2p-2h MEC response should be studied numerically,
a rough $k_F$ behavior can be extracted from simplified equations, suggesting that $R_{MEC} \sim  A k_F^3$.
As a consequence for the ratio of the two-body and one body cross sections one has $\frac{\sigma_{MEC}}{\sigma_{QE}} \sim k_F^4 $. 
Thus, as discussed in Ref.~\cite{VanOrden:1980tg} for lighter nuclei,
where $k_F$ is changing more rapidly with increasing $A$, 
the size of the MEC relative to the QE peak changes noticeably as $A$ becomes larger.
As $A$ increases toward heavier nuclei, the nuclear density saturates, causing $k_F$ to slowly approach the nuclear matter value.
This implies that for heavier nuclei all contributions will scale approximately as $A$.
Therefore, while the relative MEC contribution will be largest for heavy nuclei,
it changes most rapidly when comparing cross sections for light nuclei.
For example in Ref.~\cite{VanOrden:1980tg} it is illustrated that the size of the vector MEC
contribution relative to the QE peak increases considerably going from $^{12}$C 
to $^{58.7}$Ni but that there is very little increase in relative size when going from 
$^{58.7}$Ni to $^{208}$Pb. Similar studies should be repeated for the weak two-body currents. 

From an experimental point of view, a detailed comparison of CCQE-like flux-integrated double differential cross sections 
involving a same neutrino flux but different nuclear targets (for example C and O at T2K near detector,
or C and Ar at MiniBooNE and MicroBooNE respectively) could help to determine the $A$ dependence of 2p-2h excitations. 
Theoretical results for MiniBooNE and MicroBooNE are obtained by Gallmeister \textit{et al.}
in Ref.~\cite{Gallmeister:2016dnq} using the last version of GiBUU which includes an improved treatment of 2p-2h channel already mentioned.

We conclude this subsection by mentioning the purely theoretical work of Moreno
\textit{et al.} \cite{Moreno:2014kia} on 
the general and universal formalism for semi-inclusive charged-current (anti)neutrino-nucleus
reactions, namely those where a final-state charged lepton and some other particle
(one nucleon, or one meson, or a photon, or an alpha particle,... ) are detected in coincidence.
The semi-inclusive cross section is the sum/integral over all unobserved particles, excepting only the one that
is presumed to be detected, for example an ejected nucleon.
The formalism is general enough to allow for the presence of MEC.
The main points stressed in this work are that \textit{i)} for CC reactions 10 distinct nuclear response functions contribute to
the semi-inclusive cross section, instead of only 5 (see Eq.(\ref{m_eq_general})) when only the charged lepton is detected (inclusive process);
\textit{ii)} the semi-inclusive responses are all functions of 4 kinematic variables,
whereas the inclusive ones depend on only 2 kinematic variables;
\textit{iii)} many (essentially all) models used up to now to compare with the neutrino flux-integrated differential cross sections function of the charged lepton variables are not
applicable for semi-inclusive studies.

Modeling the coincidence reactions is in demand by the experimental community but it is a very challenging task.

\section{Pion production and inelastic channels\label{sec:pion}}
The pion production channels are traditionally important for neutrino oscillation physics for two main reasons. 
First, charged current pion productions (CC$\pi$), 
mostly charged current single charged pion productions (CC1$\pipm$), often mimic QE interactions.
In genuine QE interactions, two-body kinematics allow us to reconstruct $E_\nu$ and $Q^2$.
On the other hand, if the primary interaction is the neutrino pion production but pion is not identified,
reconstructed energy from the muon kinematics is underestimated 
and this adds a significant bias on neutrino oscillation analysis.
\footnote{This bias from the mis-reconstruction of neutrino energy can be avoided
if the oscillation analysis is based on lepton kinematics, not reconstructed neutrino energy.
This was demonstrated recently by T2K~\cite{Abe:2014tzr}.}
Second, neutral current $\piz$ productions (NC$\piz$), mainly from neutral current single $\piz$ production (NC1$\piz$)
following from $\De$-resonance, often mimics electron-like signal for $\nu_\mu\to\nu_e$ oscillation search.
The electromagnetic shower made by a gamma ray looks like electron-origin for low resolution detectors.
This can happen by either asymmetric $\piz$ decays where one gamma ray carries most of the energy,
or detector inefficiency and only one electromagnetic shower is reconstructed and detected.

Although there are a number of new data and studies,
understanding of neutrino pion production is far from satisfactory.
Currently, around 20-30\% errors are accepted for single pion production channels in oscillation analyses,
due to conflicts between different data sets and models. 
Therefore, oscillation experiments utilize internal measurements, mostly from near detectors,
to constrain errors associated to these channels.
However, such internal constraint cannot remove the all uncertainties associated to these channels. 
Furthermore, as we see from Figure~\ref{fig:flux_all}, many current and future experiments,
including NOvA~\cite{Adamson:2016tbq} and DUNE~\cite{DUNE_CDR2}
\footnote{On top of these experiments, 4-10 GeV energy region is also important for atmospheric neutrino mass ordering measurements,
such as PINGU~\cite{PINGU}, ORCA~\cite{KM3NeT}, and INO~\cite{INO}.},
are located right on the region where meson production processes are significant. 
Therefore, further understanding of these channels are an urgent program to reduce errors on current and future oscillation physics. 
In this section, we review recent data mainly from MINERvA to discuss ongoing issues of neutrino pion production physics.

\subsection{Charged current charged pion production~\label{sec:ccpip}}

In previous sections, we stressed the importance of the measurement of flux-integrated differential cross section function of
measurable variables.
The first such measurement for CC1$\pipm$ interaction was performed in MiniBooNE~\cite{MB_CCpip}.
Later, MINERvA~\cite{Eberly:2014mra} and T2K~\cite{Abe:2016aoo} performed the same measurement
with a different beam and detector.
Here, analyses require to tag an electron or positron from
$\pi^\pm\to\mu^\pm\to e^\pm$ decay chains to identify charged pions. 
However, $\pi^-$ is almost 100\% absorbed by nuclei before it decays and is not observed.
Thus, the CC1$\pipm$ sample is effectively CC1$\pip$ in these experiments even though MiniBooNE and MINERvA
do not perform sign selections. T2K ensures $\pip$ by the charge selection at the TPCs. 

\begin{figure}[tb]
  \begin{center}
    \includegraphics[width=\textwidth]{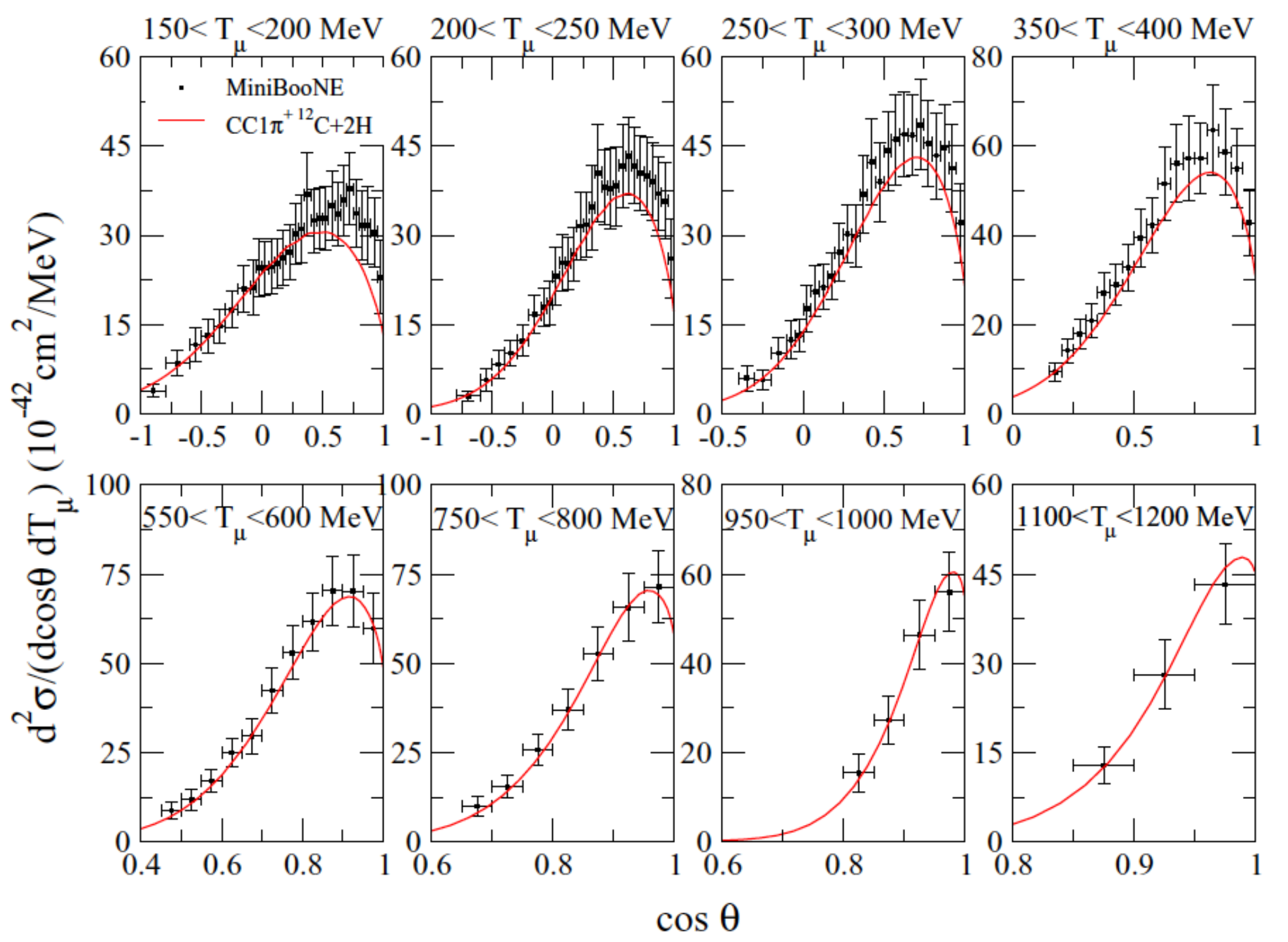}
    \end{center}
\vspace{-2mm}
\caption{
MiniBooNE CC1$\pi$ flux-integrated double differential cross section of muon kinematics ~\cite{MB_CCpip}
compared with the theoretical results of Martini {\it al.} published in Ref.~\cite{Martini:2014dqa}.
}
\label{fig:ccpip_muon}
\end{figure}

Figure~\ref{fig:ccpip_muon} compares the MiniBooNE CC1$\pi$ double differential cross section of 
muon kinematics (muon energy and scattering angle)
with the theoretical calculations  Martini {\it et al.}~\cite{Martini:2014dqa}
which, beyond the genuine QE and the np-nh excitations (see in Sec.~\ref{sec:npnh}),
allow a description of coherent and incoherent one pion production.
Although the low energy region shows a little deficit,
this theoretical approach correctly reproduces both normalization and shape of
the CC1$\pip$ double differential cross section function of muonic variables.
Reasonable agreement with MiniBooNE CC1$\pip$ differential cross sections function of muonic variables is
also obtained by the SuSA ~\cite{Ivanov:2012fm} and GiBUU ~\cite{Lalakulich:2012cj} approaches.
The GiBUU results show a good agreement in shape and a tendency to underestimate data.
However, the situation is very different for pion kinematics, as we discuss below.  

\begin{figure}[tb]
  \begin{center}
    \includegraphics[width=6cm,valign=m]{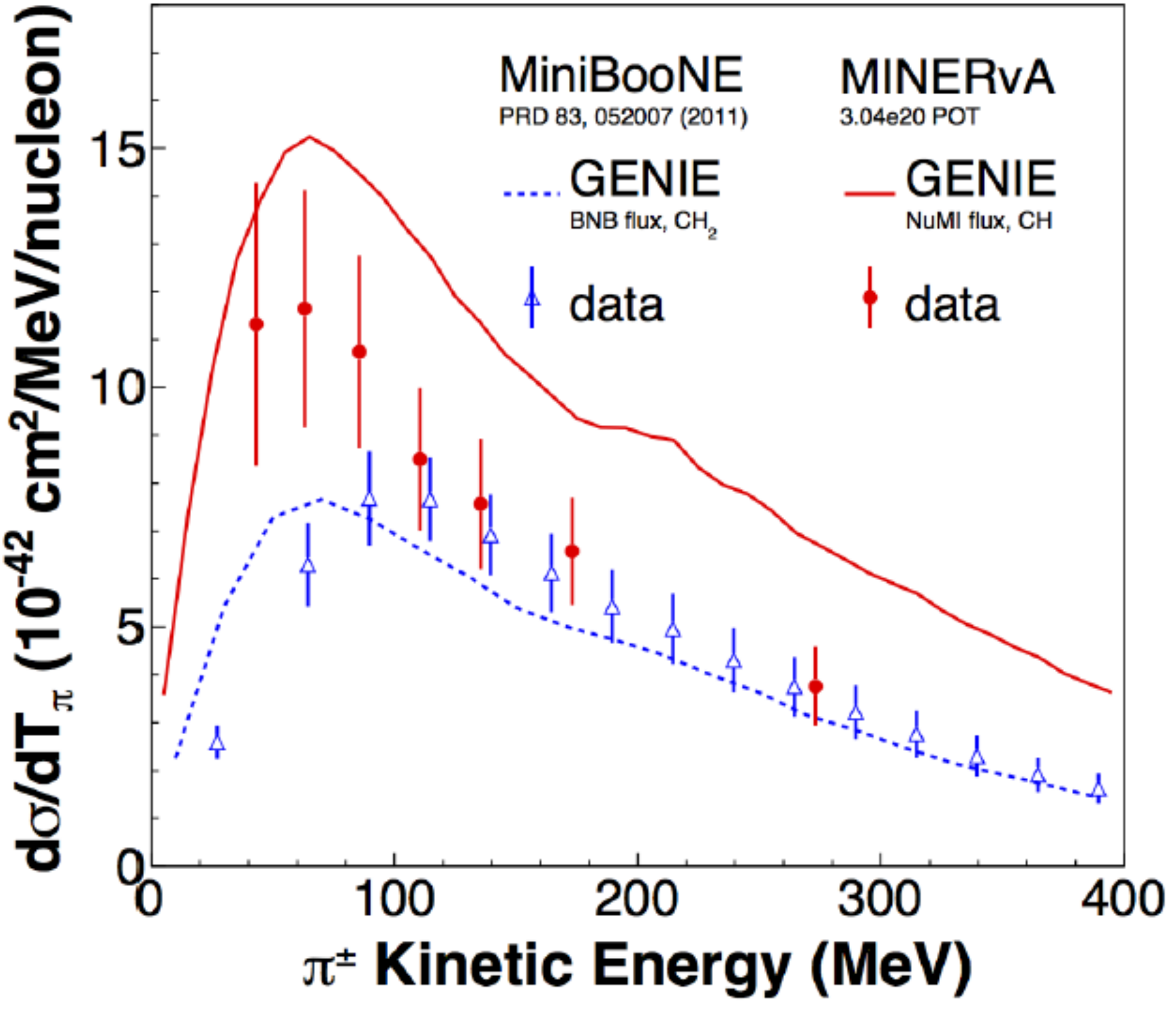}
    \includegraphics[width=6cm,valign=m]{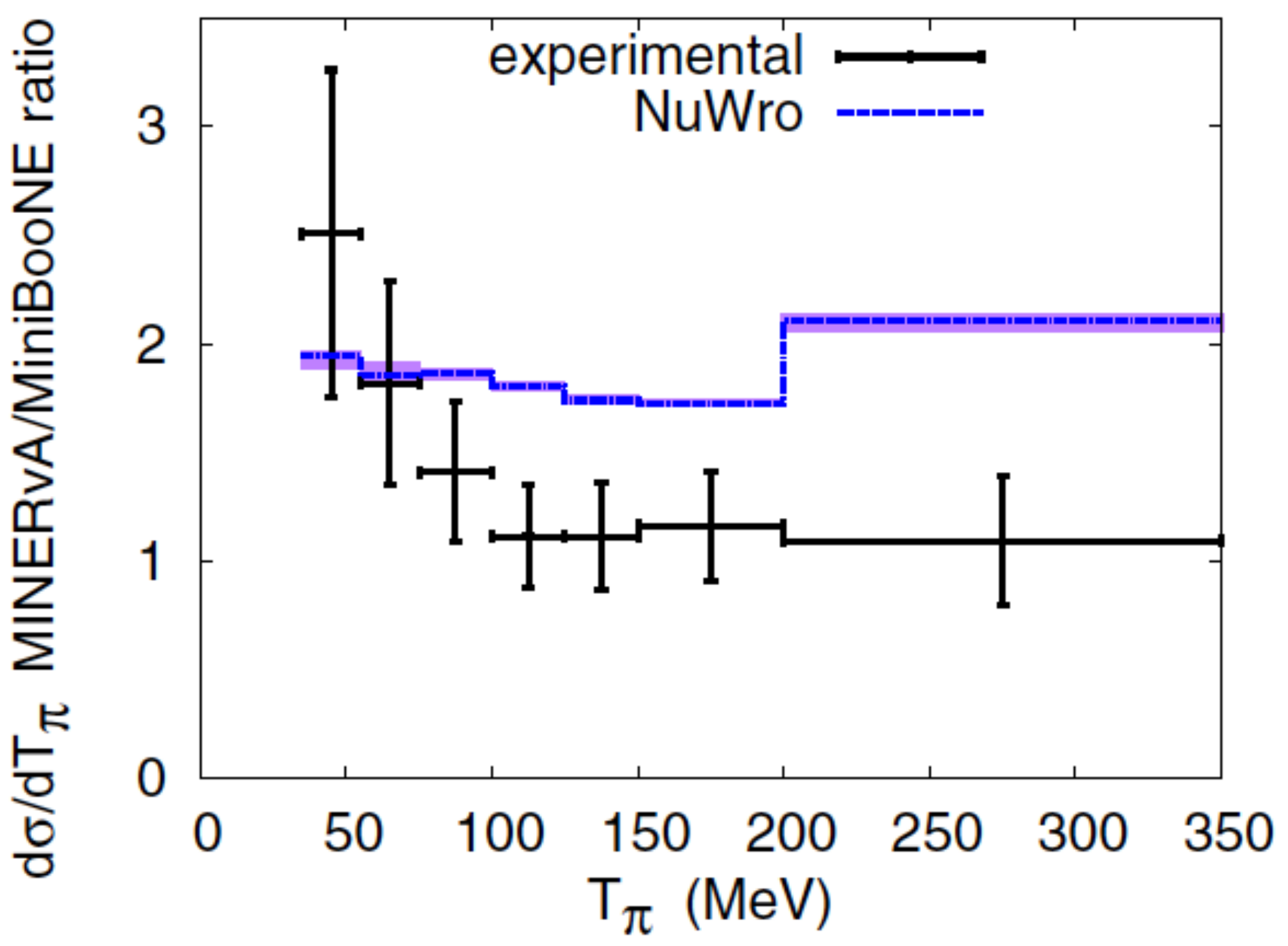}
    \end{center}
\vspace{-2mm}
\caption{
Flux-integrated differential cross section of charged pion kinetic energy from CC1$\pip$ interactions.
In the left, MiniBooNE and MINERvA data are compared with GENIE prediction~\cite{Eberly:2014mra}.
In the right, MINERvA to MiniBooNE data ratio is compared with the same ratio from NuWro~\cite{Sobczyk_MINERvApion}.
}
\label{fig:ccpip}
\end{figure}

Figure~\ref{fig:ccpip}, left, shows a comparison between the MiniBooNE and MINERvA flux-integrated differential
cross section function of the pion kinetic energy and the predictions from GENIE.
This type of plot is not easy to read, 
and here we listed few tips about how to read this plot. 

\begin{itemize}
\item[1] Topology-based signal definition: The data sample is defined from the final state particles which
include 1 muon and 1 charged pion and any number of nucleons (CC1$\pipm$). 
\item[2] Direct observables: The plot is the function of pion kinetic energy
which is a directly observable kinematic variable,
not inferred kinematic variables. 
\item[3] Flux-integrated cross section: The plot compares MiniBooNE and MINERvA data,
however, they use different neutrino beams and two data sets are not directly comparable in a same plot.
\end{itemize}

\subsubsection{Topology-based signal definition}

\begin{figure}[tb]
  \begin{center}
    \includegraphics[width=5cm]{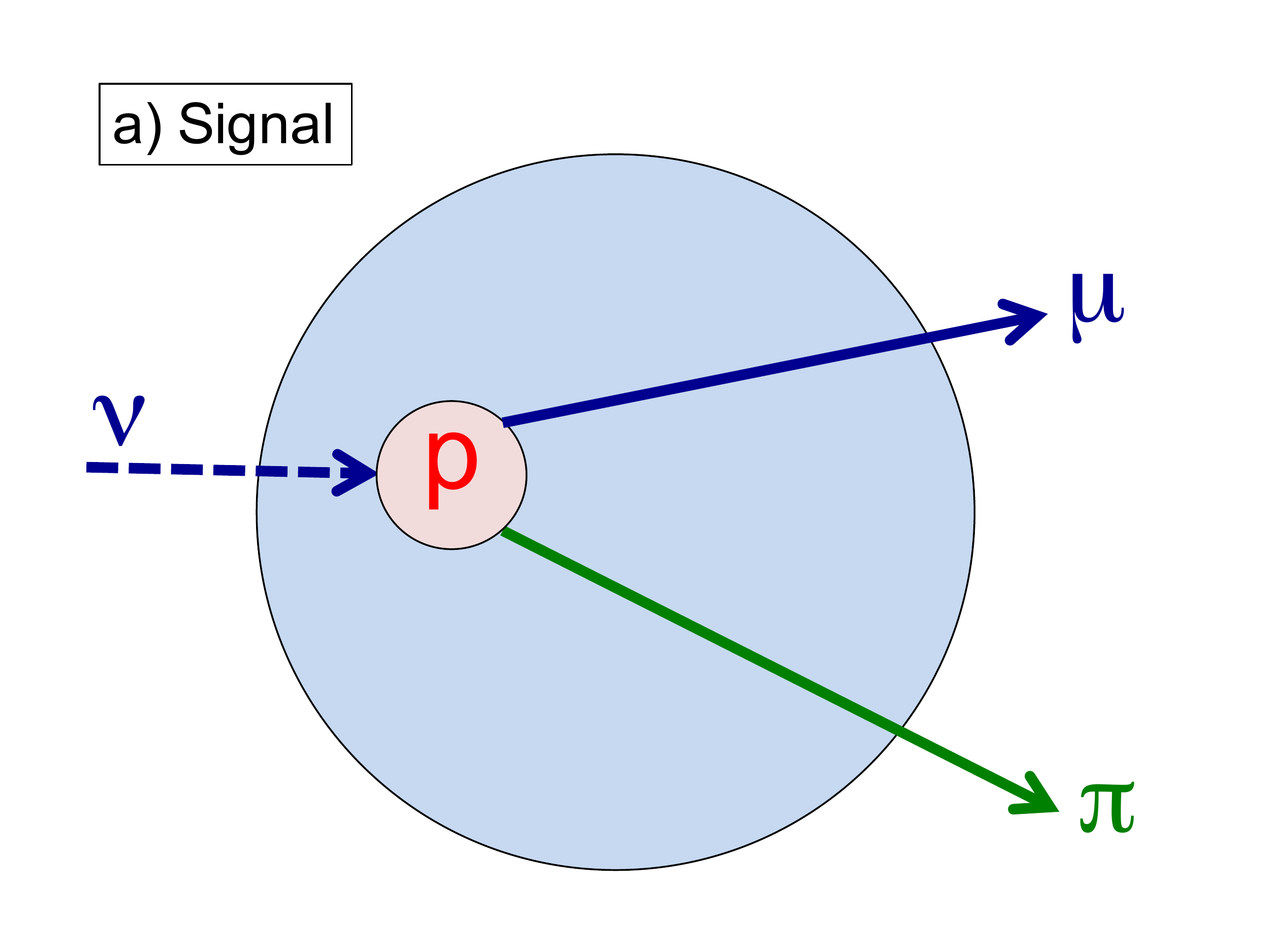}
    \includegraphics[width=5cm]{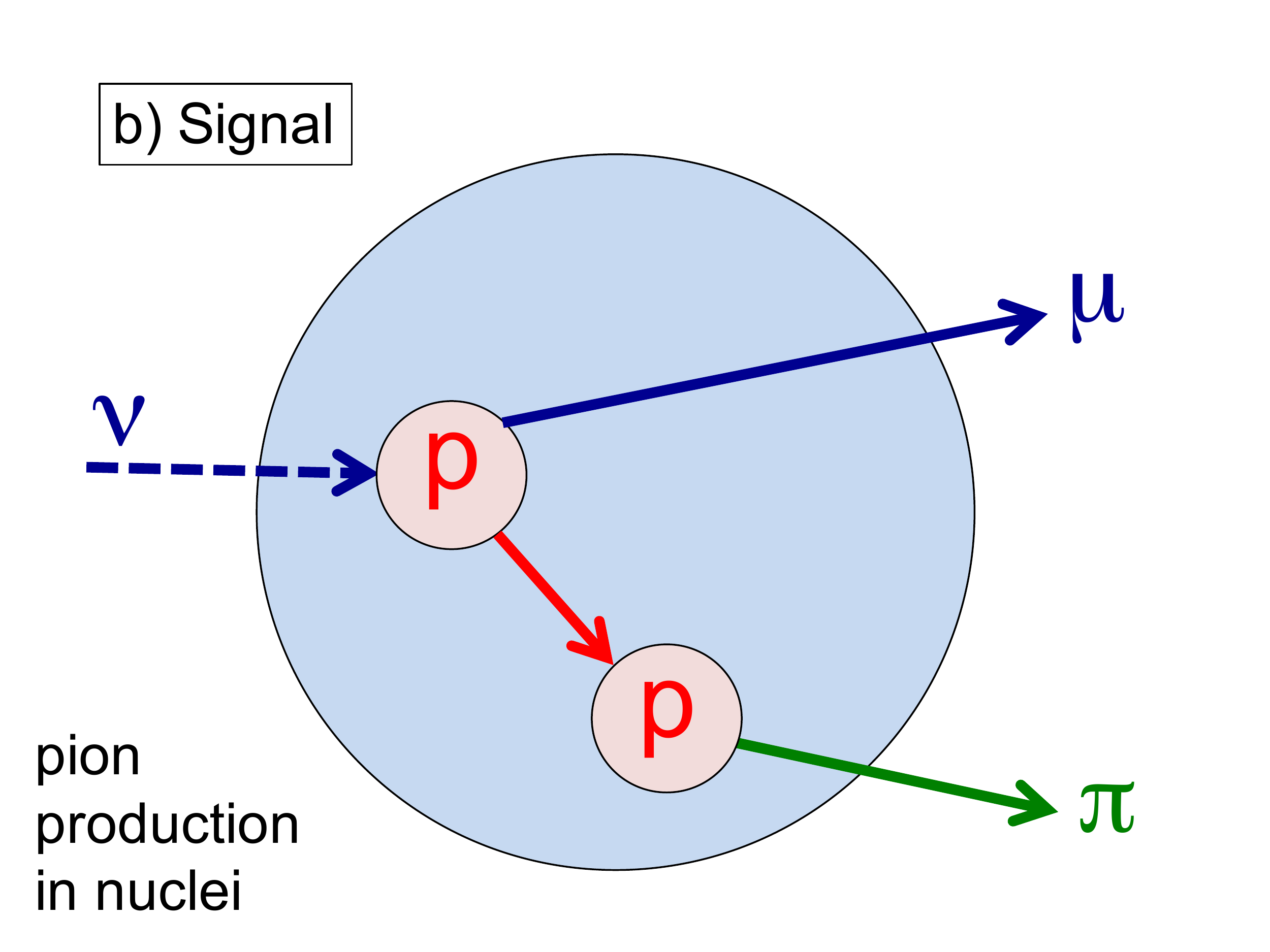}
    \includegraphics[width=5cm]{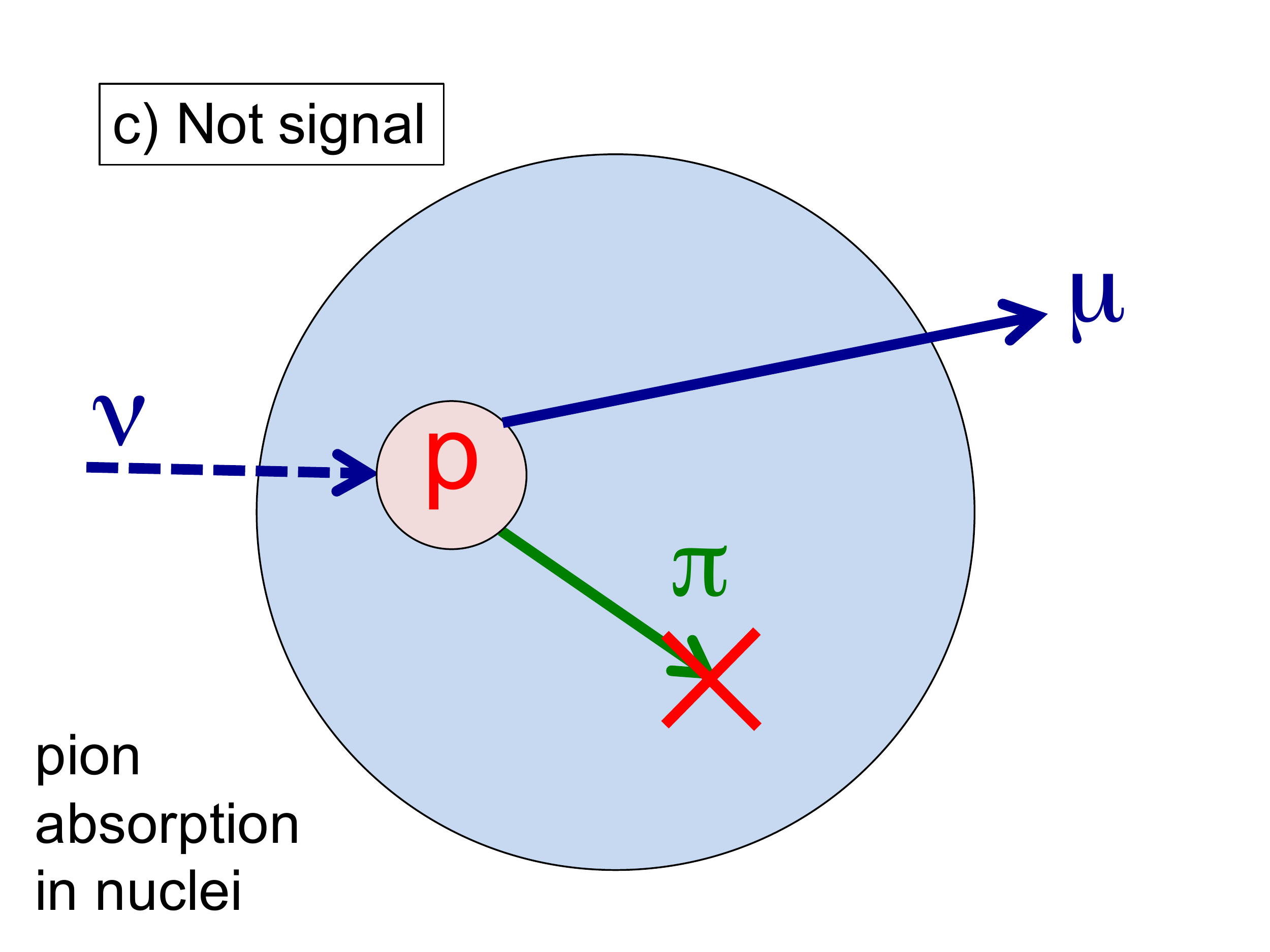}
    \includegraphics[width=5cm]{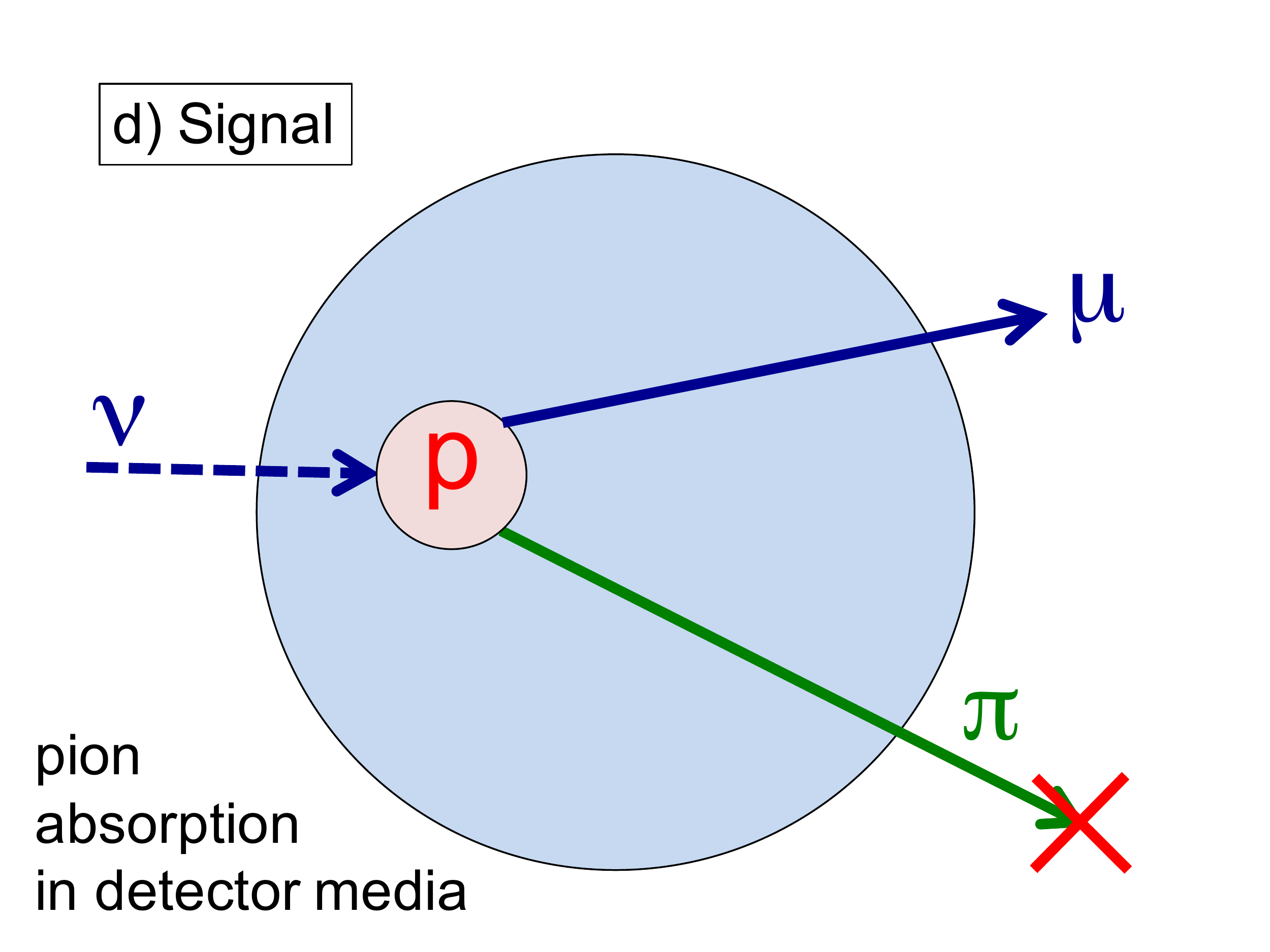}
    \end{center}
\vspace{-2mm}
\caption{Signal definition of pion production measurement.
(a), (b), and (d) are classified to be “signal”, and (c) is “not signal”.
Notice both (c) and (d) are pion adsorptions, but (c) is within the target nuclei,
and (d) is in the detector media~\cite{Teppei_MBSB}.}
\label{fig:sigdef}
\end{figure}

The signal is defined by the topology of the ``final state'', and specifically it is called ``CC1$\pipm$''. 
Here, final state refers the state of particles just outside of the target nucleus,
where particles experience nuclear effects and escape from the target nucleus.
Fig.~\ref{fig:sigdef} shows the situation. 
This sample definition does not allow a simple connection
between the data and theoretical calculations of
single pion production processes at the primary neutrino vertices.
This is cumbersome for theorists, because the data is not
directly comparable with theoretical models of baryon resonances,
but the theoretical calculation is comparable with the data only
if the calculation includes all processes with single pion in the final state. 
For example, Fig.~\ref{fig:sigdef} (b),
if the primary process does not make a pion but it is made by the FSI in the target nucleus,
this should be included in the theoretical calculation as one pion final state to compare with data. 
On the other hand, Fig.~\ref{fig:sigdef} (c),
even though the primary neutrino interaction produces a pion,   
if that disappears in the nuclear media  and it does not exit the target nucleus, 
such event is not counted as a signal in the data and the theory needs to take that into account.
The main reason for this signal definition is to avoid model-dependent background corrections.
Since the FSI simulations of hadrons are complicated and disagree between models, 
the amount of background subtracted is often suspicious. 
The best way to avoid model-dependent background subtraction is to re-define intrinsic background as signal,
and not subtract it. 
This removes the ambiguity of background subtraction associated with FSIs.  
Instead, the data becomes the tool to study FSIs. 

FSI remains one of the most important aspects of predicting hadron final events.
Over the last few years GiBUU made series of predictions on pion differential cross
sections~\cite{Lalakulich:2012cj,Lalakulich:2013iaa,Mosel:2014lja}.
As we will see, MINERvA pion data are also used directly
to constrain GENIE FSI models~\cite{Eberly:2014mra,Aliaga:2015wva,McGivern:2016bwh}.
To obtain higher precision data to tune FSI models,
there is a dedicated carbon target interaction measurement using the pion beam, DUET~\cite{DUET}.
Alternatively, Lu {\it et al.} proposed to extract FSI-free hydrogen interaction
by utilizing transverse momentum balance between $\pip$ and $p$ from from $\numu+p\to\mu^-+\pip+p$ on CH target such as
T2K near detector~\cite{Lu:2015hea}.
It is also possible to use these lepton-hadron kinematics to constrain the FSI models~\cite{Lu:2015tcr}.

As already discussed in Section \ref{sec:QE} the neutrino experimental community
prefers to classify the events in terms of observed final state particles, such as 
``1 muon + 0 pions + N protons'' (CC0$\pi$) and 
``1 muon + 1 charged pion + N protons'' (CC1$\pipm$),
as opposed to ``genuine CCQE'' or ``$\De$-resonance''. 
These classifications are already incorporated
in to the T2K oscillation analysis~\cite{T2K_nue2013}.
Section~\ref{sec:hadron} discussed 
``1 muon + 0 pions + at least 1 proton'' 2 track samples of T2K and MINERvA 
and ``1 muon + 0 pions + at least 2 protons'' (ArgoNeuT ``hammer event'' sample)
in the context of hadron information to understand neutrino interaction physics.

\subsubsection{Directly observable variables}
The direct observables, such as pion kinetic energy,
can avoid interaction model dependent detector corrections. 
The smearing of the distribution and the detection efficiency 
are direct functions of the measured kinematic variables.
This means unfolding these detector effects ($U_{ij}$ and $\ep_i$ in Eq.~\ref{eq:ddexp})
from the distributions of direct observable variables
are not involved in any inference regarding to interaction models. 
Therefore, detector effect unfolding processes have a minimum interaction model dependence on the data and
such data are often considered minimally model-dependent data.
This contrasts, for example, with differential cross section on $Q^2$.
To determine $Q^2$, the neutrino energy needs to be known {\it a priori},
but modern high statistics neutrino experiments use wide-band beam and
$E_\nu$ is a reconstructed inferred variable 
(see Sec.~\ref{sec:theo}, in particular Eqs. (\ref{enubar_muon}) and (\ref{eq_Q_rec}), and Sec.~\ref{sec:enrec}).

It is important to know that detector effects are the function of direct observables, 
and unfolding detector effects can, at best recover direct observables with a 100\% efficiency detector,
but it never recovers true kinematic variables such as $E_\nu$ which no detectors can directly measure. 
This limits the usage of kinematic variables such as  $E_\nu$, $Q^2$, $\om$, $|\nq|$, $W$, $x_{bj}$, $y$, etc,
and all of these variables need caution in modern neutrino experiments. 

\subsubsection{Flux-integrated cross section}
Although data are independent from the detectors and minimally dependent on interaction models, 
data are flux-integrated differential cross sections and 
they depend on the neutrino beam used to measure the cross section data.
Because of this, flux-integrated data from different neutrino beams are not directly comparable.
Those data are comparable only through the theoretical models convoluted with different fluxes.
We give a first illustration of this point in Section~\ref{sec:kinematics} by comparing each experiment through GENIE. 
Here, in Fig.~\ref{fig:ccpip}, left, predictions from GENIE neutrino interaction generator~\cite{GENIE}
are convoluted with NuMI flux and BNB flux to compare with MINERvA data and MiniBooNE data.
MINERvA and MiniBooNE flux-integrated data can be compared only in such a way. 
As we see, MiniBooNE and MINERvA data overlap at higher energy, 
but agreement is misleading because they use different neutrino beams.
Instead, we can see GENIE integrated with MINERvA flux is higher than MINERvA data,
and GENIE integrated with MiniBooNE flux is lower than MiniBooNE data.
Therefore, we can conclude that data exhibit serious conflict under predictions from GENIE.
This is true for other generators, such as GiBUU~\cite{Mosel_MINERvApion,Lalakulich:2012cj}
and NuWro~\cite{Sobczyk_MINERvApion}. 
Fig.~\ref{fig:ccpip}, right, is the prediction from the NuWro generator which clarifies this last point.
Here, the ratio of MINERvA to MiniBooNE data are compared with
the same ratio from the NuWro generator prediction~\cite{NuWro}, which shows a serious disagreement. 
Here, problems lay on either MiniBooNE data and/or MINERvA data and/or simulation.
We discuss this more in the later subsections.

\begin{figure}[tb]
  \begin{center}
    \includegraphics[width=6cm,valign=m]{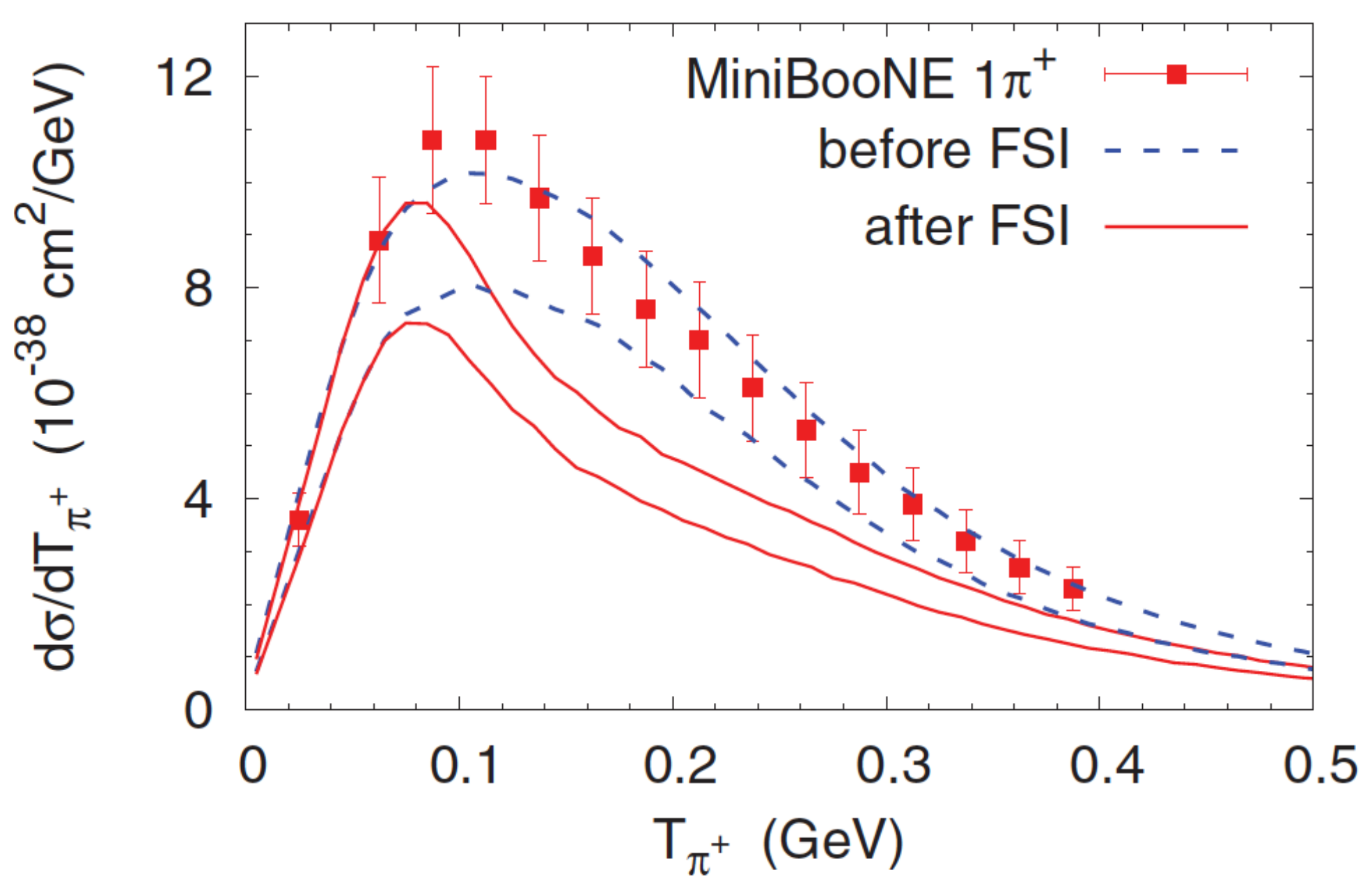}
    \includegraphics[width=6cm,valign=m]{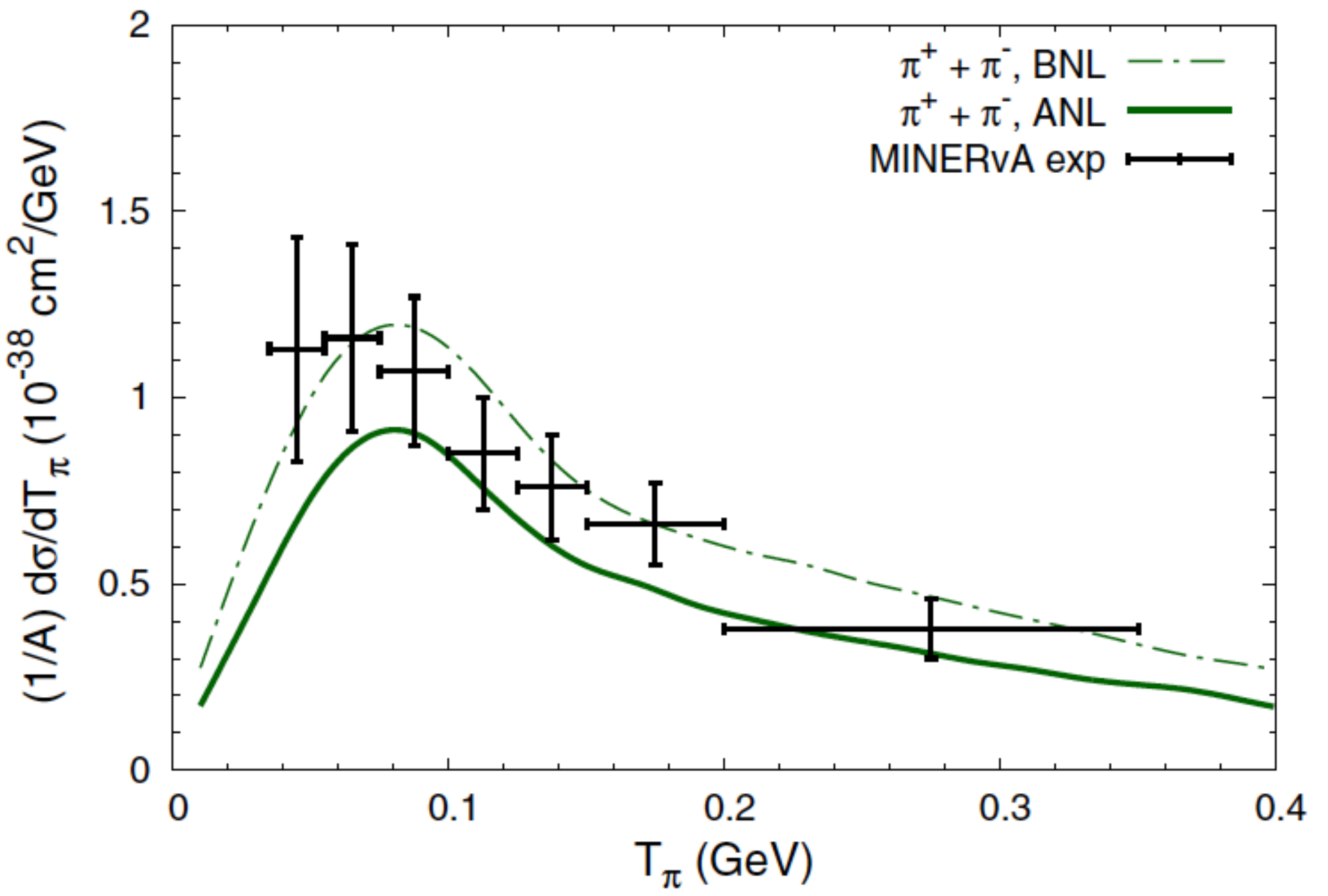}
    \end{center}
\vspace{-2mm}
\caption{
  Flux-integrated differential cross sections of CC1$\pi$ production
  charged pion kinetic energy is compared with GiBUU prediction.
  Left plot is for MiniBooNE~\cite{Lalakulich:2012cj},
  and predictions both before FSI and after FSI are shown.
  Right plot is for MINERvA~\cite{Mosel_MINERvApion}.
}
\label{fig:mosel_pionfsi}
\end{figure}

\vspace{0.5cm}

The details of these data make even more confusions.
First, the MiniBooNE data have serious shape disagreement with theoretical calculations. 
This is illustrated in Fig.~\ref{fig:mosel_pionfsi}.
Here, theoretical calculation by GiBUU is compared with MiniBooNE~\cite{Lalakulich:2012cj} (left)
and MINERvA data~\cite{Mosel_MINERvApion} (right).
From the left, apparently the calculation before FSI agrees better with data,
even though it is impossible for pions to not experience FSI when exiting nuclear targets.
A very similar conclusion is derived from Hernandez {\it et al}~\cite{Hernandez_pion},
based on a different approach, which strengthens this conclusion.
However, as discussed in Ref.~\cite{Alvarez-Ruso:2014bla},
this interpretation is oversimplified since normalization of
the primary pion production model has a large error
(which is shown as a ``band'' in Fig.~\ref{fig:mosel_pionfsi}).
Since FSIs are known to be important from other experimental results,
such as photo-production of pions on nuclei,
it is tempting to conclude that the upper limit of theoretical band including the FSIs is the right solution,
which maximizes the agreement with MiniBooNE data. 
On the other hand, the right plot from MINERvA does not select either ends of the band. 
From these exercises, squeezing this theoretical normalization error
is an important next step to understand these data.

\vspace{0.5cm}

Recently, T2K published the first flux-integrated differential cross section of charged pion production
on a water target~\cite{Abe:2016aoo}.
The measurement is different from MiniBooNE and MINERvA because of the different target material,
however, experimental differences may be even larger.
This measurement uses the water target in FGD2, where inactive water is located alternatively between
active scintillator layers. 
Because of this, many selected events are not interactions on water layers but scintillator layers,
and the subtraction of backgrounds (every interaction not on water target)
from the background dominated sample inflates systematic errors. 
However, if one increases the fiducial mass by increasing the water layers,
this necessarily reduces the scintillation layers and degrades the measurement performance.
This may be suggestive that any cross section measurements with inactive materials 
may be systematically limited for low energy neutrino experiments. 
The WAGASCI (WAter Grid And SCIntillator) detector~\cite{Koga:2015iqa} is proposed to overcome this problem
by a clever configuration of scintillator layers to retain
the tracking efficiency but maximize the fiducial volume of water. 

\subsection{Charged current neutral pion production\label{sec:ccpi0}}

\begin{figure}[tb]
  \begin{center}
    \includegraphics[width=6cm,valign=m]{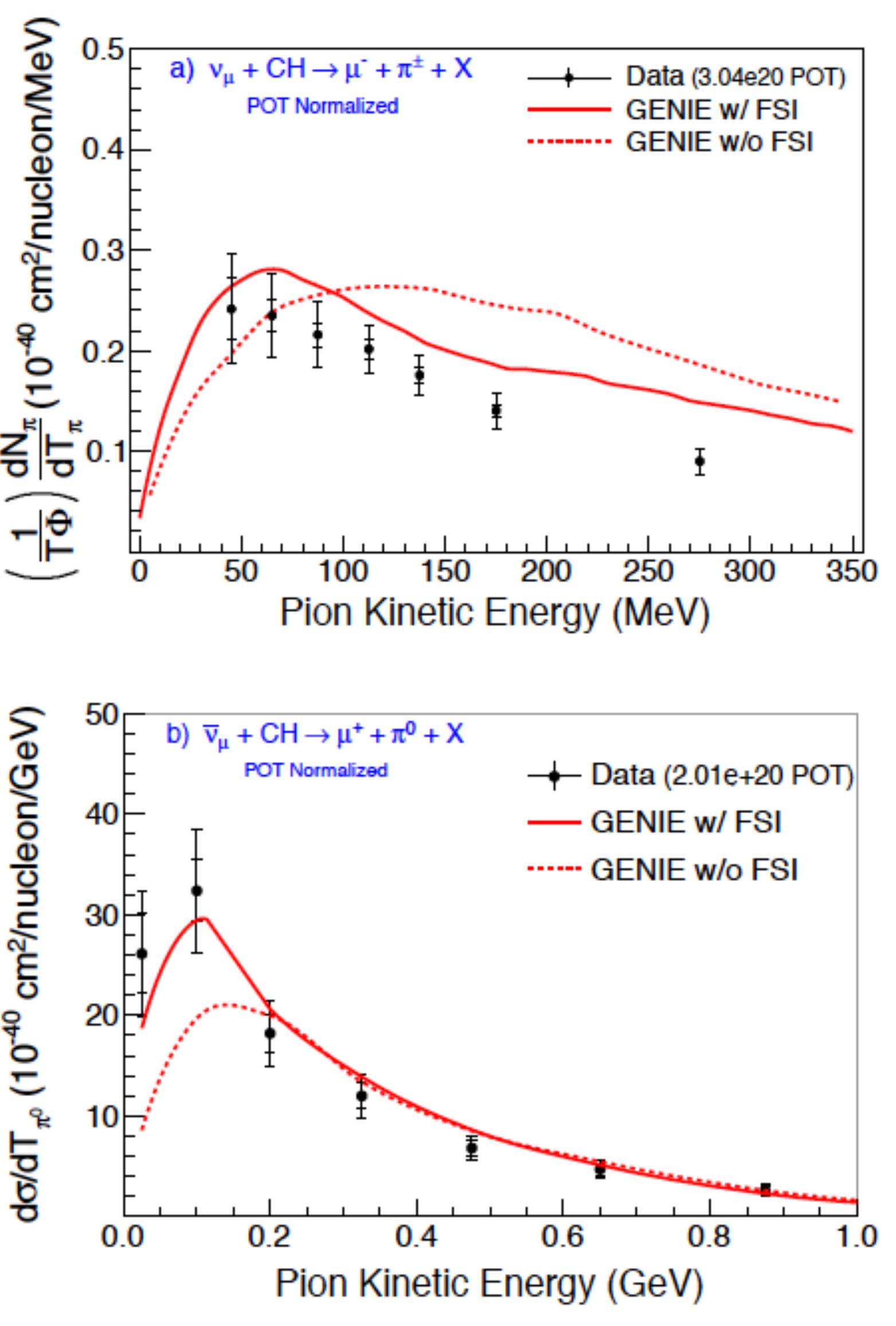}
    \includegraphics[width=6cm,valign=m]{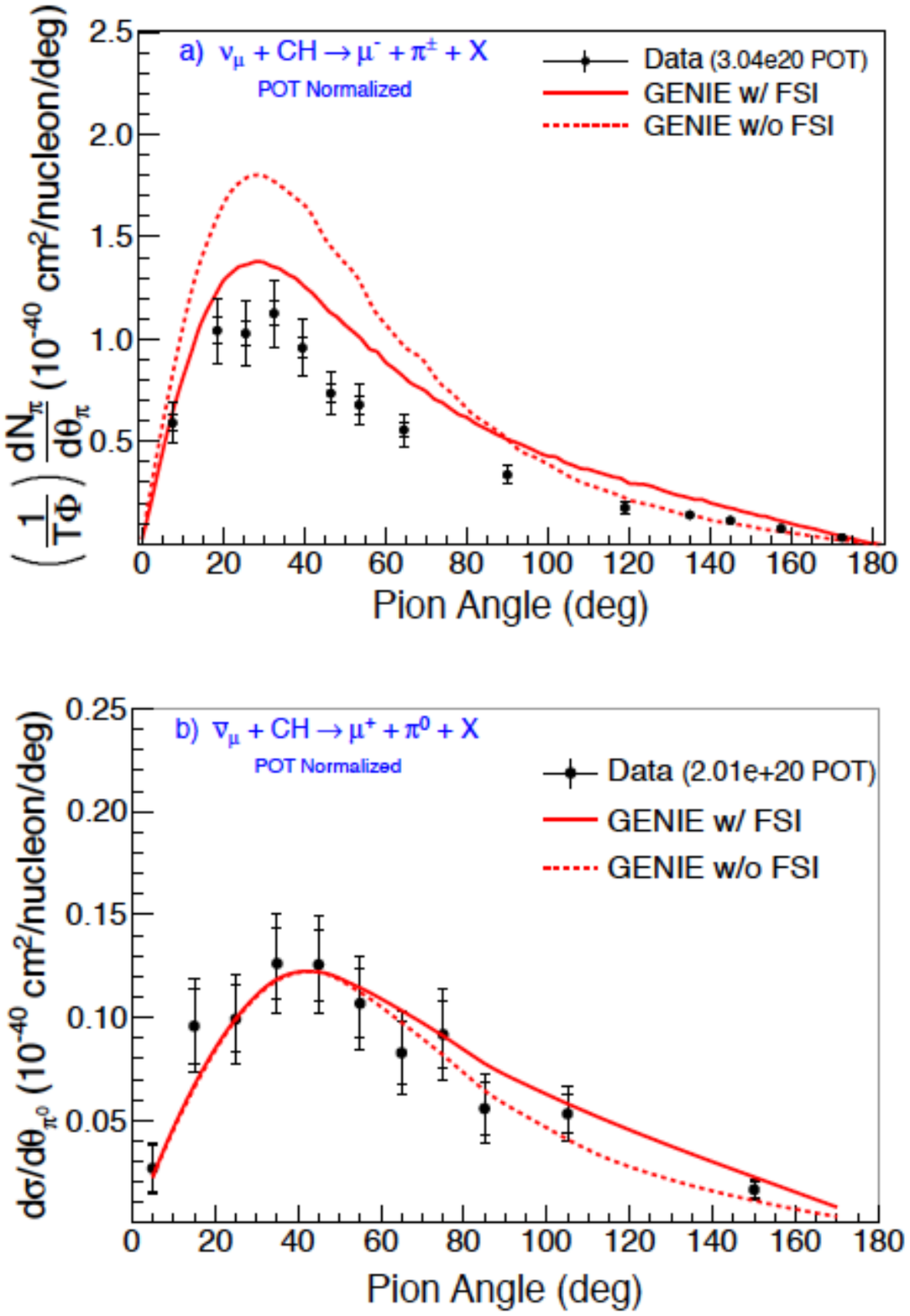}
    \end{center}
\vspace{-2mm}
\caption{
Neutrino CC1$\pip$ (top) and antineutrino CC1$\piz$ (bottom)
production flux-integrated differential cross section,
with function of $\pi$ kinetic energy (left) and $\pi$ scattering angle (right)~\cite{McGivern:2016bwh}.
Measured cross section is compared with GENIE with and without FSIs.
}
\label{fig:CCpi0_1}
\end{figure}

Charged current neutral pion production (CC$\piz$) attracts rather little interest
due to its small cross section and harder analysis to measure.
For example, MiniBooNE CC1$\piz$ differential cross section measurement~\cite{MB_CCpi0} has
the largest normalization discrepancy with their MC,
indicating difficulties to both measure and simulate this channel. 
However, theoretically this channel is key to understanding pion production mechanism,
because unlike CC1$\pi^\pm$ production it has no coherent contribution,
and this makes this channel a unique target to study.

Recently, MINERvA published the first antineutrino
CC1$\piz$ production differential cross section~\cite{Aliaga:2015wva}.
Fig.~\ref{fig:CCpi0_1} shows the differential cross section of $\pi$ kinetic energy and scattering angle.
As you see, the measured antineutrino CC1$\piz$ cross section shows a good agreement with generators (bottom plots).
Interestingly, GENIE prediction without FSI underestimates the data at low energy region,
and they agree only after FSI is included  in the prediction.
This is because the pion absorption ($\pi^\pm+\piz+X\to\piz+X'$)
and the pion charge exchange ($\pi^\pm+X\to\piz+X'$)
processes contribute to increase the cross section in this region.

\begin{figure}[tb]
  \begin{center}
    \includegraphics[width=6cm,valign=m]{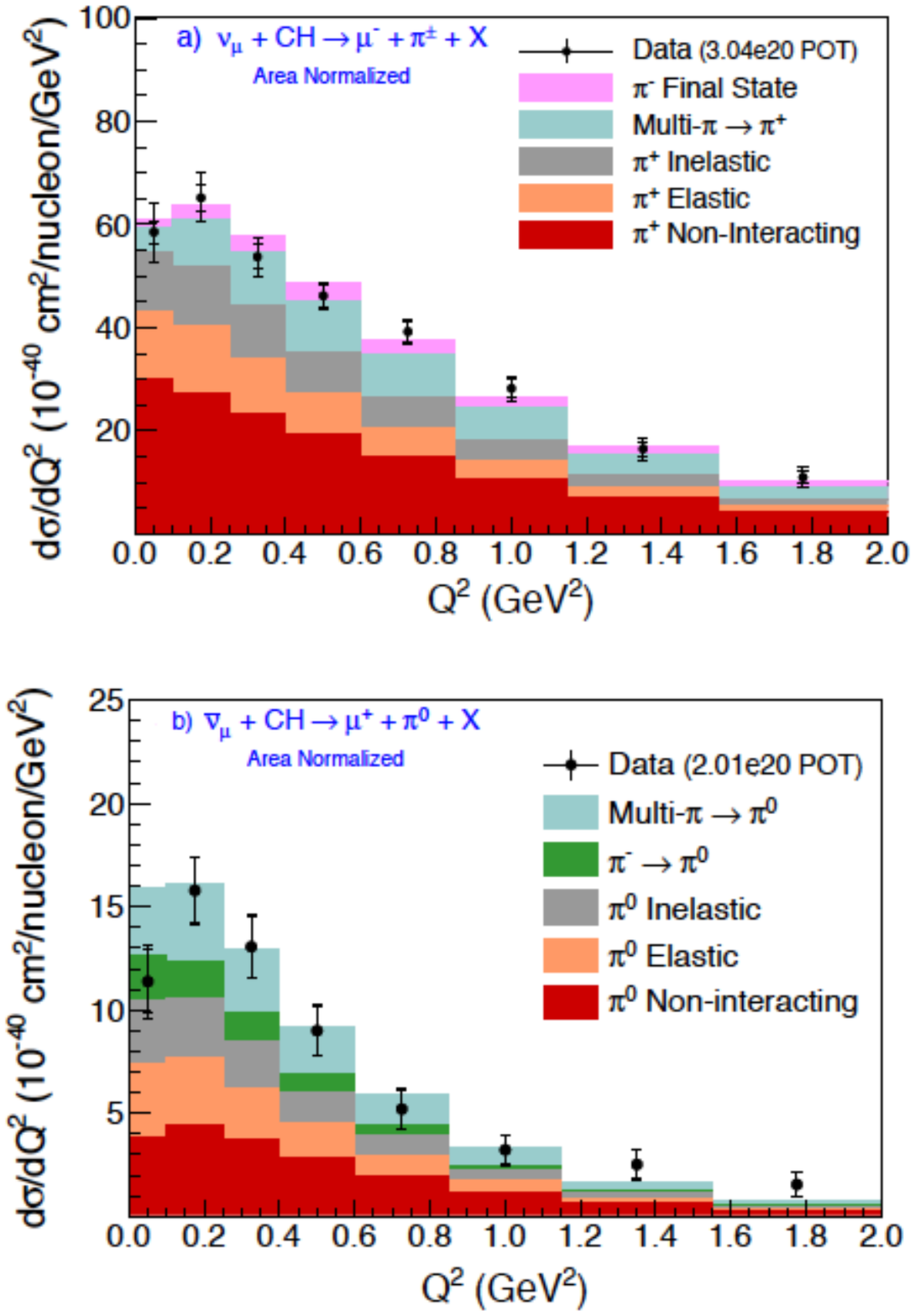}
    \includegraphics[width=6cm,valign=m]{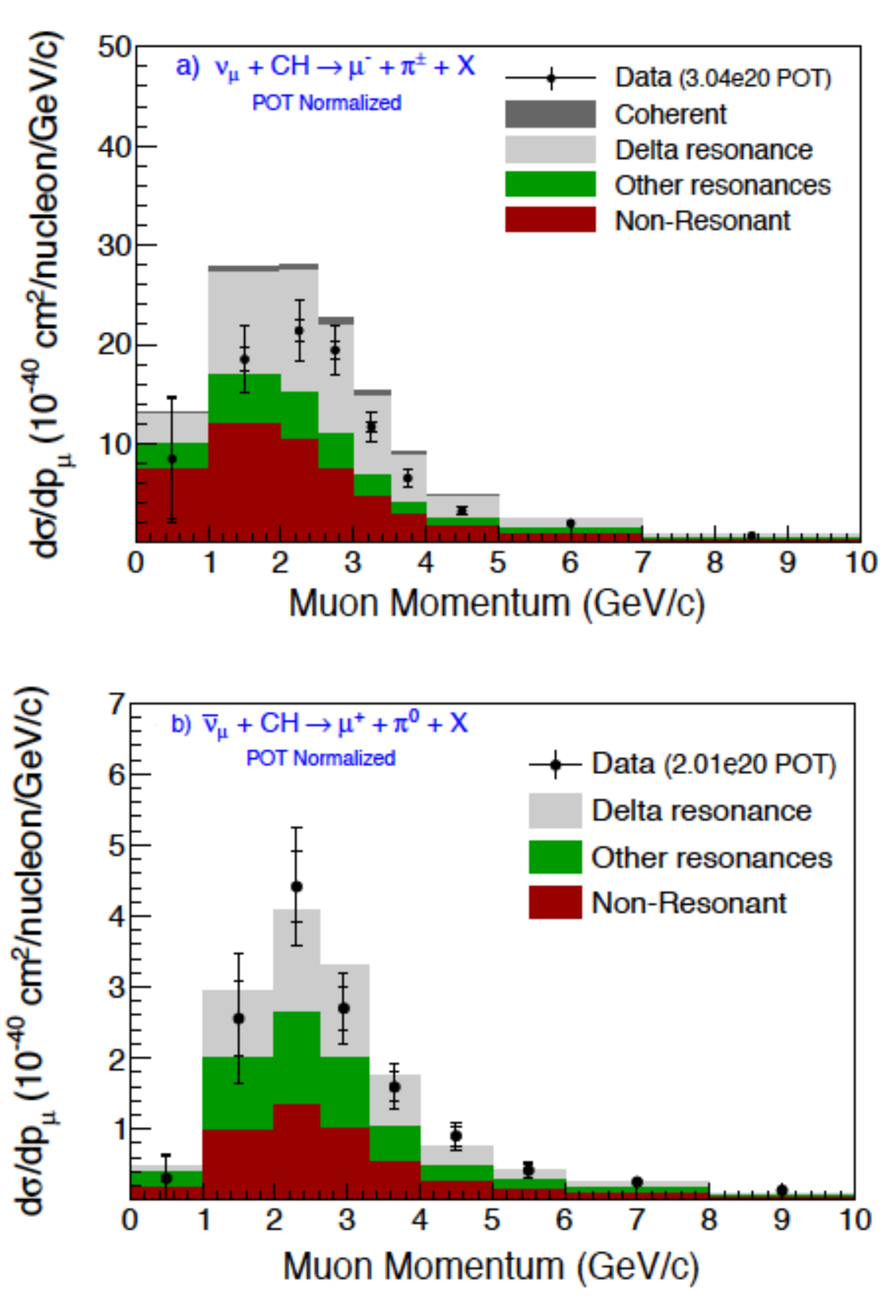}
    \end{center}
\vspace{-2mm}
\caption{
Neutrino CC1$\pip$ (top) and antineutrino CC1$\piz$ (bottom)
production flux-integrated differential cross section,
with function of $Q^2$ (left) and $\mu$ momentum (right)~\cite{McGivern:2016bwh}.
In the left plot, measured cross section is compared with GENIE prediction decomposed
with different FSI components.
In the right, measured cross section is compared with GENIE prediction decomposed
with different primary interactions.
Note left plots are area normalized. 
}
\label{fig:CCpi0_2}
\end{figure}

In Fig.~\ref{fig:CCpi0_2}, left, further details of these data are examined in the context of FSIs.
Here, predictions are broken down to different channels of FSIs in GENIE. 
Unfortunately, at this moment under GENIE there is no easy way to tune FSIs to improve data-MC agreement,
as you see from Fig.~\ref{fig:CCpi0_2}, left,
simple scaling of one of FSI processes does not dramatically improve data-MC agreement both
$\numu$CC1$\pip$ and $\numubar$CC1$\piz$ simultaneously.  
Nevertheless, these are first attempts to investigate FSIs through neutrino pion production data. 

Fig.~\ref{fig:CCpi0_2}, right, shows the detail of channels contributed to $\numu$CC1$\pip$ and $\numubar$CC1$\piz$ samples. 
One of the key points of these analysis is to understand the intrinsic background
coming from multi-pion productions where the detector fails to identify all outgoing pions.
$\numubar$CC1$\piz$ analysis accepts more background than CC1$\pip$ analysis,
and the amount of background subtraction is large. 
For the precise prediction of the background,
larger W region data is used as a control sample of the background,
and background simulation is tuned in this region to subtract from the signal region. 
The amount of correction applied to the background simulation is roughly 20\%,
leaving a little worry that subtracted background would be
a large factor to decide the data normalization. 
Rodrigues {\it et al.}~\cite{Rodrigues:2016xjj} extended
the re-analyses of bubble chamber data~\cite{Wilkinson:2014yfa} (discussed in the next section)
by including sub-dominant channels ($\nu_\mu+d\to \mu^-+\piz+p+p$ and $\nu_\mu+d\to \mu^-+\pip+p+n$),
and they found that non-resonance background in GENIE may be over-estimated.
This may be consistent with why $\numubar CC1\piz$ sample needs
a large reduction of multi-pion background by aforementioned analyses,
where in GENIE non-resonance background has higher multiplicity than resonance processes by construction. 
Indeed, as you seen in the Fig.~\ref{fig:CCpi0_2}, right, data also prefer to reduce the non-resonance background to
improve the data-GENIE agreement in CC1$\pip$ and $\numubar$CC1$\piz$ simultaneously. 

\subsection{MiniBooNE-MINERvA data comparison}

Coming back to the MiniBooNE-MINERvA data comparison (Fig.~\ref{fig:ccpip}),
here we are observing both normalization and shape disagreement between MINERvA data, MiniBooNE data, and simulation. 
One of the biggest obstacles to understand the current situation is,
we do not know who has the right normalization among those three. 
This hampers our shape analysis, which is related to our choice of FSI and baryonic resonance models.
In this section we discuss the normalization of the simulation, MiniBooNE data, and MINERvA data. 

\subsubsection{Normalization of theory}

\begin{figure}[tb]
  \begin{center}
    \includegraphics[width=6cm,valign=m]{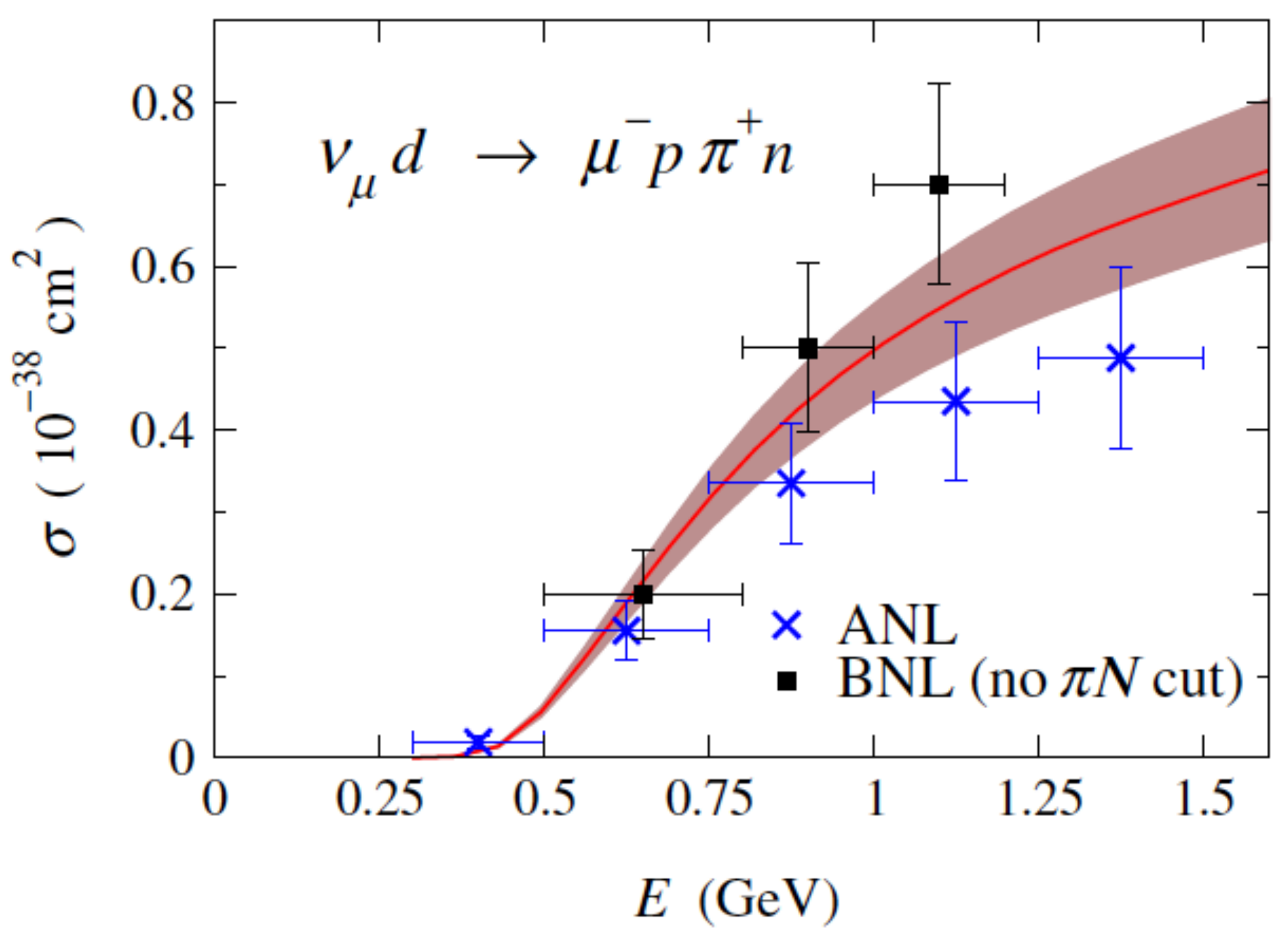}
    \includegraphics[width=6cm,valign=m]{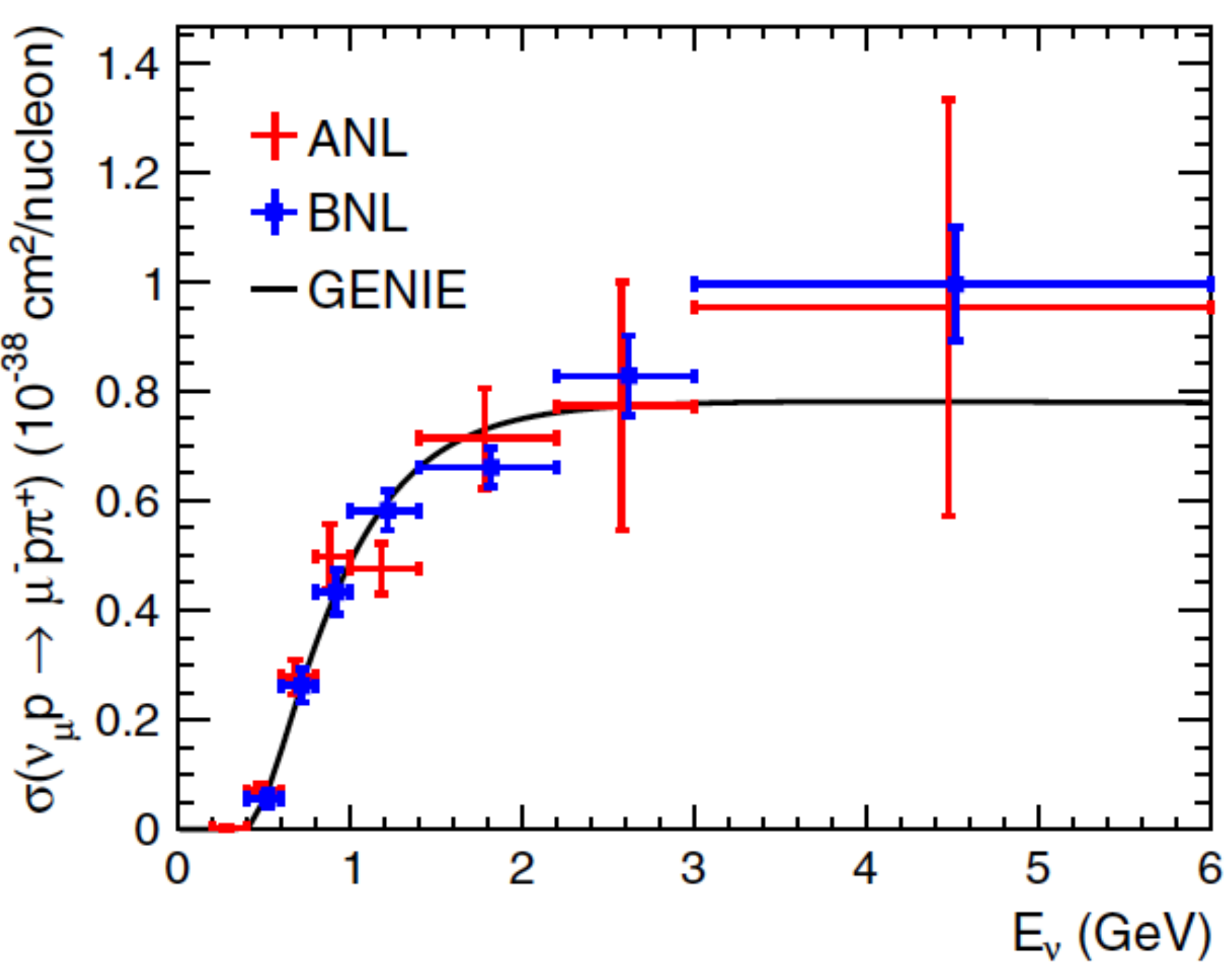}
    \end{center}
\vspace{-2mm}
\caption{
Total cross section predictions for $\nu_\mu+d\to \mu^-+p+\pi^++n$ are compared with ANL and BNL data.
In the left, the data are compared with calculation by Hernandez~{\it et al.}~\cite{Hernandez_pion}
using a dipole parameterization of $C_5^A(Q^2)$ with $\CfA (0)=1.00 \pm 0.11$.
The right plot show the re-analyzed ANL and BNL data, with prediction from GENIE~\cite{Wilkinson:2014yfa}.
}
\label{fig:ANLBNL}
\end{figure}

From Fig.~\ref{fig:mosel_pionfsi}, we observe a large shape disagreement from MiniBooNE data and models,
however, large theoretical normalization errors hamper to make any conclusions.
The origin of this theoretical error band is the different $\CfA (0)$ parameters extracted from
BNL or ANL deuteron bubble chamber data~\cite{ANL1,ANL2,BNL}. 
The $\De$-resonance hadronic current is parameterized with three vector form factors and four axial form factors.
Among them, vector form factors are precisely measured by electron scattering experiments.
For four axial form factors, $C_3^A$,  $C_4^A$,  $C_5^A$, and $C_6^A$, 
after PCAC, Adler relation, and neglecting small $C_3^A$,
only $\CfA$ axial form factor becomes the function to choose~\cite{Hernandez:2007qq,Hernandez_ANLBNL,Sobczyk_pion}. 
By using dipole parameterization, $C_5^A$ is parameterized by two parameters,
$C_5^A(0)$ and $\De$ axial mass $M_{A\De}$~\cite{Hernandez:2007qq,Sobczyk_pion},
\beq
C_5^A(Q^2)=\frac{C_5^A(0)}{\left(1+Q^2/M_{A\De}^2\right)^2}~.
\eeq
Here, $\CfA (0)$ parameter represents the $Q^2=0$ value of the $\CfA (Q^2)$ form factor, 
and it also controls the normalization of the $\CfA$ form factor,
{\it i.e.}, $\CfA (0)$ controls the normalization of axial contributions.

Figure~\ref{fig:ANLBNL}, left, shows the comparison of the total cross section of
$\nu_\mu+d\to \mu^-+p+\pi^++n$ ANL and BNL data with the theoretical calculation of Hernandez \textit{et al.}~\cite{Hernandez_pion}.
The best fit curve that can cover both data sets within the error is obtained for $\CfA (0)=1.00 \pm 0.11$,
a value below the one obtained by the Golberger-Treiman (GT) relation, $\CfA (0)_{GT}=1.15 - 1.2$.
Deviations from this value are expected to be the order of a few percents.
The model of Ref.~\cite{Hernandez_pion} could be reconciled with the GT relation
by simultaneously fitting vector form factors to the electron-proton scattering  structure  function,
leading to $\CfA (0)=1.10^{+0.15}_{-0.14}$ ~\cite{Sobczyk_ANLBNL},
nevertheless the weak point of violation of Watson's theorem still remains.
The improvement of the model of Hernandez \textit{et al.}
by imposing Watson’s theorem to the dominant vector and axial multipoles
has been presented in Ref.~\cite{Alvarez-Ruso:2015eva}. 
We postpone the discussion of the corresponding results.

Both ANL and BNL data suffer with large flux normalization error~\cite{Sobczyk_ANLBNL}. 
To improve the precision, there is also a strong desire to repeat
these bubble chamber experiments to find the $\CfA$ form factor.
However, due to recent tighter safety regularization,
hydrogen or deuteron bubble chamber experiments are not easy to be approved,
especially underground, where most of neutrino beams are located. 
Under this desperate situation,
a re-analysis on ANL and BNL pion production data was performed by Wilkinson {\it et al.}~\cite{Wilkinson:2014yfa}. 
To avoid flux normalization, they analyze the ratio of two channels,
$\nu_\mu+d\to \mu^-+p+\pi^++n$ to $\nu_\mu+d\to \mu^-+p+p$,
because this ratio is supposed to cancel the flux normalization effect.
Indeed, the extracted cross section ratio, $\frac{CC1\pi^+}{CCQE}$ from ANL and BNL data are consistent. 
Then, the CCQE cross section value calculated by GENIE is multiplied,
to recover the single pion production cross section, shown in Fig.~\ref{fig:ANLBNL}, right. 
One important assumption of this analysis is the validity of GENIE CCQE cross section in the channel $\nu_\mu+d\to \mu^-+p+p$.
If that were wrong, one couldn't recover the $CC1\pi^+$ cross-section from $\frac{CC1\pi^+}{CCQE}$ ratio.
After these studies, re-analysis on ANL data are found to be consistent with the original result~\cite{ANL1,ANL2},
while BNL data shows smaller cross section than the original BNL result~\cite{BNL}. 
Therefore, authors concluded BNL cross section data were overestimated due to incorrect flux normalization.

Apparently, this suggests  ``ANL-BNL puzzle''~\cite{Alvarez-Ruso:2014bla} can be solved by choosing the ANL data,
which naturally leads to choose the lower value of $\CfA$.  
Unfortunately, this is not the end of the story.
Recent calculations of Wu~{\it et al.}~\cite{Sato_Deuteron},
which respect GT relation and Watson's theorem by construction of the model,
showed that the suppression of pion production cross section on deuteron by NN re-scattering can be large. 
If this effect is taken into account in the extraction of $\nu_\mu+d\to \mu^-+p+p$ cross section, 
the true pion production cross section would be larger,
hence the true value of $\CfA$ could be larger,
as discussed in Ref.~\cite{Mosel_MINERvApion,Mosel:2016cwa}.
The recent improvement of the model of Hernandez \textit{et al.}
by imposing Watson’s theorem ~\cite{Alvarez-Ruso:2015eva}
leads to the following values $\CfA (0)=1.12 \pm 0.11$ and $\CfA (0)=1.14 \pm 0.07$
obtained respectively by fitting the original ANL and BNL data and the the re-analyzed ones.
These values are in agreement with the GT relation.
Without imposing Watson’s theorem,
the revisited data would also result below the GT relation.

\subsubsection{Normalization of data}

The normalization of MiniBooNE data was suspected for long time
due to the large CCQE cross sections they measured, 
and this raises the question of BNB neutrino flux normalization.
At MiniBooNE, although data from HARP~\cite{HARP} was used to improve the meson production simulation,
they do not use the data taken with the full replica target.
Instead, they used the data from the shorter target and scaled it in the simulation.
This leaves an uncertainty in the mesons from re-scattered protons,
because mesons are produced by interactions of exiting particles,
not by the primary proton-target interactions (see Fig.~6 of~\cite{Kopp_flux} taken from~\cite{Zarko_thesis}).
Such a contribution is usually hard to simulate,  
and experimentalists use the measured meson distributions from replica targets at hadron detectors (Sec.~\ref{sec:flux}). 
However, 8~GeV protons from Fermilab Booster are low energy and the majority of mesons
are produced by the primary interactions of protons and the target,
and the prediction of the neutrino flux is relatively easy. 
This was confirmed recently where the full length target data from
the HARP measurement in the simulation indicates
no significant changes of the flux prediction~\cite{Athula_thesis}.  

Also, as discussed in Sec.~\ref{sec:npnh},
theoretical models including np-nh can explain
the large normalization of the MiniBooNE CCQE-like data ~\cite{Martini:2009uj,Nieves:2011pp,Megias:2016fjk,Gallmeister:2016dnq}.
Note, the models of Martini~{\it et al.}, SuSAv2-MEC and GiBUU simultaneously reproduce
$\nu_\mu$ and $\bar{\nu}_\mu$ MiniBooNE CCQE-like and T2K $\nu_\mu$ and $\nu_e$ CC inclusive data (Sec.~\ref{sec:cc}).
Martini~{\it et al.}, Nieves~{\it et al.} and SuSAv2-MEC also reproduce the T2K CC0$\pi$ data. 
SuSAv2-MEC also reproduces the MINERvA CCQE-like data.   
This simultaneous agreement of theoretical calculations with several differential cross sections
integrated over different neutrino fluxes suggests that BNB flux normalization is within the quoted systematic error. 

However, this may not answer the questions about data normalization of all channels.
In MiniBooNE, a variety of cross section measurements were performed
and it turned out all of them have different normalization discrepancies from their predictions based on their simulation. 
For example, NC1$\piz$ cross section ($\sim$10\%)~\cite{MB_NCpi0}, CC1$\pip$ cross section ($\sim$20\%)~\cite{MB_CCpip}, 
and CC1$\piz$ cross section ($\sim$60\%)~\cite{MB_CCpi0} have all larger normalization
than their simulation with different amount indicated in the bracket,
even though the simulation used NUANCE generator~\cite{NUANCE} which uses the same model for all pion channels. 
This indicates that to understand MiniBooNE pion data normalization, 
understandings of BNB neutrino flux
and primary interaction cross section models are necessary but not sufficient.

\vspace{0.5cm}

The normalization of MINERvA data are also suspicious. 
The primary protons at the NuMI beamline are sent from the main injector,  
and meson productions from other than the primary proton-target interactions are high. 
The full NuMI target measurement is done at the MIPP experiment~\cite{MIPP} at Fermilab,
but this result is not yet fully implemented in the NuMI neutrino flux simulation.
Thus, there is a chance that NuMI absolute flux normalization could be shifted in the future.

Recent results from $\nue-e$ elastic scattering and reanalysis of flux prediction
indicate that the NuMI flux was over-estimated~\cite{Park:2015eqa,Aliaga:2016oaz}. 
Consequently, measured cross-sections from MINERvA are shifted to larger normalization~\cite{McGivern:2016bwh}. 
Since both CCQE-like~\cite{Fiorentini:2013ezn,Fields:2013zhk} and CC$\pip$~\cite{Eberly:2014mra} 
predictions tend to over-estimate MINERvA data, this new flux simulation sounds like good news.  
Does a new flux simulation of MINERvA fix the all problems? 
Unfortunately, this na\"{i}ve idea cannot resolve all mysteries,  
because in MINERvA, like the case of MiniBooNE,
data from different channels have different amounts of normalization disagreement with theories,
and especially two track sample of CCQE-like data~\cite{Walton:2014esl} and
$\numubar$CC1$\pi^\circ$ data~\cite{Aliaga:2015wva,McGivern:2016bwh} have better agreements with current simulation.
And decreasing the flux normalization as suggested from $\nu-e$ data would increase the cross sections from these data
and data-simulation agreements would be degraded.
This may require to tune the simulation model parameters,
but such tuning based on the re-analyzed MINERvA pion data with new NuMI flux has not been done. 

In summary, as we see from MiniBooNE and MINERvA cases,
differential cross section normalizations with hadronic final states vary a lot between different exclusive channels.
These diffrences are not explained by simple mismodeling of neutrino flux or baryon resonance cross sections,
but they indicate there are more problems to predict and measure hadron exclusive final states.
We further discuss this in the subsection \ref{sec:exclusivehad}.

\subsection{Neutral current neutral pion production~\label{sec:pizero}}

In MiniBooNE oscillation analyses,
NC1$\piz$ production is measured by observing two electromagnetic showers
consistent with $\piz$ decay. 
Then, $\piz$ momentum distribution simulation is corrected based on this measured $\piz$ momentum distribution.
In this way, MiniBooNE improved the background predictions for $\nue$($\nuebar$) oscillation candidate samples.
This method is valid up to the level of the correctness
to simulate a single gamma reconstruction from a $\piz$ decay~\cite{MB_NCpi0PLB}. 
T2K and NOvA measured this channel using the near detector to understand their backgrounds
for $\nue$($\nuebar$) appearance oscillation searches~\cite{SamShort_thesis,Bu:2016grw}. 

Recently, the event reconstructions in Super-Kamiokande has been dramatically improved~\cite{Abe:2015awa}.
In this new event reconstruction algorithm ``fiTQun'', 
time (T) and charge (Q) information from all PMTs are used to construct the likelihood function 
to find the best fit values of particle track information (vertex and direction)
depending on the interaction hypothesis~\cite{MB_recon}.
Furthermore, likelihood value ratio of different track hypothesis can be used as a particle ID. 
For example, the likelihood ratio of $\nue$CC to NC1$\piz$ can remove a further $\sim$70\% NC1$\piz$ background
from the oscillation sample, making NC1$\piz$ background no longer the biggest background of
$\nue$($\nuebar$) appearance oscillation measurements in Super-K. 
In the meantime, LArTPC detector (discussed in Sec.~\ref{sec:lartpc}) is expected to
eliminate the majority of NC1$\piz$ backgrounds from $\nue$ appearance search~\cite{Teppei_uB}.
Recently ArgoNeuT demonstrated the $\piz$ momentum measurement by only
using angular information of gamma rays which is related to the $\piz$ momentum~\cite{Acciarri:2015ncl}.
Although this measurement is limited with the detector size,
unprecedented resolution of the detector was demonstrated. 
Thus, these developments make NC1$\piz$ background a minor background in future $\nue$ appearance search experiments. 
However, understanding of this channel is still important as 
a part of the ``pion puzzle''~\cite{Alvarez-Ruso:2014bla,Morfin:2012kn,Garvey:2014exa,Mosel:2016cwa,Kendall_NCgamma},
where we are lacking an overall understanding of pion production data.

Note, due to the higher rejection of NC1$\piz$ background,
sub-leading order neutral current single gamma (NC1$\ga$) production becomes important.
Especially, this channel got attention as a possible source
to explain MiniBooNE low energy excess~\cite{Aguilar-Arevalo:2013pmq,Hill_3H,Hill_NCgamma,Hill_LowE}. 
However, later calculations which include nuclear effects do not support this
interpretation~\cite{Luis_NCgamma,Luis_LowE,Zhang_incoh,Zhang:2012xi,Zhang_NCgamma,Zhang_LowE}. 
Nevertheless, NC1$\ga$ may be an important background
for future oscillation experiments~\cite{Teppei_MBreview,Kendall_NCgamma,Lasorak_NuInt15}. 
Indeed, confirming that MiniBooNE low energy excess is electron-like, not gamma-like,
is one of motivations for MicroBooNE to use LArTPC technology~\cite{Teppei_uB}.

\subsection{Toward the solution of the pion puzzle~\label{sec:exclusivehad}}

So far, we see that predictions of final state hadrons require 
good models of baryon resonance cross sections and FSI.
In reality, hadrons are also generated by DIS with the hadronization process.
Moreover, the transition region between baryonic resonance and DIS,
so-called ``shallow inelastic scattering (SIS)'' region, 
is a developping field and we do not have a complete picture of how to model the cross section
in MC simulation.
To compare with exclusive hadron channel data, 
theoretical models need to take into account all of the above
to make predictions of exclusive hadron final state predictions.

On top of that, interactions of propagating particles in media may
generate more hadrons or disappear along the tracks (Fig.~\ref{fig:sigdef} (d)). 
These so-called ``secondary interactions'' add another confusion on hadron measurements. 
Since this is a part of the detector effects,
and experimentalists and theorists agree to compare their results at the instance 
when particles leave the target nuclei (to make data detector-independent),
secondary interactions should be corrected by experimentalists from their data based on their simulation.
Finally, to measure hadrons, the responses of detectors on hadrons have to be understood.
However, in general, neutrino detectors are coarsely instrumented to maximize the fiducial volume to overcome small cross section. 
They are not good at measuring short, high angle, and high $\frac{dE}{dx}$ hadron tracks
very likely produced by 1-10~GeV energy neutrino beams discussed in this article,
and the uncertainty of hadron detection efficiency is high. 

To summarize, one needs to study and analyze the following points: 
\begin{itemize}
\item[1] Baryon resonance, SIS, and DIS cross section models
\item[2] hadronization
\item[3] secondary interactions and detector efficiency correction
\end{itemize}

\subsubsection*{SIS and DIS cross section}

Nakamura {\it et al.} investigate hadron final states from neutrino interactions 
by the dynamical coupled-channels (DCC) model prediction~\cite{Nakamura:2015rta},
where a coupled equation of baryon resonance channels is solved.
By the DCC model, Delta contribution on $\numu-n$ and $\numubar-p$ single pion production are
only about half at 2 GeV~\cite{Nakamura:2015rta}.
They also found two pion productions from baryon resonances can be up to $\sim$10\% around $E_\nu\sim$2~GeV,
In this SIS region, higher resonances and non-resonance background become more important with increasing energy.
On the other hand, low $Q^2$ DIS slowly turns on and DIS reaches $\sim$50\% at 5~GeV, and $\sim$90\% at 10~GeV. 

There are several studies to apply quark-hadron duality to describe
the neutrino SIS differential cross sections~\cite{Paschos_duality,Sobczyk_duality,Lalakulich:2008tu}.
Among them, Bodek-Yang correction~\cite{BodekYang1,BodekYang2,BodekYang3} on leading order (LO) GRV98 PDF~\cite{GRV98}
reasonably describes the SIS cross sections for neutrino interaction generators~\cite{NEUT,GENIE,NuWro}.
However, these models are too na\"{i}ve compared with the aforementioned DCC model.  

DIS differential cross sections are measured precisely. For example, NuTeV~\cite{Tzanov:2005kr} measured
CC DIS double differential cross sections ($\frac{d\si^2}{dxdy}$) from 30 to 360 GeV for both $\numu-Fe$ and $\numubar-Fe$ interactions.
However, these data may not be applicable to other experiments unless
we understand the nuclear target dependence (A-dependence) of DIS cross section. 
Nuclear-dependent DIS is not well known and it is an active field for charged lepton DIS community~\cite{Morfin_shadow}.
This new subject is not studied with neutrinos very carefully, and recent results from
MINERvA target ratio of CC inclusive~\cite{MINERvA_TRatio} and CC DIS interactions~\cite{MINERvA_DISRatio}
shows strong shadowing effect for heavy elements than any models in the market.
Clearly this adds an additional concern for predictions of any heavy target nuclei. 

\subsubsection*{Hadronization}

Hadronization process is not studied carefully in neutrino physics.
It was pointed out that most generators underestimate the averaged charged hadron multiplicity measured
by bubble chamber experiments~\cite{KuzminNaumov,KevinConnolly}.

\begin{figure}[tb]
  \begin{center}
    \includegraphics[width=12cm,valign=m]{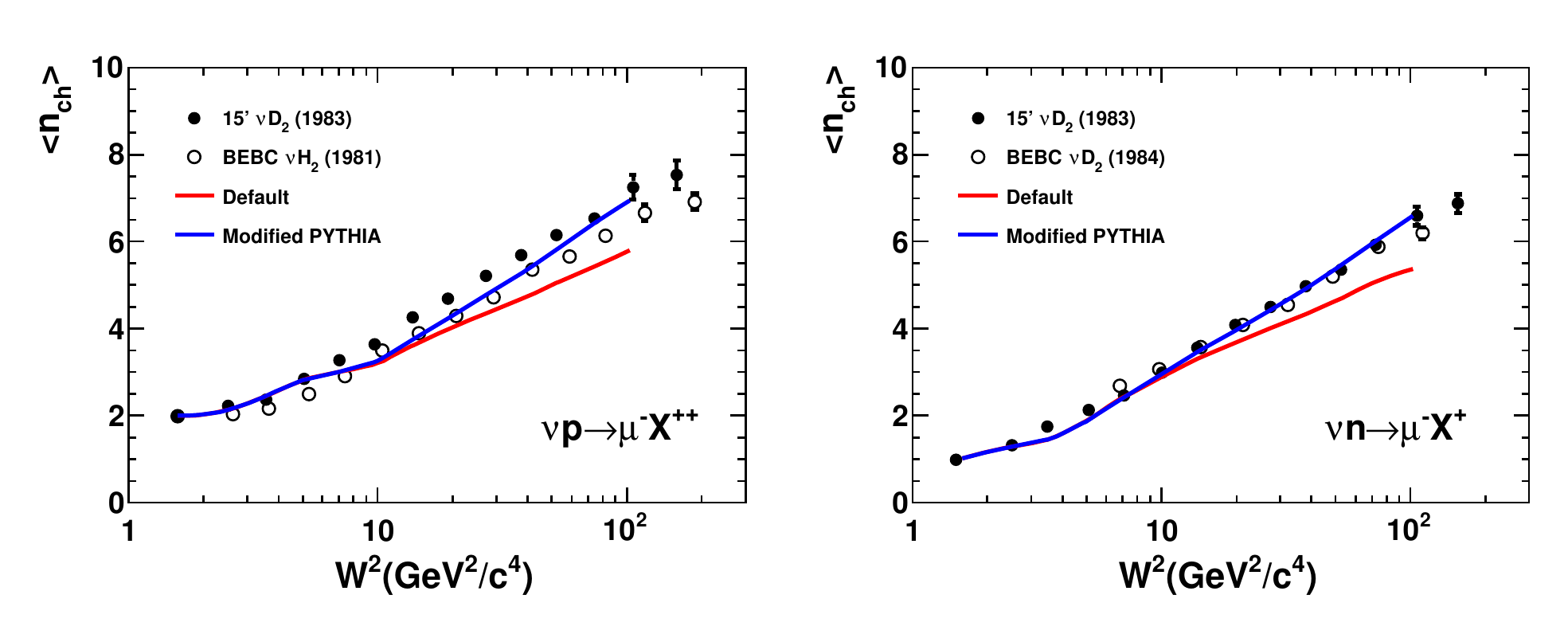}
    \end{center}
\vspace{-2mm}
\caption{
  Data-simulation comparison of the averaged charged hadron multiplicity.
  The left side is $\nu+p$ interaction, and the right side $\nu+n$ interactions~\cite{Teppei_pythia6}. 
}
\label{fig:cMulCh}
\end{figure}

The hadronization model in the neutrino interaction generators~\cite{Tingjun,GiBUU,Nowak_had,Nowak_pythia}
is often governed by the PYTHIA6 hadronization program~\cite{PYTHIA6,PYTHIA8}.
However, PYTHIA6 is by default designed for higher energy collider experiments ($\sqrt{s}\geq 35$~GeV),
and is not optimized for neutrino experiments in $1-10$~GeV region. 
Figure~\ref{fig:cMulCh} shows the data-simulation comparison of averaged charged hadron multiplicity $<n_{ch}>$
with a function of $W^2$. 
As you see, default GENIE hadronization model~\cite{AGKY} based on PYTHIA6 underestimates $<n_{ch}>$ 
comparing with the data from bubble chamber experiments~\cite{Zieminska,Allen:1981vh}.
Tuning of PYTHIA is a common practice in many experiments,
especially HERMES~\cite{HERMES_MulCh}, which is a relatively low energy gas target experiment ($\sim$27~GeV)
comparing with collider experiments, and they developed number of PYTHIA tuning
schemes~\cite{Felixtune,Hillenbrand,JoshRubin,HERMES_GluPol}.
After tuning the fragmentation function by changing PYTHIA6 parameters following the procedure by HERMES,
it was found that neutrino interaction generator can reproduce $<n_{ch}>$ data
from bubble chamber experiments~\cite{Teppei_pythia6}.
Recently, PYTHIA8 is adapted by LHC experiments.
The basic principle of tuning seems to be same with PYTHIA6
to use in neutrino experiments~\cite{Katori:2016blu}.

\begin{figure}[tb]
  \begin{center}
    \includegraphics[width=12cm]{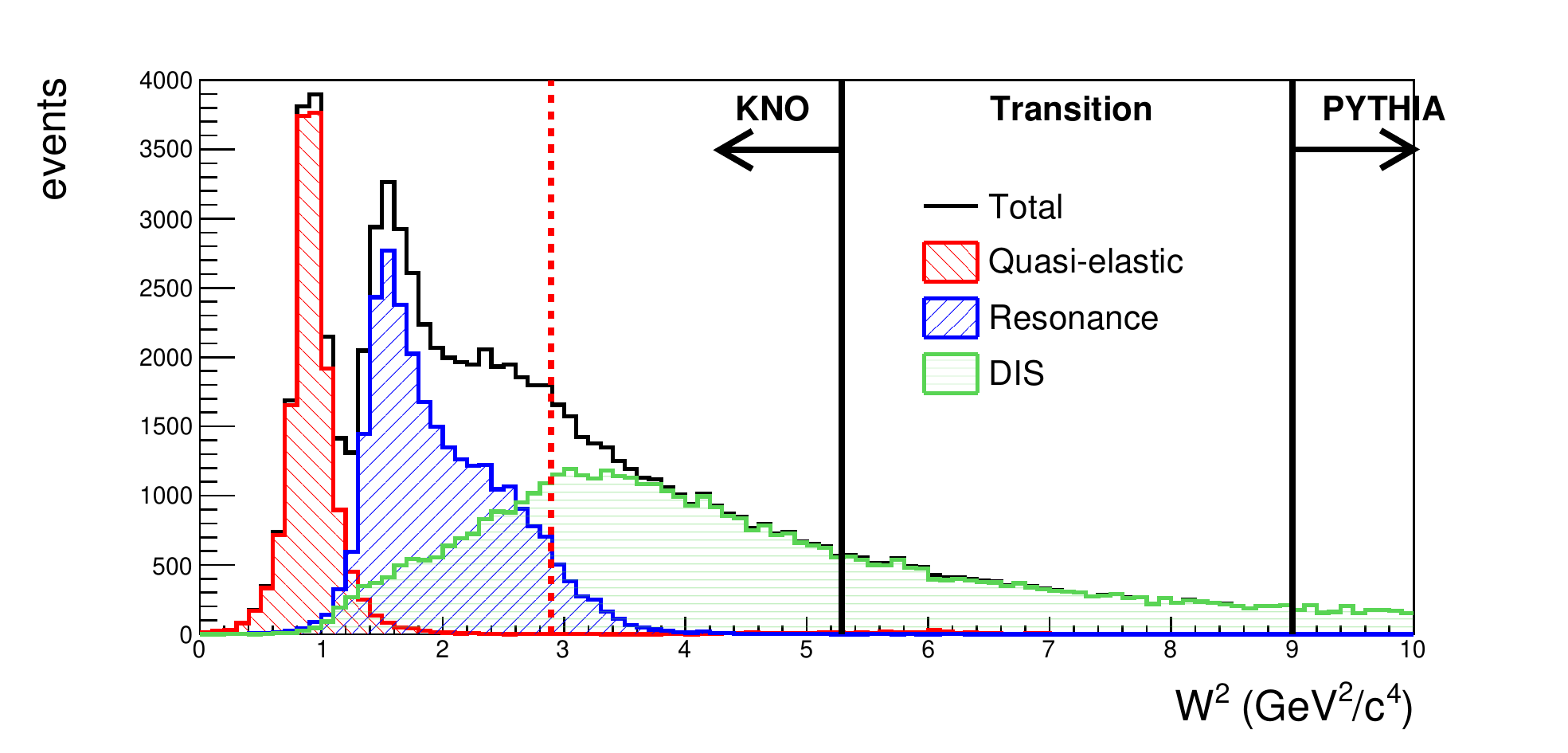}
    \end{center}
\vspace{-2mm}
\caption{
GENIE AGKY model landscape~\cite{AGKY}.
This is the $W^2$ distribution for atmospheric neutrino flux-integrated $\numu$CC interaction on water target. 
The flux is taken from ~\cite{Honda_2015}.
Here, quasi-elastic (red), resonance (blue), and DIS (green) interactions are distributed in $W^2$. 
The 3 types of interactions show how each W region is dominated by a particular interaction, especially
the red dashed line shows a theoretical transition from RES to DIS. 
The hadronization model shifts from KNO scaling-based model to PYTHIA,
where PYTHIA is turned on at $W^2=5.3 GeV^2/c^4$ and completely replace the KNO scaling-based model at $W^2=9 GeV^2/c^4$.
Here charm production channels are not included. 
}
\label{fig:had}
\end{figure}

However, PYTHIA is only valid at high W region and at low W hadronization,
generators use a KNO-scaling based data-driven model~\cite{KNO}.
For example, GENIE hadronization model (AGKY model~\cite{AGKY})
uses data to construct both averaged hadron multiplicities and their dispersions at low W events.
This model is further extended to low W region to simulate non-resonance backgorund.
Therefore, in GENIE, resonance channels and non-resonance background make an incoherent sum.
Fig.~\ref{fig:had} describes the situation.
As you see both interaction models (QE, RES, DIS)
and hadronization model (KNO scaling-based model and PYTHIA) are tried to be connected smoothly.  
Notice the large tail of ``DIS'' extends to QE region,
imitating the non-resonance background.
As we discussed in Sec.~\ref{sec:ccpi0},
re-analysis of bubble chamber data by Rodrigues~{\it et al.}~\cite{Rodrigues:2016xjj}
suggests pion contributions from the non-resonance background may be overestimated in the AGKY model. 

\subsubsection*{Secondary interaction and Detector efficiency}

There are systematics associted with secondary interactions, and these are evaluated by the detector simulation of each experiment.
Similarly with FSI, processes such as absorption and charge exchange changes multiplicity of final state hadrons.
Unfortunately experimentalists often find processes in GEANT are off from external data~\cite{AndyFurmanski_thesis},
and secondary interaction ends up with a large fraction of detector related systematics. 

In particle physics, detectors are tested with a dedicated beam line with known particles to understand the performance.
MINERvA tested their prototype detector at Fermilab Test Beam Facility (FTBF)~\cite{MINERvA_beamtest}. 
T2K tested each module of the near detector complex~\cite{Abe:2011ks}.
FGDs were tested at TRIUMF M11 secondary beamline~\cite{FGD},
and downstream ECal was tested at CERN PS T9 test beam~\cite{ECal}. 
Furthermore, detector performances are checked {\it in situ} by through going cosmic muons. 
But these calibrations often miss requirements for hadron measurements.
First, detectors, especially trackers, are not isotropic and the response depends on the angles of the tracks.
At low energy low $Q^2$ interactions, hadrons tend to go at a higher angle but they are often not studied by beam tests.
Second, low energy hadrons have large $\frac{dE}{dx}$ and
this may saturate detectors designed to measure MIPs (minimumly ionizing particles).
Third, in neutrino interactions, particles are knocked out from the inside of nuclei of the detector materials, 
but through going hadrons cannot imitate this situation. 
Therefore, we still need to rely a significant amount of detector responses on simulations.
Mismodeling of these would give incorrect efficiency correction on data.

\subsection{Kaon and strangeness production}
Atmospheric neutrino induced kaon production is a major background of the $K$-mode proton decay ($p \to \bar\nu K^+$). 
MINERvA made the first differential cross section measurement
of neutrino induced CC charged kaon production\cite{Marshall:2016rrn}.
The analysis developed a clever reconstruction of charged kaons.
Not only that, MINERvA needed to update both kaon production and propagation simulations, 
showing hadron measurements beyond pions are still very new in this community.
Recently MINERvA also announced the first evidence of neutrino-induced coherent kaon production~\cite{Wang:2016pww},
and we anticipate the results from the NC charged kaon production measurements, too. 

Neutrino induced strangeness productions offer unique opportunities to study various aspects~\cite{Alvarez-Ruso:2014bla},
including enhancement of $\pim$ productions~\cite{Singh:2006xp,Singh_antinupion} and the hyperon polarization~\cite{Akbar:2016awk}, 
however, the field is premature due to a lack of data. 
Next generation high resolution detectors (such as LArTPC discussed in Sec.~\ref{sec:lartpc})
may be able to reconstruct short hyperon tracks~\cite{Teppei_uB}.

\subsection{Coherent pion productions}

We conclude this section by reviewing the status of the coherent pion production:
production of one pion with the nucleus left in its ground state.
The CC and NC reactions are $\nu_\mu~A \to \mu^- ~ \pi^+~ A $ and
$\nu_\mu~A \to \nu_\mu ~ \pi^0~ A $, respectively.

Naive isospin ansatz gives the cross section ratio of CC and NC coherent pion production would be 2:1.
\beq
&&\frac{\left|\left<\pip ,\mu^-|0, \numu\right>\right|^2}{\left|\left<\piz ,\numu|0, \numu \right>\right|^2}
=
\frac{
\left|\left< T_{\pi},T_l;T_{\pi}^3(\pip),T_l^3(\mu^-)|0,T_l;0,T_l^3(\numu)\right>\right|^2}{
\left|\left< T_{\pi},T_l;T_{\pi}^3(\piz),T_l^3(\numu)|0,T_l;0,T_l^3(\numu)\right>\right|^2}\\
&=&
\frac{
\left|\left< 1,\frac{1}{2};+1,-\frac{1}{2}|0,\frac{1}{2};0,+\frac{1}{2}\right>\right|^2}{
\left|\left< 1,\frac{1}{2}; 0,+\frac{1}{2}|0,\frac{1}{2};0,+\frac{1}{2}\right>\right|^2}
=
\frac{
\left|\sqrt{\frac{2}{3}}\left< \frac{1}{2},+\frac{1}{2}\right|\left.\frac{1}{2},+\frac{1}{2}\right>\right|^2}{
\left|\sqrt{\frac{1}{3}}\left< \frac{1}{2},+\frac{1}{2}\right|\left.\frac{1}{2},+\frac{1}{2}\right>\right|^2}
=2\no
\eeq

Our struggle against the coherent in the 1 GeV region starts from K2K,
where they set the upper limit of CC coherent pion production cross section~\cite{K2K_CCcohpi}.
A subsequent measurement by SciBooNE supported this result~\cite{SB_CCpip},
but on the other hand according to MiniBooNE~\cite{MB_NCpi0PLB} and SciBooNE~\cite{SB_NCpi02} measurements,
NC coherent pion production seems nonzero.
This leads to following ratio
\beq
\frac{\sigma_{\textrm{CC-coh}}}{\sigma_{\textrm{NC-coh}}}=0.14^{+0.30}_{-0.28}
\eeq
measured by SciBooNE on carbon at an average neutrino energy of 0.8 GeV ~\cite{SB_NCpi02}.
It is very difficult to reconcile this value with the 2:1 ratio previously discussed and
with many theoretical predictions based on PCAC hypothesis \cite{Rein:1982pf,Paschos:2005km,Rein:2006di,Berger:2007rq} or on microscopic
calculations~\cite{Singh:2006br,AlvarezRuso:2007tt,AlvarezRuso:2007it,Amaro:2008hd,Martini:2009uj,Nakamura:2009iq,Hernandez:2010jf,Zhang:2012xi}
which obtain results between 1.5 and 2.
\footnote{for a synthetic review on these models see Ref.~\cite{Boyd:2009zz}}

\begin{figure}[tb]
  \begin{center}
    \includegraphics[width=12cm]{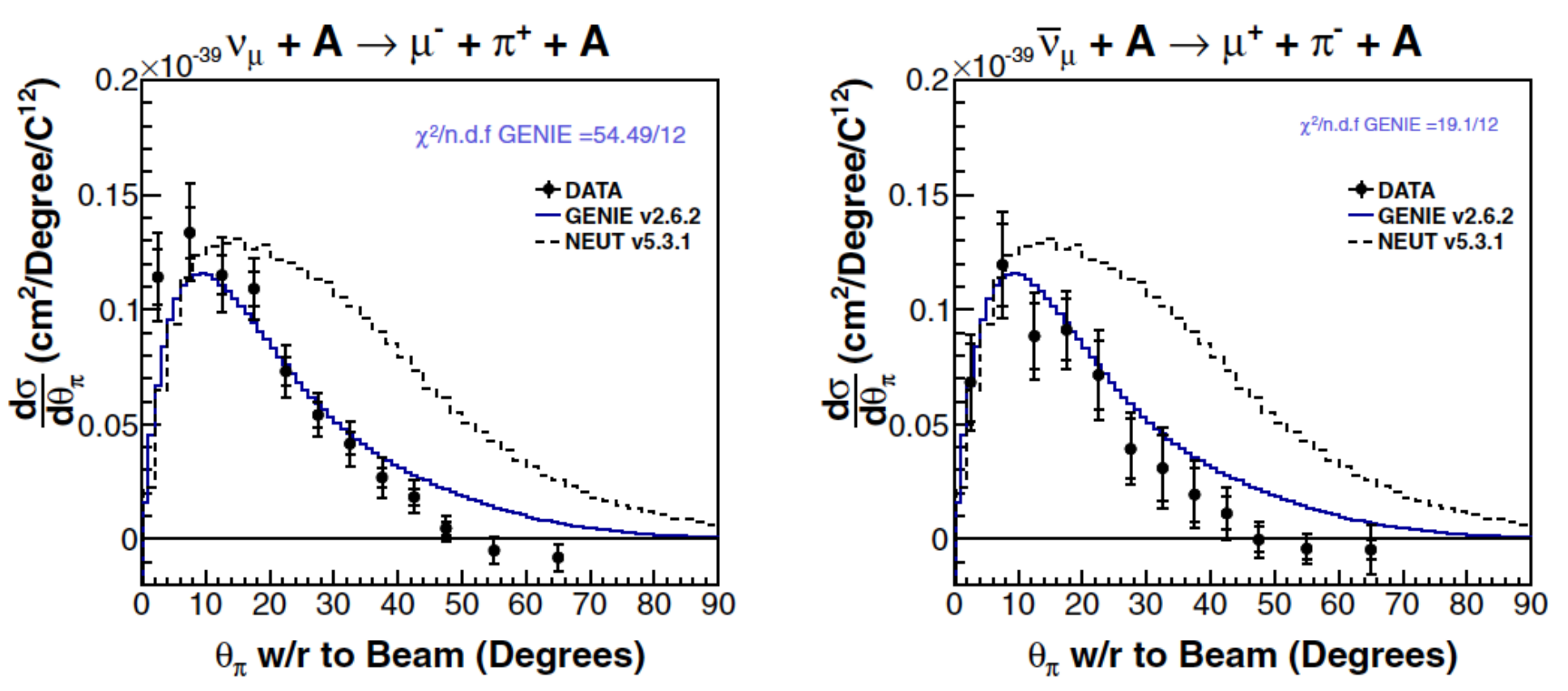}
    \end{center}
\vspace{-2mm}
\caption{
  Flux-integrated differential cross section of CC coherent charged pion production for neutrino and antineutrino mode.
  ~\cite{Higuera:2014azj}.
}
\label{fig:MINERvA_cohpi}
\end{figure}

The situation of the very elusive $\pi^+$ coherent
dramatically changed after a recent series of measurements of nonzero CC coherent pion production results. 
Figure~\ref{fig:MINERvA_cohpi} shows the MINERvA neutrino and antineutrino
CC coherent charged pion production cross sections~\cite{Higuera:2014azj}.
Clearly we see nonzero cross sections on this.   
Moreover, MINERvA measured kinematics with high precision,
which allows these data to actually test theoretical models and Monte Carlo.
For example in Fig.~\ref{fig:MINERvA_cohpi} the data are compared with GENIE and NEUT Monte Carlo predictions. 
As you see, data angle distribution is rather narrower than these models,
and the agreement with GENIE is better than NEUT. 
Surprisingly, both NEUT and GENIE use the Rein-Sehgal model~\cite{Rein:1982pf},
and the difference of distributions are made simply by how the model is
implemented~\footnote{now NEUT prediction is consistent with GENIE and NuWro after updating the code in NEUT}.
Here, the Rein-Sehgal model is based on $\pi-N$ inelastic cross section tables with analytic A-scaling,
which is expected to be inaccurate at the low pion energy ($<$1~GeV). 
On the other hand, the Berger-Sehgal model~\cite{BergerSehgal_coh} is based on $\pi-C$ elastic cross section tables
and it is more suitable for the accelerator-based neutrino oscillation experiments.
However, in order to use Berger-Sehgal model on anything other than a carbon target with a high accuracy,
one needs to come up a realistic A-scaling law. 

ArgoNeuT also shows clear evidence of CC coherent charged pion production on argon~\cite{ArgoNeuT_CCcohpi}. 
This is the first such measurement on an argon target.
On top of these measurements, the T2K off-axis detector measured nonzero coherent pion production~\cite{Abe:2016fic}
in the energy region closer to K2K and SciBooNE, contradicting the null results reported previously by these two experiments.
Furthermore, the T2K results have been compared not only to the standard Rein-Sehgal model
but also to the model of Alvarez-Ruso \textit{et al.}~\cite{AlvarezRuso:2007tt,AlvarezRuso:2007it},
representing the first example of microscopic models for the coherent pion production 
implemented in the Monte Carlo.
Initial results of the T2K on-axis detector are also consistent with nonzero coherent pion production~\cite{Kikawa_thesis}.
To summarize, from no positive results on this CC coherent channel,
the field evolved so quickly to the situation of the high precision measurement!

Although there is a strong interest and series of measurements of coherent pion production,
there is a caution because none of these experiments measure the nuclear levels.
By definition, ``coherent'' interactions require target nuclei to be the ground state after the interactions.
However none of the experiments have the ability to measure that.
Modern tracker experiments utilize so called ``vertex activity'', an energy deposit around the interaction vertex,
to select coherent events.
Lower vertex activity may be related to lower excitation of the nuclear target's final state,
but it is not clear how to check that the target nucleus remains exactly in its ground state.

Recently, there is speculation that the diffractive pion production on hydrogen may be contaminated in 
the MINERvA data of the coherent pion production on nuclear targets~\cite{Wolcott:2016hws}.
They share common kinematics, such as small momentum transfer,
yet the diffractive productions don't necessarily have small vertex activity due to the proton recoil.
We are awaiting further confirmation of this from other experiments. 

\section{Charged current inclusive cross sections\label{sec:cc}}
\subsection{$\nu_\mu$ CC inclusive\label{subsec:numuccincl}}

The charged current inclusive cross section measurement is defined as the one in which only the charged lepton is detected. 
In principle all the reaction mechanisms 
(nuclear resonances, quasielastic, multinucleon excitations, one and multi-pion production, deep inelastic scattering), 
once above a threshold, can contribute to this process.
The relative weight of the different mechanisms depend on the neutrino beam. 
The importance of inclusive measurements is related to the fact that they are less affected by background subtraction 
with respect to exclusive channels measurements,
hence they can provide useful information and tests on the neutrino fluxes and on the theoretical models. 
In the neutrino energy domain treated in this paper,
the experimental published results on $\nu_\mu$ CC inclusive cross sections on carbon are 
the ones of SciBooNE~\cite{Nakajima:2010fp}, T2K~\cite{Abe:2013jth}
and MINERvA~\cite{DeVan:2016rkm}.
The $\nu_\mu$ CC inclusive cross sections on argon have been measured
by the ArgoNeuT collaboration~\cite{Anderson:2011ce,Acciarri:2014isz}.
Gargamelle ~\cite{Blietschau:1977mu} and T2K ~\cite{Abe:2014agb}
collaborations published also the $\nu_e$ CC inclusive cross sections.
Preliminary results on the same cross section as a function of the neutrino energy
have been presented also by NO$\nu$A~\cite{Bu:2016grw}. 

The SciBooNE collaboration published in Ref.~\cite{Nakajima:2010fp}
the $\nu_\mu$ CC inclusive cross section as a function of the neutrino energy up to $E_\nu\sim$3~GeV.  
The corresponding theoretical calculations were published by Nieves~\textit{et al.} in Ref.~\cite{Nieves:2011pp}, 
by Martini and Ericson in Ref.~\cite{Martini:2014dqa}, by Ivanov \textit{et al.} and Megias \textit{et al.}
in Refs.~\cite{Ivanov:2015aya} and ~\cite{Megias:2016fjk} respectively.
All these approaches include the genuine quasielastic, the one pion production 
and the multinucleon excitations contributions.
In spite of the differences of these models in the treatment of the different channels, 
all the approaches obtain a good agreement up to $E_\nu\simeq$0.8 GeV, a small underestimation of
the data around $E_\nu\simeq$1 GeV and a larger underestimation of the data for larger $E_\nu$.
As discussed in Refs.~\cite{Nieves:2011pp,Martini:2014dqa,Ivanov:2015aya,Megias:2016fjk}, 
the natural interpretation of this underestimation is the existence of other channels which open up at high energies,
and which have not been included in the three analyses.
A likely candidate for the missing channel is the multi-pion production,
in particular the two-pion production one. 

\begin{figure}
\begin{center}
  \includegraphics[height=\textheight]{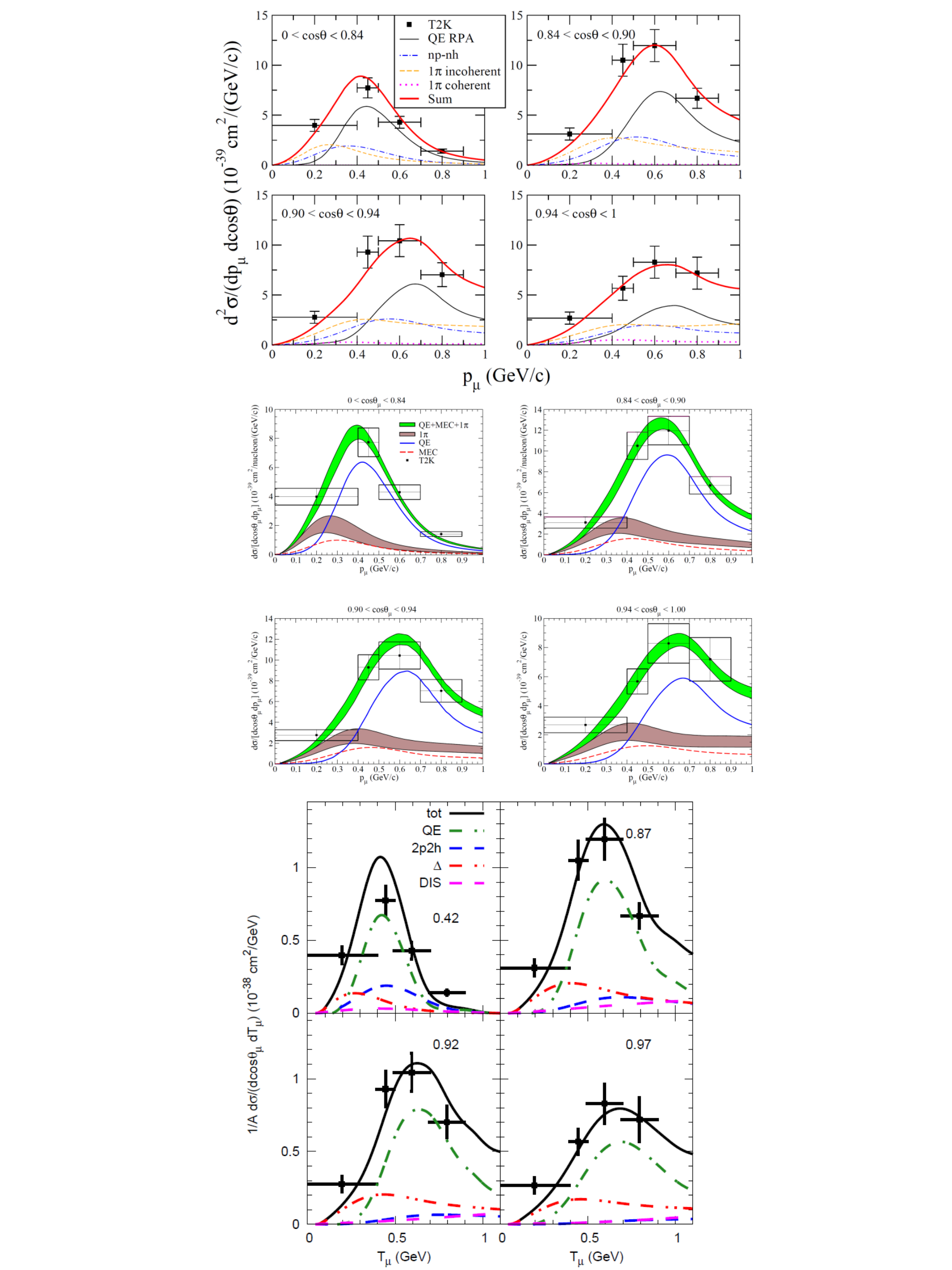}
\caption{(color online) T2K flux-integrated $\nu_\mu$ CC inclusive 
double differential cross section on carbon per nucleon
as a function of the muon momentum.
The experimental T2K points are taken from~\cite{Abe:2013jth}. 
The different contributions to the inclusive cross sections obtained in three different models are shown.
Four upper panels: Martini and Ericson results of Ref.~\cite{Martini:2014dqa};
four middle panels: SuSAv2-MEC results of Megias \textit{et al.} ~\cite{Megias:2016fjk};
four lower panels: GiBUU results of Gallmeister \textit{et al.} ~\cite{Gallmeister:2016dnq}.}
\label{fig_t2k_d2s}
\end{center}
\end{figure}

The T2K collaboration published in Ref.~\cite{Abe:2013jth}
the inclusive $\nu_\mu$ CC double differential cross section on carbon.  
The experimental data of the T2K flux-integrated double differential cross section
as function of the emitted muon momentum for the various angular bins 
are reported in Fig.~\ref{fig_t2k_d2s} and compared with the theoretical results of Martini
and Ericson of Ref.~\cite{Martini:2014dqa}~\footnote{Unlike the original figure of Ref.~\cite{Martini:2014dqa}
where the different channels were added to the genuine quasielastic,
here the different components are separately plotted.
Furthermore minor differences between the original Fig.1 of Ref.~\cite{Martini:2014dqa}
and the results reported here are due to the fact that the present calculations are
performed without cut in the neutrino energy,
while the results of Fig.~1 of Ref.~\cite{Martini:2014dqa} were obtained by considering the T2K flux only up to 3 GeV.},
the ones of the SuSAv2-MEC approach of Megias \textit{et al.}~\cite{Megias:2016fjk}
and the GiBUU results of Ref.~\cite{Gallmeister:2016dnq}. 
In Fig.~\ref{fig_t2k_d2s} the different components of
the theoretical cross section in the three different approaches are shown:
the genuine quasielastic channel, the multinucleon component,
the single pion production cross section and then the total one.
Other components are also shown: the coherent pion production in the case of Martini and Ericson ~\cite{Martini:2014dqa}
and the DIS one in the GiBUU case ~\cite{Gallmeister:2016dnq}.
Both contributions in this $\nu_\mu$ CC inclusive cross section are too small to be singled out.
All the theoretical evaluations are compatible with the T2K inclusive data.
As in the previous analysis of the MiniBooNE QE-like cross sections
the multinucleon component is needed in order to reproduce the experimental results,
especially in the Martini and Ericson ~\cite{Martini:2014dqa} and in the SuSAv2-MEC ~\cite{Megias:2016fjk} cases.
These results represent another successful test of the necessity of the multinucleon emission channel
in an experiment with another neutrino flux (but in the same neutrino energy domain) with respect to the one of MiniBooNE. 

In spite of the relative similarity of the theoretical results in the different approaches for the inclusive cross sections,
differences appear when one separately compares the different channels.
For example, the genuine quasielastic contribution of Megias \textit{et al.} and GiBUU
is larger with respect to the one of Martini and Ericson. 
RPA suppression effects which characterize the approach of Martini \textit{et al.} are absent in the SuSAv2 and in GiBUU.
In GiBUU the np-nh contribution is of only minor importance for most of the angles, except for the largest one.
This is not the case of the SuSAv2-MEC and the Martini and Ericson results,
the ones presenting the largest multinucleon contributions.
We remind that the SuSAv2 is based on a relativistic mean field approach
in which many of the NN correlation effects are already
present via strong scalar and vector meson exchanges, while in the case of Martini \textit{et al.}
the NN short range correlation contributions are part of the np-nh channel.
Furthermore NN correlations-MEC interference contributions are present only in the Martini and Ericson case.
As often happens, a one-to one correspondence between the different exclusive
channel's contributions of different theoretical calculations is hardly possible.

Interestingly, the calculations of Martini and Ericson and the ones of Megias \textit{et al.} show that
there is a region where the multinucleon excitations dominate with respect to the genuine quasielastic:
this is the low $p_\mu$ one. Unfortunately this is also the region where one pion production contribution is also important.

Other contributions not included in the three descriptions mentioned above could be also relevant for the calculation 
of the inclusive cross sections such as excitations of
low-lying giant resonances at low energy and the 2p-2h1$\pi$ excitations 
and two-pions production at high energy.
Several studies have been made on the excitation of low energy collective states in neutrino
interactions~\cite{Kolbe:1995af,Volpe:2000zn,Jachowicz:2002rr,Botrugno:2005kn,Martini:2007jw,Samana:2010up,Pandey:2013cca,Pandey:2014tza}. 
Their role in the T2K inclusive double differential cross section,
in particular in the forward bins, has been analyzed by Pandey \textit{et al.} in Ref.~\cite{Pandey:2016jju}.
Concerning the high energy excitations, for the moment theoretical calculations of 2p2h1$\pi$ 
and two-pions production contributions to neutrino flux-integrated double differential cross sections on nuclei are absent.
We mention however the theoretical calculations of the neutrino
two-pions production total cross sections on nucleon of Hernandez \textit{et al.}~\cite{Hernandez:2007ej}
and Nakamura \textit{et al.}~\cite{Nakamura:2015rta}.
2p-2h1$\pi$ excitations have been calculated for sake of illustration only
for one value of the neutrino energy ($E_\nu=1$ GeV) by Nieves \textit{et al.} in Fig. 19 of Ref. ~\cite{Nieves:2011pp}.  

Martini \textit{et al.}, Megias \textit{et al.}, and Gallmeister \textit{et al.} have studied
the inclusive differential cross sections by isolating the different contributions
not only in connection with the T2K $\nu_\mu$ data but also with T2K $\nu_e$ ones.
The $\nu_e$ inclusive cross section is discussed in the next section.
A comparison with the inclusive cross sections has also been performed
by Meucci and Giusti using the Relativistic Green's function model
which turned to underestimate the $\nu_\mu$ and $\nu_e$ CC T2K data ~\cite{Meucci:2015bea}.
The theoretical calculations within the same Relativistic
Green's function approach are compared in Ref.~\cite{Meucci:2013gja}
with the $\nu_\mu$ CC inclusive cross section on argon measured by ArgoNeuT~\cite{Anderson:2011ce,Acciarri:2014isz}.
Also in this case an underestimation of the data appears.
Theoretical predictions with GiBUU of $\nu_\mu$ CC flux integrated inclusive double differential cross sections on argon
by considering the BNB (MicroBooNE) and the NUMI (DUNE) beams are presented in Ref.~\cite{Gallmeister:2016dnq}.

\subsection{$\nu_e$ cross sections \label{subsecc:nue}}

A precise and simultaneous knowledge of $\nu_\mu$ and $\nu_e$ cross sections
is important in connection to the $\nu_\mu \to \nu_e$ oscillation experiments aiming at the 
determination of the neutrino mass ordering and the search for CP violation in the lepton sector.
The wealth of experimental and theoretical results on muon-neutrino cross sections
contrast with the few published results on electron-neutrino cross sections.
This is essentially due the relatively small component of electron-neutrino
fluxes with respect to the muon-neutrino ones hence to small statistics.
For this reason the electron-neutrino experimental published results essentially concern inclusive cross sections:
Gargamelle ~\cite{Blietschau:1977mu}, T2K ~\cite{Abe:2014agb} and NOvA preliminary results ~\cite{Bu:2016grw} 
(a prominent exception is represented by the quasielastic measurement of MINERvA \cite{Wolcott:2015hda}).

\begin{figure}
\begin{center}
  \includegraphics[width=12cm,height=8cm]{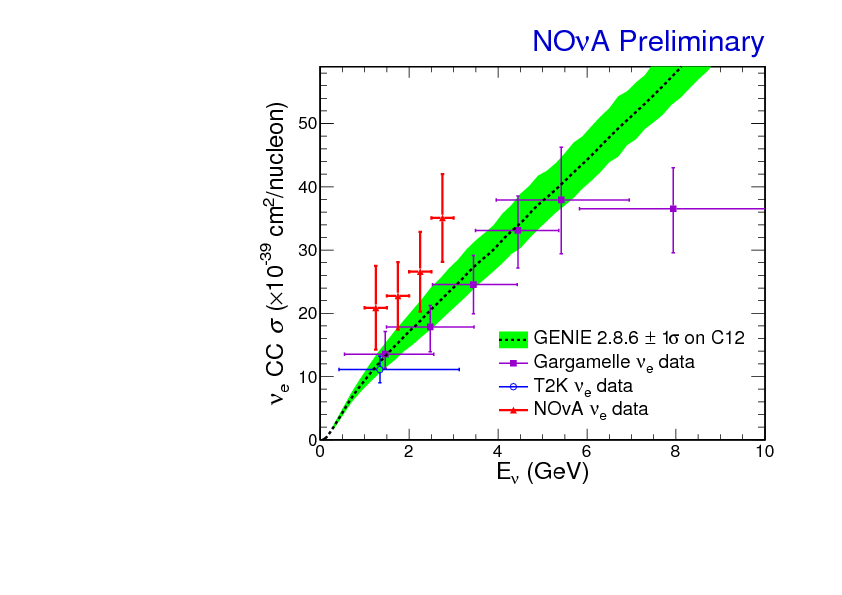}
\caption{(Color online) The $\nu_e$ CC inclusive cross sections as a function of the neutrino energy measured by NOvA ~\cite{Bu:2016grw},
Gargamelle ~\cite{Blietschau:1977mu}and T2K ~\cite{Abe:2014agb}. 
The figure is taken from Ref.~\cite{Bu:2016grw} where the preliminary results of NOvA are shown for the first time. }
\label{fig_incl_nova}
\end{center}
\end{figure}

The three $\nu_e$ CC inclusive cross sections of Refs.~\cite{Blietschau:1977mu,Abe:2014agb,Bu:2016grw}
as a function of the neutrino energy are shown in Fig.~\ref{fig_incl_nova} taken from Ref.~\cite{Bu:2016grw}.
By comparing the NOvA preliminary results with the Gargamelle ones,
one can observe the general trend of the NOvA data to be larger than the the Gargamelle ones.
We remind that the Gargamelle cross section was measured on bubble chamber,
while the NOvA one on Carbon.
Nuclear effects such as multinucleon excitations could be the responsible of the difference between the two measurements.
Also the T2K cross section refers to carbon but the only published point of T2K hardly can help.

\begin{figure}
\begin{center}
  \includegraphics[height=\textheight]{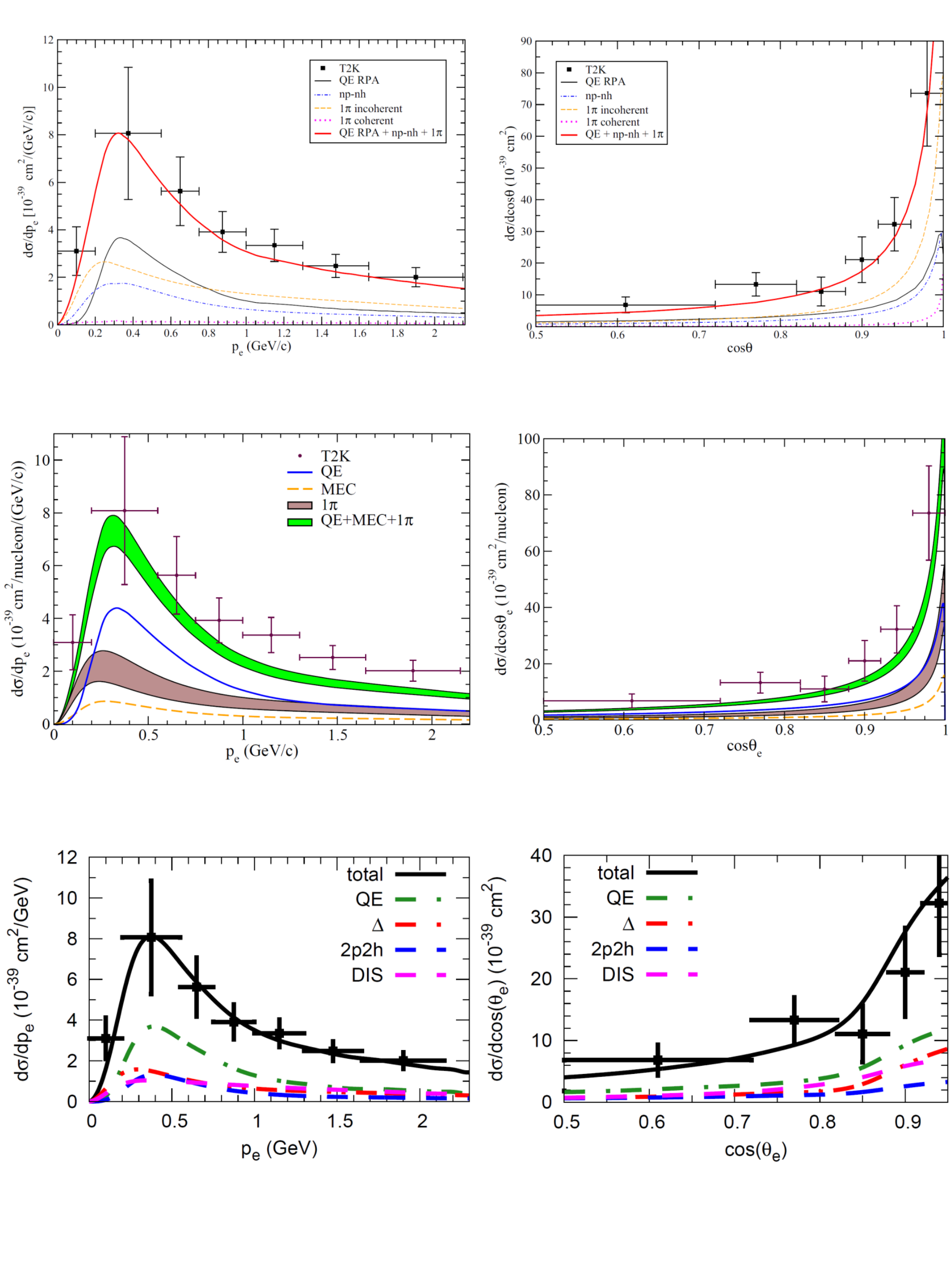}
\caption{(color online). T2K flux-integrated $\nu_e$ CC inclusive 
differential cross sections on carbon per nucleon
as a function of the electron momentum (left panels) and lepton scattering angle (right panels).
The experimental T2K points are taken from~\cite{Abe:2014agb}. 
The different contributions to the inclusive cross sections obtained in three different models are shown.
Upper panels: Martini \textit{et al.} results of Ref.~\cite{Martini:2016eec};
middle panels: SuSAv2-MEC results of Megias \textit{et al.} ~\cite{Megias:2016fjk};
lower panels: GiBUU results of Gallmeister \textit{et al.} ~\cite{Gallmeister:2016dnq}.}
\label{fig_comp_nue_incl}
\end{center}
\end{figure}

On the other hand the T2K measurement is especially important since beyond the total cross section,
the flux-integrated differential cross sections $\frac{d \sigma}{d p_e}$ and $\frac{d \sigma}{d \cos \theta_e}$
are published in Ref.~\cite{Abe:2014agb}.
These differential cross sections are shown in Fig.~\ref{fig_comp_nue_incl} where they are compared
to the theoretical calculations of Martini \textit{et al.} published in Ref.~\cite{Martini:2016eec},
to the ones of the SuSAv2-MEC approach of Megias \textit{et al.} ~\cite{Megias:2016fjk}
and to the GiBUU results of Ref.~\cite{Gallmeister:2016dnq}. 
This comparison parallels with the one of Fig.~\ref{fig_t2k_d2s}
related to the $\nu_\mu$ CC inclusive cross T2K cross sections.
Also in this case the three theoretical calculations of
the $\nu_e$ CC inclusive differential cross sections substantially agree with data.
Differences appear in the separate channel contributions. The DIS cross section,
present only in the GiBBU results ~\cite{Gallmeister:2016dnq} could reduce
the underestimation of Martini \textit{et al.} and Megias \textit{et al.},
which however is very small in the case of Martini \textit{et al.}
Once again the agreement with data needs the presence of the np-nh excitations.

The same conclusion holds also for the only $\nu_e$ CCQE-like differential cross sections on hydrocarbon
published by MINERvA in Ref.~\cite{Wolcott:2015hda} and compared with the SuSAv2+MEC approach
by Megias \textit{et al.} in Ref.~\cite{Megias:2016fjk}. 
In the MINERvA paper ~\cite{Wolcott:2015hda} the ratio of the $\nu_e+\bar{\nu}_e$
$Q^2$ distribution over the corresponding MINERvA $\nu_\mu$ one
is also shown and compared with the GENIE calculation.
The same ratio is calculated also by Megias \textit{et al.} in Ref.~\cite{Megias:2016fjk}.
The error bars of the experimental data seem too large to allow any conclusion
when compared with GENIE and SuSA2+MEC predictions. 

\subsection{$\nu_e$ \textit{vs} $\nu_\mu$ cross sections}

\begin{figure}
\begin{center}
  \includegraphics[width=12cm,height=8cm]{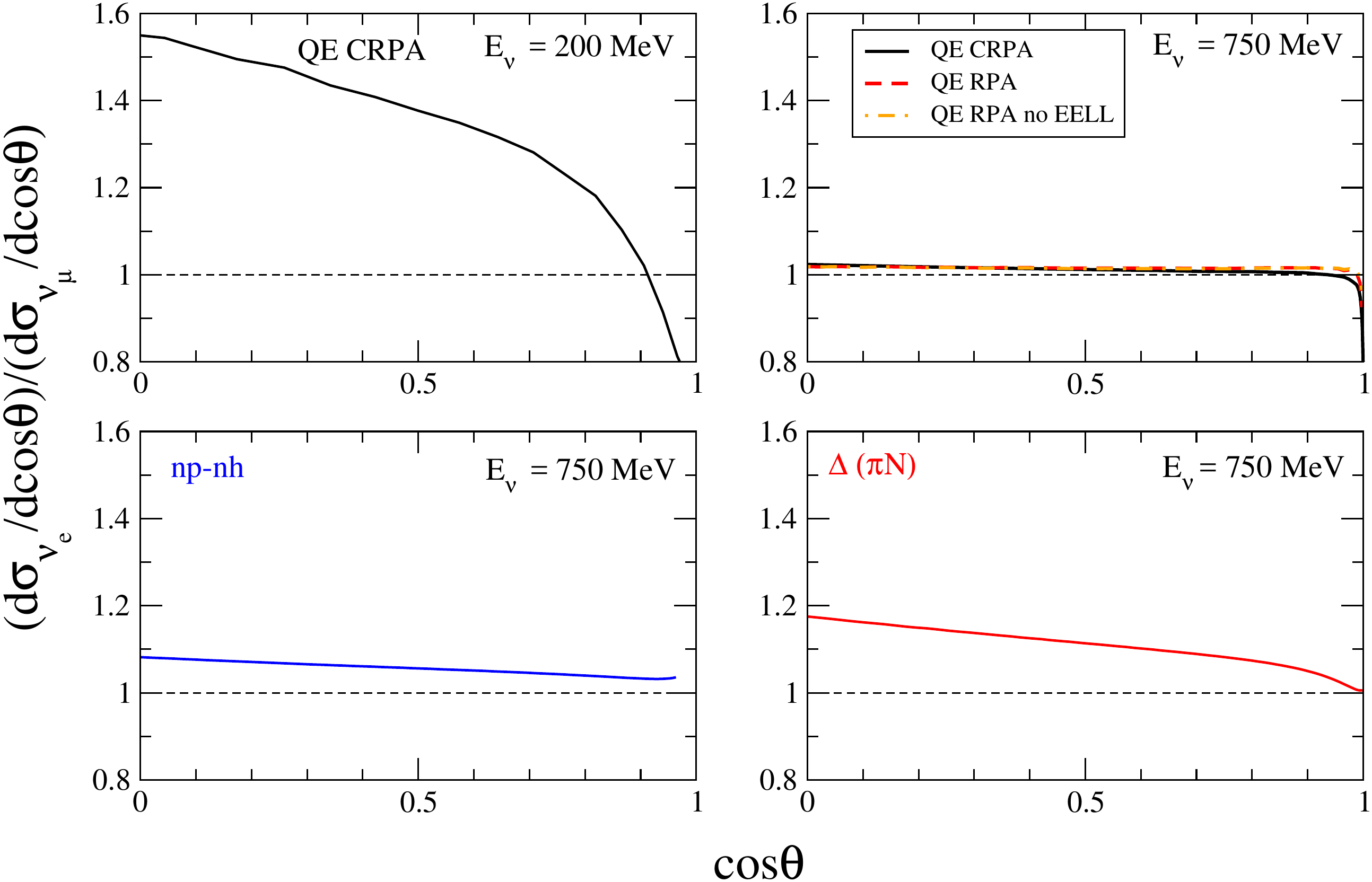}  
\caption{(color online) Ratio of the $\nu_e$ over $\nu_\mu$  differential cross section on Carbon calculated for
two fixed values of incident neutrino energies as a function of the cosine of the lepton scattering angle.
The 1p-1h results in the CRPA approach are shown for $E_\nu$=200 MeV and $E_\nu$=750 MeV. 
The 1p-1h results in the RPA approach, the np-nh excitations and
the one pion production (via $\Delta$ excitation) results are shown for $E_\nu$=750 MeV.
The figure is taken from Ref.~\cite{Martini:2016eec}.}
\label{fig_nue_numu_crpa_ratio_E200_final}
\end{center}
\end{figure}

In principle the ratio and the difference between the $\nu_e$ and $\nu_\mu$ cross sections are interesting quantities to
study the differences among $\nu_e$ and $\nu_\mu$ scattering.
A theoretical  comparison of the  $\nu_\mu$ and $\nu_e$ cross sections was performed
by Day and McFarland in Ref.~\cite{Day:2012gb}
and by Akbar \textit{et al.} in Ref.~\cite{Akbar:2015yda}. 
They analyzed the influence of kinematic limits due to the different final lepton-mass, of radiative corrections,
of uncertainties in nucleon form factors and of second class currents on
the cross sections as a function of the neutrino energy and of $Q^2$.
In Ref.~\cite{Martini:2016eec} Martini \textit{et al.} analyzed  the influence of the final lepton-mass difference focusing
on the $\nu_\mu$ and $\nu_e$ differential cross sections.
They considered two different approaches. 
On one hand the one of Martini \textit{et al.}~\cite{Martini:2009uj} based on nuclear response functions, 
treated RPA including the quasielastic excitation,
multinucleon emission and coherent and incoherent single pion production.
On the other hand, the model of Jachowicz \textit{et al.} \cite{Jachowicz:2002rr}
based on the continuum random phase approximation (CRPA),
originally developed to study electroweak reactions in the giant resonance region and then extended
by Pandey \textit{et al.} \cite{Pandey:2013cca,Pandey:2014tza} to the quasielastic regime.
Among the different  $\nu_\mu$ and $\nu_e$ results of Ref.~\cite{Martini:2016eec},
for the sake of illustration we show in Fig.~\ref{fig_nue_numu_crpa_ratio_E200_final}
the ratio of the single differential cross section 
$\frac{d \sigma_{\nu_e}}{d \cos \theta}/\frac{d \sigma_{\nu_\mu}} {d \cos \theta}$ for $E_\nu$=200 MeV and $E_\nu$=750 MeV.
In the low energy case of $E_\nu$=200 MeV the ratio for the 1p-1h channel calculated in
CRPA (the more appropriate method for low energies)
deviates very appreciably from 1 while at larger neutrino energies, such as $E_\nu$=750 MeV,
it gets closer to 1 in the CRPA as well as in the RPA case.
In Fig.~\ref{fig_nue_numu_crpa_ratio_E200_final} this quantity
$\frac{d \sigma_{\nu_e}}{d \cos \theta}/\frac{d \sigma_{\nu_\mu}} {d \cos \theta}$ at $E_\nu$=750~MeV
is given also for two other channels,
the pion production and multinucleon excitations in the RPA approach of Martini \textit{et al.}
This $\frac{d \sigma_{\nu_e}}{d \cos \theta}/\frac{d \sigma_{\nu_\mu}} {d \cos \theta}$ ratio,
always larger than 1, is characterized by a smooth decreasing behavior. 
For the pion emission channel (via $\Delta$ excitation) this ratio is
larger than the one for the np-nh and 1p-1h excitations.
Due to the different kinematic limits, the $\nu_e$ cross sections are
in general expected to be larger than the $\nu_\mu$ ones.
However for forward scattering angles this hierarchy is opposite.
This appears for the 1p-1h excitations (genuine QE and giant resonances) at low neutrino energies.  
This behavior is related to a non-trivial dependence of momentum transfer on lepton mass and scattering angle,
and to a subtle interplay between lepton kinematic factors and response functions.
As it is known, the different contributions to the cross sections
(Coulomb, Longitudinal, Transverse,..., see Eqs.(\ref{m_eq_general}) and (\ref{eq1:cross_section}))
are characterized by different angular and energy dependence.
Further details are given in Ref.~\cite{Martini:2016eec}.

In the precision era of neutrino oscillation physics the $\nu_e$ cross sections should be known with the same accuracy as the $\nu_\mu$ ones. 
Trying to deduce the $\nu_e$ cross sections from the experimental $\nu_\mu$ ones can be considered only as
a first approximation in the study of the $\nu_e$ interactions. The first experimental
observation of electron neutrinos and anti-neutrinos in the ArgoNeut Liquid Argon Time Projection Chamber, in the energy range
relevant to DUNE and the Fermilab Short Baseline Neutrino Program has been presented in Ref.~\cite{Acciarri:2016sli}.

\section{Neutrino energy recostruction and neutrino oscillation analysis\label{sec:enrec}}

The neutrino energy reconstruction problem has been already mentioned in the introduction. 
In accelerator-based experiments the neutrino beams 
(at difference with respect to electron beams, for example) are not monochromatic but they span a wide range of energies,
hence the incoming neutrino energy is reconstructed from the final states of the reaction. 
One possibility, for which the MINOS and DUNE collaborations has opted, is
the calorimetric energy reconstruction method whose main limitations are
the acceptance and the efficiencies of the detector~\cite{Ankowski:2015kya}. 
The other possibility, the one chosen by lower-energy experiments and
in particular by Cherenkov detector based experiments like MiniBooNE and Super-Kamiokande,
is the quasielastic kinematics-based method.
In the following we analyze this second option.
In this case the determination of the initial neutrino energy is done through
the charged current neutrino-nucleus quasielastic-like events, where only the charged lepton is observed.

The measured charged lepton variables (energy and scattering angle)
are used to reconstruct the neutrino energy via a two-body formula by assuming that
the neutrino interaction in the nuclear target takes place on a nucleon at rest. 
In this case, as discussed in Sec.~\ref{sec:theo}, the quasielastic condition gives
the reconstructed neutrino energy $\overline{E_\nu}$ (cf. Eq.~\ref{enubar_muon}).
A binding energy correction can be introduced in Eq.~(\ref{enubar_muon}) but it is irrelevant for our discussion. 
Several nuclear effects, going from Pauli blocking and Fermi motion to more complicated many-body effects,
can influence the determination of the reconstructed neutrino energy. 
Benhar and Meloni ~\cite{Benhar:2009wi} have dealt with the influence of the nuclear spectral function. 
Martini \textit{et al.}~\cite{Martini:2012fa,Martini:2012uc} and Nieves \textit{et al.}~\cite{Nieves:2012yz}
have investigated the impact of the collective nature of the nuclear response in the framework of the RPA.
They also investigated the impact of the multinucleon excitations,
a topic treated also by Lalakulich \textit{et al.}~\cite{Lalakulich:2012hs}.
Ankowski \textit{et al.}~\cite{Ankowski:2014yfa} have analyzed the relevance of the final-state interactions
between the struck nucleon and the residual nucleus.
Another process which can simulate the quasielastic interaction is the charged current pion production
if the real pion produced is absorbed on its way out of the nucleus. 
This case has been considered by Leitner and Mosel in Ref.~\cite{Leitner:2010kp}. 
In the following we focus on np-nh excitations, largely discussed in Sec.~\ref{sec:QE},
since until recently this initial reaction mechanism was neglected
in all the Monte Carlo employed in the experimental analyses,
but its inclusion produces important modifications of the neutrino energy reconstructed distributions.
Indeed the np-nh events have no reason to fulfill the quasielastic relation. 
This means that for a given set of lepton variables, $E_l$ and $\theta$,
a variety of neutrino energy values is possible, 
instead of the unique quasielastic value implemented in the neutrino energy reconstruction formula,
as illustrated in Sec.~\ref{sec:theo}. 

\begin{figure}
\begin{center}
\includegraphics[width=10cm]{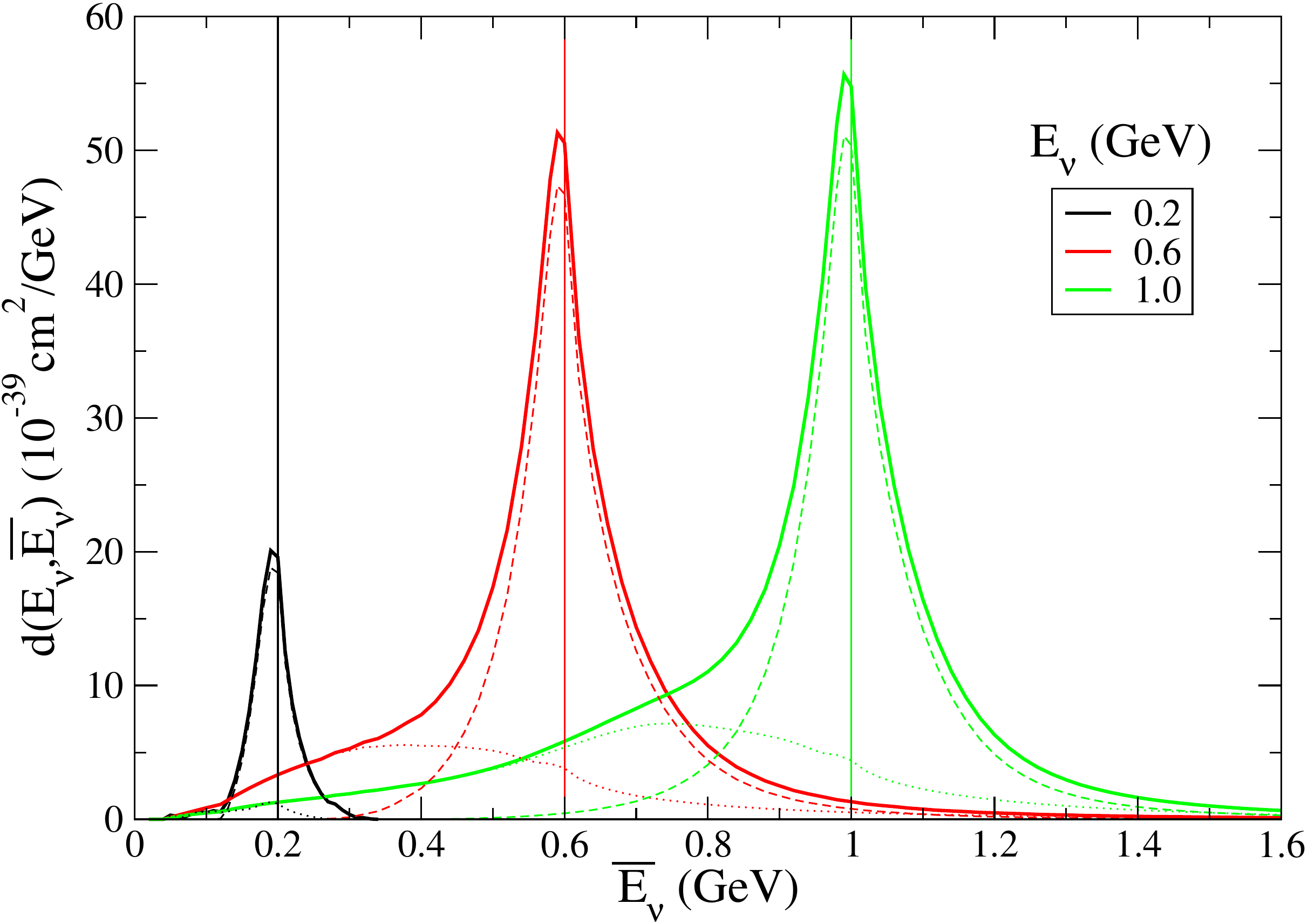}
\caption{\label{fig_integral_vs_erec} The spreading function per neutron of $^{12}$C evaluated
for three $E_{\nu}$ values. The genuine quasielastic (dashed lines) and
the multinucleon (dotted lines) contributions are also shown
separately. The figure is taken from Ref.~\cite{Martini:2012uc}. }
\end{center}

\end{figure}

The corrections corresponding to the transformation from real to reconstructed energy and
vice versa once the multinucleon excitations contribution is taken into account are discussed in detail in
Refs.~\cite{Martini:2012fa,Martini:2012uc,Nieves:2012yz,Lalakulich:2012hs} to which we refer the reader. 
Here we just summarize some of the main results following the approach of Refs.~\cite{Martini:2012fa,Martini:2012uc}.
Starting from a theoretical distribution expressed with real energies,
a smearing procedure to deduce the corresponding distribution of the events, ${D_{rec}(\overline {E_{\nu} })}$,
in terms of the reconstructed energy can be performed. 
This distribution can be expressed in terms of the double differential neutrino-nucleus cross section, according to 
   \begin{eqnarray}
  \label{rho_enubar}
 {D_{rec}(\overline {E_{\nu} })} &= &  \int  
 d E_{\nu}   \Phi (E_{\nu})  \int _{E_l^{min}}^{E_l^{max}} d E_l\frac{ME_l - m_l^2/2}{ \overline E_{\nu}^2 P_l}\left[\frac{d^2 \sigma}{d \omega  ~d\cos \theta }\right]_{\omega=E_{\nu}-E_l,~\mathrm{cos}\theta=\cos \theta(E_l, \overline {E_\nu })}\nonumber\\
 &= &  \int  
 d E_{\nu}   \Phi (E_{\nu}) d(E_{\nu},\overline {E_{\nu}}). 
\end {eqnarray}
We denote the cosine of the charged-lepton angle solution of Eq.~(\ref{enubar_muon})
for a given set of values, $E_l$ and $\overline{E_\nu}$, by $\cos\theta(E_l , \overline{E_{\nu}})$.
(Eq.~(\ref{enubar_muon}) refers to muons but the same expression holds for all the leptons by replacing $\mu$ with $l$). 
The expression of Eq.~(\ref{rho_enubar}) involves the neutrino flux distribution $\Phi(E_{\nu})$,
hence the neutrino oscillation parameters.
The second integral on the r.h.s. of Eq.~(\ref{rho_enubar}),
denoted as $d(E_{\nu},\overline {E_{\nu}})$,
represents the spreading function and depends on $E_{\nu}$ and $ \overline {E_{\nu} }$. 
Some examples of its $ \overline {E_{\nu} }$ dependence for several $ E_\nu$
values are given in Fig.~\ref{fig_integral_vs_erec},
published in Ref.~\cite{Martini:2012uc}. As one can observe this spreading function is not symmetrical around $ E_\nu$.
The multinucleon excitations play a crucial role since they create a low energy tail.
Similar results have been obtained in Refs.~\cite{Nieves:2012yz,Lalakulich:2012hs}.
In Ref.~\cite{Lalakulich:2012hs} the zero pion events are considered, hence
the stuck reabsorbed pion contribution is also included.
It produces an additional low-energy bump in the reconstructed energy distribution,
as already shown in Ref.~\cite{Leitner:2010kp}. 

\begin{figure}
\begin{center}
\includegraphics[height=\textheight]{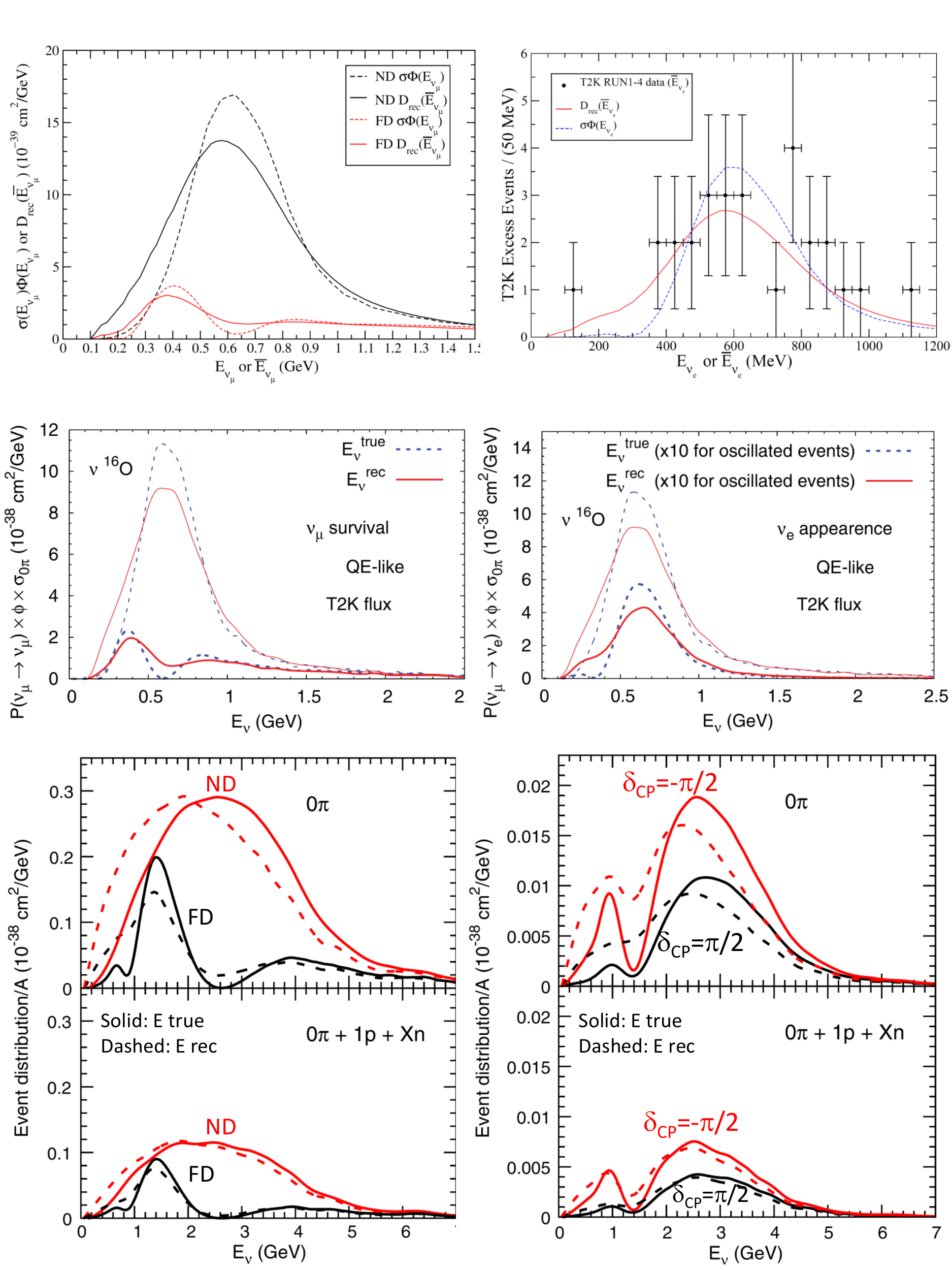} 
\caption{\label{fig_comp_en_rec}
Left panels: distributions of charged current muon neutrino events before and after the energy reconstruction correction in the near (i.e. before oscillation)
and far (i.e. after oscillation) detector. Right panels: the same for $\nu_e$ appearance CC events. 
Top four plots are for T2K and bottom four plots are for DUNE. 
The top two plots are taken from Martini \textit{et al.}~\cite{Martini:2012uc}. 
The next two plots are taken from Lalakulich \textit{et al.}~\cite{Lalakulich:2012hs},
and the bottom four figures are taken from Mosel \textit{et al.}~\cite{Mosel:2013fxa}.
}
\end{center}
\end{figure} 

In Fig.~\ref{fig_comp_en_rec} the impact of the energy reconstruction correction
on different neutrino oscillation distributions is shown. 
The left panels refer to the $\nu_\mu$ disappearance channel;
the right ones to the $\nu_e$ appearance one.
The top four panels refer to T2K distributions, as analyzed by Martini \textit{et al.}
(top two figures)~\cite{Martini:2012uc}
and Lalakulich \textit{et al.} (next two figures)~\cite{Lalakulich:2012hs} using the GiBUU model. 
GiBUU was also used by Mosel \textit{et al.}~\cite{Mosel:2013fxa} to investigate the DUNE distributions,
reported in the bottom four panels of Fig.~\ref{fig_comp_en_rec}. 
In this figure the results taking into account the energy reconstruction correction are compared with the products  
$\sigma (E_{\nu}) \Phi_{\nu}(E_{\nu})$ which represent the distributions of muon CC events before reconstruction
in the near detector (ND) and far detector (FD) and the electron CC events before reconstruction at far detector. 
The salient features of the results obtained by applying the smearing procedure are the broadening effects.
In the near detector there is a clear low energy enhancement. 
In the far detector, where the unsmeared $\nu_\mu$ disappearance distribution displays a pronounced dip,
the smeared one acquires a low energy tail and the middle hole is largely filled.
In the $\nu_e$ appearance distributions the reconstruction correction tends to make events leak outside the high flux region,
in agreement with the observed experimental behavior.
All these smearing effects can be described as a trend to escape the regions of high fluxes,
with a tendency to concentrate at lower energies, when one goes from true to reconstructed energies.
These effects are largely due to the multinucleon component of the quasielastic-like cross section.
It is interesting to notice how the Martini \textit{et al.} results~\cite{Martini:2012uc}
and the GiBUU ones~\cite{Lalakulich:2012hs} are very similar, hence robust, 
in spite of the differences among the two approaches (different treatment of np-nh excitations,
inclusion of stuck reabsorbed pion contribution in the case of GiBUU). 

Taking into account the np-nh excitations as possible initial reactions,
mechanisms beyond the genuine QE can affect the determination of neutrino oscillations parameters.
The theoretical studies of Fernandez-Martinez and Meloni~\cite{FernandezMartinez:2010dm} 
and of Meloni and Martini~\cite{Meloni:2012fq} focused on this point, 
in the context of the beta beams and of the T2K $\nu_\mu\to\nu_\mu$ and $\nu_\mu\to\nu_e$ oscillation data respectively, 
by considering as input neutrino-nucleus cross sections
as a function of the neutrino energy including or not the multinucleon emission channel. 
Further theoretical analises~\cite{Martini:2012uc,Coloma:2012ji,Coloma:2013tba,Ankowski:2016bji},
mainly in the $\nu_\mu$ disappearance channel have also taken into account the additional neutrino energy reconstruction problem. 
In Refs.~\cite{Coloma:2013tba} two event generators, GiBUU and GENIE
are considered and detector effects such as thresholds and energy resolution are also included. 
We refer to these works, as well as to Ref.~\cite{Ankowski:2016jdd}, for more quantitative details. As a general comment, in an analysis which takes into account the np-nh excitations the smearing effect due to their contribution 
is likely to lead to some increase of the oscillation mass value and decrease of the mixing angle. 

In Ref. ~\cite{Abe:2015awa} the T2K collaboration performed an analysis of the $\nu_\mu$ disappearance results
by taking into account for the first time the multinucleon emission channel. 
The conclusion of this T2K analysis was that for the considered exposure,
the multinucleon effect can be ignored, but future analyses will need to 
incorporate these effects in their model of neutrino-nucleus interactions~\cite{Wilkinson:2016wmz}.
There are few remarks about this conclusion.
First of all, this analysis has been performed by considering
the multinucleon contributions as calculated by Nieves \textit{et al.} which,
as it has been shown in Sec.~\ref{sec:npnh}, are smaller with respect to the calculations of Martini \textit{et al}.
Second, the full model by Nieves \textit{et al.} is not implemented yet in the Monte Carlo used by T2K,
and there are some delicate points in the present implementation ~\cite{Wilkinson:2016wmz}.  
For example, part of non-genuine quasielastic events, the ones related to the non pionic decay 
of the Delta in the nuclear medium, was already included in the Monte Carlo (NEUT) employed for the T2K analysis. 
Finally, neither  Nieves \textit{et al.} nor Martini \textit{et al} provide theoretical errors for the np-nh component,
and T2K collaborators implemented only normalization of these channels as tunable parameters. 
But the shape uncertainties of models are necessary for a precise fit.
This point was already discussed in Sec.~\ref{sec:qe_minerva} as a reason of strong tension between MiniBooNE and MINERvA data.
Clearly, further studies are needed for both theory and experiment. 

Another example of neutrino oscillation analysis which quantitatively
takes into account the effects of the multinucleon emission channel
is the one performed by Ericson \textit{et al.}~\cite{Ericson:2016yjn}
in the context of the MiniBooNE $\nu_\mu\to\nu_e$ low energy anomaly~\cite{AguilarArevalo:2008rc,Aguilar-Arevalo:2013pmq},
related to the possible existence of sterile neutrinos. 
Ericson \textit{et al.} shown that taking into account the multinucleon effects
in the analysis of MiniBooNE data allows a slightly better fit
of the MiniBooNE low-energy excess and induces a shift of the allowed region in the $\sin^2 2\vartheta$--$\Delta{m}^2$
plane towards smaller values of $\sin^2 2\vartheta$ and larger values of $\Delta{m}^2$ in the framework of two-neutrino oscillations.
In Ref.~\cite{Ericson:2016yjn} Ericson \textit{et al.} also performed a global fit of short-baseline neutrino oscillation data
in the framework of 3+1 neutrino mixing showing that taking into account the multinucleon interactions
in the analysis of MiniBooNE data lead to a decrease of the appearance-disappearance tension.
However, this effect is not enough to solve the problem of the appearance-disappearance tension in the global fit
of short-baseline neutrino oscillation data,
because the value of the appearance-disappearance parameter goodness-of-fit is still too small.

Mosel~\textit{et al.} discussed in Ref.~\cite{Mosel:2013fxa}
a possible way to reduce the systematic uncertainties due to the multinucleon emission channel
in the neutrino oscillation events by maintaining
the QE-based energy reconstruction method instead of the calorimetric one in connection with DUNE distributions~\cite{Mosel:2013fxa}. 
This is illustrated in the bottom four panels of Fig.~\ref{fig_comp_en_rec}.
They suggest to consider ``CC0$\pi$1p'' sample, {\it i.e.} final state events with one charged lepton,
0 pions, and 1 proton, instead of traditional ``CC0$\pi$'' sample where the final state particles include one charged lepton and 0 pions,
and any number of protons.
These events in CC0$\pi$1p sample are primarily due to an original QE event 
(about 80\% of contribution with respect to a 50\% in the CC0$\pi$ case). 
As it appears from Fig.~\ref{fig_comp_en_rec},
the more restrictive requirement of 0 pions and exactly 1 proton allows to
obtain the true and reconstructed results quite close to each other.
The shift of the oscillation peaks and the tendency to concentrate
the events at lower energies are drastically reduced. 
The price to pay is that one loses a factor 3 in the number of events.
This example shows a power of information from the hadronic system,
especially the multiplicity of nucleons in the final state.  
We discuss this point further in Sec.~\ref{sec:nmulti}. 
Furmanski and Sobczyk proposed to include full energy-momentum conservation on CC0$\pi$1p sample to
improve the CCQE data sample and energy reconstruction~\cite{Furmanski:2016wqo}.
However, in order to utilize these ideas in real experiments, we need carefully evaluation of proton measurement systematics,
including FSIs and detection efficiency uncertainties.

In the recent study by Ankowski~{\it et al.}~\cite{Ankowski:2015kya} on comparison of calorimetric and QE-based energy reconstruction
in $\numu$ disappearance oscillation measurement with DUNE setup, it is found that the results obtained with the QE-based method exhibit
less sensitive to the detector efficiency error with respect calorimetric method.
This suggests that even if the detector has a capability to reconstruct neutrino energy calorimetrically,
there might be an advantage to use QE-based energy reconstruction.

\section{Future experiments\label{sec:future}}
\subsection{Detectors}

\begin{figure}[tb]
  \begin{center}
    \includegraphics[width=6cm,valign=m]{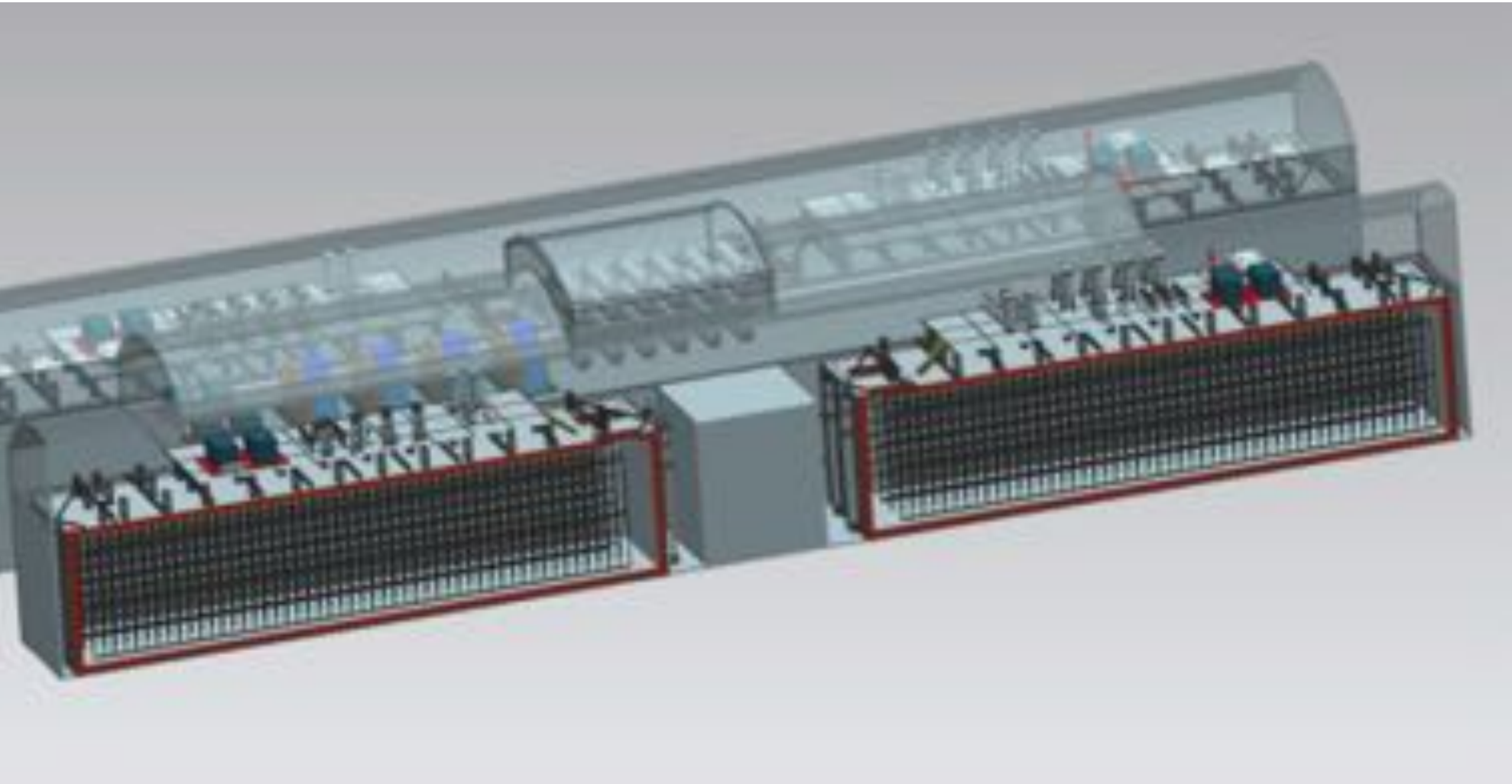}
    \includegraphics[width=6cm,valign=m]{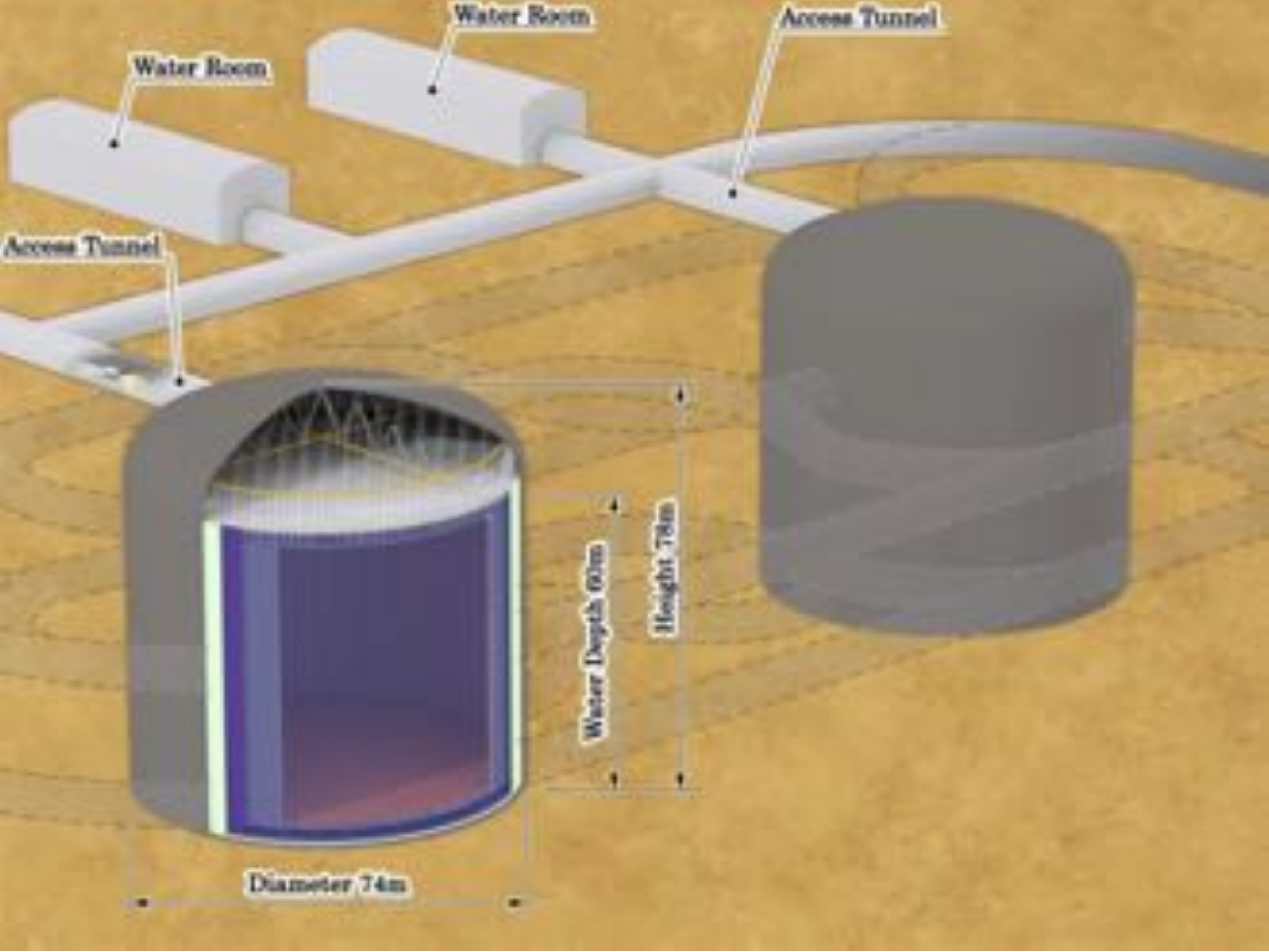}
    \end{center}
\vspace{-2mm}
\caption{
  Detector designs of DUNE (left, circa 2016)~\cite{DUNE_CDR4}
  and Hyper-Kamiokande (right, circa 2016)~\cite{TITUS}.
}
\label{fig:future}
\end{figure}

To discuss the future of this field, 
it is important to realize we selected two targets to be the most important ones 
for future accelerator-based neutrino experiments,
\begin{itemize}
\item argon, for liquid argon time projection chambers (LArTPCs)
\item water, for water/ice-Cherenkov detectors
\end{itemize}
Not surprisingly, near future experiments focus on R\&D of these detectors. Figure~\ref{fig:future} shows the planned detectors of DUNE (Deep Underground Neutrino Experiment)~\cite{DUNE_CDR4}
and Hyper-Kamiokande~\cite{TITUS}.
Both collaborations are working on optimization of design details and the detector details may be changed, 
but the basic designs will not be changed. 

\subsubsection{Liquid Argon Time Projection Chamber (LArTPC)~\label{sec:lartpc}}

Fermilab will host the beam (Fig.~\ref{fig:flux_all}) of the DUNE~\cite{DUNE_CDR3} experiment.
In this 1300~km baseline long baseline neutrino oscillation experiment,
four of 10~kton membrane cryostat LArTPC detectors are planned to be installed at SURF
(Sanford underground research facility), South Dakota, USA. 

The Fermilab short baseline neutrino (SBN) program will host three LArTPC detectors
at the BNB beamline~\cite{ICARUS_FNAL}, MicroBooNE (2015-), SBND (2018-), and ICARUS (2018-), as 
a R\&D of the detector technology and physics including
1eV sterile neutrino search and cross section measurements. 

The concept of LArTPC is a ``modern bubble chamber''.
The high voltage liquid argon volume is the fiducial region
(and hence, the neutrino interaction target is the argon nucleus).
The charged particles created in the fiducial volume ionize argon atoms along the tracks,
and ionized electrons are drifted to anode wire planes. 
First, wire planes see these electrons through induction,
and wires from the last wire plane collect these electrons.
Utilizing the different orientations of wires in each plane and drifting time information of electrons, 
in principle LArTPC allows to reconstruct all three dimensional charged particle tracks (Fig.~\ref{fig:argoneutdisplay}).
Although LArTPC has no timing information of interactions,
it often companies with a scintillation photon detection system
which has $\sim$ns precision to identify neutrino interaction time.
Intensive R\&D programs are being performed to understand the detector technology~\cite{LArTPC_FNAL1,LArTPC_FNAL2}.
The main detector challenge includes the design of the high-voltage feed-through~\cite{LArTPC_HV},
purification and circulation to keep the highest purity liquid argon~\cite{Pordes_filter,Pordes_luke,LAPD},
process of thousands of wire signals in LAr~\cite{LArTPC_ColdFEB}, 
detection of 128~nm scintillation light from argon~\cite{uB_PMTtest,uB_PMTsystem,Chiu,Ben_nitrogen,Ben_methane,Ben_benzophenone},
understanding of properties of membrane cryostat~\cite{LBNE35ton},
and studies of alternative dual phase LArTPC technology~\cite{DUNE_CDR4}, etc.
The R\&D of this detector is a world-wide effort.
Not only in USA, but both Japan~\cite{KEKTPC} and Europe~\cite{LBNO,ARGOTUBE,DUNE_CDR4} own LArTPCs to study different aspects. 

Besides detector challenges,
the biggest software challenge of this detector is the event reconstruction. 
Since the amount of information from each interaction is so large,
reconstructing particle tracks and extracting physics information,
such as particle type, energy, and momentum, becomes more complicated.
The electromagnetic shower events are difficult ones to reconstruct and measure kinematics~\cite{Szelc_Nu14}, 
but these are especially important because they are candidate events of 
$\numu\to\nue$($\numubar\to\nuebar$) oscillation signals. 
We expect a significant amount of input from the LArIAT beam test experiment at Fermilab~\cite{LArIAT}
and future protoDUNEs at CERN~\cite{DUNE_CDR4}. 
In these experiments, LArTPC detectors are installed in
the dedicated beamlines with beams of known particles
to understand detector responses carefully.

LArTPC detectors are expected to provide superior results on 
$\numu\to\nue$($\numubar\to\nuebar$) oscillation measurements.
In these measurements, single gamma ray from neutral current interactions,
such as NC1$\piz$ and NC1$\ga$ (Sec.~\ref{sec:pizero}),
is background for typical tracking and Cherenkov detectors
which cannot distinguish electromagnetic showers by an electron and a gamma. 
In LArTPC, they can be separated by either measuring a gamma conversion length and/or $\frac{dE}{dx}$~\cite{Teppei_uB}.
If the shower is made by an electron or a positron, the track starts directly from the interaction vertex,
while a gamma ray has no activities between the interaction vertex and pair production points.
Also, gamma rays decay to a $e^+-e^-$ pair, which has twice the $\frac{dE}{dx}$ of an electron or a positron,
before the electromagnetic shower is developed.

\subsubsection{Water Cherenkov detector}

On the other hand, water Cherenkov detector is a fully established technology. 
Hyper-Kamiokande (Hyper-K) detector~\cite{HK_2016} has $\sim$ half megaton mass or
$\sim$17 times bigger fiducial volume than the Super-Kamiokande (Super-K) detector.
This colossal water Cherenkov detector serves to both beam physics and astrophysics. 
Although, the basic technology and the photo-cathode coverage are the same, 
higher efficiency photo sensors~\cite{Hayato_HPD}
make the Hyper-K to outperform the Super-K.
As we discussed in Sec.~\ref{sec:pizero},
recent improvements of reconstruction made the NC$\pi^\circ$ a minor background for the T2K experiment~\cite{Abe:2015awa}.
However, neutrino interaction systematics are continuously the most important systematics for oscillation analyses.

The R\&D projects for the future water detectors are also very active.
One approach is to improve traditional scintillation trackers with the water layer,
by introducing an alternative configuration of plastic scintillators
to keep high tracking performance with large fiducial volume, such as the WAGASCI detector~\cite{Koga:2015iqa}.
Other approaches involve dopings in the water.
EGADS (Evaluating Gadolinium's Action on Detector System)
is a 200~ton R\&D pilot detector of GADZOOKS!~\cite{GADZOOKS} experiment.
The basic concept is to dope the gadolinium compound in the water to capture low energy neutrons from neutrino interactions.
The main goal is to detect low energy ($\leq$30~MeV) diffused supernova neutrino background (DSNB).
However, this technology is also useful for the neutrino interaction physics as we discuss later. 
The neutron capture can be performed by hydrogen atoms in the water~\cite{SK_NCapH},
but gadolinium neutron captures cause multiple photon emissions from higher Q-value ($\sim$8~MeV), 
and it is easier for water Cherenkov detectors to tag neutrons. 
The concept of the technology was demonstrated~\cite{SK_NCapGd}, 
and EGADS is currently running at the Kamioka mine to study water purification and neutron capture physics.
Another doping idea is the water-based liquid scintillator (WbLS)~\cite{ADSC}.
This allows the water Cherenkov detector to tag low energy charged particles below the Cherenkov threshold,
which include low energy protons. 
In short, doping of gadolinium compound and WbLS in principle makes
neutrons and low energy charged hadrons to be visible by water Cherenkov detectors. 
The R\&D of these technologies are very active, 
and all of the future accelerator-based neutrino experiments with water target,
such as ANNIE~\cite{ANNIE}, TITUS~\cite{TITUS}, 
and $\nu$PRISM~\cite{nuPRISM} consider either gadolinium and/or WbLS doping. 

The neutrino telescopes, such as IceCube~\cite{IC_osc2013} and ANTARES~\cite{ANTARES_atmonu}
are another type of successful ice and water Cherenkov detectors,
but now the interaction media are natural ice or sea water.
Both are planning low energy physics for the future, 
PINGU~\cite{PINGU} (precision IceCube next generation upgrade)
is the low energy part of IceCube's future extension ``IceCube-Gen2''~\cite{G2_2015},
and ORCA (Oscillation Research with Cosmics in the Abyss) is the low energy part of future KM3NeT~\cite{KM3NeT}.
These experiments (together with DUNE and Hyper-Kamiokande) utilize atmospheric neutrinos around 2-10 GeV to
look for matter oscillations to find the neutrino mass ordering (NMO) through 
$\numu$($\numubar$) and $\nue$($\nuebar$) measurements.

\subsection{Higher precision hadron information}

\subsubsection*{Hadron energy deposit}
The total hadron energy deposit is the necessary information for the calorimetric neutrino energy reconstruction
\beq
E_\nu = E_{lepton}+E_{had}~,~E_{had}=\sum_i E_{had}^i=\om 
\eeq
and already actively used by experiments.  
However, it is impossible to measure the energies of all hadrons, notably energy deposits by neutrons are always hard to measure.
Therefore, experiments, such as MINERvA, compromise to measure  ``$E_{avail}$'', or the available hadronic energy,
which is the sum of kinetic energies of protons and charged pions, and the total energy of other hadrons except neutron.
This is a quantity related to energy transfer $\om$, but their connection is only given through the simulation.
Once $\om$ is obtained, together with lepton kinematics, $|{\bf q}|$ is reconstructed.
This allows MINERvA to measure the first $E_{avail} - |{\bf q}|$ double differential cross section ~\cite{Rodrigues:2015hik}
(Sec.~\ref{subsubsec_minerva_evail}). 
It is important to study the np-nh contribution in the lepton kinematics.  
However, understanding hadronic energy deposits from 2p-2h is important for oscillation experiments such as NOvA~\cite{Adamson:2016tbq}. 
Recently, the NOvA data-simulation agreement of the hadronic energy measurement was improved
by incorporating the latest GENIE 2p-2h simulation~\cite{NOvA_Nu2016} motivated from MINERvA $E_{avail} - |{\bf q}|$ data.
This is a confirmation that 2p-2h channel is important not only for lepton kinematics, but also for the hadronic system prediction. 

$E_{had}$ may be a key parameter to determine NMO by atmospheric neutrino oscillations. 
Since mass ordering has roughly opposite effect for neutrinos and anti-neutrinos~\cite{Akhmedov_PINGU},
charge separation is desired for atmospheric neutrino interactions.
Assuming we know the atmospheric neutrino flux with sufficient precision,  
particle kinematics could be used to separate neutrino and anti-neutrino interactions. 
More specifically, ``inelasticity ($y$)'' can be used for the charge separation~\cite{Smirnov_inel}. 
Here, inelasticity is the fraction of hadron energy from the given neutrino interaction ($y=\frac{E_{had}}{E_{had}+E_l}$),
and such measurement requires precise hadronic system simulations. 
Improvement of hadron shower reconstruction is underway by PINGU~\cite{PINGU} and ORCA~\cite{KM3NeT} for this purpose. 

\subsubsection*{Nucleon multiplicity\label{sec:nmulti}}
A natural direction of future experiments is to measure more ``details'' of interactions.  
Among them, ``nucleon multiplicity'' may be a next extension to go beyond
the existing analyses which exploit all lepton kinematics.
As we see in Sec.~\ref{sec:npnh}, it is still possible for several theoretical models
to explain the CCQE data from MiniBooNE, MINERvA, and T2K.
The problem is that the data only show lepton phase space.
The nucleon multiplicity is the next accessible parameter to further classify events to smaller sub-groups.

The power of nucleon multiplicity is demonstrated by the GiBUU~\cite{Lalakulich:2012ac} and TITUS collaborations~\cite{TITUS}.
As we see in Fig.~\ref{fig_comp_en_rec}, bottom right,
the difference of $\de_{CP}$ values are the same size of true and reconstructed neutrino energy,
suggesting nuclear effects can be a dangerous systematics to measure $\de_{CP}$ effect.
However, if the LArTPC detector can measure outgoing proton multiplicity, namely,
if the analysis is based on a ``CC0$\pi$1p'' sample, instead of a ``CC0$\pi$'' sample,
energy reconstruction is under control. 

An equivalent approach with proton counting in LArTPC
may be possible for the water Cherenkov detectors by counting neutrons.   
TITUS~\cite{TITUS} is a proposed water Cherenkov detector for the near detector of the Hyper-K~\cite{HK_2015},
and the collaboration is actively studying this ``neutron multiplicity'', $\left< n\right>$. 
Na\"{i}vely, in neutrino-neutron CCQE interaction one can expect there is no outgoing neutron.
On the other hand, antineutrino-proton CCQE interaction has one outgoing neutron 
\begin{description}
\item[$\numu$CCQE] $\left< n\right>\sim 0$, $\numu+n\to\mu^-+p$~,
\item[$\numubar$CCQE] $\left< n\right>\sim 1$, $\numubar+p\to\mu^++n$~.
\end{description}
By assuming neutron-proton pair is the dominant type
($\sim$90\%),
interactions with correlated nucleon pairs have different neutron multiplicities than genuine CCQE. 
\begin{description}
\item[$\numu$CC2p-2h] $0<\left< n\right> <1$, $\numu+(np)\to\mu^-+p+p~$ ($\sim$90\%),~$\numu+(nn)\to\mu^-+n+p$ ($\sim$10\%), 
\item[$\numubar$CC2p-2h] $1<\left< n\right> <2$, $\numubar+(np)\to\mu^++n+n$ ($\sim$90\%),~$\numubar+(pp)\to\mu^++n+p$ ($\sim$10\%).
\end{description}
This idea can be extended to other channels, including inelastic interactions and NC interactions.
Clearly, this is a too na\"{i}ve picture, first, as we discussed in Sec.\ref{subsubsec_theonppp},
the real fraction of neutron-proton pair is not a constant but kinematics dependent, and second,
the presence of FSIs in the target nuclei and secondary interactions by particle propagation in detector media
will modify the final state nucleon multiplicity. 
Nevertheless, it is possible to use nucleon multiplicity
to statistically extract information about primary interactions within systematic errors. 

\begin{figure}[tb]
  \begin{center}
    \includegraphics[width=14cm,valign=m]{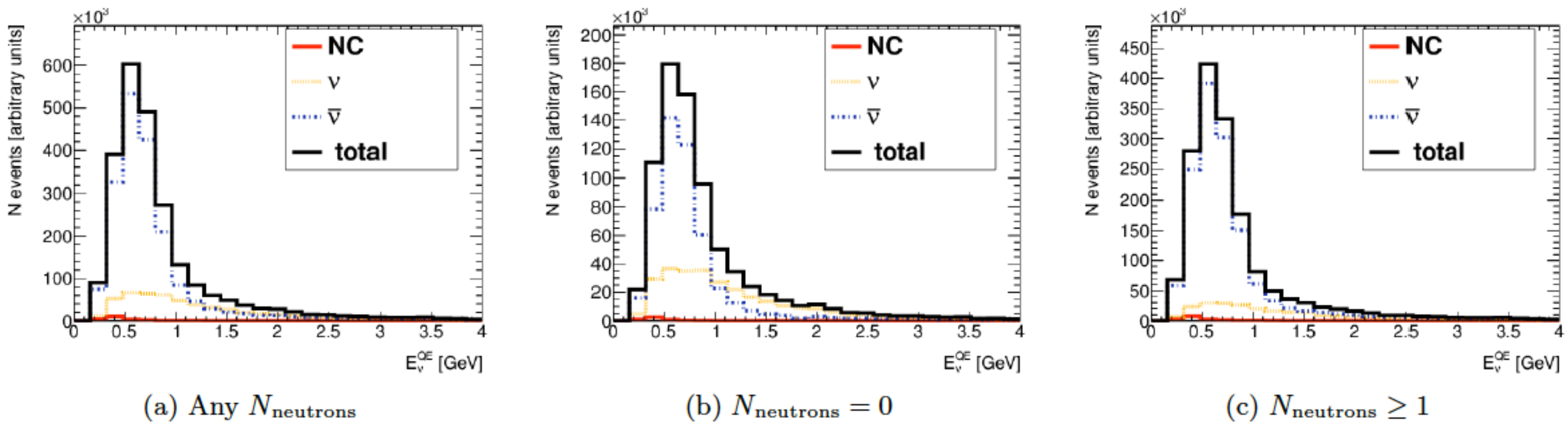}
    \includegraphics[width=14cm,valign=m]{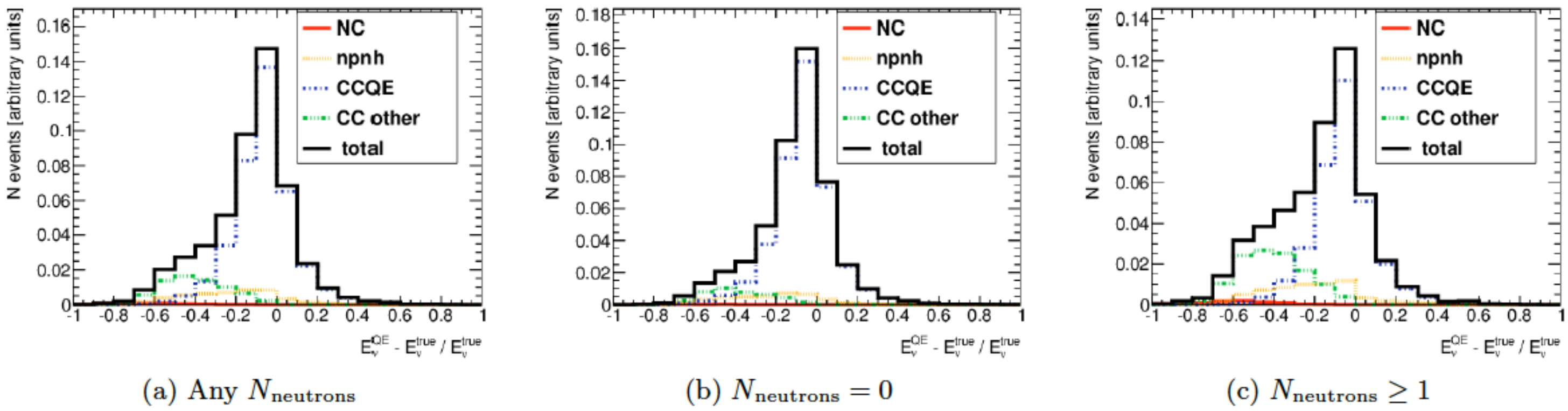}
    \end{center}
\vspace{-2mm}
\caption{
  Top three plots shows the reconstructed neutrino energy from the TITUS simulation in antineutrino mode.
  The neutrino energy is reconstructed from reconstructed muon kinematics by the TITUS simulation. 
  Left plot is the CC0$\pi$, the standard data sample for oscillation analysis. 
  As you see, contamination of wrong sign background ($\numu$ interaction) can be suppressed by requiring 
  neutrino number $\geq$ 1 (right, ``CC0$\pi\geq$1n'' sample).
  On the other hand, neutron number=0 (middle, CC0$\pi$0n sample) is used as a control sample of background channels.
  Bottom three plots are resolutions of neutrino energy reconstruction for neutrino mode.
}
\label{fig:nucmulti}
\end{figure}

Figure~\ref{fig:nucmulti} shows the simulation of the TITUS Gd-doped water Cherenkov detector in the antineutrino mode.
Here, in the top three plots, outgoing anti-muon tracks are reconstructed,
and measured kinematics are used to reconstruct neutrino energy. 
In the antineutrino mode, where the cross section is lower and the wrong sign (WG) flux background (Sec.~\ref{sec:flux}) is higher, 
the oscillation measurement such as $\numubar\to\nuebar$ is in general harder than $\numu\to\nue$.
The left plot shows the reconstructed neutrino energy from CC0$\pi$ sample. 
The right plot shows the reconstructed neutrino energy with more than 1 neutron in the final state (``CC0$\pi\geq$1n'' sample).
An extra constraint of detecting neutrons suppresses the neutrino WS background.
On the other hand, the middle plot is the same distribution from CC0$\pi$0n sample
which can be used to understand backgrounds. 
Further classification is possible (CC0$\pi$1n, CC0$\pi$2n, etc)
to simultaneously constrain different interaction channels,
and these studies are ongoing by collaborators.
Bottom three plots are resolutions of neutrino energy reconstruction for neutrino mode.
In this case, CC0$\pi$0n sample (middle) can be used to suppressed unwanted backgrounds,
where CC0$\pi\geq$1n sample (right) serves as a background control.
  
\subsubsection*{Exclusive hadronic channels}
More details of hadronic information beyond the simple nucleon counting should be useful to understand further details. 
At this moment, among modern high statistics wide-band beam experiments,
ArgoNeuT is the only experiment to show the ability to measure two proton tracks
on top of the muon track from CC interactions~\cite{Acciarri:2014gev} (Sec.~\ref{sec:npnh}).

\begin{table}[tb]
    \centering
    \footnotesize
    \
    \begin{tabular}{c|c|c|r}
        T2K & SBND & reaction & expected \#evts \\
        \hline \hline
        CC0$\pi$ &CC0$\pi^\pm$Np    &$\numu+N\to\mu+Xp$ & 3,552k \\
                 &CC0$\pi^\pm$0p    &$\numu+N\to\mu+0p$ &   793k \\
                 &CC0$\pi^\pm$1p    &$\numu+N\to\mu+1p$ & 2,028k \\
                 &CC0$\pi^\pm$2p    &$\numu+N\to\mu+2p$ &   359k \\
                 &CC0$\pi^\pm\geq$3p&$\numu+N\to\mu+\geq 3p$ & 371k \\
        \hline
        CC1$\pi$ &CC1$\pi^\pm$      &$\numu+N\to\mu+1\pi^\pm+Xp$       & 1,162k \\
        \hline
        CC others&CC$\geq$2$\pi^\pm$&$\numu+N\to\mu+\geq 2\pi^\pm+Xp$   &    98k \\
                 &CC$\geq$1$\pi^\circ$&$\numu+N\to\mu+\geq\pi^\circ+Xp$ &   498k \\
        \hline \hline
    \end{tabular}
    \caption{Comparison of current T2K event~\cite{T2K_nue2013} classification and
      SBND event classification plan~\cite{ICARUS_FNAL}.
      For SBND, GENIE v2.8 is used to simulate events.
      The estimated total event number in the active volume through the entire run is shown.
      Note proton detection threshold is assumed to be 21~MeV~\cite{Acciarri:2014gev}.}
    \label{tab:SBND}
\end{table}

But future ambitious experiments, such as SBND~\cite{ICARUS_FNAL} (Short Baseline Neutrino Detector),
are ready to measure more hadronic final states.
Table~\ref{tab:SBND} shows the simulated number of events in each final topology in the SBND LArTPC detector. 
Notice that high statistics of SBND events are classified to further sub-groups comparing to T2K.
For example, CC0$\pi$ category of T2K can be further classified depending on the number of protons. 
SBND is also expecting sub-groups depending on the number of charged pions. 
However, as discussed in Sec.~\ref{sec:exclusivehad}, 
the prediction of final state hadrons require good models of baryon resonance, SIS, and DIS cross sections,
FSIs, hadronization process, and secondary interactions, on top of a good understanding of detector performance. 
Also, theoretical understanding of nuclear target is always challenging,
especially for large non-isoscalar targets, such as argon nuclei.
Therefore, instead of believing the numbers from Table~\ref{tab:SBND} seriously, 
SBND data will find these numbers from their data and
they will be used to improve many features of hadron physics.

\subsection{Highly controlled neutrino beam~\label{sec:futurebeam}}

The precision era of neutrino oscillation physics requires not only
an excellent knowledge of the final states of the neutrino reactions
(via very good detectors and Monte Carlo) but also a robust knowledge of initial neutrino fluxes and beam contamination. 
It is pointed out that the current flux normalization error is unlikely to be reduced
less by than $\sim$5\% by just improving hadron production errors from external measurements~\cite{Garvey:2014exa}. 
This forces the community to invent new ideas to produce or monitor neutrino beams
to further reduce the flux systematic error.
We list below the major new possibilities.

\subsubsection*{Neutrino factory}
The neutrino factory~\cite{NF} is a planned future neutrino beam for this purpose.
The traditional neutrino beams, created by meson DIF is called ``super beam''.
On the other hand, neutrino factory makes the neutrino beam by muon decays,
especially muon decays in the muon storage ring. 
The neutrino distributions from muon decays are predicted by the Standard Model and well known,
and this allows higher precision neutrino interaction measurements.  
The problem is the high cost of such machine, 
and even a small scale version, such as $\nu$STORM (Neutrinos from STORed Muons)~\cite{nuSTORM}, 
is still considered expensive and the possibilities to build this type of machine in the USA or Europe are for the moment unknown. 
Recently, a project based on this concept is proposed in China~\cite{MOMENT}.
If this is realized, it would be the first step towards the neutrino factory. 

\subsubsection*{Beta beam}
Another popular idea of future neutrino beams is the ``beta beam''~\cite{betabeam}, 
which produces nearly mono-energetic neutrino beams from the decays of isotopes. 
Recently, a slightly different version is studied carefully by IsoDAR collaboration~\cite{IsoDAR}. 
In IsoDAR, a high intensity proton beam produced by a high power cyclotron, 
such as the injector designed for the DAE$\de$ALUS experiment~\cite{DAEdALUS_1,DAEdALUS_2},  
hits a $^9$Be target in a $^7$Li sleeve and $^8$Li produced by the neutron capture makes a $<E_\nu>\sim$6.4~MeV $\nuebar$ beam. 
These neutrinos are useful for oscillation experiments,
but they can be also used for low energy cross section measurements. 

\subsubsection*{Tagged-neutrino beam}
Associated particles from the neutrino beamlines are often monitored to understand the neutrino beam better.
Both NuMI~\cite{NuMI} and J-PARC neutrino beam lines~\cite{T2K_flux} are equipped with series of muon monitors in their beam dumps
to understand dominant $\numu$ flux.
One advanced idea,  ENUBET (Enhanced NeUtrino BEams from kaon Tagging),
is to measure $\piz$ from $K_{e3}$~($K^+\to e^++\nue+\piz$) in the decay pipe to monitor $\nue$ flux~\cite{Longhin:2014yta}.
Such an idea is not only useful to suppress the intrinsic backgrounds for $\numu\to\nue$ oscillation search,
but it also provides precise $\nue$ flux for $\nue$ cross section measurements.  
This may be the first step to future tagged-neutrino beam facility~\cite{Pontecorvo:1979zh}, 
where leptons from DIF mesons are measured (``tagged'') in the beamline simultaneously with neutrino interactions in the neutrino detector  
so that one can reconstruct neutrino energy event-by-event for the neutrino interactions measured at the neutrino detector. 

\subsubsection*{DAR neutrino beam}
Comparing with high intensity meson decay-in-flight (DIF) neutrino beam,
neutrinos from decay-at-rest (DAR) mesons are low intensity, but spectrums are well understood.
Low energy pion DAR neutrinos ($\leq$60~MeV) are often thought to be a useful tool
to look for sterile neutrinos~\cite{Aguilar:2001ty,KARMEN_osc} 
or neutrino-nucleus coherent scattering~\cite{CSI,COHERENT,SCENE}. 
Also, two-body decay DAR neutrinos are mono-energetic.
Especially, higher energy kaon DAR neutrinos (235.5 MeV) may be useful to study CCQE interactions~\cite{KDAR_xs}.
The combination with high resolution detector, such as LArTPC, may be able to measure the giant resonances
induced by neutrino scattering~\cite{Kolbe:1995af,Volpe:2000zn,Jachowicz:2002rr,Botrugno:2005kn,Martini:2007jw,Samana:2010up,Pandey:2013cca,Pandey:2014tza,Martini:2016eec,Pandey:2016jju} on argon.
High intensity pion DAR neutrinos are expected from many non-neutrino facilities as free byproducts. 
Both Ork Ridge SNS (Spallation Neutron Source)~\cite{OscSNS,CSI,COHERENT} 
and J-PARC MLF (Materials and Life Science and Experimental Facility)~\cite{KPIDAR}
do not have a neutrino beamline, but they have proposals to perform neutrino experiments. 
A space constraint is often an obstacle since none of these facilities were 
designed to install large neutrino detectors.

\subsubsection*{$\nu$PRISM} 
\begin{figure}[tb]
  \begin{center}
    \includegraphics[width=6cm,valign=m]{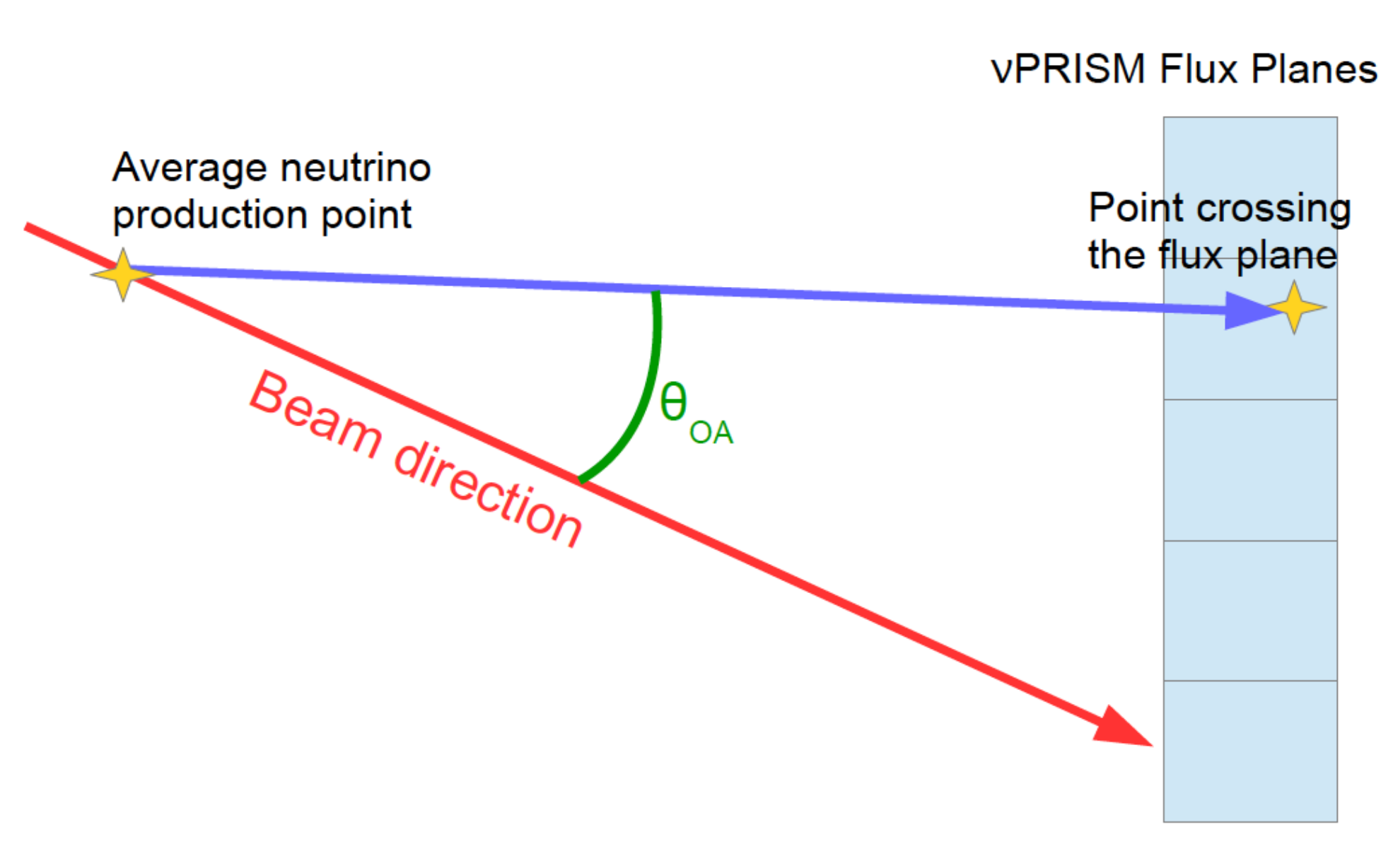}
    \includegraphics[width=6cm,valign=m]{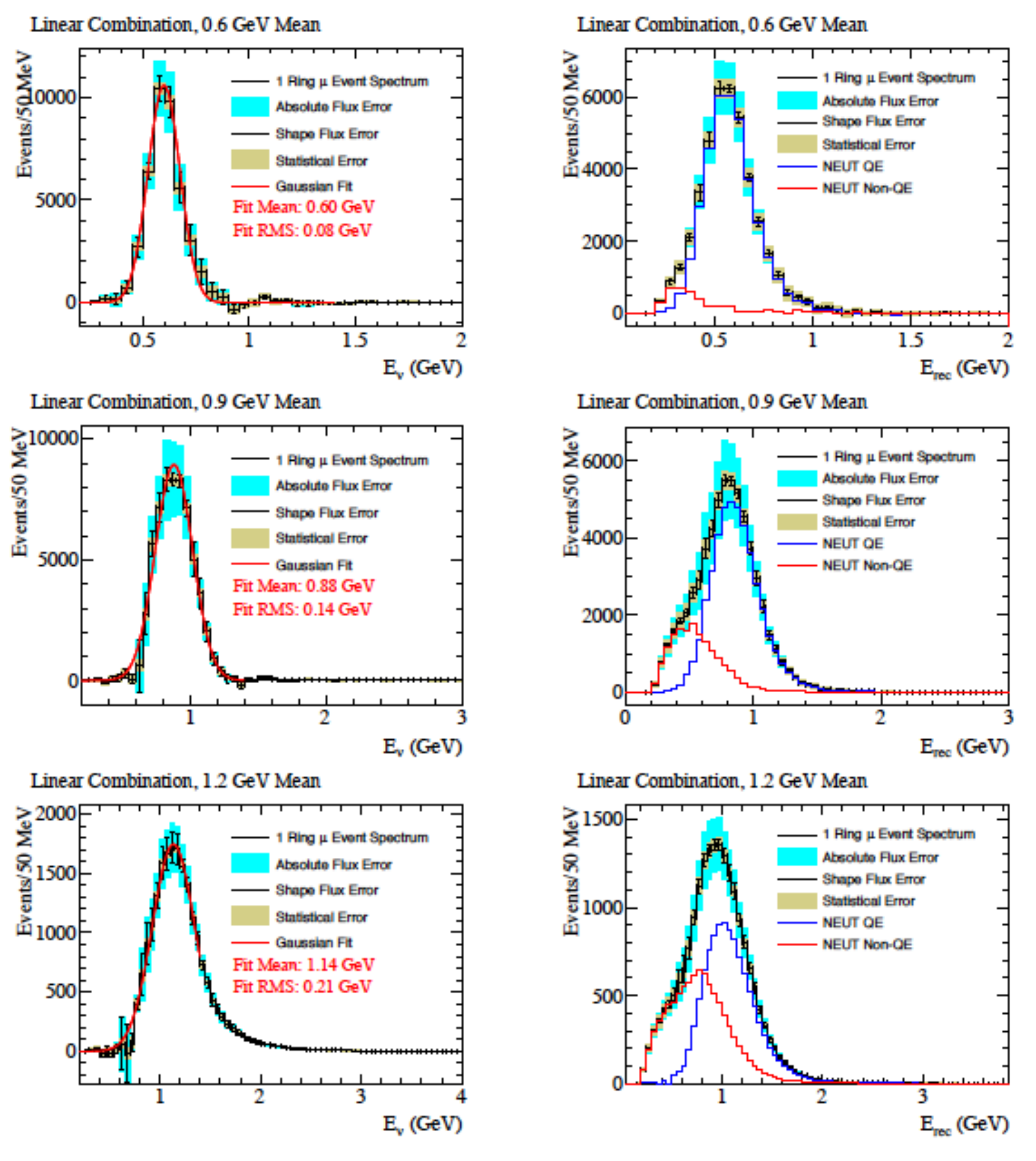}
    \end{center}
\vspace{-2mm}
\caption{
  Left cartoon shows the concept of $\nu$PRISM detector.
  A simultaneous measurement of neutrino beam
  with different angles is performed by a elongated water Cherenkov detector.
  This allows to find pseudo-monochromatic neutrino beam at some angles (middle).
  Right shows reconstructed neutrino energy~\cite{nuPRISM}. 
}
\label{fig:nuPRISM}
\end{figure}

Another approach was proposed in Japan, as a potential near detector concept for Hyper-K. 
The $\nu$PRISM~\cite{nuPRISM} is a proposed water Cherenkov detector
that does not to plan to measure details of exclusive channels to understand interactions. 
Instead, it focuses on simultaneous measurements of different off-axis angles from a given neutrino beam
to produce a pseudo-monochromatic neutrino beam. 
Figure~\ref{fig:nuPRISM} shows a concept. 
The neutrino spectrum changes with angles, 
and in general at a higher angle than the beam axis, 
neutrino spectrum becomes softer, and narrower~\cite{Kopp_flux,offaxis}. 
By combining measurements at separated off-axis angles,
$\nu$PRISM can find a reasonably narrow spectrum of neutrino spectrum for any energy between $\sim$0.4-1.0~GeV,
and lepton kinematics measured in such detector can be used as precise test of different QE or np-nh models.  

\subsubsection*{Internal constraint}
It is also common to constrain the neutrino flux by using the neutrino data from the own detector.  
NOMAD~\cite{Lyubushkin:2008pe} checked the flux normalization in two ways, by utilizing 
known DIS and inverse muon decay (IMD) cross sections.
For lower energy neutrino experiments which we focus on this review,
neither DIS nor IMD are very practical for such a purpose. On the other hand, MINERvA~\cite{Park:2015eqa}
measures $\numu-e$ elastic scattering to constrain the flux.
This process also has a theoretically well-known cross section,
and a distinctive experimental signature (forward going electromagnetic shower) allows them to be selected efficiently. 
Thus, by measuring $\numu-e$ elastic scattering events, 
MINERvA can effectively measure the neutrino flux.
Using this method, MINERvA found that the predictions of NuMI flux are consistently higher than the measurements.
This is confirmed by improved NuMI flux simulation~\cite{Aliaga:2016oaz}.

For DIS dominant experiments, such as CCFR~\cite{Seligman:1997fe}, NuTeV~\cite{Tzanov:2005kr},
so called low-$\nu$ method~\cite{Bodek_lowEflux} is often used to find the shape of flux.
This flux prediction takes advantage of the fact that neutrino and antineutrino interactions
with low nuclear recoil energy ($\nu$) have a nearly constant cross section as a function of incident neutrino energy.
The technique was extended to the lower energy region to apply to MINOS~\cite{MINOS_CC}.  
Recently MINERvA~\cite{DeVan:2016rkm} used the technique to make an {\it. in situ} prediction of
the shape of the flux as a function of neutrino energy from 2-50
GeV to measure the inclusive charged-current neutrino and antineutrino
cross sections on CH in the NuMI low-energy mode. 
This measurement is the lowest energy application of the low-$\nu$ flux technique. 
One needs to be careful since the technique is strictly applied only on DIS,
where $Q^2$ dependence from QE and resonance interactions must be corrected,
also the flux found by low-$\nu$ method needs to be normalized, either by using simulation or
the world average isoscalar cross section where A-dependence is ignored.  


\subsection{Near future neutrino experiments}

From Fig.~\ref{fig:flux_all} and a quick overview of this section, 
there is one trend for future neutrino experiments ---
For neutrino oscillation physics, a slightly higher energy region,
say 2-10 GeV, will become very important in the next few years.  
In the last $\sim$10 years, there was a stronger interest
to understand neutrino interaction systematics around 1 GeV or less
for MiniBooNE~\cite{Aguilar-Arevalo:2013pmq} and T2K~\cite{Abe:2015awa}.
However, in the near future, precise knowledge of 2-10 GeV region will become equally important for neutrino oscillation experiments,
including NOvA~\cite{Adamson:2016tbq}, PINGU~\cite{PINGU}, ORCA~\cite{KM3NeT}, Hyper-K~\cite{HK_2015}, 
INO~\cite{INO}, and DUNE~\cite{DUNE_CDR1}.
This will be challenging for the theoretical cross section models
particularly developed and tested in the 1 GeV region.

\section{Conclusions}
In the past, the role of neutrino-nucleus interactions was realized to be very important systematics of neutrino oscillation experiments,
but there was not much effort to understand it. 
This motivated the community to start the NuInt conference series; 
it is impressive how far we have come from NuInt 01~\cite{Lipari_NuInt01}!
``Supposedly well-known'' neutrino cross-section measurement was
a fringe of the neutrino oscillation program. 
However, there were growing doubts in neutrino interaction models implemented in the Monte Carlo generators used
by the experiments starting from K2K, MiniBooNE, and MINOS,
but there were not many things for theorists to do,
because there was no data to compare with theoretical models,
and theoretical models were simply compared among them and with Monte Carlos~\cite{Sobczyk_NuInt09_1,Boyd:2009zz}.

It was at NuInt 09 where a large amount of data from MiniBooNE became available for the first time.
Two major points raised great interest and debate:
\textit{i)} the CCQE data appeared too large with respect to theoretical calculations using the standard value of the axial mass and  
\textit{ii)} it was realized that full reconstruction of kinematics of the neutrino-nucleus reactions is impossible,
hence cross sections in terms of traditional kinematic variables like $Q^2$ and $E_\nu$ are available only within assumptions.

Concerning the point \textit{i)}, it took a short time to realize that the missing component
of describing neutrino MiniBooNE data falsely called ``quasielastic'' was the initial reaction mechanisms of
multinucleon excitations (or 2p-2h, or np-nh). 
To overtake the point \textit{ii)}, neutrino flux-integrated differential cross sections on direct observables
(like muon variables, see Eq.~\ref{eq:ddexp})
were used as main format to publish data.

This led the community to the following protocol:
focus on flux-integrated differential cross sections (the ``meeting spot''
of neutrino experimentalists and nuclear theorists) in terms of the final state topology of the reactions, e.g. CC0$\pi$ instead of CCQE. 
Since then, 
a variety of flux-integrated differential cross section data
from many experiments has been compared with theoretical models.
Nowadays there is an agreement among theorists to describe CC0$\pi$ sample,
and this information is used for flagship particle physics experiments, such as T2K and NOvA.
The interactions and the integration of the communities
(theory-experiment; particle-nuclear; interaction-oscillation) have been accelerated.
The particle data group has included a neutrino interaction section since 2012~\cite{PDG}.
A number of new collaborations between experimentalists and theorists had been formed, too.
The number of workshops and conferences is in continuous expansion. 

Flux-integrated cross section in terms of leptonic variables, which are far to be known at $\sim 5\%$ level
(see for example the multinucleon emission contributions),
provide in any case only partial information of kinematics,
and there are many attempts to break-though this uncomfortable situation.
Precise predictions and measurements of hadronic final states are clear next steps. 
This allows to fix the kinematics of given neutrino interactions, to further constrain interaction models, 
which eventually allows the higher precision neutrino oscillation experiments.
The community is moving toward to this path, 
and some experiments already produced once-thought-impossible data,
including the MINERvA double differential cross sections in $E_{avail}$ and $|{\bf q}|$, 
and the ArgoNeuT CC interaction with two proton final states.

We are still many steps away from the golden goal,
the full understanding of kinematics of neutrino interactions.
Probably the most urgent program is the understanding of neutrino-induced pion production channels. 
This is the key test to confirm our understandings, not only baryonic resonances, 
but also, pion final state interaction, hadronization, and SIS to DIS cross sections. 
It seems precise measurements and predictions of hadronic final states
may bring the next break-through in this community.
Also in this high precision era, both data and theory necessarily pass the rigorous statistical test.
T2K neutrino interaction working group is moving in this direction by comparing theories against external data. 

The future of the long-baseline neutrino oscillation community is the LArTPC and the water Cherenkov detector.
Therefore, understanding of neutrino-argon and neutrino-water interactions is very important.
Near future experiments will provide large amounts of data.  
However, without theoretical understanding and adequate Monte Carlo implementation
of all the reaction channels in the whole 1 to 10 GeV neutrino energy range,
the interpretation of these data will be difficult.

New intriguing results like CP violation in the leptonic sector (and maybe other surprising results!) necessary 
passes through a precise knowledge of neutrino-nucleus interaction.

\ack{ 
  We thank Magda Ericson for the interest on this review paper and for numerous discussions.
M.M. thanks Magda Ericson also for the long and continuous
collaboration on the research field of this paper.  
We appreciate help and hospitality by the members of the following groups:
T2K neutrino cross section group,
MINERvA collaboration, 
NuSTEC collaboration,
Queen Mary University of London neutrino group, 
Neutrino SPP and Nuclear Structure SPhN at CEA/Saclay, Theoretical Physics Department at CERN, 
Theoretical Nuclear and Statistical Physics at Ghent University, 
Theoretical Physics at IPN Lyon, 
Theoretical Many-Body and Nuclear Physics at Torino University and INFN.   
We appreciate the help by Dipak Rimal and Mateus Carneiro (MINERvA), Xinchun Tian (NOvA), Laura Fields (DUNE)
for providing neutrino flux files. 
We also thank to Jon Paley, \v{Z}arko Pavlovi\'{c},
Kevin McFarland, Tammy Walton, Sara Bolognesi, Paul Martins, Carl Stanley for their critical comments.
We thank also all the participants of the ``Two-body current contributions in neutrino-nucleus scattering''
ESNT workshop for useful discussions. 
Finally, we thank for active participation on discussions of various topics
the subscribers of the Neutrino Cross-Section Newsletter~\cite{nuxsec}.

T.K.'s work is supported by STFC (UK). M.M. acknowledges the support and the framework of the ``Espace de Structure et
de r\'eactions Nucl\'eaire Th\'eorique'' (ESNT, \url{http://esnt.cea.fr} ) at
CEA. 

}

\bibliography{only_marco,rev_nu_teppei}

\end{document}